\pdfoutput=1 % only if pdf/png/jpg images are used
\newcommand{\mus}{~\rm \upmu s}
\newcommand{\mum}{~\rm \upmu m}
\newcommand{\neqcm}{~\rm n_{eq}/cm^2}

\newcommand{\krad}{~\rm krad}
\newcommand{\C}{~{}^{\circ}\rm C}
\newcommand{\e}{~e~\rm ENC}
\newcommand{\pccm}{/\rm cm^3}
\newcommand{\ohmcm}{\rm~ \Omega \cdot cm}
\documentclass[a4paper,11pt]{article}

\usepackage{jinstpub}
\usepackage{upgreek}
\usepackage{multirow}% http://ctan.org/pkg/multirow
\usepackage{subcaption}
\usepackage{sidecap}
\usepackage{placeins}
\usepackage{lineno}

\title{Progress on the radiation tolerance of CMOS Monolithic Active Pixel Sensors}

\author[a,1]{M. Deveaux}
\affiliation[a]{Institut f\"ur Kernphysik, Goethe University Frankfurt,\\
  Max-von-Laue-Str. 1, 60438 Frankfurt am Main, Germany}

\emailAdd{deveaux@physik.uni-frankfurt.de}

\abstract{CMOS Monolithic Active Pixel Sensors (CPS) are ultra-light 
and highly granular silicon pixel detectors  suited for highly sensitive charged particle tracking. Being 
manufactured with cost efficient standard CMOS processes, CPS may integrate sensing elements
together with analogue and digital data processing circuits into one monolithic chip. This turns into
$50 \mum$ thin sensors, which  provide an outstanding typical spatial resolution of few$\mum$ and a detection efficiency for minimum ionizing particles above $99.9\%$.  The radiation tolerance of CPS was
initially constrained by the limits of the CMOS processes used for their production but has been improved by orders of magnitudes
during the last years.  

This work reviews the related R\&D on the radiation tolerance of traditional CPS with partially depleted active medium as pioneered by the MIMOSA-series developed by the IPHC Strasbourg.
Procedures for assessing radiation damage  in those non-standard pixels are discussed and the major mechanisms of radiation damage are introduced. Techniques for radiation hardening 
are shown.  Moreover, recent results on next generation CPS featuring fully depleted sensors based on improved CMOS processes are
summarized.

}


\begin{document}
\maketitle
\flushbottom
\setcounter{secnumdepth}{5}
\setcounter{tocdepth}{5}
%\setpagewiselinenumbers
%\linenumbers

\section{Introduction}

CMOS Monolithic Active Pixel Sensors (CPS or MAPS) are silicon pixel sensors, which are produced with standard CMOS processes.
By definition, active pixels are formed from a sensing element and a pre-amplifier. In addition, the pixels of modern CPS
may integrate a full shaper and discriminator chain. 

CPS were initially developed for a use as cost efficient optical imagers and became increasingly successful in the field of digital photography.
Their main economical advantage consists in the fact that they may be realized as a cheap monolithic CMOS chip while more traditional imagers like CCDs required
their sensing and readout units to be placed on two separate chips. The latter was done to solve a technological conflict: Photo diodes are preferably 
fully depleted, which calls for lowly doped silicon and high voltages. Building the highly integrated circuits needed for amplifying and processing the diode's signal calls for the use of CMOS technology, which bases typically on comparably highly doped silicon and is compatible  with low voltages only. Building a system from two separate
chips allows to use the most suited technology for both, sensors and amplifiers. However, it comes also with the costs and complexity of building and interconnecting both subsystems.

CPS integrate both building blocks to a single CMOS chip, which reduces cost and complexity. However, the sensing elements are designed with CMOS processes, which 
were initially neither intended nor optimized for this purpose. Among the compromises introduced by this design choice were the need to use non-depleted, or only partially depleted, sensing elements, 
which are less performing than their 
optimized, fully depleted counter parts. Moreover, the use of PMOS transistors in the pixels was initially not possible, which constrained the design of the on-pixel amplifier circuits. Despite those
compromises, very satisfactory performances were reached.

Already short time after the invention of CPS, it was recognized that those sensors are sensitive to minimum ionizing particles and thus suited
for a use in tracking detectors for relativistic heavy ion and particle physics \cite{TurchettaPaper}. Despite its above mentioned limitations, the technology
was considered to provide an interesting potential in this field: The size of pixels of CPS may amount
few$\mum$ only. In
charged particle detection, small pixels measure the impact point of minimum
ionizing particles with outstanding precision. Moreover, CPS are by themselves very thin, energy efficient and do not require supporting readout chips.  
This allows to integrate them into very light sensor systems, which minimizes the multiple coulomb scattering of minimum ionizing particles
in the sensor material and thus the feedback of the instrument on the particle trajectories to be measured.

Indeed, first CPS prototypes showed a single point resolution of \mbox{$1 - 2 ~\rm \upmu$m} and a detection efficiency close to 100\% \cite{Paper_Deptuch_DevelopmentOfMAPS} in a $\sim 100 \rm~ GeV/c$ pion beam at the CERN-SPS. Those excellent results were reproduced with a full size CPS up to 10$^6$ pixels on active areas as large as \mbox{$4~$cm$^2$} \cite{Paper_Deptuch_DevelopmentOfMAPS,TestPaperMimosa-5}. Soon after, the feasibility of thinning the sensors to
a thickness of $50\mum$ without loss of performance was demonstrated.
On the negative side, the time resolution of early CPS was in the order of $10 ~\rm ms$, which compares with the $25 ~\rm ns$ of classical hybrid pixel sensors. The radiation tolerance of CPS was initially fully unknown and came out to be rather moderate for the first sensor generation.

Their weakness in terms of time resolution and radiation tolerance hampered first generation CPS from reaching the rate capability of the very fast and radiation hard hybrid pixel sensors as used e.g. at the early LHC. However, given their strong points, CPS came out to complement those pixels and to form the technology of choice for experiments, which call for highest precision of the trajectory measurement but allow for compromises in terms of rate capability. This holds e.g. for the vertex detector systems of electron/positron collider experiments and of relativistic heavy ion experiments
aiming for the measurement of open charm particles.

%The benefit of the technology with respect to elder silicon pixel detector technologies like CCDs consists in the use of commercial
%CMOS wafers, which allows to benefit from the widely available and cost efficient tools and installations of the CMOS industry.
%On the other hand, the choice of a non-optimized technology turned into the need for compromises in terms of sensor design.
%Due to their low price and modest quality, CPS were initially used for low budget digital cameras. 
%Responding to the increasing market share of the devices, the CMOS industry provided consecutively production processes, which
%allowed to overcome the initial weak points of the technology. Thanks to those improvements, CPS can be considered as one
%of the leading technologies in optical imaging nowadays. 

 One of the driving forces pushing this idea forward was the PICSEL group \cite{PICSELGroup} of the 
IPHC Strasbourg (previously IReS and LEPSI), which designed and tested the above mentioned first CPS prototypes. 
This work aimed initially for the TESLA experiment \cite{TeslaTDR,TeslaDetector}, which became the Future International Linear Collider (ILC) \cite{ILCTDR} later on. The potential of the technology was rapidly recognized
outside the ILC community and a first working vertex detector based on CPS was finally realized in the heavy ion experiment
STAR, which is carried out at the RHIC collider \cite{PaperSTARPixel}. In parallel, an R\&D program aiming for a use of CPS in the Micro Vertex Detector of the Compressed Baryonic Matter experiment (CBM) \cite{Paper:CBM-MVD1,Paper:CBM-MVD2,CBMMvdPrototype} was started. The related prototype was converted into the  Small Acceptance Vertex
Detector of NA61, which realized a first CPS vertex detector in a fixed target geometry \cite{Paper:NA61LeadAddendum, Paper:NA61Annual2016,PaperNA61ShineSQM}. 

%Among the future applications of the technology are the upgrade of the Inner Tracking System (ITS) of the ALICE experiment \cite{Paper:AliceItsTDR} and the  Moreover, a use of CPS in upgraded trackers of LHC experiments has been proposed (see e.g. \cite{CMOSInATLAS}).

%While the capability of CPS to do precision particle tracking was almost immediately demonstrated, improving their radiation tolerance came out as a major research project. This is as the commercial CMOS processes used for building CPS are not intended for a use in radioactive environments. However, significant progresses were made in the recent years in this field. This work reviews the results of the related R\&D carried out at the IPHC Strasbourg and the Goethe University Frankfurt am Main and gives a brief insight on the status of CPS research of other groups.

Using CPS in the above mentioned applications required to improve their rate capability and time resolution by orders of magnitude. Moreover, it was necessary to establish their radiation tolerance and to extend it substantially. This task was 
complicated by the specific features of CPS. Conventional CPS use a non-depleted, low resistivity, p-doped active medium and collect their signal charge dominantly by means of thermal diffusion. Due to those properties, CPS feature a potentially unique device physics and  most established knowledge on radiation tolerance as provided e.g by the CERN RD48 and RD50 collaborations could not be applied in a straight forward way. Moreover, the active medium of CPS is as thin as $\sim 10 \mum$, which comes with a particularly small signal charge of few $100~e$ only. This makes the sensors vulnerable to particularly subtle effects injecting a noise of $\sim 10 ~e ~\rm ENC$, which were previously rarely in the focus of research. The system-on-a-chip concept of CPS does not allow to access sensing elements and amplifiers directly or separately and any information on radiation damage had to be extracted from the output of the potentially damaged, integrated amplification chain of the device. Finally, it became only recently possible to request modifications of the CMOS processes used for device production. Therefore, any modification aiming for higher radiation hardness had to be done within the strict limits of the design rules of the non optimized commercial processes.

%which is their particularly 
%While being used in vertex detectors, CPS are exposed to radiation and have to resists to the related radiation doses without loss of performances. To asure this,
%it was necessary to understand and to extend the radiation tolerance of CPS.
%This task was complicated by the specific features of the technology. Thanks to the efforts of the LHC community, e.g. the CERN-RD48 (ROSE) collaboration \cite{Rose}, 
%rich knowledge was available on the radiation tolerance of silicon detectors. However, the results applied first of all for mildly N-doped and 
%fully depleted sensor cells with a size of $\sim 100 \times 100 \mum^2$ and 
%a thickness of few $100\mum$, which formed the LHC baseline technology. In contrast, the sensing element of a typical CPS was non-depleted and relied on an, in comparison, highly and 
%P-doped silicon. It had a typical size of few $\sim 10 \times 10 \mum^2$ and a thickness of $\sim 10 \mum$, which turned into a by one order of magnitude
%reduced signal charge. Moreover, the highly integrated design of CPS did not allow for a direct access to the sensing diodes (more precisely speaking,
%results relying on such a direct access were found unreliable). This feature hampered applying test procedures to the novel technology. 

The focus of this work consists in reviewing the test procedures suited for testing the radiation tolerance of CPS and discussing the mechanisms limiting this tolerance in CPS with non- and partially depleted sensitive medium. Moreover, it introduces the technological approaches, which allowed to extend the radiation tolerance of CPS based on classical, double well CMOS processes by more than one order of magnitude and such to reach the performances required for using them in a first set of heavy ion experiments.

%This review is complemented by a summary of recent research projects on CPS aiming for high rate experiments at the LHC and FAIR, which rely on next generation CMOS processes. Those processes were made available by the CMOS industry few years ago and provide among others the 
%necessary features to use PMOS transistors in on-pixel pre-amplifiers and to deplete the active volume of the devices. This can be considered as a technological break-through, which will, according to first results, extend the performance and radiation tolerance of CPS once more a by significant step.  

The document is structured as follows: In Sec. \ref{Sec:MechOfRadDamage}, a brief introduction on established radiation damage effects will provided. 
Sec. \ref{Sec:TechOfCMOS} introduces the technological approach of CPS relying on 
traditional CMOS processes. An overview over the methods for testing the radiation tolerance of those devices is given in Sec. \ref{Sec:Methods}. 
Sec. \ref{Sec:IonizingDamage} and \ref{SectionNonIonizingRadiation} discuss the effect of integrated ionizing and non-ionizing radiation doses respectively
and introduce device hardening strategies. Sec. \ref{Sec:ThermalAnnealing} covers the effects 
of thermal annealing of ionizing and non-ionizing radiation damage. In Sec. \ref{Sec:SpecificRadDamage}, we will discuss specific radiation damage effects,
including Random Telegraph Signal, the effects of cold (thermal) neutrons and heavy ions.  

The impressive commercial success of commercial CMOS cameras motivated industry to provide optimized CMOS processes. The appearance of those processes 
eliminate the limitations of the first generation CPS. This triggered the interest of a broader community and turned into the start of a number of promising new projects aiming for next generation detector systems.
The so far most prominent of those next generation devices is the ALPIDE-sensor \cite{ALPIDE1}, which was chosen as technology for the upgrade of the
ALICE ITS \cite{Paper:AliceItsTDR}. Other promising R\&D programs aim to extend the radiation hardness and rate capability of CPS to a 
level, which allows to use them in the ATLAS experiment. An outlook in Sec. \ref{CPSOtherGroups} provides an overview about this R\&D.
 Finally, the major results on mechanisms of radiation tolerance are summarized.

\section{Mechanisms of radiation damage in CMOS devices}
\label{Sec:MechOfRadDamage}
Radiation damage in CMOS devices can be subdivided into two major groups, which are denoted as ionizing and non-ionizing. 
Ionizing radiation damage occurs by definition due to a radiation induced displacement of electrons from their atom while
non-ionizing radiation damage\footnote{Note that this definition differs from the one used in the context of life-science and radiation protection.} is consequence of a displacement of the full atom e.g. out of its position in a crystal lattice.
Both kinds of radiation damage may be reversible, e.g. if the lost electron is replaced by another electron obtained from the conduction band of conductive materials; or irreversible, e.g. if
a chemical binding in a molecule is permanently broken. 
 
Most particles cause both kinds of radiation damage. Therefore, irradiated devices will typically show superimposed radiation effects.
However, an understanding of the individual mechanisms is required to improve the radiation tolerance of the devices.
 
\subsection{Ionizing radiation damage}

Ionizing radiation damage is caused by electro-magnetic interactions, which are created by all kinds of charged particles and photons with an energy above $10~\rm eV$. The energy deposit of ionizing radiation is measured in the SI-unit Gray ($1~\rm Gy= 1~J/kg$). The elder unit $1~ \rm rad = 0.01~ Gy$ is still widely used in literature and will also be used in this work. 

An ionizing energy deposit in bulk silicon is typically reversible and not of worry. Lasting damage is caused by the irradiation of SiO$_2$ structures, which are widely used as isolators in CMOS chips. The precise mechanisms of the radiation damage generated in those structures is reviewed in \cite{PaperTotalDoseEffects}, which is summarized in the following:

Ionizing radiation excites electron/hole pairs in SiO$_2$, which are separated in the presence of an electric field $E_{ox}$. The separation probability increases with the electric field. Weakly ionizing particles like photons show a substantially higher charge yield $f(E_{ox})$ than strongly ionizing particles  (e.g. $\alpha$-particles).  The total number of free holes\footnote{The free electrons do in first order not contribute to the relevant radiation damage effects and are ignored in the following.} per $\rm cm^2$ Si-SiO$_2$ interface, $N_h$, is given with:

\begin{equation}
N_h=f(E_{ox}) \cdot g_0 \cdot t_{ox} \cdot D
\label{Eqn:HoleYield}
\end{equation}

Here, $g_0=8.1\times 10^{12}/\rm cm^3$ per rad for SiO$_2$ denotes the density of initially produced holes, $t_{ox}$ the thickness of the oxide and $D$ the dose.  

In CPS, the electric fields pushes the holes slowly toward the interface between the SiO$_2$ structures and the bulk silicon. They are trapped in oxygen-vacancies, which are found in great number nearby the Si-SiO$_2$ interface. The number of traps depends strongly on the precise production process of the device. Consequently, the fraction of trapped holes may vary between few percent and close to 100\%. A saturation at a positive surface charge of $dQ/dA \approx 3\times10^{12}/\rm cm^2$ is reported for good silicon strip detectors \cite{Lutz} . The trapping cross-section scales with ${E_{ox}}^{-1/2}$, which counteracts to some degree the field dependent increase of the charge yield. 

One may distinguish two major groups of trapped holes: Holes trapped near the interface in the non-conductive SiO$_2$ are referred to as oxide charge. Their charge is always positive. Their number is reduced starting from the moment of irradiation by room temperature annealing but a significant fraction of traps may remain even after several months of annealing. Holes trapped on the conductive Si-side may create defects in the band gap of the silicon and are referred to as interface traps. Depending on their energy and the local Fermi-level (doping), their charge may be positive, neutral or negative. 
%Remarkably, interface traps are formed only slowly and the formation process continuous after the irradiation. According to measurements done with a $47~\rm nm$ thick oxide layer and a field of $1~\rm MeV/cm$, 50\% of all traps are formed after about $35 ~\rm s$ but the full build-up requires as long as $10^5~\rm s$. 
Interface traps are not annealed at room temperature. 
The charge generated of both kinds of traps adds and the sum may become negative if negatively charged charged interface traps outnumber the oxide charges. 

\iffalse =================================== Comment on ===============================

\begin{figure}[t]
 \centering
\begin{subfigure}{.45\textwidth}
  \centering
 \includegraphics[viewport=1.0cm 1.0cm 19.4cm 19.8cm, clip,width=1\columnwidth]{Figures/DefectDensityTox.png}
    \caption{Surface defect density as a function of the oxide thickness.}
    \label{fig:DefectDensity}
\end{subfigure}
%
\hspace*{0.05\textwidth}
%https://www.overleaf.com/6458423ktgndt#%1
%
\begin{subfigure}{.45\textwidth}
  \centering
   \includegraphics[viewport=1.0cm 0.95cm 19.4cm 19.75cm, clip,width=1\columnwidth]{Figures/FlatBandVoltageTox.png}
			
		\caption{Flat band voltage shift as a function of the oxide thickness. }
    \label{fig:FlatBandShift}
\end{subfigure}%
%
%2
%
\caption{Generation constants of surface damage and radiation induced flat band voltage shift as a function of the thickness of the SiO$_2$ layer and the electric field at $T=295~\rm K$ and ``as-grown'' SiO$_2$.
Data from \cite{IonizingRadiation3}. See original work for additional data.}
\label{fig:IonizingRadiationEffectsOnSiO2}
\end{figure}

\fi  %============================ End Comment ======================================

In \cite{IonizingRadiation3}, measurement results on the defect densities are shown. Similarly to oxide traps (see equation \ref{Eqn:HoleYield}), the number of interface traps $D_{it}$ scales linearly with the dose for a given oxide configuration and electric field. For doses well above $1~\rm Mrad$,  the onset of a saturation turning into a sub-linear dependence of $D_{it} \propto D^{2/3}$ is observed. As expected, the number of interface traps increases with the thickness $t_{ox}$ of the oxide 
% (see figure \ref{fig:DefectDensity})
. 
For thick oxides, this increase scales with $D_{it} \propto t_{ox}^{1.6}$ \cite{IonizingRadiation3} and thus stronger than expected from a naive scaling of equation \ref{Eqn:HoleYield}. More importantly, $D_{it}$ shrinks dramatically for oxide thicknesses of $\lesssim 12 \rm ~ nm$.  This effect is explained by the elimination or compensation of trapped charge by tunneling of electrons from
the silicon into either oxide traps or electron traps associated with trapped holes \cite{PaperTotalDoseEffects}. Due to the limited range of the tunnel effect, this process is most efficient for thin oxide structures.

\begin{figure}[t]
 \centering
\begin{subfigure}{.45\textwidth}
  \centering
 \includegraphics[width=0.8\columnwidth]{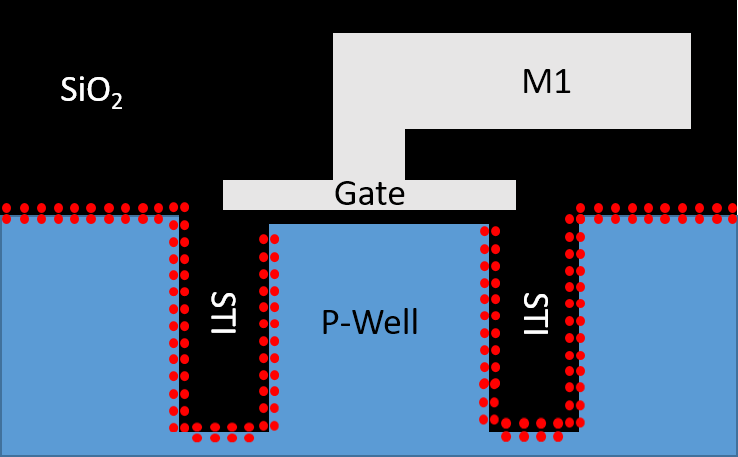}
    \caption{Cross-section of a CMOS transistor along line A in Fig. \ref{fig:irradiatedTransistors3}. Structures vulnerable to surface damage are indicated with red dots.}
    \label{fig:irradiatedTransistors1}
\end{subfigure}
\hspace*{0.05\textwidth}
%https://www.overleaf.com/6458423ktgndt#%1
%
\begin{subfigure}{.45\textwidth}
  \centering
   \includegraphics[width=0.8\columnwidth]{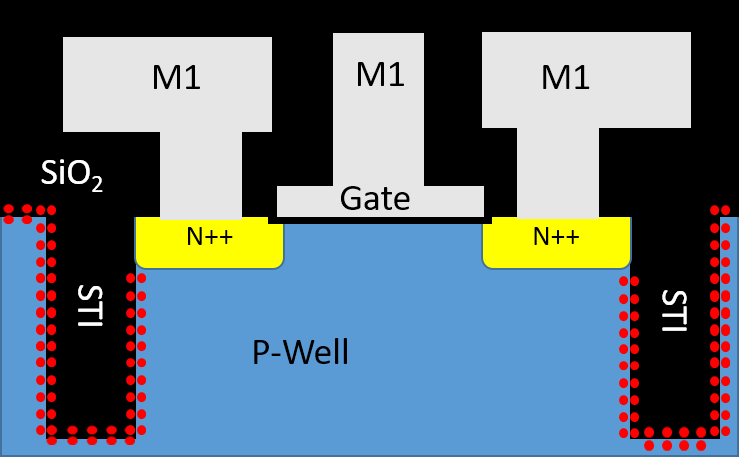}
			
		\caption{Cross-section of a CMOS transistor along line B in Fig. \ref{fig:irradiatedTransistors3}. Structures vulnerable to surface damage are indicated with red dots.}
    \label{fig:irradiatedTransistors2}
\end{subfigure}
%		\newline
	\begin{subfigure}{.45\textwidth}
  \centering
 \includegraphics[viewport=-1.5cm 0cm 14.2cm 10cm, clip, width=1\columnwidth]{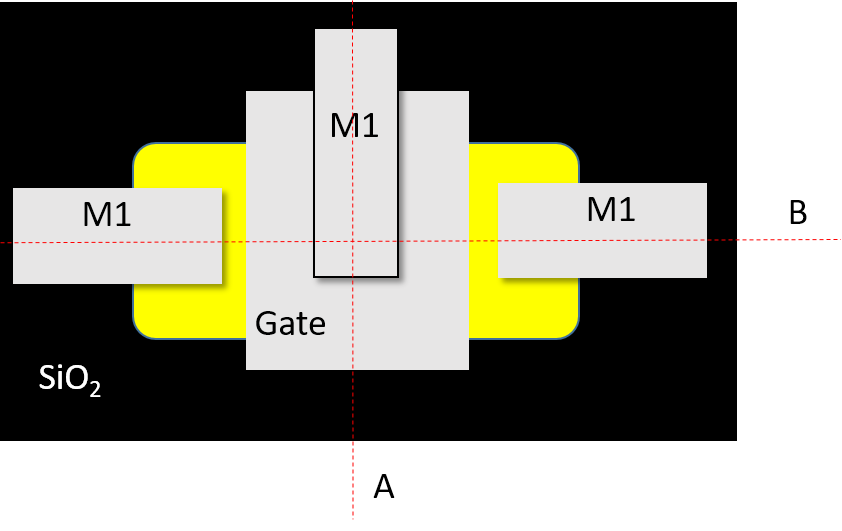}
    \caption{Top view of a CMOS transistor. The lines of the cross-sections in panel a and b are indicated.}
    \label{fig:irradiatedTransistors3}
\end{subfigure}
\hspace*{0.05\textwidth}
%https://www.overleaf.com/6458423ktgndt#%1
%
\begin{subfigure}{.45\textwidth}
  \centering
   \includegraphics[viewport=-1.5cm -1.5cm 14.2cm 8.5cm, clip,width=1\columnwidth]{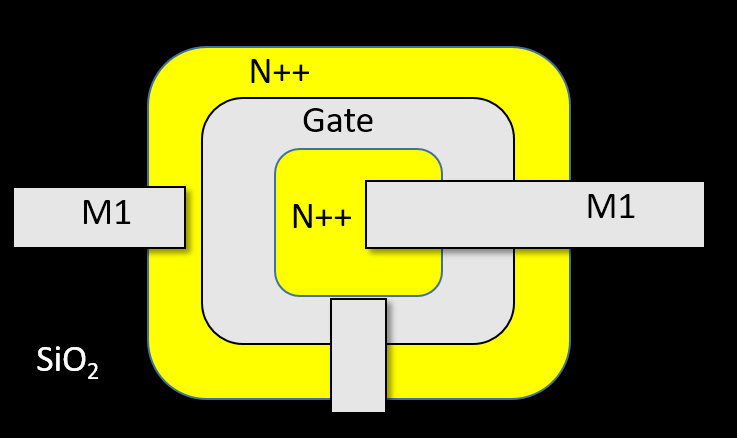}
			
		\caption{Top view of an enclosed transistor. The geometry eliminates the STI aside the channel.}
    \label{fig:irradiatedTransistors4}	
\end{subfigure}
\caption{Structures vulnerable to surface damage in linear transistors carried out in modern CMOS processes and top view of a radiation hard enclosed transistor.}
\label{fig:irradiatedTransistors}
\end{figure}

The relevance of the radiation induced charge build up is due to the related fields. Those fields may act similar to the steering fields of field effect transistors (FET) and deform the bands nearby the interface. This may generate unwanted conductive paths, e.g. in case the flat band voltage is modified such that p-silicon separating two n-doped structures becomes locally n-conductive. Moreover, the fields generated by the radiation induced charge add literally to the steering fields of FETs and modify their threshold, which is referred to as voltage shift. 
%The degree of this voltage shift as a function of the oxide thickness is shown in figure \ref{fig:FlatBandShift}. 
One observes that the voltage shift scales with ${t_{ox}}^3$ for ``thick'' SiO$_2$ and that it reaches values of few $\rm V/Mrad$ for the $47~\rm nm$ thick silicon. In \cite{PaperTotalDoseEffects}, a weaker voltage shift scaling with ${t_{ox}}^{1.45}$ to ${t_{ox}}^{1.85}$ is reported. Again, very thin silicon is less vulnerable than predicted by this scaling. 

\iffalse ================ Comment ========================0
\begin{figure}[t]
 \centering
\begin{subfigure}{.45\textwidth}
  \centering
  \includegraphics[viewport=1.5cm 0cm 19cm 17cm, clip, width=\textwidth]{Figures/ThresholdShiftHillemanns.PNG}
    \caption{Threshold voltage shift of transistors with $0.18 \mum$ gate length as a function of the gate width and the total ionizing dose. The point at
"$20~\rm Mrad$" represent a measurement of sensors, which were irradiated to $10~\rm Mrad$ and hereafter annealed for 24 hours.}
    \label{fig:ThresholdShifts}
\end{subfigure}
%
\hspace*{0.05\textwidth}
%https://www.overleaf.com/6458423ktgndt#%1
%
\begin{subfigure}{.45\textwidth}
  \centering
  \includegraphics[viewport=1.3cm 0cm 19cm 17cm, clip, width=\textwidth]{Figures/TransistorenHillemanns2.PNG}
			
		\caption{Threshold voltage shift of high voltage transistors as a function of the gate gate geometry and the total ionizing dose. The points at
"$20/30~\rm Mrad$" represent a measurement of sensors, which were irradiated to $10~\rm Mrad$ and hereafter annealed for 24 hours / one week.}
    \label{fig:ThresholdShifts2}
\end{subfigure}

\caption{Threshold shifts of NMOS transistors implemented in a $0.18\mum$ CMOS process used for CPS as a function of the radiation dose. Data from \cite{HillemannsTransistoren}. Mind the scale.}
\label{fig:ThresholdShiftsFull}
\end{figure}
\fi %========================================= End Comment ====================

In practical terms, oxide and interface charges in PMOS transistors show both a positive charge and their fields cause a negative voltage shift. In NMOS transistors, the interface charge is dominantly negative. Consequently, their fields tend to cancel out to some extent and even an inversion of the initially negative voltage shift may occur at high doses as the negatively charged interface traps start to outnumber the positively charged oxide traps. For both kinds of transistors, the voltage shift depends on the thickness of the gate oxide of the transistors. In modern deep sub-micron CMOS technologies, this thickness is typically $t_{ox}\ll 12 ~\rm nm$ and thus sufficiently thin for recovering radiation damage by tunnel effect. Therefore, deep sub-micron CMOS processes are often considered as intrinsically radiation hard. This statement should however be taken with caution. The thickness of isolation structures others than the transistor gates is usually not reduced. Therefore, most structures remain vulnerable.

This is illustrated in Fig. \ref{fig:irradiatedTransistors}, which shows a simplified cross-section of a FET in CMOS and indicates structures, which are typically affected by surface damage. Those structures include a thick shallow trench isolation (STI) aside the intended conductive channel of the transistor. A charge build up in those STI may
create fields, which act from the side to the conductive channel of the transistor and such generate a threshold voltage shift, which is independent of the gate oxide. The magnitude of the related threshold shifts was recently studied in detail for a CMOS process with $0.18\mum$ feature size, which is frequently used in CPS construction \cite{HillemannsTransistoren}. 
Some results of this study are displayed in Fig. \ref{fig:ThresholdShiftsFull}. The figure shows the threshold voltage shift for regular NMOS transistors as a function of the width of the transistor and the radiation dose. Transistors with narrow gates show a significantly higher threshold shift than wide transistors. Moreover, the increase of the threshold shift does not scale linearly with the dose, which reflects the saturation effects for doses beyond $1 ~\rm Mrad$ discussed above. Note that besides the voltage shifts, irradiated STI may also cause significant source to drain leakage currents, which may persist within the specified steering voltage range of the device.

\begin{figure}[t]
 \centering

\begin{minipage}{.55\textwidth}
  \centering
     \includegraphics[viewport=0.5cm 0.5cm 9.0cm 8.0cm, clip, width=\textwidth]{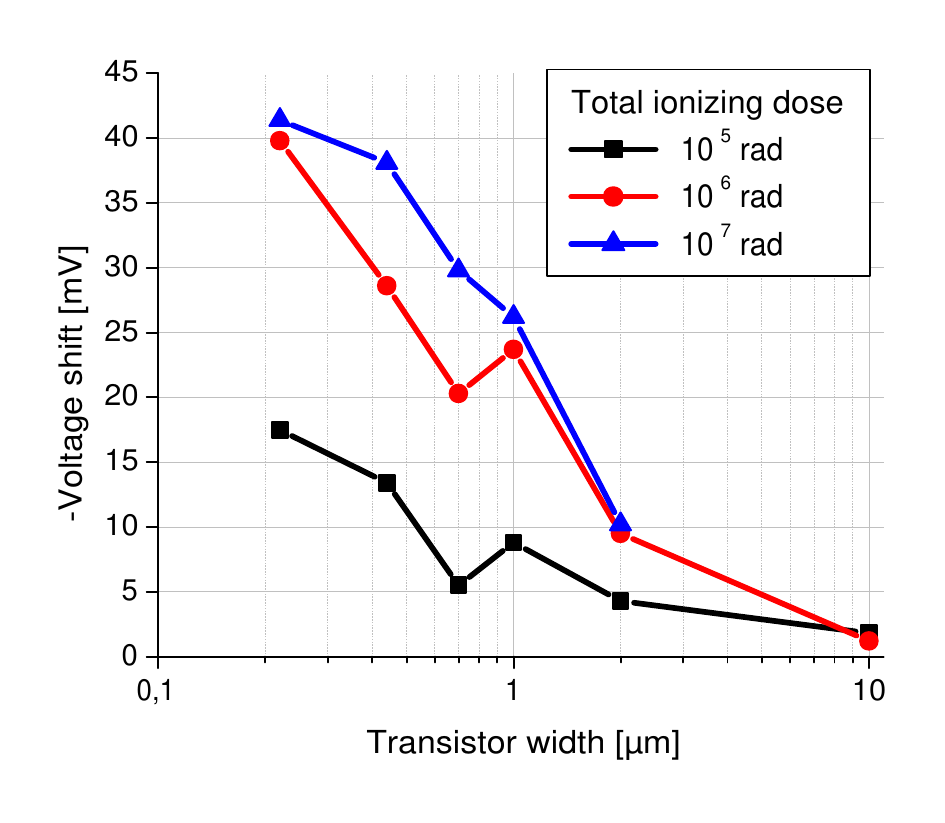}
	\end{minipage}
\hspace{0.05\textwidth}	
\begin{minipage}{.35\textwidth}
  \centering
  \caption{Threshold shifts of NMOS transistors with $0.18\mum$ gate length as a function of the gate width and the radiation dose.
	The transistors were implemented in a $0.18\mum$ CMOS process used for CPS. Data from \cite{HillemannsTransistoren}, see 
	original work for additional data.} \label{fig:ThresholdShiftsFull}
	\end{minipage}
				
\end{figure}

To suppress threshold shifts and source to drain leakage currents in transistors, one may consider the use of enclosed transistors (ELT) \cite{EnclosedTransistors1,EnclosedTransistors2}, which separate source and drain by a circular gate (see Fig. \ref{fig:irradiatedTransistors4}). Thanks to this geometry, no thick and vulnerable SiO$_2$ is located aside the gate and the radiation effects are substantially reduced. This advantage has to be balanced against the higher capacitance of the gate and the substantially increased size of the transistor.

\subsection{Non-ionizing radiation damage}
\label{SectionNonIonizingTheory}
Non-ionizing radiation doses are caused by an electromagnetic or strong interaction between the impinging particle and the nuclear cores of the atoms of the detector material. Once the energy transfer of an interaction exceeds $25~\rm eV$ \cite{Lutz}, the silicon atoms are displaced out of their position in the crystal lattice. Due to momentum conservation, a minimum kinetic energy of  $190~\rm eV$ and $260~\rm keV$ is required for generating this energy transfer with protons/neutrons and electrons, respectively \cite{Lutz}. Hard $\gamma$-rays create point defects indirectly via a generation of fast electrons, e.g. by means of Compton scattering \cite{GammaRadiationPaper}. The latter effect is however of second order in most particle physics environments. Groups of defects are created if the recoil atom gains enough energy to initiate secondary displacements. The threshold for this defect cluster generation is given with $2~\rm keV$ (Si-ion), $15~\rm keV$ (proton/neutron) and $4.6~\rm  MeV$ (electron) \cite{Lutz}. 

The dosimetry on non-ionizing radiation is done in the framework of the Non-Ionizing Energy Loss model \cite{NIEL-Modell}, which assumes the radiation damage to scale with the non-ionizing energy deposit. This energy deposit is usually expressed as multiples of the damage caused by $1~\rm MeV$ reactor neutrons and the related unit is $1 ~ \rm n_{eq}/cm^2$. This unit is equivalent to a displacement cross-section of $D=95~\rm MeV~mb$ or a NIEL of $2.03~\rm keV~cm~g^{-1}$ \cite{LindstromNonIo}. Tables \cite{NIEL-compilation} may be used for computing the equivalent damage of particles other than $1~\rm MeV$ neutrons. The NIEL model assumes that microscopic aspects of the radiation damage can be ignored, which is a useful but sometimes too rough simplification.

\iffalse
Based on this normalization, the NIEL caused by a radiation field is frequently expressed in the unit $\rm n_{eq}/cm^2$, which is rather convenient to compute doses from particle fluxes. The radiation damage caused by the most frequent hadrons in silicon can be normalized to units of $\rm n_{eq}/cm^2$ by means of tables . The related data for protons, neutrons, electrons and pions are plotted in Fig. \ref{Fig:NIEL-Factor}. 
\begin{figure}[t]
\begin{minipage}{0.7\textwidth}
 \includegraphics[width=\columnwidth]{Figures/NIEL.png}
\end{minipage}
\hspace{0.02\textwidth}
\begin{minipage}{0.26\textwidth}
 \caption{NIEL per particle for different particle species. Data from \cite{NIEL-compilation}.}
    \label{Fig:NIEL-Factor}
\end{minipage}
\end{figure}
\fi

Radiation induced bulk defects generate states in the band gap of the silicon. Those states ease the generation and recombination of free charge carriers in the material and thus increase the thermal leakage current of charge collection diodes. This increase is found to scale linearly with the NIEL and not to depend on the specific type of the silicon  \cite{LindstromNonIo}. A diode with a depleted volume $V$ does thus generate a radiation induced additional leakage current of:
\begin{equation}
\Delta I= \alpha(T) \cdot \Phi \cdot V
\label{Eqn:VolumeCurrent}
\end{equation}
Here, $\Phi$ represents the NIEL in units of $\rm n_{eq}/cm^2$ and the scaling factor $\alpha(T)$ depends exponentially on the temperature. It is usually given for $T=20^{\circ}\rm C$ and amounts $\alpha (20^{\circ}\rm C)= 4.0 \times 10^{-17} ~\rm A/n_{eq}/cm$ after the recommended annealing of 80 min at $\rm 60^{\circ}C$.  Note that $\alpha$ is also reduced by room temperature annealing \cite{MollPhD}. The leakage currents for other temperatures may be derived like:
\begin{equation}
I(T_R)=I(T)\cdot R(T)
\end{equation}
\begin{equation}
R(T)=\left( \frac{T_R}{T} \right)^2 \exp \left (- \frac{E_g}{2 k_B} \left[ \frac{1}{T_R} - \frac{1}{T} \right] \right)
\end{equation}
Here, $T_R=293~\rm K$ is the reference temperature, $T$ the device temperature and $E_g=1.12~\rm eV$ the band gap.

%P-doped silicon does not show a type inversion. Instead, one observes an initial decrease of the effective doping of the samples, which is referred to as acceptor removal \cite{NeutronIrradiationPaperWermes}. Due to this acceptor removal, the effective doping shows a minimum  after neutron doses of $10^{14} - 10^{15} \neqcm$ and increases hereafter. For doses below $10^{15} \neqcm$, a low initial doping is of advantage. At higher radiation doses, samples with initially 10, 20 and $2000~\rm \Omega \cdot \rm cm$ resistivity show a similar effective doping of $\sim 10^{14}~\rm N/cm^3$.  Now, the initially low resistivity silicon shows a by a small factor lower effective doping. 

In addition, bulk damage is observed to create a loss of minority charge carriers as excited by the impinging particle. This loss may be caused by a recombination (the charge carrier is permanently lost) or by trapping (the electron is captured by a defect state but thermally released after some time). While the first case turns into a loss of signal charge, the second case may delay the arrival of the charge such that it arrives to late for contributing to the sensor signal. For non-irradiated, moderately p-doped silicon as used in CPS, the lifetime of minority charge carriers amounts $\tau_{\rm eff_{e,h}}=1~\rm ms$. If the doping exceeds few $p=10^{15}/\rm cm^3$, $\tau_{\rm eff_{e,h}}$ shrinks inverse to the doping concentration \cite{SolarzellenPaper}. 
For irradiated sensors, the lifetime of the charge carriers until recombination or trapping scales with: 
\begin{equation}
\frac{1}{\tau_{\rm eff_{e,h}}} =const \cdot \Phi
\label{Eq:ChargeCarrierLifeTime}
\end{equation}
For recombination, the constant is identified with the recombination damage constant $K$. A value of $K=2.5 \times 10^{-6} \rm ~cm^2/s$ is given in \cite{MinorityChargePaper} for electrons in P-doped silicon, which was exposed to $1~\rm MeV$ neutrons. For trapping, the constant is identified with the trapping time parameter $\beta_{e}$ and numerical values ranging from $3.4 \pm 0.3 \times 10^{-7}~\rm cm^2/s$ to  $4.7 \pm 0.4 \times 10^{-7}~\rm cm^2/s$ (depending on the detailed P-doped material) are reported for $T=20\C$ in \cite{PTrappingPaper}.

Non-ionizing radiation modifies the effective doping of silicon, $N_{\rm eff}=|N_d-N_a|$. Both, n- and p-dopants are being absorbed and neutralized by radiation induced defects  \cite{AcceptorRemovalPaper}. In parallel, radiation induced defects acting as effective p-dopants, are generated. N-doped silicon looses its initial doping and may become effectively p-doped silicon at some point. This effect is referred to as charge inversion. The generation rate of the radiation induced p-dopants is reduced by about a factor of three in silicon with strong oxygen doping \mbox{($>\rm 10^{17}/cm^3$)} \cite{PaperOxiSilicon}.  This holds for proton and pion radiation but not for neutron radiation, which forms a prominent exception from the NIEL scaling hypothesis. 
%This observation forms a prominent exception from the NIEL scaling hypothesis. It is likely explained by the fact that the charged particles generate dominantly point defects while neutrons produce dominantly defect clusters. It appears that the oxygen acts exclusively on the point defects. 
The effective p-doping of irradiated silicon is observed to increase further once the silicon is annealed, which is referred to as reverse annealing. Reverse annealing is reduced by oxygen doping and may be suppressed by cooling irradiated sensors permanently to moderate negative temperatures.
%\subsubsection{Radiation induced modification of the effective doping}
\label{Sec:Neff}

For p-doped silicon, the acceptor removal is typically faster than the generation of novel states. Therefore, the effective p-doping shows a minimum at some point. Measurements for the p-doped silicon of CPS were carried out recently. 
According to the results, the effective doping of p-doped silicon can be described with (see e.g. \cite{WermesNeffMessungBonn}):
\begin{equation}
N_{\rm eff} = N_{\rm eff0} - N_C \cdot (1-\exp(-c \cdot \Phi_{eq})) + g \cdot \Phi_{eq}
\end{equation}
Here, $N_{\rm eff0}$ is the initial acceptor concentration of the substrate, $N_C$ the concentration of the removed acceptors, $c$ the removal constant and $g$ the introduction rate of stable deep acceptors for neutron irradiation. 
\begin{figure}[t]
\begin{minipage}{0.6\textwidth}
\includegraphics[viewport=1.3cm 1.3cm 25.5cm 19.5cm, clip,width=\columnwidth]{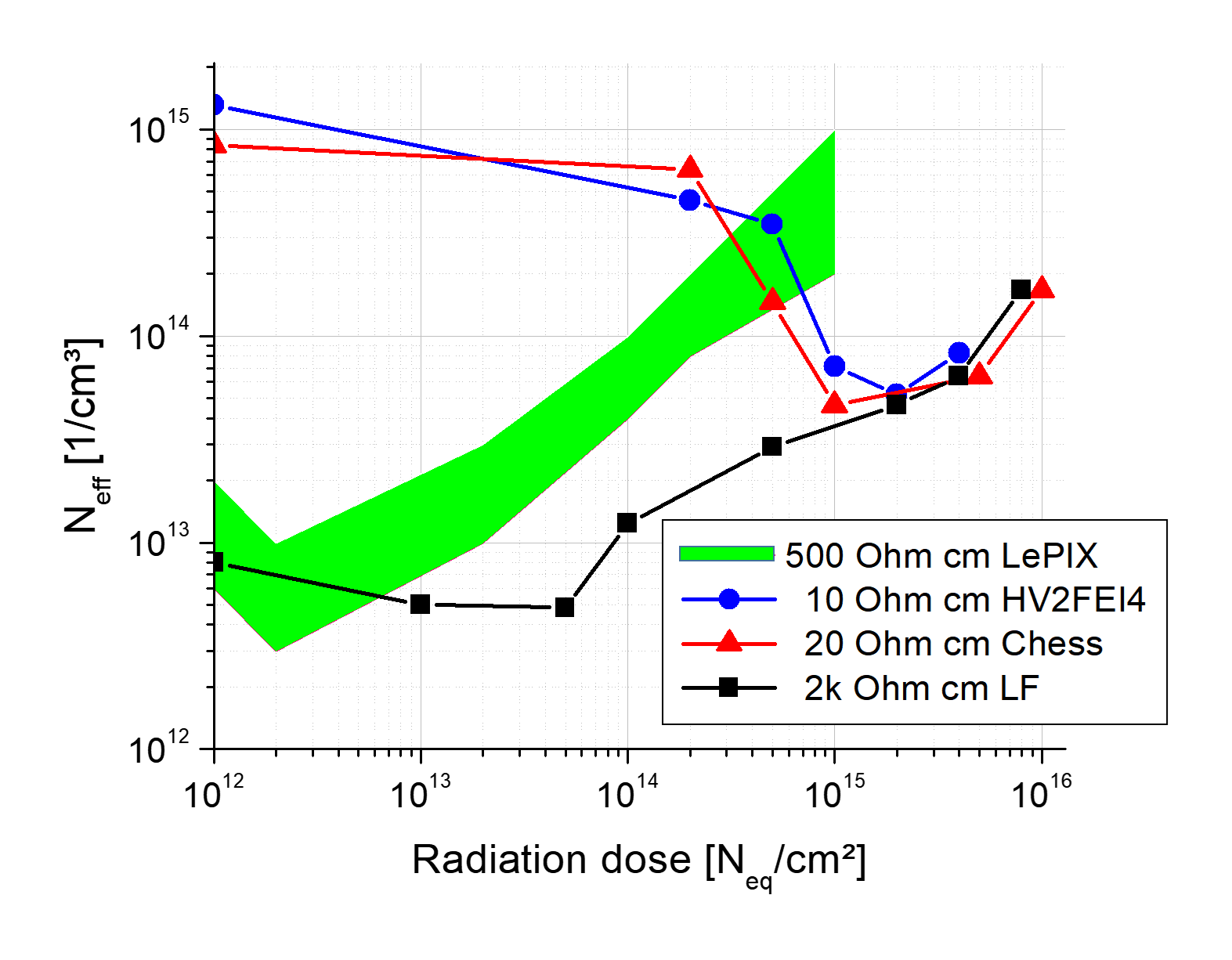}
\end{minipage}
\hspace{0.02\textwidth}
\begin{minipage}{0.36\textwidth}
 \caption{Effective doping of different processes used for CMOS sensors. The points at $10^{12} \neqcm$ represents 
the non-irradiated devices. Data on HV2FE14, Chess and LF from \cite{WermesNeffMessungBonn,NeffPaper2}. The LePIX - data was extracted from Fig. 4 of \cite{LePixPaper2}. See original works for details
on measurement strategies and uncertainties.}
    \label{fig:N_eff}
\end{minipage}
\end{figure}

Measured data on the effective doping of the active volume of different CMOS processes is compiled in Fig. \ref{fig:N_eff}. 
One observes that the high initial doping of low resistivity silicon is consecutively reduced and the reduction amounts 
more than one order of magnitude for doses up to $\sim 10^{14} \neqcm$.
A minimum of the effective doping is observed at few $\sim 10^{15} \neqcm$. 
For high resistivity silicon, the minimum of $N_{\rm eff}$ is reached at a lower radiation dose. For higher doses, the $N_{\rm eff}$ of the
silicon presented in \cite{WermesNeffMessungBonn} is independent of the initial doping within error bars. For the HV2FE14, CHESS and LF data sets,
a common parametrization of $c=10^{-15}/\rm cm^2$, $N_C/N_{\rm eff0} = 0.6$ and $g_c=0.047\rm cm^{-1}$ with uncertainties of around 30\% are reported.
The data reported for the
LePIX wafer follows qualitatively the observations of the other lines. The discrepancy may be because both measurements of the depletion depth 
underlying the $N_{\rm eff}$ were done
with rather different approaches (LePIX: C/V - measurements, others: TCT). 
At the time given, it appears that the effect of acceptor removal and acceptor generation is qualitatively present in various p-doped materials.
However, the 
precise removal and generation rates seem dependent to some extent on the details of the material and the radiation source (protons/neutrons) \cite{HitiPaper, MandicRD50}.
This is subject of ongoing research. %fixme Update papers sichten und einfügen

\section{Technology of CPS based on standard CMOS processes}

\label{Sec:TechOfCMOS}

\subsection{Classical CPS}
\subsubsection{Fundamental design considerations}
\label{Sec:FundamentalDesign}
Charged particles and photons of sufficient energy generate electron/hole-pairs in silicon. Any kind of silicon radiation detector aims to separate the electrons and the holes,
to collect their charge by means of a charge collection diode and to amplify this charge in order to obtain a signal. This signal has to exceed the noise of the amplification chain. In average, a radiation energy of $3.6 ~\rm eV$ is needed to generate an e/h pair \cite{Lutz}. 

X-rays and fast charged particles interact with silicon in different ways. X-rays tend to concentrate their full energy in a tiny volume and thus create a localized charge cloud. Fast charged particles (minimal ionising particles, MIPs) are typically not stopped in the sensor and generate $\sim 80$ e/h pairs per $\rm \upmu m$ along their trajectory in silicon.
The amplitude of the signal scales therefore with the length of the particle trajectory in the active silicon of the sensor. A minimum thickness of few $100\mum$ of the sensors was traditionally considered as appropriate for obtaining a good number of electron/hole-pairs. The separation of electrons and holes is done by electric fields as found in the depleted volume of the charge collection diode. Conventional sensor designs extend this depleted zone over the full active sensor volume.

The width $W$ of a reverse biased one-sided abrupt flat junction is given, e.g. in \cite{BookSze}, with:
\begin{equation}
W= \sqrt{\frac{2 \, \epsilon_s\, (V_{bi}-V)}{ e \, N_B}}
\label{Eqn:DiodeThickness}
\end{equation}
Here, $e$ denotes the elementary charge, $\epsilon_s$ the electrical permitivity in silicon, $V_{bi}$ the build-in-voltage of the PN-junction and $N_B$ the
doping concentration at the lowly doped side of the junction. $V$ is the negative, externally applied reverse biasing voltage of the diode\footnote{One assumes $N_B$ to be negligible w.r.t the doping concentration of the highly doped side. $V_{bi}$ amounts $<1~ \rm V$ and can thus be neglected in many cases.}. From this equation, one learns that reaching a $W>100\mum$ requires a combination of lowly doped silicon ($N_B \lesssim 10^{13}/\rm cm^3$) and high reverse bias voltages of $\sim 100~\rm V$.
Due to missing commercial interest, this low doping was initially not found in wafers used for CMOS electronics. Moreover, the guaranteed voltage tolerated by CMOS chips does typically not exceed $5~\rm V$. Until both limitations vanished thanks to the appearance of optimized CMOS processes (see section \ref{CPSOtherGroups}), it was not possible to realize charge collection diodes with thick depleted volume in CMOS technology.

%The traditional solution for handling the limitations consists in realizing the charge collection diode and the amplifiers in two different chips, which are electrically connected hereafter. The diode is placed on a dedicated, very pure and lowly doped wafer, which does not host transistors but may tolerate high voltages instead. The amplifiers are put on a CMOS chip, which is optimized for analogue and digital electronics. This concept allowed for building successful designs like silicon strip detectors or the hybrid pixel detectors used e.g. at the LHC\footnote{See e.g. \cite{Lutz} for a comprehensive introduction.}. However, the solution comes with limitations. The material of the sensor chips and the readout chips tends to scatter the charged particles of interest and thus to change their trajectory. Moreover, the necessary interconnections between both chips are expensive, somewhat vulnerable and constrain the size of the sensor cells. Therefore, hybrid pixel and strip detectors are not the first choice, once a physics case requires to measure particle trajectories with highest precision.

%C{\scriptsize MOS} monolithic active Pixel Sensors  for charged particle tracking (CPS) were developed in order to overcome this limitation. To do so, they integrate the charge collection diode and its amplification chain on one single CMOS chip, which reduces the material and eliminates the interconnections. However, the above mentioned restrictions concerning doping and voltage apply and force a number of compromises in the design. 
The history of CPS is best understood as a series of increasingly successful trials to bypass and overcome those restrictions. This was among others possible as the CMOS Monolithic Active Pixel Sensors for optical imaging, from which the CPS were initially derived, had a great commercial success, e.g. as cameras of smart phones. This triggered the development of improved CMOS processes, which became also accessible to the particle physics community. 

\subsubsection{The classical sensing element}

\begin{figure}[t]
  \begin{center}
   \includegraphics[viewport=1.0cm 1cm 25cm 17cm, clip,width=0.8\columnwidth]{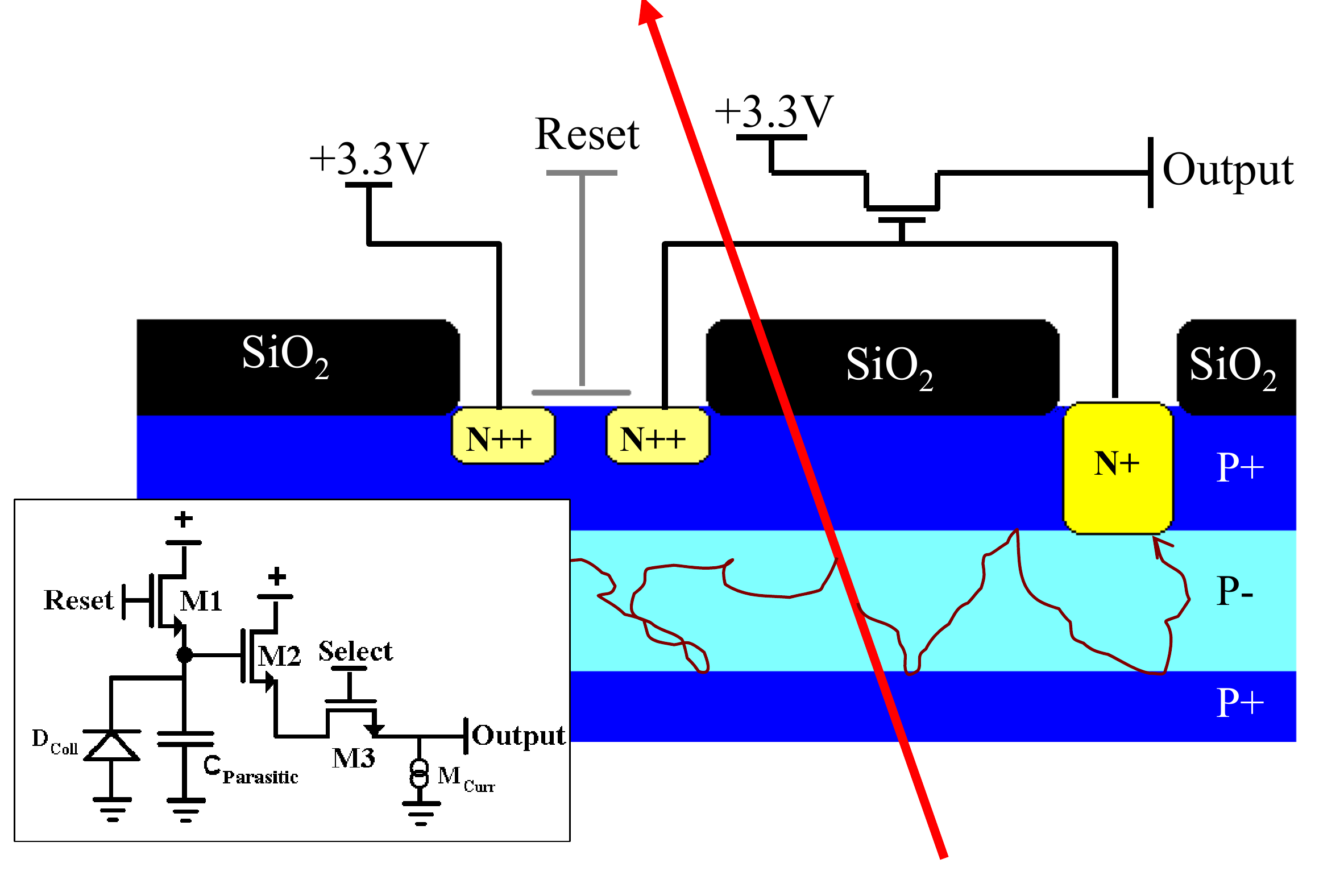} 
    \caption{Fundamental layout of a CMOS-sensor. The trajectory of an impinging minimum ionizing particle and the diffusion paths of two free electrons is shown. The diagram of the preamplifier shown is displayed in the lower left box.}
    \label{fig:sandwich}
  \end{center}
\end{figure}

The classical sensing element of a CPS was designed for being compatible with low resistivity double well CMOS processes relying on wafers with epitaxial layer.  A simplified sketch of a CPS implemented in this process is shown in Fig. \ref{fig:sandwich}.  The lowest layer of the device is formed from a silicon wafer with a doping of about $p=10^{19}/\rm cm^3$. On top of this wafer, one processes an epitaxial layer, which has a typical doping of few $p=10^{15}/\rm cm^3$ (P- in the figure) and may show a typical thickness of few$\mum$ up to $20 \mum$. Above, the designers may add so-called wells (P+ and N+ in the figure), which may have a p- or n-doping of something like $p=10^{17}/\rm cm^3$. This layer may embed p- or n-diffusions (P++ and N++), which show a doping roughly $p=10^{19}/\rm cm^3$. Note that the thickness of the structures as much as the precise doping concentrations vary among different manufacturers. They are, if known at all, subject to non-disclosure contracts.

According to equation \ref{Eqn:DiodeThickness}, a charge collection diode with $N_B=3 \times 10^{15}\pccm$ and $V=5~\rm V$ will show \mbox{$W=1.5 \mum$}, which is not sufficient to obtain a reasonable signal from an impinging MIP. The limitations of the early CMOS processes in terms of voltage limitation and high doping concentration did not allow for extending this depleted volume. Instead,  a typically $\sim 15\rm ~\upmu m$ thick epitaxial layer is used as an extension of the diode. This epitaxial layer is not depleted and therefore free of electric fields. The $\sim 1200$ signal electrons generated by a MIP in this layer are therefore not attracted by drift but diffuse to the thin charge collection diodes. This time consuming process succeeds thanks to the $\sim 1~\rm ms$ long life-time of the free electrons. 

The full sensing element is formed from the epitaxial layer, which is surrounded by the substrate and a p-well layer. The doping gradients between the lowly doped epitaxial layer and each of the highly doped p-well and substrate generate a small potential drop of $\sim 100~\rm mV$, which hampers the electrons from leaving the epitaxial layer\footnote{See Sec. \ref{ChapterDopingGradients} for further information.}. The photo-diode is formed from n-well implantations, which penetrate the p-well layer, and the epitaxial layer. 

Despite the charge is collected by diffusion, non-irradiated classical CPS pixels may show a rather high charge collection efficiency of fairly above 90\%. However, the diffusion creates a significant charge sharing and only about  25\% of the electrons are found in the diode located closest to the impact point of the impinging particle (seed pixel). Most of the remaining signal charge is spread over a cluster of $3 \times 3$ pixels while few electrons recombine before reaching a pixel diode and are lost. 

The sensing elements of CPS provides therefore a rather tiny signal charge of typically some 100 electrons. To distinguish this signal from noise, one requires for a very low noise pre-amplifier, which must moreover be suited for being integrated into the individual pixel. 

\subsubsection{The 3T-amplifier}

In the standard double-well CMOS processes available in the early days of CPS development, the on-pixel amplifiers had to rely exclusively on NMOS transistors. This is as building a PMOS (p-n-p) transistor requires the use of a n-well implantation. This n-well combines with the epitaxial layer to an unwanted second charge collection diode. This diode competes 
with the primary charge collection diode in terms of charge collection and such reduces the signal sensed by the latter diode in a rather dramatic way. Still, PMOS transistors could be placed on the same chip in non-sensitive areas aside the pixel matrix. This limitation vanished in modern CMOS processes as the n-well of the PMOS transistor and the active volume may be separated by a deep p-well implantation (see Sec. \ref{CPSOtherGroups}). 
However, this feature was not available for the design of the first CPS generations, which limited the design options substantially. This is as building blocks like high gain amplifiers and discriminators require PMOS transistors
and could thus not be used in the on-pixel electronics. 

The on-pixel amplifier of the traditional 3T-pixel forms a rather successful solution to overcome this limitation. The diagram of the charge-to-voltage amplifier is shown in the box of \mbox{Fig. \ref{fig:sandwich}}. It is formed by three transistors, the parasitic capacitance of the pixel diode and a current source, which is typically located outside the sensor-chip. The amplifying process is twofold: First, a charge-to-voltage conversion is performed by charging the collected electrons into the tiny capacitance and generating such a voltage drop seen by the gate of the transistor M2. This transistor and the current source $M_{\rm curr}$ form a source follower with a voltage gain slightly below one. The latter buffers the signal and such reduces the output impedance\footnote{The output impedance of a signal source is a measure for the amount of current, the source may deliver without showing a voltage drop. This is modelled in a most simple picture by a resistor/impedance, which is put into series to an ideal voltage source. As a good rule of thumb, the output impedance of a signal source should remain by one order of magnitude below the input impedance of the next amplification stage. 

The capacitance of a CPS pixel stores only few 100 electrons signal charge, which does not allow to deliver any current and thus shows a very high output impedance. The source follower does however not consume current from its input (very high input impedance) and delivers a macroscopic current at its output (low output impedance). This does not amplify the voltage signal but stabilizes it to the level needed to drive the voltage amplifiers of the next amplification stage. This reduction of the impedance is denoted as buffering.} 
of the pixel.
The output signal $U_S$ of the 3T-pixel is given with:
\begin{equation}
U_S=g \cdot \frac{Q_s}{C}
\end{equation}
Here, $Q_S$ is the signal charge, $C\approx 10 ~\rm fF$ the capacitance of the pixel and $g$ the summed gain of the source follower and possible further voltage amplification stages, which are located outside the pixel. 

For non-irradiated pixels, the noise of the pixel is dominated by the amplification chain. The signal-to-noise ratio can be increased by reducing $C$ while increasing $g$ by means of adding voltage amplifiers located outside the pixel amplifies both, the signal and the noise. 
Reducing $C$ suggests to reduce the size of the diode and the transistor gate of M2. This strategy has to be balanced against 
the bad charge collection capabilities of tiny collection diodes and the occurrence of Random Telegraph Signal (see Sec. \ref{sectionRTS}) in transistors with too small gates.
The typical noise of a good and non-irradiated 3T-pixel amounts $\sim 10~ e ~\rm ENC$ (equivalent noise charge). 

Besides the signal charge, the leakage current of the collection diodes discharges the pixel capacitance. To avoid a saturation of the pixel, this capacitance is recharged by opening the reset switch M1. The recharging is process somewhat history dependent \cite{DoktorarbeitDeptuch} and, due to kTC-noise, not very accurate. A substantial improvement in the noise performance of the pixel can be achieved by means of correlated double sampling (CDS), which consists in taking a reference for the voltage $U_1(t_1)$ of the pixel before starting a particle measurement but after performing the reset cycle. After the integration time of this particle measurement is completed, a sample $U(t_2)$ is taken and the voltage drop in the pixel is derived by subtracting $U_S=U(t_2)-U(t_1)$. This eliminates the signal offset introduced by the reset noise. However, the reset of the pixel and the reference measurement introduce a dead time.

The leakage current of the diode can be considered as constant in time. It thus provides a DC-pedestal of $U_S$, which can be measured and subtracted. 
The capability of 3T-pixels to indicate the leakage current of its diode down to a fraction of $\rm fA$ makes those pixels a valuable tool for understanding radiation damage in both, the depleted and non-depleted active volume of the pixel. 

\subsubsection{The rolling-shutter readout scheme}
\label{sec:RollingShutterReadout}
Experiments of sub-nuclear physics aim usually to reach high counting rates in order to provide the statistics needed to observe rare particles and subtle effects. 
To match this goal, one would preferably connect each pixel with an individual, fast readout chain, which delivers the pixel signal immediately to the outside world.
This is however not feasible as the number of electrical connections between the chip and the outside world is restricted to $\sim 100$ while a full size CPS
hosts typically $\sim 10^6$ pixels. The simple solution used in first generation CPS consists in connecting all pixels consecutively to an analogue output line (multiplexing) and to process the data stream obtained with external resources like an external ADC. To connect and to disconnect the pixel from this common readout line is the mission of 
the ``select'' switch of the 3T-pixel (transistor M3 in Fig. \ref{fig:sandwich}). As  the start and end of the integration time of a frame is given by this moment of
pixel readout, it varies as a function of the pixel number $i$, which is referred to as rolling shutter. The precise time of the 
beginning ($t_1(i)$) and the end ($t_2(i)$) of the integration time of pixel $i$  is given with:
\begin{eqnarray}
t_1(i)&=& t_0 + i / f\\
t_2(i)&=& t_1(i) + N / f
\end{eqnarray}
Here, $f$ stands for the readout frequency of the readout line, $N$ for the number of pixels connected to this line and $t_0$ for time of the readout of the first pixel.
The integration time of the detector is given with:
\begin{equation}
t_{int}=N/f
\end{equation}
As the frequency of the readout $f$ is restricted to $\sim 50~\rm MHz$ for an analogue readout, the time resolution of this kind of readout remains at $\sim 10~\rm ms$.
The readout may be accelerated by using a couple of readout lines in parallel, which is however limited in practice by the number of external channels (e.g. high speed ADCs) required.

\subsection{CPS with accelerated time resolution and rate capability}
\subsubsection{Design challenges and approaches for solutions}

The time resolution and the rate capability of the 3T-pixel is limited by two major issues. On one hand, the pixel reset and the need to obtain a reference frame before initiating the particle measurement generates a significant dead time, which is not acceptable for modern particle physics experiments. Moreover, the speed of the over-all system is limited by the data bus transporting the signals obtained from the pixel to the outside world. 

The first issue was solved by replacing the reset switch by a system providing a continuous leakage current compensation. Thanks to this, the reset cycle became obsolete and the related dead time was eliminated. This feature is implemented in pixel structures, which are denoted as Self-Bias (SB-)pixels.
The second issue was addressed by digitizing the data if possible already on the pixel, to compress it on the chip and to submit only compressed digital data to the outside world. The compression is in simple terms done by exclusively submitting the address of the few firing pixels instead of charge information of all pixels, which reduces the data volume of frames with low occupancy dramatically. 

Realizing this concept is complicated by the fact that the analysis of CPS data includes a number of steps like pedestal correction and correlated double sampling, which were initially done offline and had to be integrated into the ASIC. Copying the traditional digital processing came out as unrealistic as the memory required could not be placed to the sensors. Instead, the functionalities had to be realized with analogue electronic circuits.
Designing circuits doing the analogue processing of few 100 electron signals in CMOS technology came out to form a tough engineering challenge. 
%This is as the individual components show a significant production tolerances, which hampered the production of pixels with equal dark potential and amplifier gain and comparators with similar discrimination level. 
This is as the individual CMOS components show significant production tolerances, which generate offsets in the dark signal of the devices. Compensating those offsets e.g. by means of placing trim DAQs into highly granular pixels is excluded by space constraints. Therefore, all amplification and discrimination circuits have to be designed with internal analogue offset compensation. 

% cannot be compensated by dedicated trim . On the other hand, 
%Moreover, the pixels are too small to hold trim DACs, which could be used to compensate the production tolerances. Instead, the discriminators of groups of numerous pixels were steered by one common DAC setting the threshold. Finally, different solutions based on elaborated, self-calibrating electronic circuits were and are being designed to overcome this obstacle.

%\subsection{Pre-amplifiers and strategies for improvement}
%The readout of CPS is traditionally done by means of a rolling shutter. The signal of multiple pixels is multiplexed to one ADC, which may be located outside the chip. Due to this readout, all pixels have the same integration time but this time starts and stops with a specific delay, which motivates 
%the naming. The readout frequency may reach $\sim 10 ~\rm MHz$ (pixels/s) per readout line and only few lines per chip can be used in practical terms.
%Consequently, this approach is restricted to a readout time of few $\rm ms$. Improved readout concepts were consecutively realized in order to accelerate this readout time and the related time resolution of the device. This development will be briefly summarized as the different amplifiers show specific responses to radiation damage.

\subsubsection{Eliminating the dead-time - The Self-Bias-Pixel}
\label{SectionSBPixel}
\begin{figure}
  \begin{center}
  \includegraphics[viewport=1.3cm -1cm 9.3cm 7.5cm, clip,width=0.35\columnwidth]{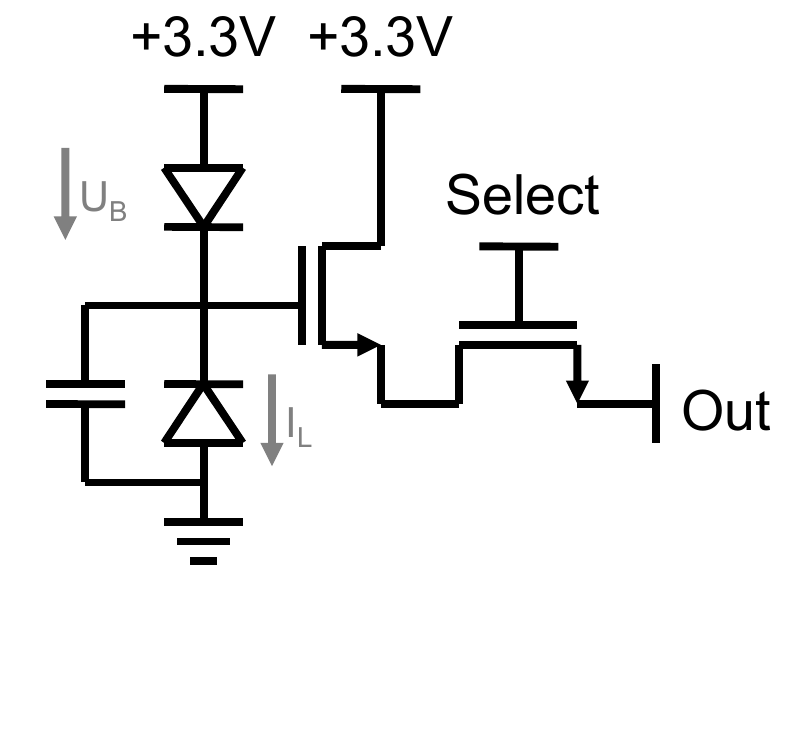}
	\hspace{0.025\columnwidth}
  \includegraphics[viewport=0.5cm 0cm 25cm 19cm, clip,width=0.55\columnwidth]{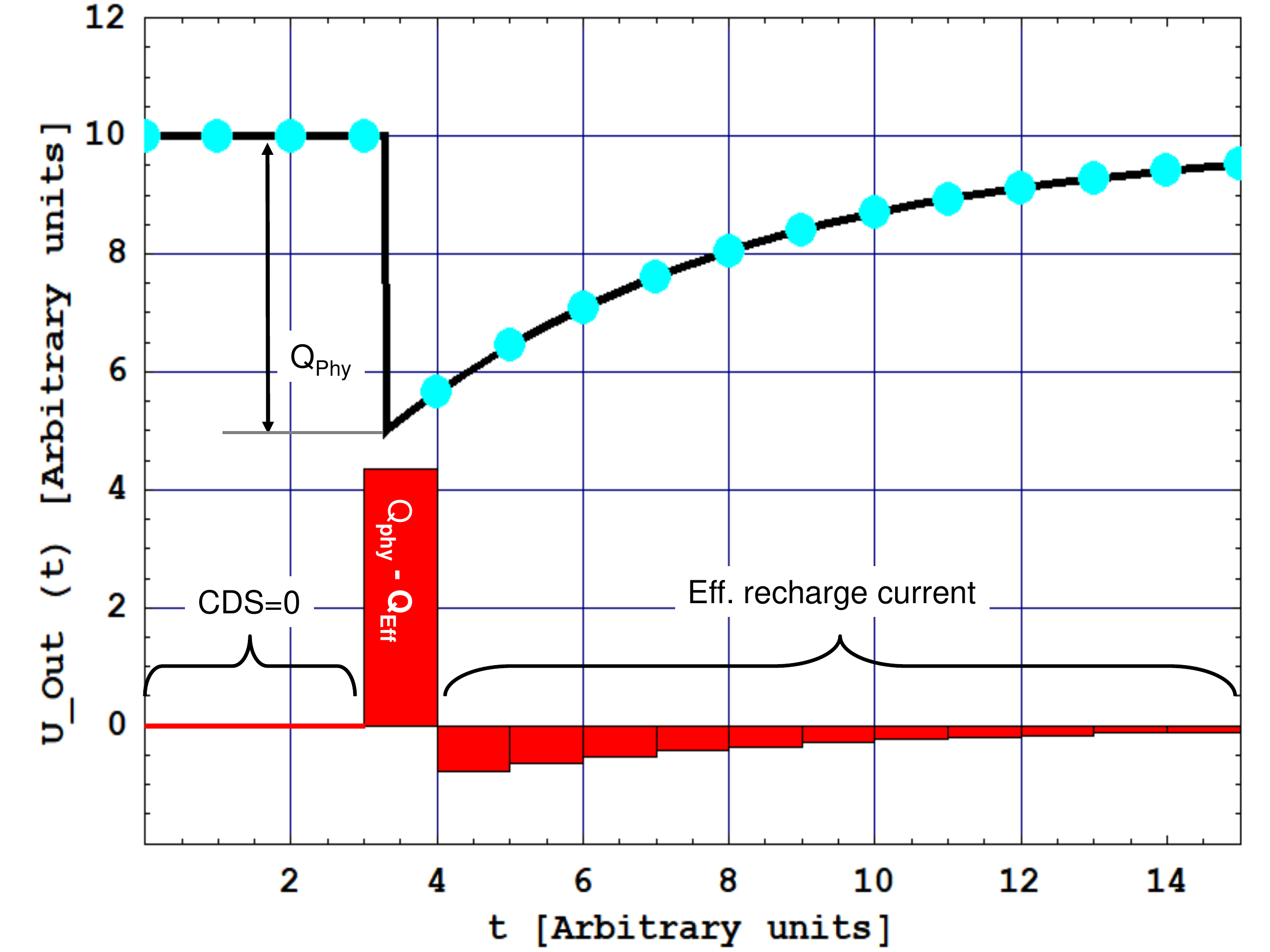}
    \caption{Left: Diagram of a SB-pixel. Right: Response of a SB-pixel to a particle signal. The signal $Q_{phy}$ is cleared slowly after the hit impinges. Performing CDS shapes the signal.}
    \label{fig:SB-readoutSchema}
  \end{center}
\end{figure}

In the Self-Bias-pixel (SB-pixel, see Fig. \ref{fig:SB-readoutSchema}, left), the reset transistor is replaced by a high-ohmic, forward biased diode. The current passing through this diode replaces continuously the charge, which was lost to the leakage current of the collection diode. Therefore, the reset cycle and the related dead time become obsolete. Moreover, the pedestals of the pixels disappear after performing CDS (see Fig. \ref{fig:SB-readoutSchema}, right). 
In case the pixel is hit, the signal voltage $U_S$ increases the voltage drop at the biasing diode and the current passing this diode increases. This increase removes the signal charge from the pixel capacitance. This clearing process can be well modelled with an exponential decay of the signal charge: 
\begin{equation}
Q_{S}(t)= Q_{Phy} \cdot \exp \left(- \frac{t}{\tau}\right)
\end{equation}
Here, $Q_{Phy}$ denotes the signal charge created by the particle impact at $t=0$ and $\tau$ the time constant of the clearing. This constant should be kept as long as feasible by design. It decreases dynamically in case a fast series of hits creates a sizeable signal charge in the pixel capacitance. Therefore, SB-pixels don't saturate in practical terms and the pixel remains sensitive to further hits during the clearing process. However, the charge injected by the clearing reduces the amplitude of the consecutive particle hit.  
Moreover, in the case of high occupancies and/or high leakage currents, $\tau$ approaches the integration time of the pixel and a significant fraction of $Q_{S}$ is cleared before the pixel is read out. This and specific radiation damage effects will be discussed in more detail in Sec. \ref{ChapterIonizingRadDamageSB}.

%\begin{itemize}
%\item Goal: Charge to Voltage conversion
%\item Impact of diode capacitance
%\item Design constraints for diode size - Charge collection efficiency and RTS
%\item 3T-design - optimal for spectroscopy
%\item SB - design for continuous readout
%\item Column parallel readout and MIMOSA-26
%\item Pixel parallel readout and ALPIDE
%\end{itemize}

\subsubsection{From sequential to column parallel - The clamping-pixel}
\label{Sec:ClampingPixel}
While the SB-pixel features a leakage current compensation, it is not yet suited to perform CDS and signal discrimination.
Those building blocks were added in the clamping pixel (CP-pixel) \cite{Mimosa8Paper}. 
The CP-pixel is designed for standard CMOS technology, which prohibits the use of PMOS transistors and 
thus discriminators in the pixel area. Instead, the discriminators are placed on the same chip aside the pixel matrix. 

The readout is typically organized 
in a column parallel way, which means that the signal of all pixels of a column is multiplexed to one common 
discriminator. This discriminator supports typically a pixel readout frequency
of $\sim 5~\rm MHz$ and is thus one order of magnitude slower than the related external devices. 
However, as about 1000 readout blocks are operated in parallel, one obtains a speed-up of 
two orders of magnitude with respect to the traditional readout. 

\begin{figure}[t]
\begin{minipage}{0.7\textwidth}
 \includegraphics[viewport=0cm 0cm 25.5cm 20cm, clip,width=\columnwidth]{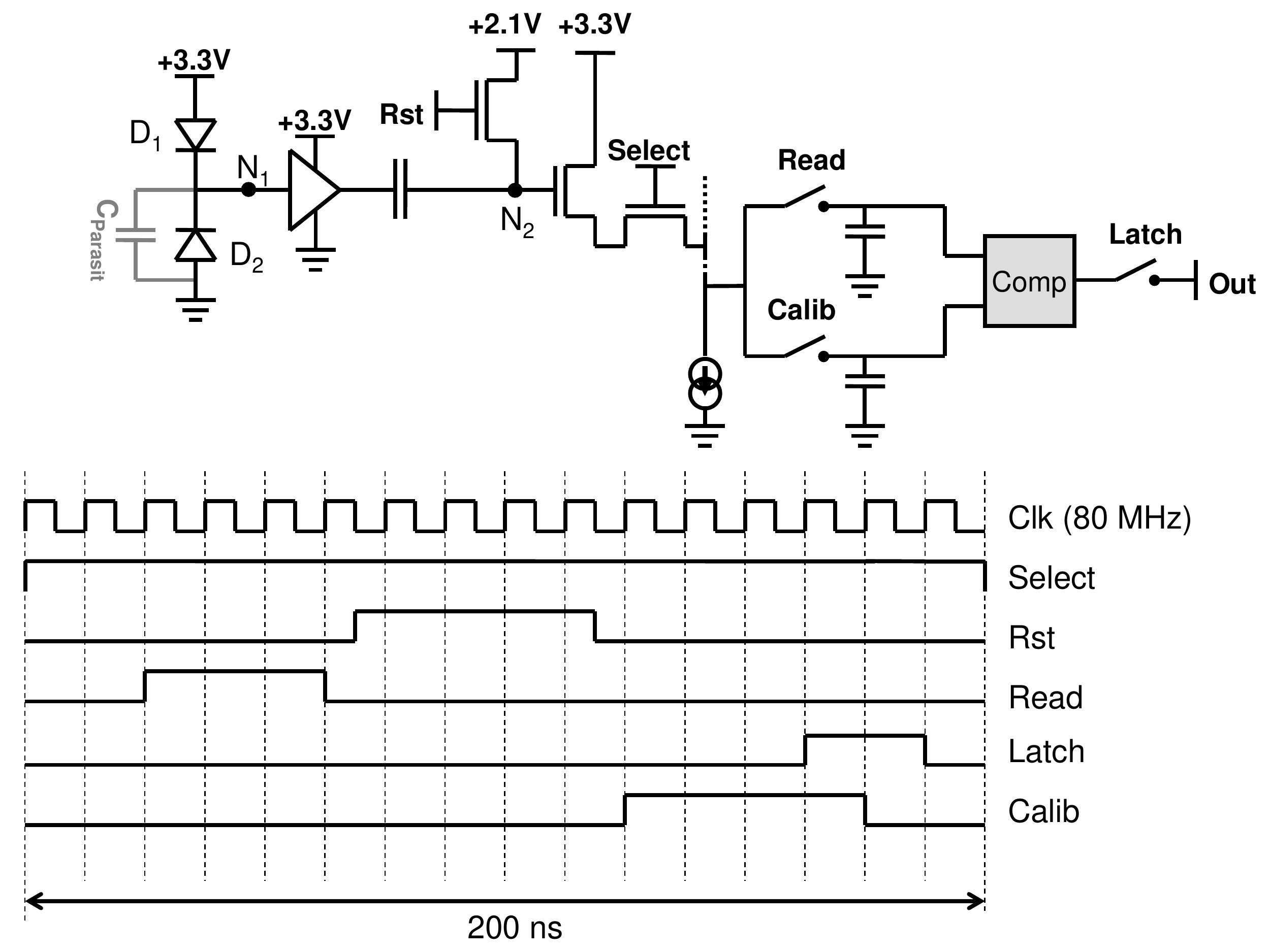}
\end{minipage}
\hspace{0.02\textwidth}
\begin{minipage}{0.26\textwidth}
 \caption{Diagram of the DC coupled SB-version of a clamping pixel and its readout sequence. In the 3T-version, the diode $\rm D_1$ is replaced by a reset transistor,
in the AC-coupled version, a capacitor is added at the point $\rm N_1$.  After \cite{Mimosa8Paper, RadToleranceDev}.}
    \label{fig:ClampingPixel}
\end{minipage}
\end{figure}

A simplified diagram of a CP-pixel and a switching diagram\footnote{Note that the diagram starts at the end of a pixel integration time and ends at the beginning of the consecutive integration time.} of a pixel readout is shown in Fig. \ref{fig:ClampingPixel}. The sensing element may be similar to the one of a 3T- or SB-pixel and shares the related properties. The source follower transistor is replaced by a AC or DC coupled voltage amplifier. This amplifier sends its signal via a second capacitor to the clamping node $N_2$. The potential of the latter is fixed prior to each readout cycle by opening the \emph{Rst} switch. Hereafter, it is buffered by a source follower and forwarded via the \emph{Calib} switch to an analogue storage capacitor located in the end-of-column discriminator block. After the integration time, the potential is sent via the \emph{Read} switch to the second capacitor and the potential of both capacitors is compared by the comparator. Thanks to the AC-coupling between the amplifier and the node $N_2$, the difference obtained is equivalent to the signal after CDS. If this signal exceeds a threshold, a digital hit indication is sent to the output. 

The clamping pixel was found decisively less vulnerable to the production tolerances of the CMOS process than earlier design attempts \cite{Mimosa6Paper}, which aimed to integrate the storage capacitors and differential amplifiers needed for doing CDS into the individual pixels. In good designs, the offsets caused by those tolerances remained substantially below the thermal noise of the pixels. After adding a digital data compression stage, the architecture was successfully used by multiple users. 

% in the MIMOSA-26 sensor known from the EUDet - telescope \cite{PaperChristineMi26, Mi26Ref} and in the ULTIMATE sensor of the STAR PXL detector \cite{PaperUltimateSensor}. 

%The functionality of the CDS procedure can be understood from the pixel readout cycle displayed on  the lower right panel of Figure \ref{fig:ClampingPixel}. The cycle starts with reading the potential of the pixel \emph{after} the integration time and charging it into one of the storage capacitors located at the end of the column by closing the ``read''-switch. Next, the clamping node is reset and the ``calib''-switch is closed in order to store the respective potential as reference voltage into the second capacitor. Given 
%the AC-coupling of the preamplifier to the clamping node, the voltage stored in the ``read''-capacitor represents the sum of the signal after CDS-processing and the reset potential of the clamping node. To eliminate latter contribution,  the voltage stored in the ``calib''-capacitor is subtracted  from the one stored in the `read''-capacitor in a second step. Provided the reset potential is arbitrary but well reproducible, the output of this procedure is a pedestal corrected CDS-signal, which is ``computed'' during the readout of the individual pixel. This signal is compared with a threshold, which is common for all pixels. To provide zero suppression, the location of an individual active pixel (or sets of up to four contiguous fired pixels in a row) is encoded in a data word, which is later sent to the outside world via a digital data connection. 

\subsection{Performances of selected CMOS sensors based on standard CMOS processes}
\label{Sec:Performances}

Classical CPS designed for charged particle tracking showed a surprisingly good performance from the start. The first first prototypes of the MIMOSA-series demonstrated a single point resolution of \mbox{$1 - 2 ~\rm \upmu$m} and a detection efficiency close to 100\% \cite{Paper_Deptuch_DevelopmentOfMAPS} in a $\sim 100 \rm~ GeV/c$ pion beam at the CERN-SPS. Those excellent results were soon reproduced with a full size CPS (MIMOSA-5) with up to 10$^6$ pixels on active areas as large as \mbox{$4~$cm$^2$} \cite{Paper_Deptuch_DevelopmentOfMAPS,TestPaperMimosa-5}. Moreover, it became clear that the devices can be thinned to $50 \mum$ without loss of performance. 
The charge sharing of the pixels allowed to apply centre-of-charge methods, which improved the spatial resolution to $\sim p/10$, where $p$ is the pixel pitch \cite{BessonPaper}. This has to be compared with a value of $p/\sqrt{12}$, which holds for pixels showing no charge sharing. 

The properties of selected detectors of the MIMOSA-series\footnote{The properties of other sensors, which typically rely on advanced CMOS processes, are discussed in Sec. \ref{CPSOtherGroups}} is shown in Tab. \ref{Tab:Sensors}. 
The first sensor listed, MIMOSA-5 was mostly a technology demonstrator, which relied on 3T-pixels. It was not really used for measurements due to its long integration time and the dead-time required for resetting the device. Instead, a prototypes relying on SB-pixels
were designed. This R\&D helped a lot to improved our understanding on the radiation hardness of CPS and yielded into sensors for the early EUDET-telescope (MimoTEL) \cite{EarlyEUDET}. However, the research line was abandoned as fully integrated sensors with clamping pixels became available. 

The first of those sensors, MIMOSA-26  \cite{PaperChristineMi26, Mi26Ref}  was a fully integrated detector featuring column parallel readout, internal data discrimination, internal data sparsification and could be configured by a JTAG interface. 
The readout is binary but the spatial resolution is better than $p/\sqrt{12}$ as the charge sharing between the pixels turns into different cluster shapes, which may be analysed by
means of centre-of-gravity methods. The sensor was available in a version with standard epitaxial layer but  also in  version with more radiation tolerant high resistivity epitaxial layer.
The sensor is being used for the present EUDET-telescope. Moreover, it was used for the prototype of the CBM-Micro Vertex Detector \cite{Paper:CBM-MVD1, Paper:CBM-MVD2} and the first generation (so called small  acceptance)
vertex detector of the NA61/SHINE experiment \cite{PaperNA61ShineSQM}. The MIMOSA-28 (ULTIMATE) \cite{PaperUltimateSensor} sensor is a larger and slightly slower device, which essentially relies on the same architecture of 
MIMOSA-26. It can be considered as the most prominent sensor as it was used in the STAR-PiXeL detector \cite{StarSummaryPaper}. The FSBB-M0 was the most developed sensor relying on clamping 
pixels. It was developed aiming for a use in the ALICE inner tracker and was optimized for fast readout, among others by reading out two lines in parallel. The sensor relied already
on a next generation process (Tower 180nm) but did not use specific features of this process. Still, the novel process allowed for a substantially improved tolerance to ionizing radiation.
Despite of reaching its design goals, the FSBB architecture was abandoned once the feasibility of realizing the more powerful, next generation sensor ALPIDE (see Sec. \ref{Sec:ALPIDE})  was demonstrated.

\begin{table}[t]
\label{tab:CPSPerformance}       % Give a unique label
%\begin{tabular}{p{3.3cm} p{2.5cm} p{2.5cm}| p{2.5cm} p{2.5cm}}
\small
\begin{tabular}{| l ||c|c|c|c| }

\hline
 Sensor	name				& MIMOSA-5 				 & MIMOSA-26 				& MIMOSA-28 					& FSBB-M0   \\
						&  			 &   				& (ULTIMATE) 					&    \\
\hline
Year				& 2001 					 & 2008 					& 2011 						& 2015	\\
Technology			& AMS $0.6\mum$	 		 & AMS $0.35\mum$ 			& AMS $0.35\mum$ 			& Tower 180 nm	\\
Epitaxial layer		& Low res.				 &	High res.				&	 High res.				&	 High res. \\
\hline
User			    & None	 				 & EUDET 					& STAR 						& None\\
				    &       		 		 & NA61/SHINE 				& 							&  	\\

\hline
Pixel  number			& $1024 \times 1024$	     & $1152\times 576$ & $960 \times 928$	& $416\times 416^a$	\\
Pixel size 			& $17 \times 17 \mum^2$  & $18.4\times 18.4 \mum^2$ & $20.7\times 20.7 \mum^2$	& $22 \times 33 \mum^2$	\\
Pixel type & 3T & Clamping & Clamping & Clamping \\
\hline
Spatial resolution 	& $\sim 2 \mum$			 & $3.5\mum$ 				& $3.8\mum$ 				& $\sim 4.5 \mum$ \\
Readout time 	& $6.5 ~\rm ms$ 		 & $115.2 \mus$  			& $185.6 \mus$ 				& $41.6 \mus$\\
\hline
Discriminator 		& External ADC 			 & End of column  			& End of column				& End of column\\
Readout	parallelism			& 4 pixel (analogue)           	 & Single line 		& Single line			& Double line\\

Power dissipation   & 		N/A				 &	$ 250~\rm  mW/cm^2 $	& $160~\rm mW/cm^2$			& $<160 ~\rm mW/cm^2$\\
Max. data rate   & 		$4 \times 40~\rm MHz^a$			 &	$ 160~\rm Mbps $	& $320~\rm Mbps$			& $2\times320 ~\rm Mbps$ \\
\hline
Ionizinig rad. tol. & $\sim 100~\rm krad$	 &	$150~\rm krad$	& $150~\rm krad$			& $> 1600~\rm krad$\\
Non. Io. rad. tol.  & $<10^{12}\neqcm$	 &	$>10^{13}\neqcm$	& $>10^{13}\neqcm$			& $>10^{13}\neqcm$\\

\hline
\end{tabular}

\vspace{2mm}

{\footnotesize $^a$ Three FSBB were intended to form a full reticle size sensor.  }

\caption{Performances of reticle size sensors relying on standard CMOS processes (MIMOSA-series) and partially depleted pixel diodes.}
\label{Tab:Sensors}
\end{table}
\section{Methods for observing radiation damage}
\label{Sec:Methods}
%\subsection{Fundamental appraoch and observables}
In the most simple approach, the radiation tolerance of a device is evaluated by measuring its properties before and after irradiating it. As any radiation dose creates microscopic radiation damage, one expects the properties to change during irradiation. However, the related modification may remain without impact on the macroscopic properties of the device.
A device is considered as tolerant to a specific radiation dose, if it still operates according to its specifications after irradiation.
Typical mistakes in radiation tolerance studies consist in not defining the specification to be fulfilled and
in confusing a missing instrumental sensitivity to radiation damage 
with the absence of this damage.

Besides estimating the radiation tolerance of a given device, there is an interest in spotting the damaging mechanism limiting this tolerance. 
The latter allows to fix the weak points of the design. Those kind of studies are particularly
complex in highly integrated devices like CPS. This is because the sensing elements of CPS can only be accessed via the internal amplification
chain\footnote{One may of course consider to build e.g. isolated charge collection diodes connect them with external high precision instruments. However,
the diode capacities and leakage currents are typically too small for obtaining reliable results. Moreover,
identical diodes showed different leakage currents if placed isolated or into a pixel 
(see Sec. \ref{Section:3TLeakageCurrent}).}, which complicates the measurements and limits the number of accessible parameters. Consequently, studies on radiation tolerance are
preferably carried out with sensors with a simple readout chain, e.g. with 3T- and SB-pixels. A brief overview over the accessible parameters for 
those pixels will be given in the following. More detailed information can be found e.g. in \cite{MichaelPhD}.

\subsection{Setting up the testing hardware}

As CPS are sensitive to light and many radiation effects are temperature sensitive, they should be tested in a temperature controlled dark chamber. 
The dark chamber should be air tight to avoid a destruction of the 
sensor by condensed water. Unless dry nitrogen is injected, the sensors should be started at room temperature and cooled only while running. 
%If this protocol is followed, the power dissipation of the electronics increases its temperature with respect to the primary cooling installations (e.g. coolant pipes). A potential condensation of air humidity will thus concentrate to the cold, primary cooling installations which are typically reasonably water tolerant. 
Few tens of minutes are typically required to reach a stable thermal equilibrium.
% The installation
%has to be warmed up above room temperature before opening the dark chamber. If this protocol is used, the dissipated power of the electronics keeps its temperature above
%the one of the primary cooling installation (e.g. coolant pipes). Thus, potential humidity condenses at this installation while electronics and sensors remain free of water/ice. 

In 3T-pixels and SB-pixels, the readout of the pixel properties is typically done via a linear, analogue amplification chain. The signal is hereafter
digitized by means of an ADC. Due to the relatively wide spread of the dark signal of the pixels as compared to the pixel noise, this ADC 
should provide at least a 12-bit resolution. After applying CDS, the pixel amplitude may be indicated negative. The readout chain has thus
to support negative numbers. 
%
%The processing of the data includes typically the so-called Correlated Double Sampling (CDS). 
%This CDS consists in measuring the output voltage of the individual pixel prior and after the integration time and in subtracting both values.
%This operation eliminates the individual pedestal of the pixels and the result reflects the integrated charge of the pixel.
%
%Note that due to noise, the signal value of the pixel after CDS may be negative. Therefore, the processing chain has to be able to handle negative 
%numbers.
Besides of showing a wide spread from pixel to pixel, the dark signal of the pixels changes significantly with temperature.
Caution is required to keep the (non CDS processed) signal in the linear range of the readout chain as saturated pixels may generate rather funny
electronic effects. The symmetry of the charge of the signal clusters is worth checking as an asymmetry in the average charge of all pixels in the order of readout is an indicator for an issue in the 
analogue readout chain.
 
%Before interpreting the data obtained, one should check the readout chain. This holds in particular as this chain tends 
%to provide close-to-reasonable data despite of major issues. A frequent mistake consists in operating a fraction of the pixels
%in saturation in the sense that their output signal leaves the linear range of the external amplification chain. This occurs because 
%the dark signal of the pixels varies substantially from pixel to pixel. The variation depends in general on the temperature and
%increases in case the sensors are irradiated. Note that cooling decreases the variation for 3T-pixels but may increase it for
%non-irradiated SB-pixels. As a good rule of thumb, one has exclude pixel saturation prior to each individual measurement. 
%Otherwise, the saturated pixels may cause all kinds of funny effects. 

%Computing the average charge of all pixels in a cluster may help to identify potential problems in the signal
%transmission. An asymmetric average shape has found to point to issues as missing line termination resistors, an accidential replacement of those resistors by capacitors and broken signal cables. A missing ``calibration peak'' in the $^{55}$Fe-spectrum (see Sec. \ref{SectionFe55Spectrum}) of a sensor, which should normally show this peak,
%was caused by an insufficient current of the external current source biasing the source follower of the related pixel matrix. In SB-pixels, 
%it is however a frequent signature of radiation damage.

\subsection{Irradiating the sensors}

A good choice of irradiation sources may ease spotting the origin of radiation damage. Most charged particles generate ionizing and non-ionizing radiation at the 
same time and generate therefore both, bulk and surface damage. However, X-rays with $\sim10\rm ~keV$ have too small momentum to displace atoms and to create bulk damage. Neutrons are missing the charge needed to ionize electrons and generate therefore almost exclusively bulk damage. Using $\sim 10\rm ~keV$ X-rays and $\sim 1~\rm MeV$ neutrons does thus allow to separate the effects of bulk and surface damage. 

Note that $\sim 1 ~\rm MeV$ neutrons deposit their energy in nuclear interactions, which transfer high average recoil energies to the silicon atoms. The recoil atoms obtain sufficient
energy for secondary interactions and create dense accumulations of defects (defect clusters) at the end of their trajectory. For charged particles, the additional Column interactions 
lead to low average recoil energies, which rather create isolated point defects. Some radiation effects, e.g. the beneficial effect of oxygen enrichment in silicon \cite{PaperOxiSilicon} are sensitive
to this. Therefore, while neutron irradiation is suited for creating isolated bulk damage, additional irradiations with charged hadrons should be considered
for obtaining final numbers on the tolerance to those particles.

\subsubsection{Irradiating CPS with X-rays}

Irradiations with X-rays are typically done with medical X-ray tubes, which are complemented with a dosimetry system. The typical dose rates of $\gtrsim 100~\rm krad/h$ allow for fast irradiations but let room for potential dose rate dependent effects. 
The device under irradiation has to be biased and operated during irradiation to provide the fields required for separating the electron/hole pairs in the $\rm SiO_2$ structures. Not biasing is equivalent to reducing the dose by a significant factor (see \cite{IonizingRadiation3} for data on the impact of biasing voltages to surface damage). The biasing is best done by mounting the device on a test PCB and operating it. Obviously, it is of advantage to test the device before irradiation. Note that the active components of the PCB are typically not radiation tolerant and have to be shielded against the X-rays, e.g. with $\gtrsim 1~\rm cm$ brass.

As a part of the surface damage is reduced by room temperature annealing, the dose rate and the temperature during irradiation \cite{DetteIrradationTemperaturePaper} may have an impact on the results. To avoid unwanted/uncontrolled annealing, the irradiated device should be tested already during or short after the irradiation. Moreover, one should avoid to expose the sensors to high temperatures as for example caused by direct sun light. This is of particular concern if the irradiated device has to be transported prior to testing it.

\subsubsection{Irradiating CPS with neutrons}

Neutron irradiations are typically performed with non-moderated reactor neutrons as provided by suited research reactors. In case the neutron spectrum contains a sizeable amount of additional thermal neutrons, one should consider that P-doped CPS might receive additional damage via the neutron capture reaction $\rm ^{10}B(n,\upalpha)^7Li$ (see Sec. \ref{SectionBorIrradiation}). 

Test PCBs may become activated by neutrons, so it is usually more practical to irradiate bare chips. The missing biasing of the chips is likely beneficial as it has the potential to reduce the unwanted surface damage caused by the $\gamma$-ray background of the neutrons. However, the sensors cannot be tested before or during irradiation. Instead, one compares the properties of irradiated sensors with non-irradiated devices of the same series. This imposes
uncertainties due to production tolerances. 

The CPS may be wrapped in aluminum or held by Gelpacks during irradiation. The latter were found neither activated nor degraded by doses of $10^{14}\rm~n/cm^2$ (lower limit). However, it is mostly impossible to remove $50 \mum$ thin CPS from the packs. Therefore, thick chips should be chosen. Due to limited production yields, some of the non-tested sensors may be broken already before irradiation. One should thus irradiate several sensors per dose level in order to obtain at least one (initially) working device.

Non-ionizing radiation damage in CPS was found to show less room temperature annealing than ionizing radiation damage. However,
few days were typical required for placing and bonding the sensors on PCB, which lets room for overlooking such effects.

\subsection{Accessible quantities and their measurement}

\subsubsection{Chip temperature}

\begin{figure}[t]
\begin{minipage}{0.5\textwidth}
 \includegraphics[width=\columnwidth]{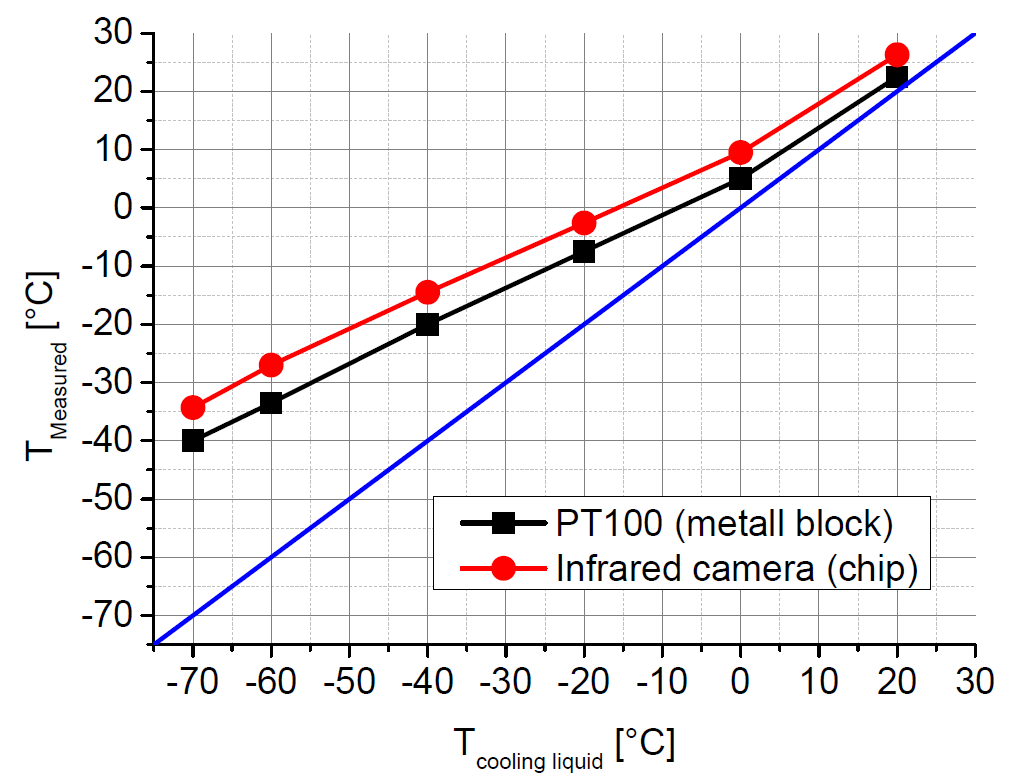}
\end{minipage}
\hspace{0.02\textwidth}
\begin{minipage}{0.36\textwidth}
 \caption{Temperature of a MIMOSA-18 sensor as measured with an IR camera in comparison to the temperature of the
supporting metal block as a function of the coolant temperature of the chiller. The ideal curve of identical temperatures is indicated
by a blue line. See text. From \cite{PaperPhD-Dennis}.}
    \label{fig:TempCompare}
\end{minipage}
\end{figure}

Measuring and controlling the temperature of the chip appears simple but comes out to be tricky as the chip is self-heating
during operation and as the heat contact via the supporting PCB toward the water cooled cooling support is far from perfect.
Moreover, it is hard to install a temperature sensor on a CPS prototype without endangering the bonds. 
The use of IR-thermometers is hampered by the high reflectivity of CMOS chips and indicates the arrangement and density
of the metal lines of the chip rather than its temperature \cite{Jungermann}. 
This issue was solved by spraying a MIMOSA-18 sensor with non-conductive, black colour (TETENAL 105202 camera varnish) \cite{PaperPhD-Dennis}. 
Figure \ref{fig:TempCompare} shows the temperature measured for this chip in comparison with the coolant temperature of the cooling system
and the indication of a PT100 sensor placed nearby the sensor PCB on the cooling support. The plot can be considered as representative
for most MIMOSA sensors relying on serial analogue readout. It should be mentioned that, in the absence of better options, the coolant temperature was used as a temperature reference in a couple of early publications
on the radiation tolerance of MIMOSA series CPS. As the IR-measurement is too complex for a daily use, later works indicate the
reading of the PT100 sensor.

\subsubsection{Leakage currents}

The leakage current of the charge collection diodes of CPS is typically too small for a direct measurement by means of an ammeter. 
The procedure may only be applied with large groups of diodes connected in parallel, which increases the current proportional to the 
number of diodes. Doing this with isolated diodes, which are not surrounded by the remaining pixel structures, was found
non-reliable: A mismatch of a factor 20 was observed when comparing the directly measured leakage current of isolated charge collection diodes
with the one of their counter parts, which were integrated in a pixel (see \cite{DoktorarbeitDeptuch}, P. 213).
Still, direct measurements may be used to estimate the leakage current of SB-pixels. This is feasible provided that the biasing of the pixel diodes is separated
from the general bias lines of the chip. 

The suggested measurement procedure for leakage currents consists in extracting the leakage current of the pixel diodes from the dark charge being indicated by 3T-pixels after
CDS processing. Knowing the gain of the pixels from measurements with a $^{55}$Fe-source (see below), this dark signal can be translated from voltage to charge in units of electrons. The current is computed accounting for the known integration time of the pixel. In order to reduce the impact of the noise, one
does typically take $\sim 100$ samples and computes the mean value. Note that the leakage current of CPS increases during a few seconds warm up phase after the start
of the sensor and should thus only be measured with a suited delay. 

The procedure yields the leakage current  of the individual pixel diodes, which varies from pixel to pixel due to significant production tolerances. 
The result is typically displayed as the mean leakage current of all pixels. The width of the typically asymmetric distribution is indicated as 
error bars, which should not be confused with the substantially lower uncertainty of the mean value measurement. The leakage current measurement of the 
diode may be biased by i) the drain-to-source leakage current and ii) the bulk-to-source leakage current of the reset transistor.

\subsubsection{Pixel noise (analogue readout)}

CPS exhibit multiple kinds of noise, which are caused by the noise of their intrinsic components (high frequency noise), pick-up from external sources (common mode noise) and Random Telegraph Signal (RTS, see Sec. \ref{SectionRTS}). Moreover, the pixel-to-pixel fluctuation of the dark signal of the pixels, despite technically not a noise, is sometimes referred to as fixed pattern noise. It is of relevance in case the signal discrimination thresholds cannot be tuned for the individual pixels. The different noise components may be measured as follows:

\paragraph{High frequency noise} 

The measurement of high frequency noise is done in analogy to the measurement of the leakage currents of the pixels. 
Instead of the mean value of multiple pixel signals after CDS, the related standard deviation is measured. As for the leakage current, 
the high frequency noise varies from pixel to pixel and the variation follows a rather broad distribution, which is often expressed in the
error bars of the related measurements. Note that a low high frequency noise is not necessarily turning into a low dark rate as the latter is often determined by 
hot pixels being affected by RTS (see below). This 
is of particular importance as 
strategies for reducing the capacitive, average noise of the pixels tend to increase the number and noise of hot RTS-pixels (and vice versa).
%
\iffalse
This holds in particular, if pixels are affected by RTS in the input stage of the pre-amplifier. 

The best way to express the noise depends on the context. In the easiest case, one may communicate the mean value and the 
standard deviation of the noise distribution. This is perfectly appropriate for imagers. It gets debatable once the R\&D aims for
particle tracking sensors with integrated discriminator. Due to limited space, only few voltage generators for setting the discriminator
threshold can be integrated into a chip. Therefore, the discriminators of sizeable groups of pixels are operated with
a common threshold and only some individual ``hot'' pixels may be masked without loosing efficiency. The threshold 
has to be set sufficiently high to suppress fake hits from the remaining most noisy pixels of the group. In this scenario,
the average noise of the pixels is an irrelevant quantity and the noise of the ``hot'' pixels should be considered. 
\fi
%
Besides of RTS, the noise measurement may be biased by the noise of the external readout chain including the discretization noise of the ADC
and common mode noise.

\iffalse
Those noise sources are not a property of the CPS and thus bias the measurement. Their impact may be reduced by
amplifying the output signal as upstream as possible in the readout chain. As a good rule of thumb, the gain should be chosen
such that the most probable noise remains above 2 ADU. The combination of this recommendation with the wide bandwidth of 
the dark signal of the pixels before applying CDS determines the recommended $\geq 12$ bit resolution of the ADC.

The indicated noise may be reduced by filtering common mode noise (see below) and 
by masking hot pixels. Both steps are easy to do for sensors with analogue output but hard to realize in sensors with integrated
discriminator. Again, one should  reflect well if the results obtained remain relevant for the real application.
\fi

\paragraph{Common mode noise}

For 3T- and SB-pixels with serial analogue readout, a low frequency pick up (e.g. 50 Hz from the electric grid) manifests itself as a slow modulation
of the pixel signal as function of the pixel number in the order or the readout.  To some extent, the pixel dark signal (before and after CDS) samples the pick up somewhat similar 
to an oscilloscope, where the y-axis is the amplitude of the pick up and
the x-axis corresponds to the pixel number and thus to the time of sampling. This width of the signal is determined by relation between the frequency of the noise
and the frequency of the readout clock. As the pick up injects a somewhat similar signal into multiple
pixels, it is often referred to as common mode noise.

 The common mode noise may be estimated by computing the average pixel signal after CDS
of a group of neighbouring pixels (in the sense of readout) within one frame. A filtering of the noise can be done in software by subtracting this average from the signal of the pixels in the group. A significant common mode noise points to a missing filtering of sensitive biasing lines and may be reduced in hardware by adding
a low pass filter (e.g. a capacitor) to this line.

In pixels with digital rolling shutter readout, common mode noise manifests itself as groups of firing, consecutive pixels or lines. 
This effect can only be alleviated by hardware improvement. Note that under bad circumstances, CPS generate common mode by themselves. 
This was observed for chips with integrated data sparsification circuits, which were biased with long flex print cables.
Once the sparsification circuit obtained a high data load, its current consumption increased and the resistance in the mass bias line modified the common digital and analogue mass potential. This indirectly increased the threshold, which reduced the data load. As an outcome, the system started to oscillate. 

\paragraph{Random Telegraph Signal (RTS)}

The term RTS reflects the observation that the current passing diodes and MOSFETs may be modulated between different discrete states
. To observe this effect, the dark signal of the individual pixel has to be sampled for up to several minutes
and to be plotted as the function of time.
Sampling the signal after CDS is particularly sensitive to RTS in the sensing diode of 3T-pixels, sampling the raw pixel signal may exhibit
RTS in the input stage of the on-pixel amplifyier (typically the source-follower transistor). More details on RTS are found in Sec. \ref{SectionRTS}.

\paragraph{Fixed pattern noise}
The term fixed pattern noise describes the spread of the dark signal of multiple pixels and is thus technically not a noise. This spread is rather excessive in CPS but is reliably eliminated by CDS processing
for pixels with analogue output. For pixels with in-pixel CDS circuits, the quantity forms an important performance parameter and may be extracted from the spread of the discrimination thresholds of the individual pixels by means of standard S-courve measurements.  

\subsubsection{The $^{55}$Fe amplitude spectrum - gain and charge collection efficiency}
\label{SectionFe55Spectrum}

The term charge collection efficiency (CCE) denotes the number of signal charge carriers collected as compared to the number of signal charge carriers created in the active medium of the sensor. In CPS, signal charge carriers may escape from the charge collection diode of the pixel penetrated by a particle by i) recombination and ii) by the diffusion of charge carriers towards neighbouring pixels. The second effect is typically dominant and the full charge is only seen if the signal of all pixels of a pixel cluster is added. Any statement on the related CCE is only meaningful if the number of pixels used for building this sum is stated. 

The CCE of CPS is measured by injecting a known number of signal charge carriers by means of a monochromatic X-ray source. The most suited source is $^{55}$Fe, which emits dominantly  $5.9~\rm keV$ photons and thus injects 1640 signal electrons. The related hits are recorded and the signal amplitudes are histogrammed to an amplitude spectrum, which is sometimes also called charge collection spectrum. 
\begin{figure}[t]
 \centering
%1
\begin{subfigure}{.45\textwidth}
  \centering
	\vspace{-1.3cm}
   \includegraphics[viewport=1.5cm 1.2cm 24cm 18cm, clip, width=1\columnwidth]{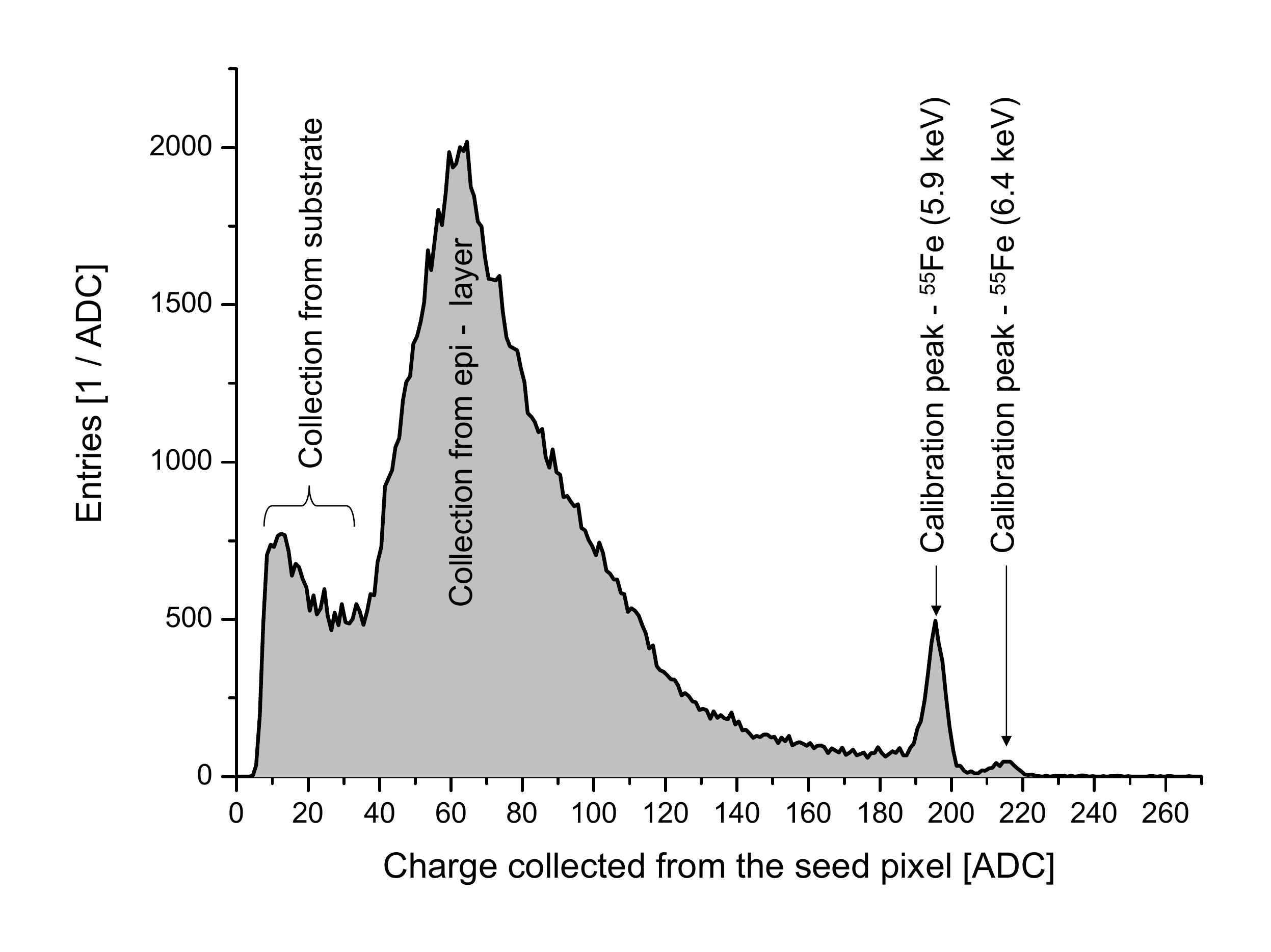}
    \caption{Signal from the seed-pixel.}
    \label{fig:Fe55SeedSpectrum}
\end{subfigure}%
\hspace*{0.05\textwidth}
%2
\begin{subfigure}{.45\textwidth}
  \centering
   \includegraphics[viewport=1.5cm 1.2cm 24cm 18cm, clip,width=1\columnwidth]{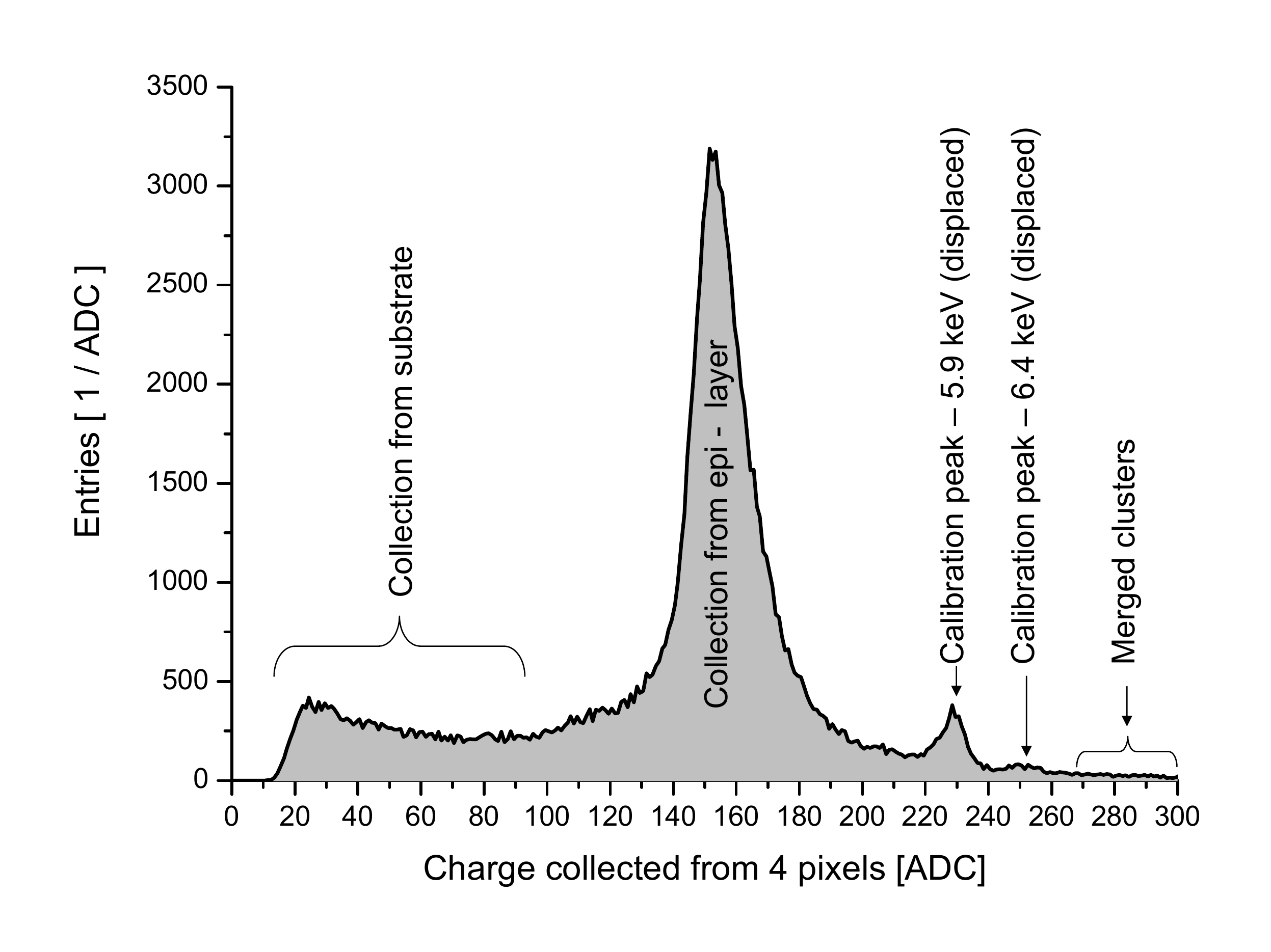}
    \caption{Summed signal of the 4 pixels with highest signal of a signal cluster. Summing the signal of all pixels would move
		the main ``collection'' peak to the ``calibration'' peak.}
    \label{fig:Fe55FourSpectrum}
\end{subfigure}%https://www.overleaf.com/6458423ktgndt#
\caption{Typical amplitude spectrum as recorded by illuminating a CPS with low resistivity epitaxial layer with photons from a $^{55}$Fe-source.}
\label{fig:Fe55SpectrumTextbook}
\end{figure}
A textbook example of such spectra is shown in Fig. \ref{fig:Fe55SpectrumTextbook}. The x-axis of those spectra represents the amplitude of a signal as recorded by a pixel. The y-axis defines, how often this amplitude was recorded. It is instructive to assume that the number of X-ray interactions in a sub-volume of the active medium of the pixel does not depend on the location of this sub-volume\footnote{Note that the attenuation length of X-rays in silicon is in the order of the thickness of the epitaxial layer of CPS. Therefore, this assumption forms a rather rough approximation.}. If this assumption of a uniform illumination is made, one may interpret the spectrum as a measure of the volume (y-axis) providing a certain charge collection efficiency (x-axis). 

The spectrum in Fig. \ref{fig:Fe55SeedSpectrum} represents the amplitudes recorded by the central, so-called seed pixel of a pixel cluster. This pixel collects by definition the highest fraction of the signal charge. The spectra was taken with a sensor relying on 3T-pixels and a low resistivity epitaxial layer. Only a small fraction of the active volume is depleted and most charge is collected by thermal diffusion. The spectrum of such sensors shows two distinguished peaks, the so-called ``charge collection peak'' and the so-called ``calibration peak''. Those peaks are generated by two major sub-volumes of the active medium, which show each an about uniform charge collection efficiency (CCE).  
The ``collection peak'' is associated with a charge collection from the non-depleted epitaxial layer, which forms the major active volume of the sensor. The ``calibration peak`` is associated with the depleted volume of the sensor. The related CCE is commonly accepted to amount 100\%. This is as signal electrons cannot escape from the strong electric fields found in this depletion zone.

The ``calibration peak'' is used to measure/calibrate the gain of the pixels and the readout chain of the CPS. In the example, the charge of 1640 electrons corresponds to an amplitude of 195 AD-units. The position of the ``collection peak'' is found at $65~\rm ADU$. Comparing it with 
the value for the calibration peak, one obtains the average CCE$_{\rm Seed}$ (roughly 30\%) of the epitaxial layer. Fig. \ref{fig:Fe55FourSpectrum} shows the spectrum as obtained from the summed amplitude of four pixels. Here, the ``collection peak''
is found at a position of about $155~\rm ADU$, which points to a CCE$_4$ of about 80\%. By summing up the signal of all, typically less than $5 \times 5$ pixels
of a cluster, one would typically obtain a CCE$_{25}$ of 100\% for good, non-irradiated sensors. Note that the calibration peak position found in the summed amplitude spectra
is not reliable. For SB-pixels, the position of all peaks are biased by pixel charge clearing (see Sec. \ref{SectionSBPixel}) and the reading is only reliable if this effect is suppressed e.g. by cooling. 

The charge collection spectra may also be used to obtain an indicative measure of the depleted volume of the pixel. To do so, the number of entries of the calibration peak ($N_1$) and the number of entries of the charge collection peak ($N_{25}$) are extracted from the charge collection for the seed pixel and the one for 25 summed pixels respectively. The depleted volume $V_{depl}$ of the pixel 
is then given with:
\begin{equation}
V_{depl} \approx \frac{N_1}{N_{25}} ~ V_{pixel}
\label{Eqn:Depletion}
\end{equation}
The active volume $V_{pixel}$ of the pixel has to be extracted from the known pixel pitch and thickness
of the epitaxial layer.

%, besides a simple measurement of the CCE, the amplitude spectra may be used for a more detailed understanding of different sensor effects. This reading of the spectra relies on interpreting it in terms of volumes with given CCE. A number of instructive spectra together with a brief explanation of the underlying detector physics effects is given in Appendix. 
%Figs. \ref{fig:TypicalFe55Signatures} and \ref{fig:TypicalFe55Signatures-2}.  

\subsubsection{The $^{90}$Sr amplitude spectrum - signal charge and SNR}

The best laboratory measure for the signal charge and the SNR of the sensors is provided by illuminating 
the sensors with hard $\beta$-rays as for example emitted by a $^{90}$Sr. The $\beta$-rays interact
approximatively like minimum ionizing particles and the related amplitude spectrum follows a Landau-distribution.
The related seed pixel amplitude spectrum may be used to estimate the SNR of the sensor, which 
is then defined as the most probable charge according to the Landau distribution divided by the average high frequency noise.
As a good rule of thumb, an SNR $> 15$ is required for a good particle detection.
A spectrum of the summed signal of typically 25 pixels provides an estimate of the full charge collection from the 
active medium.

\subsubsection{Pixel clearing constant}
\label{SectionClearingConstant}
Pixels with SB-structure (see Sec. \ref{SectionSBPixel}) rely on a controlled clearing of the pixel capacitance of the signal $U_S$, which is in the simplest case done by a biasing diode. This clearing follows in good approximation an exponential function:
\begin{equation}
U_S(t)=  U_S(t=0) \cdot \exp \left(- \frac{t} { \tau} \right)
\label{SB-RechargeEquation}
\end{equation}
%The clearing time constant $\tau$ decreases with the inverse of the leakage current of the pixel. As the latter increases with radiation, an strong acceleration of $\tau$ is frequently observed in irradiated sensors. If $\tau$ approaches the integration time of the sensor, a substantial part of the signal is cleared before the sensor is readout.
%The latter limits frequently the radiation tolerance of the device.
%Measuring the clearing time is complicated by the fact that the effect is steered by an interplay between the charge collection diode and the entrance node of the on-pixel pre-amplifier. The relevant scale of signal charges is $\sim 100~e$. Both aspects exclude a test of the amplifiers with external instruments.

The time constant $\tau$ may be measured by pulsing the pixel and observing the signal decay by means of the standard readout chain. In the frequent case that the pixels don't integrate a dedicated pulse generator, 
pulsing may be done by detecting photons from a $^{55}$Fe-source, which provide a signal amplitude corresponding to the one of the calibration peak. Hereafter, the indicated signal after CDS
is recorded for multiple integration times. This signal shows negative sign as it corresponds to the charge injected by the clearing mechanism. Fitting the results over time provides a good estimate of
$\tau$, provided that this constant is reasonably slower than the integration time.

\section{Effects of Total Ionizing Doses}
\label{Sec:IonizingDamage}
\subsection{Increase of leakage currents and shot noise}
\subsubsection{3T-pixel}
\paragraph{Leakage current}
\label{Section:3TLeakageCurrent}

\begin{figure}[t]
\begin{minipage}{0.6\textwidth}
 \includegraphics[width=\columnwidth]{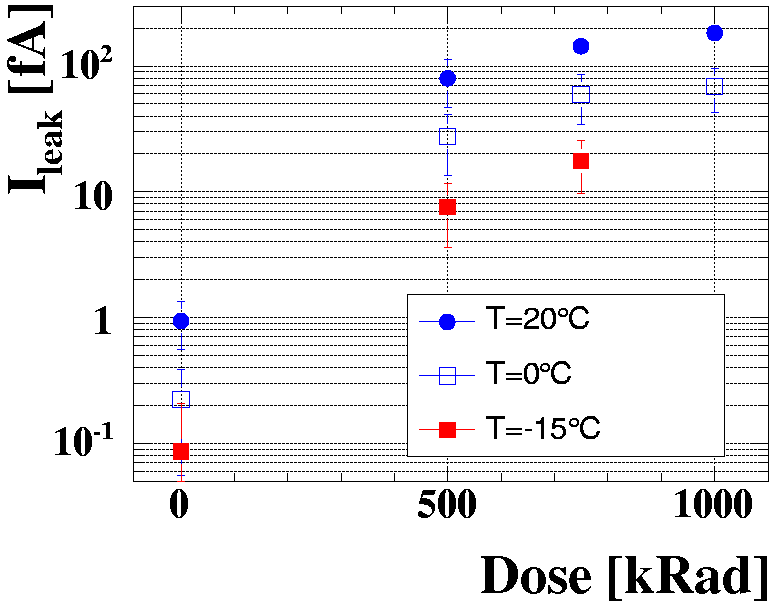}
\end{minipage}
\hspace{0.02\textwidth}
\begin{minipage}{0.36\textwidth}
 \caption{Leakage current of SUCCESSOR-1 a CPS imager with low resistivity epitaxial layer as function of total ionizing dose and temperature ($T_{Coolant}$). From \cite{Su1-Paper}.
                See original work for details on the device.}
    \label{fig:Su1LeakageCurrent}
\end{minipage}
\end{figure}

The most prominent effect of ionizing radiation in CPS consists in an increase of the 
leakage currents of the pixel. This was first reported in a study with CPS used
for optical imaging, which were exposed to $30~\rm Mrad$ \cite{Paper30Mrad}. 
The amount of the increase is in general specific to the individual CPS design and the production
process. As illustrated in Fig. \ref{fig:Su1LeakageCurrent}, the current may scale a fraction of a  $\rm fA$ to few $\sim 100~\rm fA$
per diode after a dose of $1~\rm Mrad$ \cite{Su1-Paper}. Despite this is not easy to see in the logarithmic
plot, the leakage current increases, within an anticipated $10\%$ dosimetry uncertainty, linearly with the total ionizing dose. A proportionally factor of
$1.9~ \rm aA/krad$, $6.6 ~ \rm aA/krad$ and $17.5 ~ \rm aA/krad$ may be extracted for a $T_{Coolant}$ of
$-15\C$, $0\C$ and $20\C$ respectively. Note that those values are considered as device specific and provided without any claim of 
generality.

In 3T-pixels, the additional leakage current may discharge 
the pixel capacitance prior to the readout and thus saturate the pixel. This is not of worry 
if the integration time of the pixel is sufficiently fast to reset it
in time. Alternatively, the current may be reduced by cooling. Note
that also the pixel-to-pixel fluctuations of the current increases. This feature may challenge
the dynamic of the amplification chain. Moreover, it makes it impossible
to discriminate the signal of irradiated 3T-pixels with a common threshold.
\begin{figure}[t]
 \centering
\begin{subfigure}{.75\textwidth}
  \centering
 \includegraphics[width=1\columnwidth]{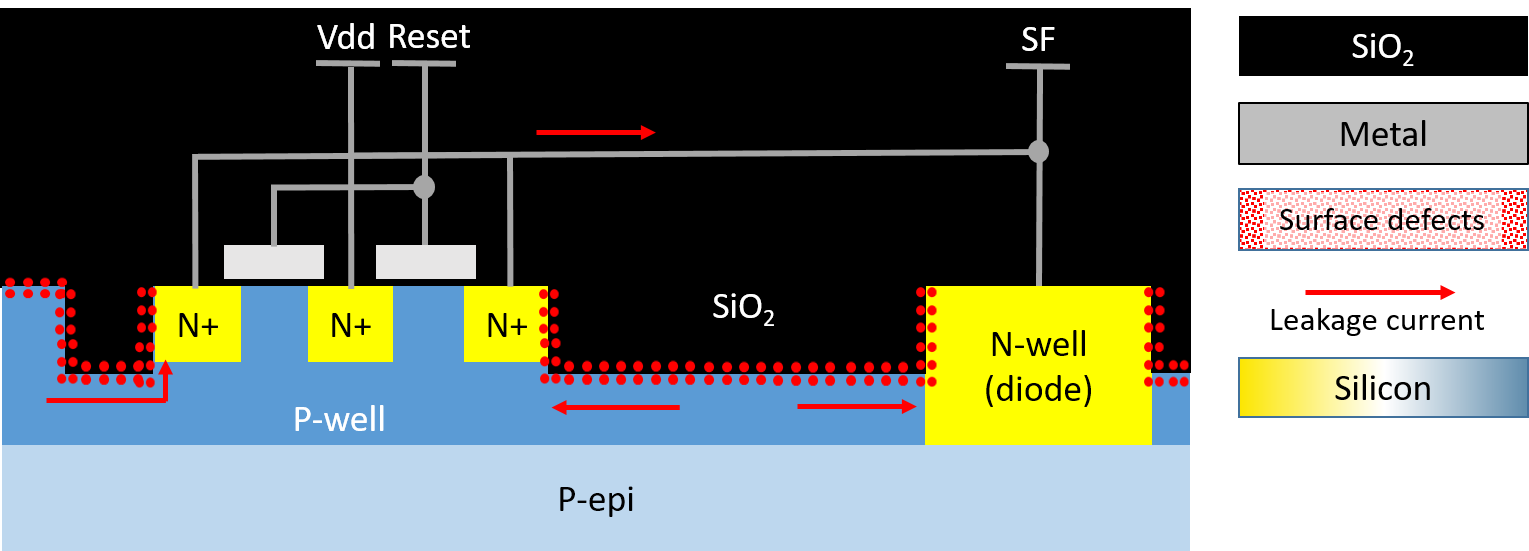}
    \caption{Radiation soft pixel: A cross-section of an enclosed reset transistor (left, see also Fig. \ref{fig:irradiatedTransistors4})
		and of a charge collection diode (right) is shown. Vulnerable, thick SiO$_2$ structures
		are placed nearby the diode. The radiation induced defects at the surface of those structures generate leakage currents, which are 
		collected by the diode. Additional leakage current is collected by the outer ring of the
		reset transistor and forwarded to the pixel node.}
    \label{fig:RadSoft3TPixel}
\end{subfigure}
%
%\hspace*{0.05\textwidth}
%https://www.overleaf.com/6458423ktgndt#%1
%

\begin{subfigure}{.75\textwidth}
\vspace{0.2cm}
  \centering
   \includegraphics[width=1\columnwidth]{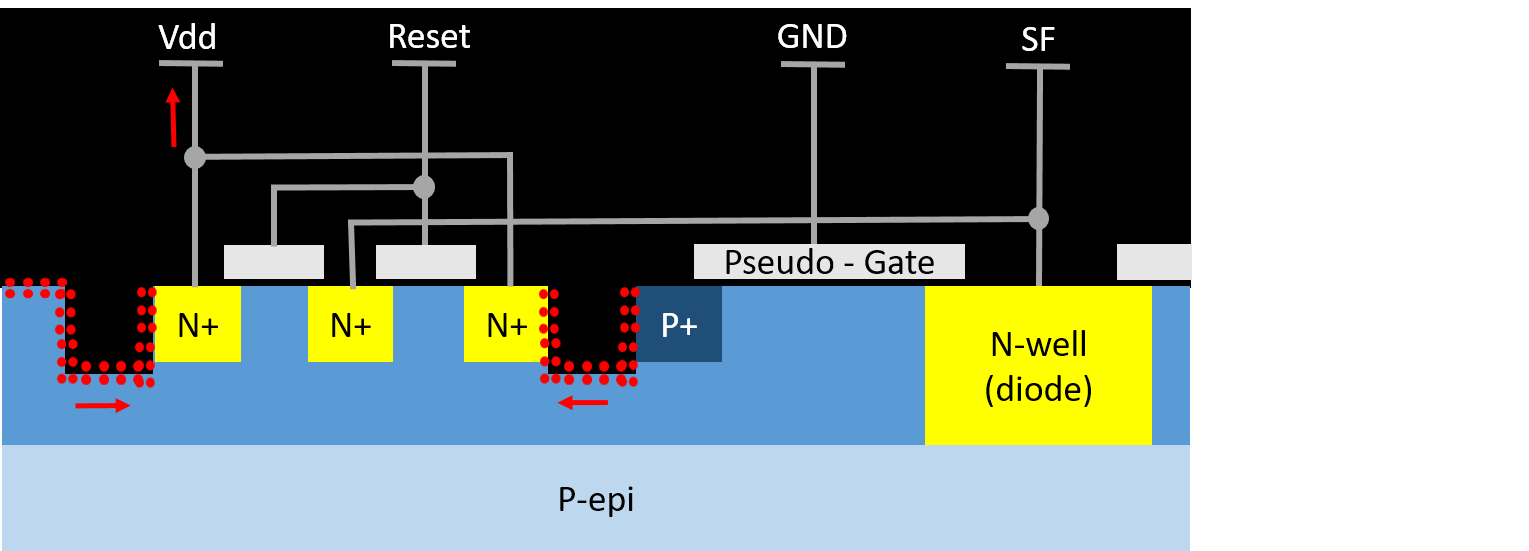}
			
		\caption{Radiation hardened pixel: Here, the protected inner ring of the reset transistor is connected
		to the pixel node. Leakage currents collected by the outer ring are thus evacuated to the 
		power supply. A p-diff (P+, alternatively n-diff may be used) guard ring is placed around the n-well diode.
		A ``pseudo gate'' guard ring forces the generation of thin and thus radiation hard SiO$_2$
		between this guard ring and the N-well diode.
		}
    \label{fig:RadHard3TPixel}
\end{subfigure}
%		\newline
	
\caption{Cross-section of a radiation soft and a radiation hardened 3T-pixel with enclosed reset transistor. Surface damage is indicated
by red dots, radiation induced leakage currents by red arrows, the black surfaces represent SiO$_2$, the dark blue surfaces the p-well.}
\label{fig:RadHardness3TPixel}
\end{figure}
\begin{figure}[t]
\begin{center}   
\includegraphics[viewport=1cm 0.2cm 26cm 20cm, clip, width=7.1cm]{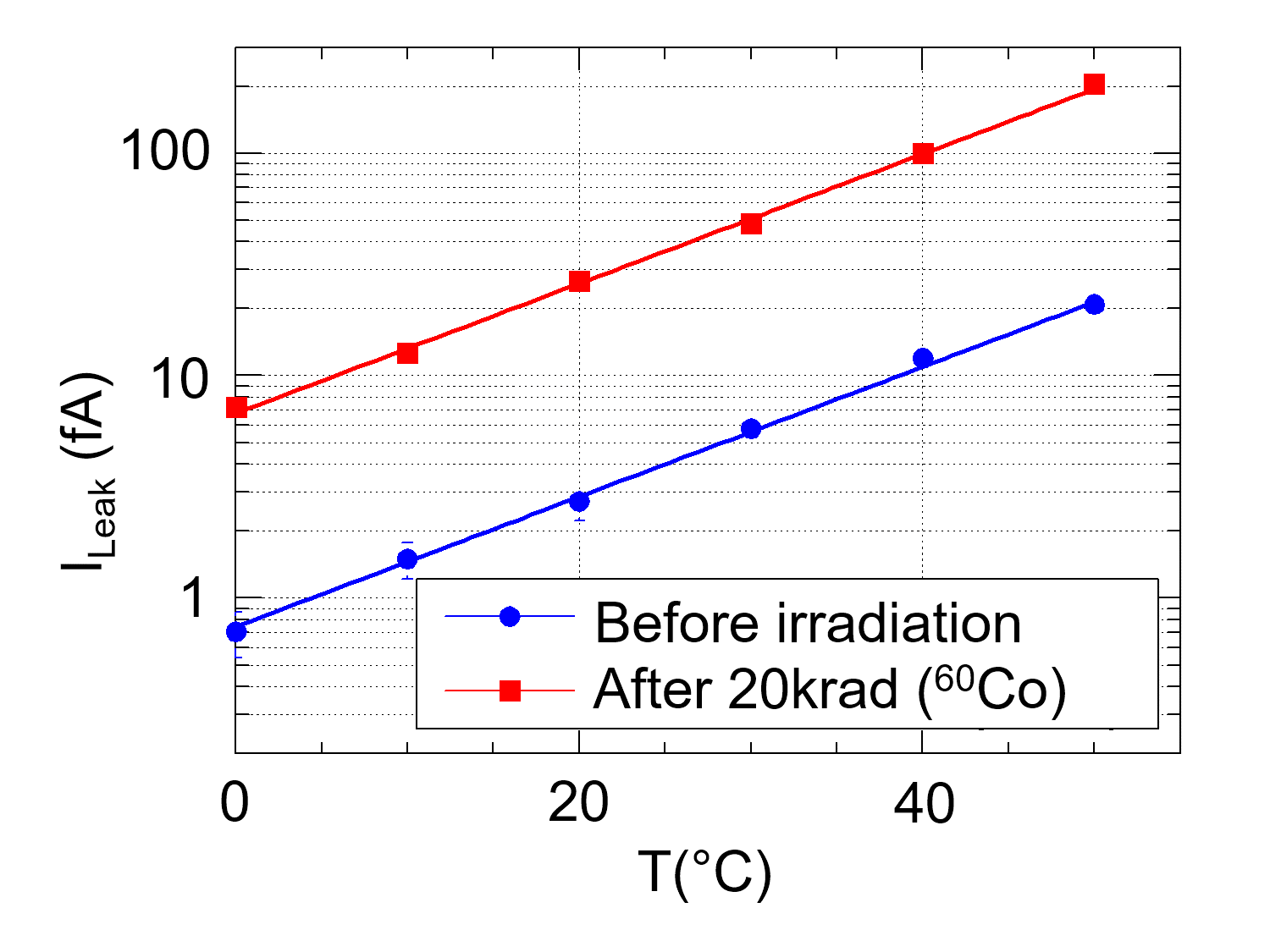}
\includegraphics[viewport=1cm 0cm 26cm 20cm, clip, width=7cm]{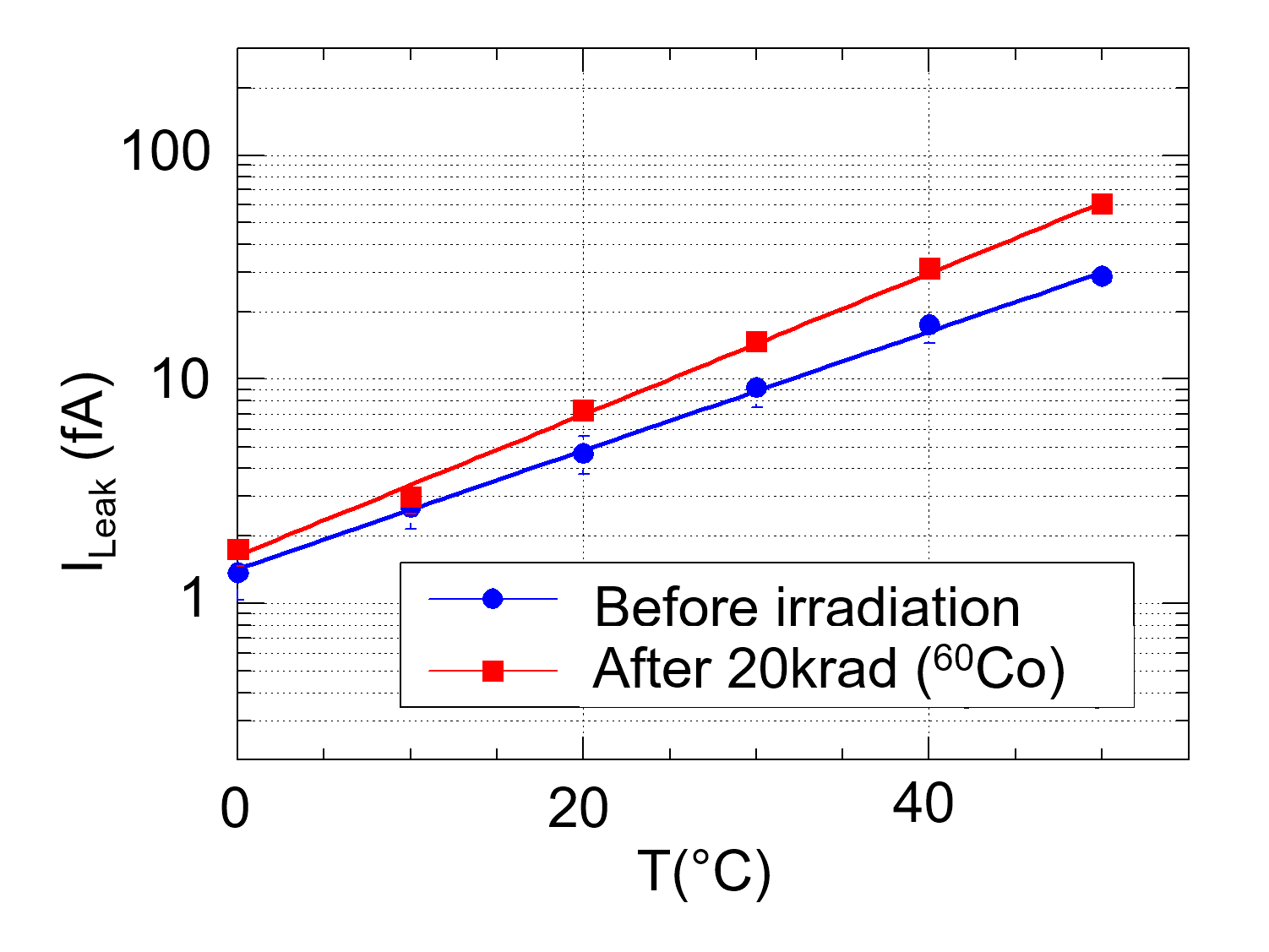}
\end{center}
  \caption[Leakage current of the radiation soft and the radiation hard 3T-pixel of MIMOSA-9 before and after an irradiation with \mbox{$\sim 20 \ kRad$}.]
	 {Leakage current on the radiation soft (left) and the radiation hard (right) 3T-pixel of MIMOSA-9 before and after an irradiation with \mbox{$\sim 20 \ \rm kRad$} $\gamma$ - rays from a $^{60}$Co - source. One observes the increase of the leakage current to be substantially reduced in the radiation hard design. From \cite{MichaelPhD}. }
 \label{Mi9RadSoftAndHard}
 \end{figure}	

A plausible source of the additional leakage current is generated by the surface defect states
located at the interfaces between Si and thick SiO$_2$ located nearby the diode. This is 
illustrated at the right side of Fig. \ref{fig:RadSoft3TPixel}), which displays a simplified
design of a related charge collection diode and indicates the location of the N-well implantation and of the 
thick oxide in a qualitative way. One observes that the radiation induced surface defects 
found at the interface between the thick SiO$_2$ and the Si generate thermal currents,
which may be collected by the n-well implantation. This would be further eased in case
the positive charge build-up related to the defects deforms the bands and creates thus a
conduction channel along the surface toward the diode.

Design efforts to reduce the leakage current of the collection diodes aimed to remove thick SiO$_2$ structures like the STI from the vicinity of the diode \cite{SB-PixelPaper}. A successful design is illustrated in the right part of Fig. \ref{fig:RadHard3TPixel}.
It consists in surrounding the n-well of the diode by a
circular gate, which is terminated in a p-diff guard ring.  By doing so, the thick silicon located nearby the diode is replaced by the thin, radiation hard gate oxide. The doping gradient
generated with the p-diff guard ring generates a built in voltage, which repulses free electrons from the diode. Alternatively, a n-diff guard ring may be used, which absorbs the currents. 
In the most successful attempt to realize this structure, the gate voltage 
was set to the internal GND of the sensor. Applying other voltages by means of external voltage sources was tested but found to inject first of all an unacceptable pick-up noise. 
Note that the implementation of the above mentioned strategy with a standard CMOS process may require to bend design rules,
which caused issues in some cases. A comparison between the measured leakage currents of a 3T-pixel with standard and radiation hardened diode is shown in Fig. \ref{Mi9RadSoftAndHard}. In this case, the guard ring was realized as n-diff. One observes that the radiation induced increase of the leakage current is reduced by factors.

Besides the diode, the reset transistor of a 3T-pixel collects (bulk-to-source) leakage currents. This is as the source of the reset transistor and the surrounding 
p-doped bulk form another pn-junction, which may be larger in size than the diode. This effect is illustrated at the left side of Fig. \ref{fig:RadSoft3TPixel}, which shows
the cross-section of a non optimized enclosed transistor: The outer ring of the transistor is placed in direct contact to thick SiO$_2$ structures and 
collects leakage currents, which are generated by the surface defects located there. Those leakage currents are guided to the pixel node and add 
to the ones of the diode. Fig. \ref{fig:RadHard3TPixel} illustrates an improved design. Here, the pixel node is connected to the protected inner ring
of the transistor. Indeed, a pixel using a radiation protected diode
and an optimized reset transistor layout was found to have a leakage current $31.6\pm 0.2~ \rm fA$ after $1~\rm Mrad$ and at $T_{\rm Coolant}=10\C$ while an otherwise identical
pixel with non-optimized reset transistor with showed a leakage current of $286.9\pm 1.7 ~\rm fA$ under the same conditions (see \cite{MichaelPhD}, P. 148). This suggests that the leakage current
of a non-optimized transistor may dominate the one of a charge collection diode by far. 

\paragraph{Noise}
The radiation induced increase of the leakage current $I_l$ may increase the shot noise of the
pixel.  This noise is caused by the fact that the current is transported by discrete charge carriers.
The charge $Q_L$ collected during one integration time $t_{int}$ of the pixel is composed from
$N$ individual electrons. Therefore:
\begin{equation}
Q_L= N \cdot e=\int^{t_{int}}_{0} I_l ~dt
\end{equation}
According to  Poisson statistics, the uncertainty of N amounts $\Delta N =\sqrt{N}$.
The electron equivalent noise $Q_L$ created by this current can be considered as the
fluctuation of the collected charge. It amounts:
\begin{equation}
\Delta Q_L = e \cdot \sqrt{N} =
% e \sqrt{\int^{t_{int}}_{0} \frac{I_l}{e} ~dt} = 
\sqrt{ e \cdot I_l \cdot t_{int}}
\label{3TShotNoiseEquation}
\end{equation}
%In more general terms, the shot noise may be expressed as fluctuation $\Delta I$ of the current $I$ passing 
%an electrical component as:
%\begin{equation}
%\Delta I = \sqrt{2 e \cdot |I| \cdot f_{max} - f_{min}}
%\label{Eqn:ShotNoiseGeneral}
%\end{equation}
%Here $f_{max} - f_{min}$ expresses the bandwidth of the measurement. 
The shot noise adds quadratically to the initially dominating 
noise $\Delta Q_A\approx 10~ e$ of the pre-amplifier:
\begin{equation}
\Delta Q\approx \sqrt{ Q_A^2 +  e \cdot I_l \cdot t_{int}}
\label{EqShotNoise3T}
\end{equation}
It is suited to push the total noise of the pixel to an unacceptable value, which restricts the functional radiation tolerance
of the devices. As $\Delta Q_L$ depends on the leakage current and on the integration time, the radiation hardness limit
of a device may be extended by cooling and accelerating the readout. Despite both strategies are restricted by 
practical constraints, the radiation tolerance of a given sensor thus may vary by more than one order of magnitude depending
on the operation conditions. 

\subsubsection{SB-pixel}
\label{ChapterIonizingRadDamageSB}
\begin{figure}[t]
\begin{minipage}{0.6\textwidth}
 \includegraphics[width=\columnwidth]{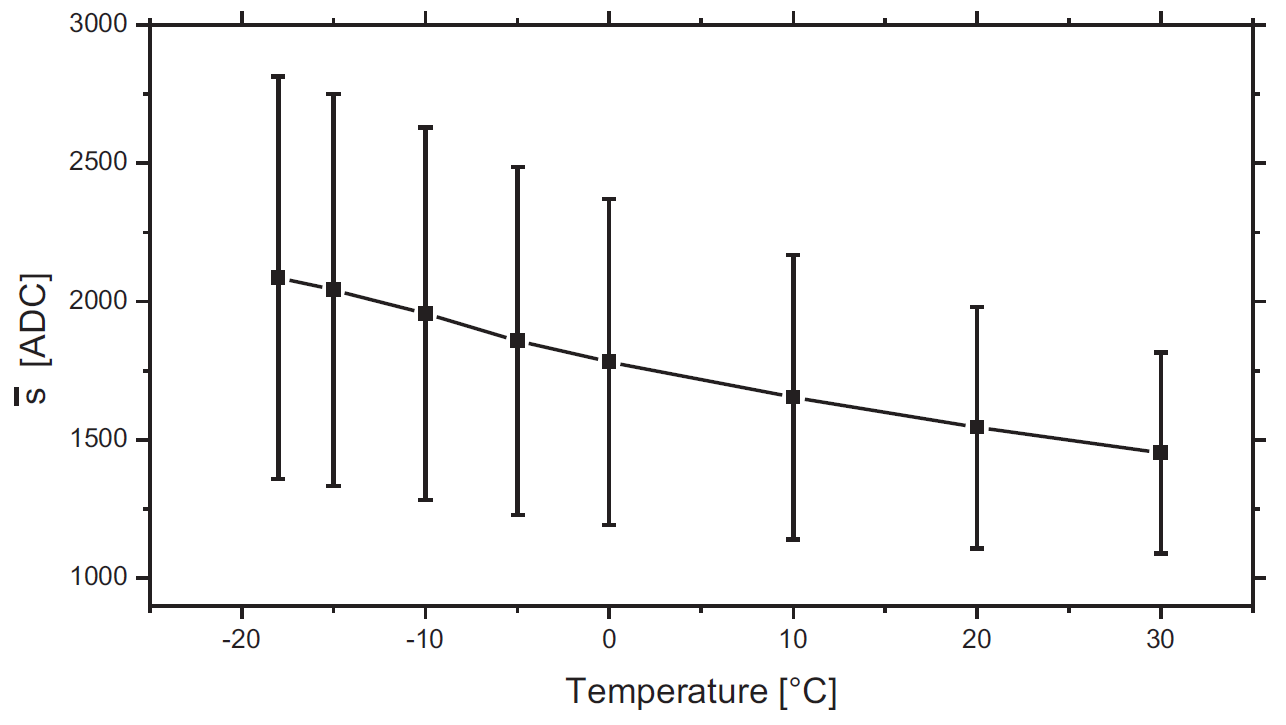}
\end{minipage}
\hspace{0.02\textwidth}
\begin{minipage}{0.36\textwidth}
 \caption{Mean value and RMS (error bars) of the dark signal distribution of SB-pixels on a non-irradiated CPS based on a $0.35\mum$ process. 
 $I_L$ increases with $T$. The shift of the working point and the reduction of the RMS is visible. Saturated pixels were removed. 1 ADC unit $\approx 4~ e$. From \cite{PhD-Samir}.}
    \label{fig:SB-WorkingPoint}
\end{minipage}
\end{figure}

\paragraph{Leakage current}
In first order, SB-pixels react rather friendly to radiation induced increases of the leakage current $I_L$. The current is compensated by the biasing diode, which cancels it from the output signal of the pixel after applying CDS and hampers the pixel from saturating. It is thus possible to discriminate the CDS processed signal of multiple, irradiated SB-pixels with a common threshold.
However, the variations in $I_L$ create a number of rather relevant, indirect effects, which should be considered. 

The best way to approach the indirect leakage current effects of a SB-pixel is to consider it as a combination of the forward biased diode and a current source, which is realized by the reversed biased collection diode. Both components form a voltage divider\footnote{See schematics in Fig. \ref{fig:SB-readoutSchema}.}. The working point is set by the voltage drop $U_B$ caused by $I_L$ in the biasing diode. As the $I_L$ of cooled and non-irradiated diodes amounts the order individual electrons per readout cycle, $U_B$ tends to be smaller than the $\sim 0.7~\rm V$, which are usually considered for macroscopic currents passing a forward biased diode. This working point moves once $I_L$ increases due to increasing temperature or irradiation (see Fig. \ref{fig:SB-WorkingPoint}), which should be considered in the design of the amplification chain. Moreover, it is subject to important pixel-to-pixel fluctuations in $I_L$. In contrast to the situation in the 3T-pixel, this variation tends rather to decrease with increasing $I_L$. The origin of this decrease is unknown. Potentially, the electrical voltage drop caused by tiny ($< 1 \rm fA$) currents at the forward biased diode is influenced by badly controlled parasitic effects (e.g. individual crystal defects), which become negligible at higher currents.

Few data on the leakage current of SB-pixels is available as this quantity is hard to measure. As similar diodes are used, their leakage currents should match the numbers known from 3T-pixels and the related radiation hardening strategies apply. Quite obviously, no leakage current from a reset transistor is expected in SB-pixels. 

\paragraph{Signal clearing}

\begin{figure}[tp]
 \centering
\begin{subfigure}{.45\textwidth}
%\vspace{0.2cm}
  \centering
   \includegraphics[viewport=3cm 1.2cm 26cm 20cm, clip, width=1\columnwidth]{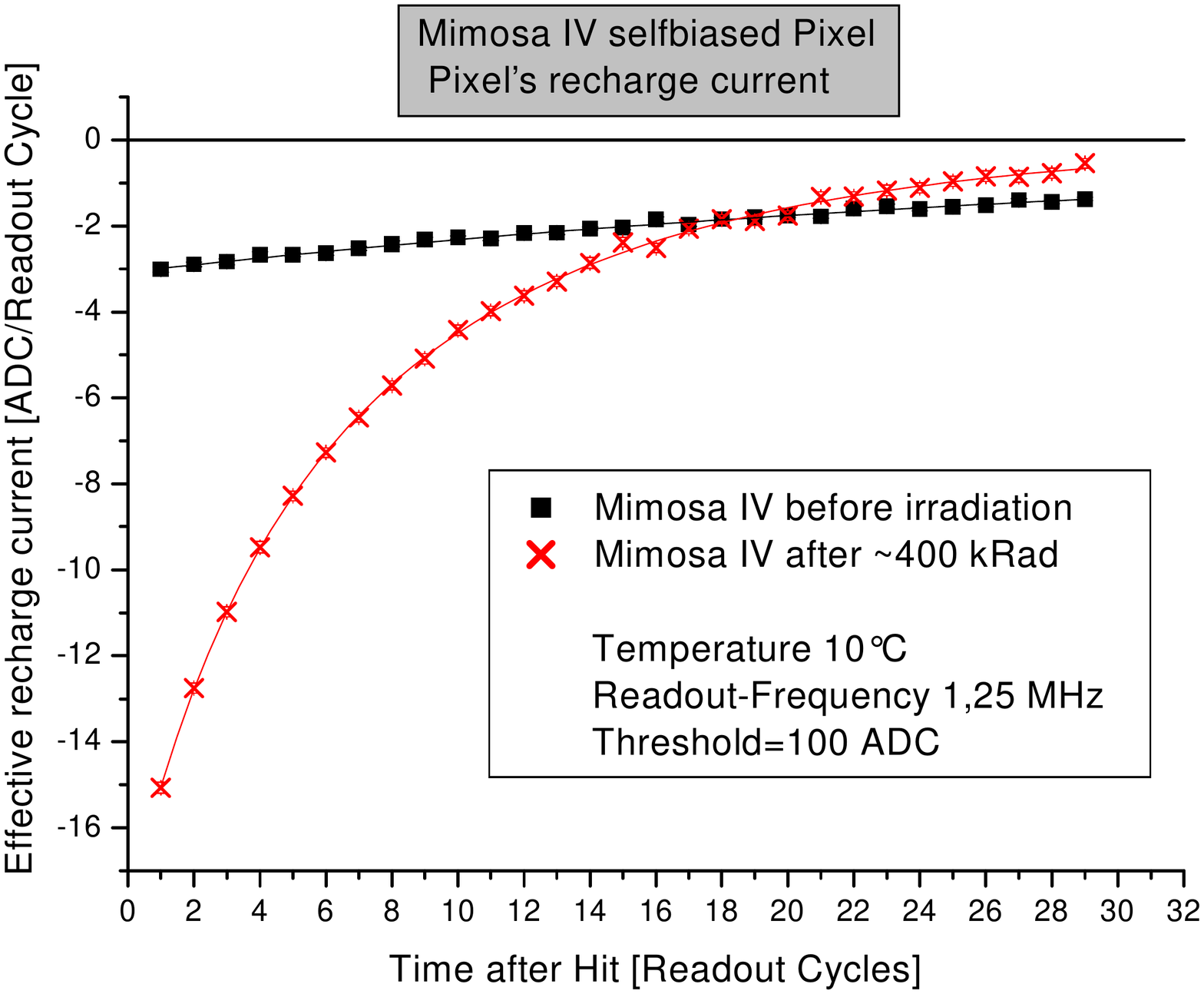}
			
		\caption{Measured recharge current of a non-irradiated and irradiated SB-pixel as
		function of the radiation dose.}
    \label{fig:SB-PixelRechargeCurve}
\end{subfigure}
\hspace{0.05\textwidth}
\begin{subfigure}{.45\textwidth}
  \centering
 \includegraphics[viewport=1.5cm 1.5cm 23cm 18cm, clip, width=1\columnwidth]{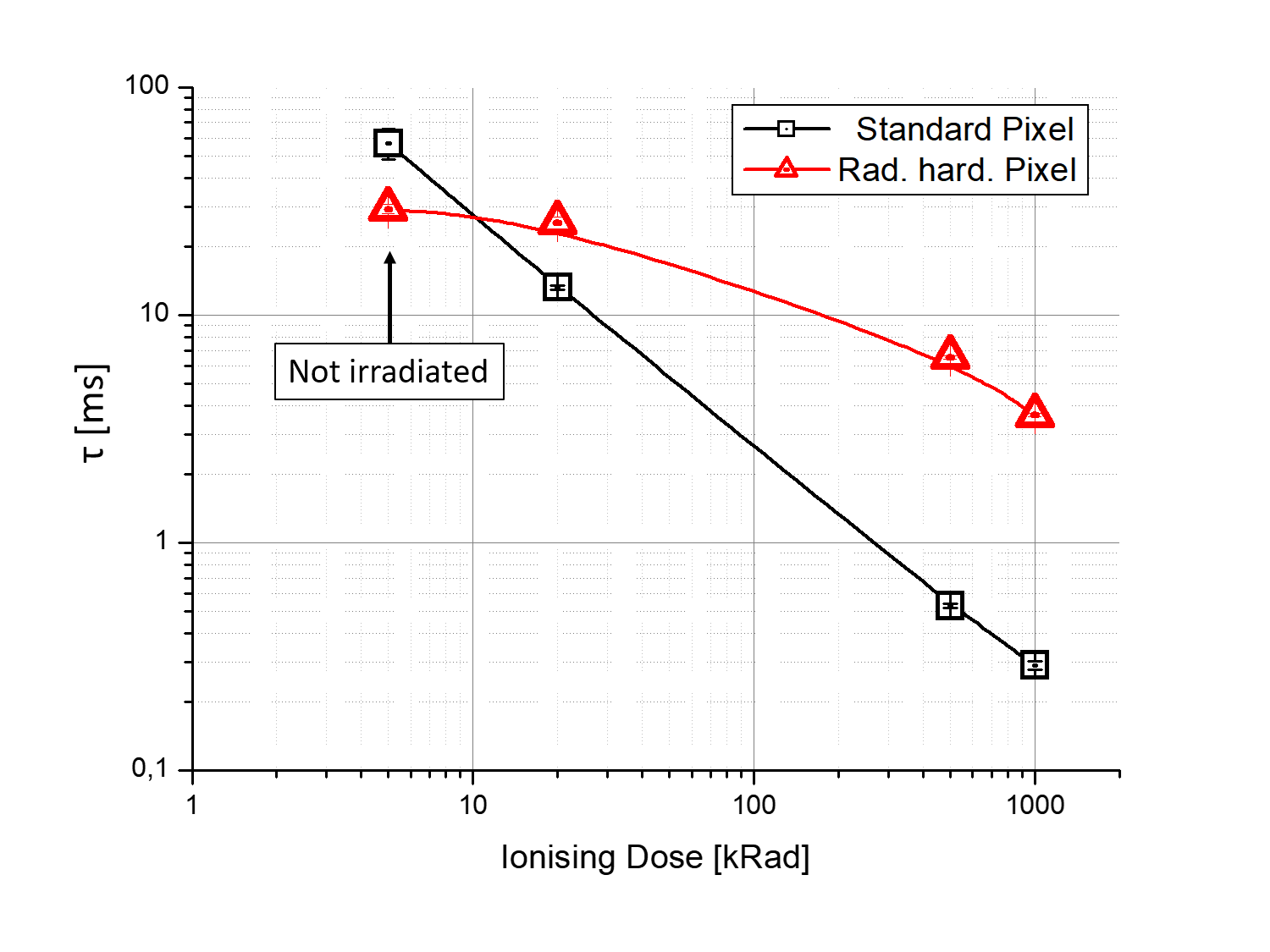}
    \caption{Clearing time constant $\tau$ as a function of the radiation dose at $T=10\rm ^{\circ}C$ for a standard SB-pixel
		on a LR epitaxial layer.}
    \label{fig:Mi11IonizingTau}
\end{subfigure}

%https://www.overleaf.com/6458423ktgndt#%1
%
\begin{subfigure}{.45\textwidth}
\vspace{1cm}
  \centering
	   \includegraphics[viewport=1.5cm 1.2cm 24cm 18cm, clip, width=1\columnwidth]{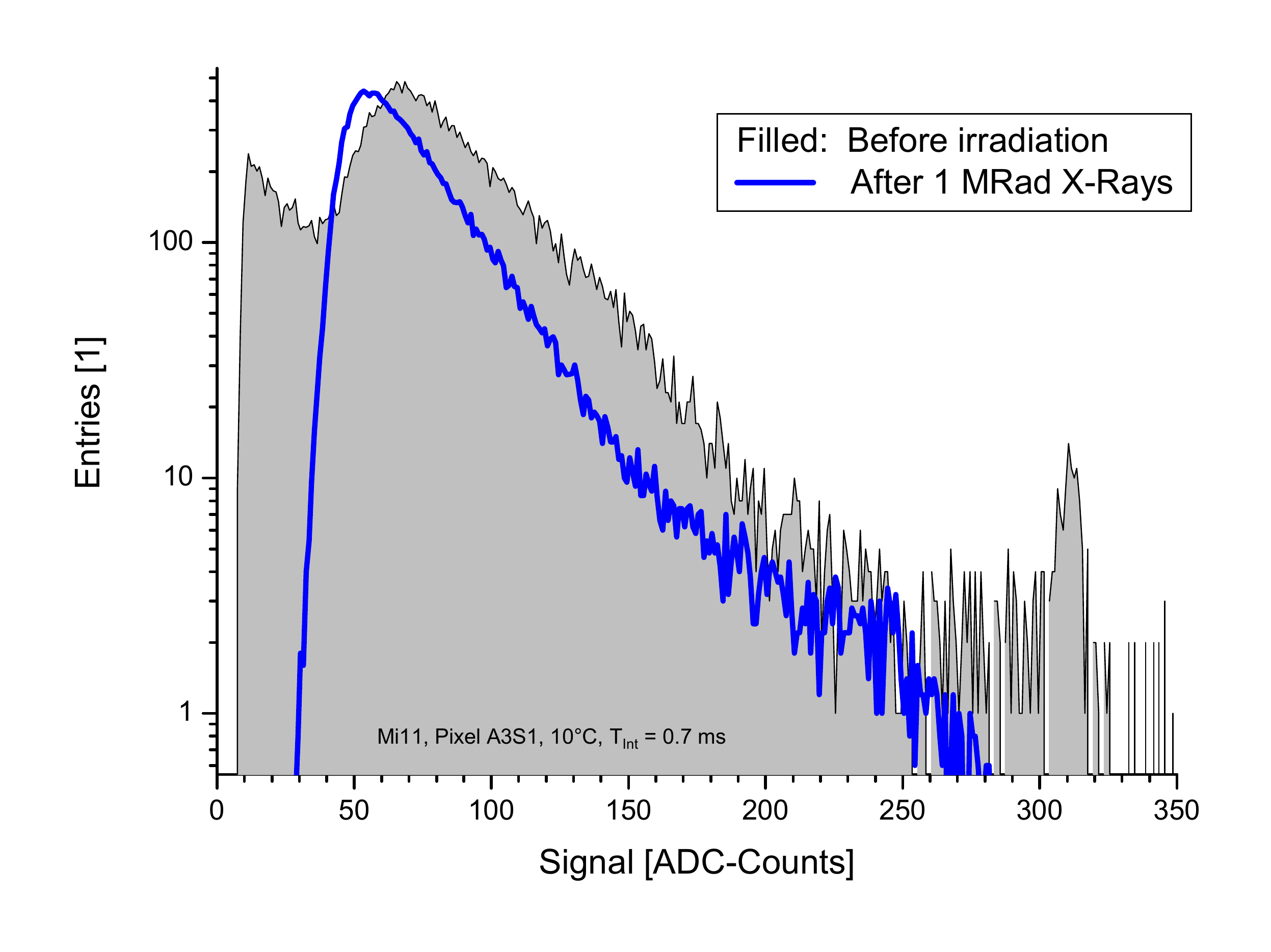}
    \caption{Seed spectrum ($^{55}$Fe) of a SB-pixel before and after irradiation with X-rays. A mild accelerated clearing smears the calibration peak and reduces the signal amplitudes.}
    \label{fig:Fe55SeedSpectrum-SBPixel2}
\end{subfigure}%
\hspace{0.05\textwidth}
\begin{subfigure}{.45\textwidth}
\vspace{0.2cm}
  \centering
\includegraphics[viewport=2.5cm 1.0cm 26.8cm 20cm, clip,width=1\columnwidth]{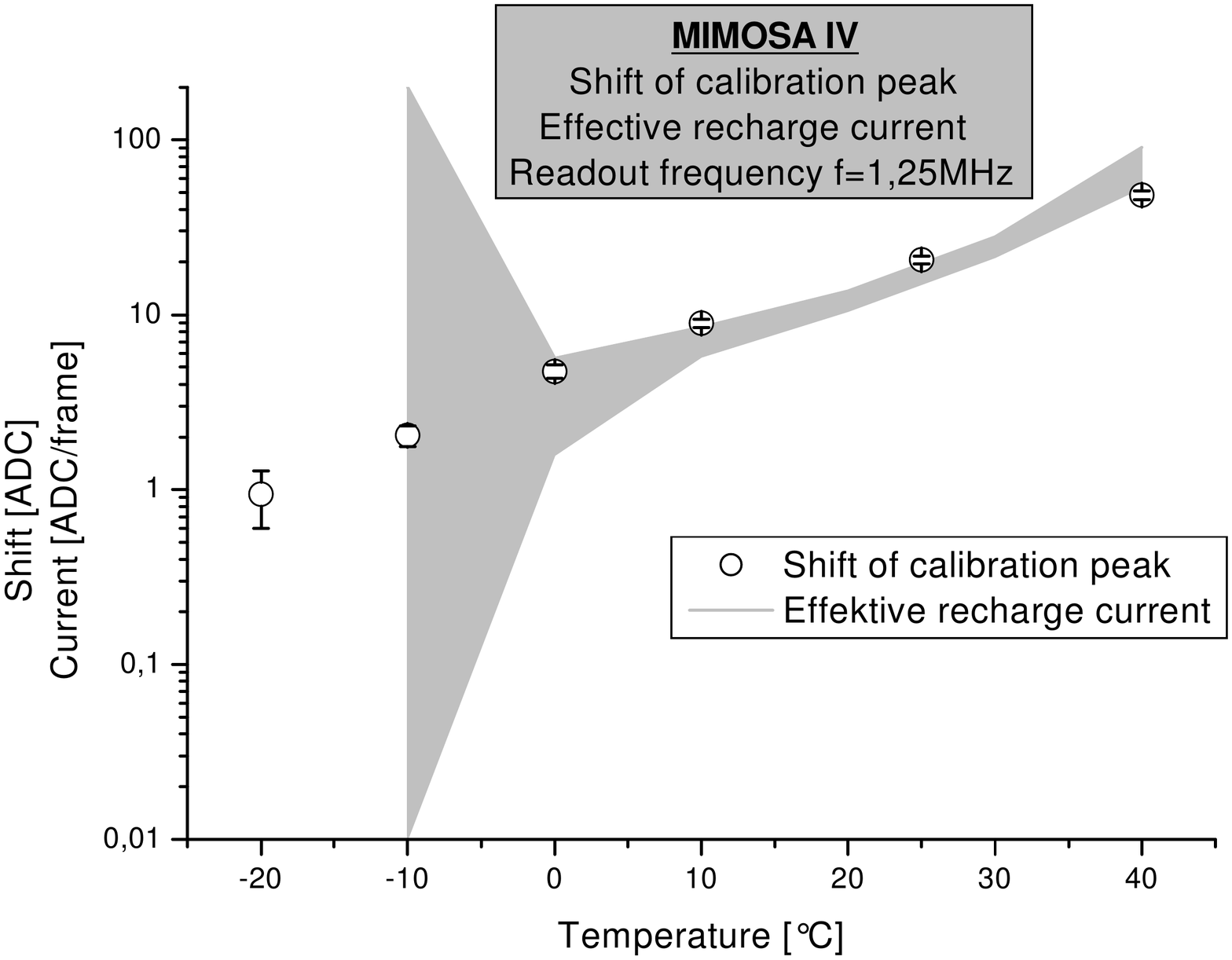}			
		\caption{Correlation between the displacement of the calibration
		peak and the amplitude of the recharge current. See text. From \cite{MichaelPhD}.}
    \label{fig:SB-pixel-Clearing-PeakShift}
\end{subfigure}
%		\newline
	
\caption{Illustration of the major effects caused by accelerated signal clearing. The plots are based on measurements taken with different sensors, more details are found in \cite{MichaelPhD,SB-PixelPaper}. See text.}
\end{figure}

An radiation induced increase of the leakage current of the collection diode $I_L$ may accelerate the signal clearing in SB-pixels. This clearing of the signal charge $Q_S$ follows in good approximation \cite{SB-PixelPaper}:
\begin{equation}
Q_S(t)=  Q_S(t=0) \cdot \exp \left(- \frac{t} { \tau} \right)
\end{equation}
In this equation, $t$ is the time after the signal generation and $\tau$ the time constant of the clearing process, which is given with:
\begin{equation}
\tau \approx \frac{n \cdot k_B \cdot T \cdot C}{ e} \cdot \frac{1}{I_L+I_S}
\label{Eqn-SB-Tau}
\end{equation}
Here, $I_S$ is the saturation current of the biasing diode, $I_L$ the leakage current of the collection diode, $1<n<2$ represents the emissivity of the biasing diode, $e$ the elementary charge, $k_B$ the Boltzmann constant and $T$ the temperature. For ideal diodes with the size and doping of a typical biasing diode, $I_S$ is roughly in the order of $1 ~\rm fA$, which can be neglected with respect to the leakage current of irradiated pixels.

%\cite{Paper_Deptuch_DevelopmentOfMAPS, SB-PixelPaper}

The clearing process of the pixel can be measured with the procedure discussed in Sec. \ref{SectionClearingConstant}. A result of this measurement is shown in Fig. \ref{fig:SB-PixelRechargeCurve}, which shows the related curve for a SB-pixel before and after irradiation. As expected, the higher leakage current of the irradiated collection diode turns into an increase of the amplitude of the recharge current and a shortening of the clearing constant $\tau$. The evolution of the clearing time with increasing radiation dose is illustrated in  Fig.~\ref{fig:Mi11IonizingTau}. One observes the clearing time $\tau$ of the non-irradiated sensor to amount several $10~\rm ms$ before irradiation. The SB-pixel with radiation hardened diode shows a slight disadvantage with respect to the standard design prior to irradiation. After irradiation, the time constant is accelerated by orders of magnitude and the radiation hardened pixel shows better performance already after small doses. This is as the increase of leakage current is suppressed in this pixel. 

Once $\tau$ approaches the integration time $t_{int}$ of the pixel, a part of the signal is cleared prior to the readout.
This is illustrated in Fig. \ref{fig:Fe55SeedSpectrum-SBPixel2}, which shows the charge collection spectrum of an affected pixel. One observes a shifting of small calibration peak toward smaller values\footnote{See Sec. \ref{SectionFe55Spectrum} on how to read $^{55}$Fe charge collection spectra.}. This reflects the clearing of a fraction of the signal between the time of the particle hit and the time of readout.
As the time of particle impact is random, this time span avaliable for signal clearing follows a flat distribution, which ranges from zero to $t_{int}$. This creates the smearing of the peak.

The modification of the spectrum due to signal clearing is sometimes confused with a loss of signal amplitude due to reduced CCE. However, both are separate processes.
This is illustrated in Fig. \ref{fig:SB-pixel-Clearing-PeakShift}, which shows an example of the correlation between the observed displacement of the calibration peak (points) and the effective recharge current
generating the clearing. Both quantities were recorded as a function of the temperature in order to modify the leakage current and thus $\tau$. The effective recharge current
is displayed as a grey shaded area. The width of this area which reflects the uncertainties of the related measurement. One observes that the amplitude of the clearing current is reduced with decreasing temperature. This is as $\tau$ becomes longer while the integral of the exponential, which relates to the charge to be cleared, remains constant. The observed and expected clearing are in good accordance, 
which supports the above mentioned understanding. Doing this test allows to separate signal clearing from reduced CCE, mostly as effects reducing the CCE are typically not strongly temperature dependent. 
Note that besides cooling, reducing the integration time forms an obvious mean to reduce unwanted signal clearing.

\paragraph{Noise}

\begin{figure}[t]
\begin{minipage}{0.6\textwidth}
 \includegraphics[viewport=1.5cm 1.8cm 23cm 17cm, clip,width=\columnwidth]{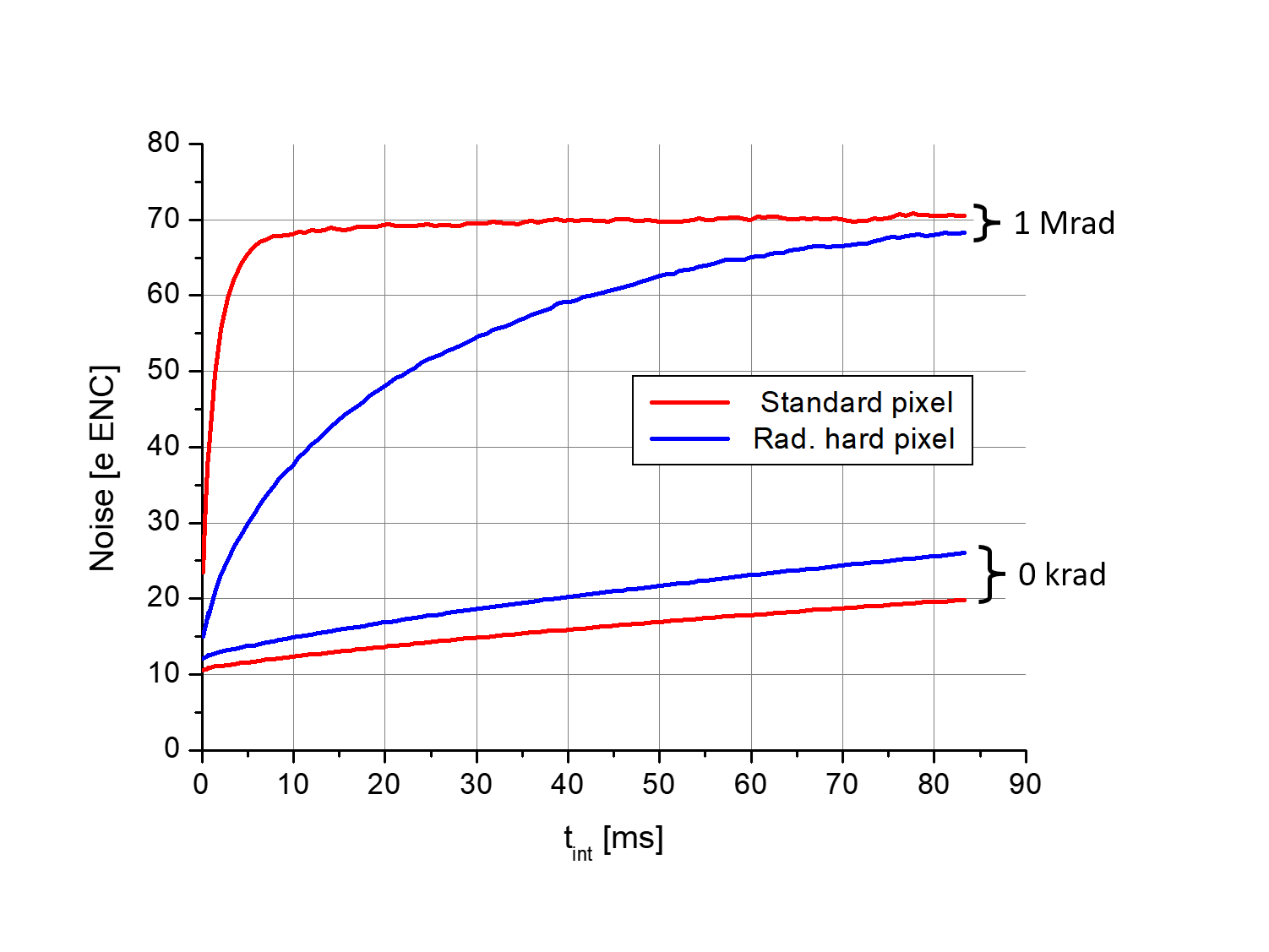}
\end{minipage}
\hspace{0.02\textwidth}
\begin{minipage}{0.36\textwidth}
 \caption{Noise of a standard and a radiation hardened SB-pixel as function of the radiation dose and the integration time. The data 
was taken at $T=\rm 10^{\circ}C$. After \cite{SB-PixelPaper}.}
    \label{fig:SB-PixelNoise}
\end{minipage}
\end{figure}

As for the 3T-pixel, the radiation induced increase of the leakage current $I_L$ generates shot noise in SB-pixels. Due to the continuous bias, the shot noise is 
not described with the model used for the 3T pixel (Eq. \ref{EqShotNoise3T}).  However, one may still assume that the leakage current $I_L$ removes a charge
$Q_{out}=I_L\cdot t_{int}$ out of the pixel capacitance during the integration time $t_{int}$. At the same time, this charge is replaced by the biasing diode, which 
injects $Q_{in}=I_L\cdot t_{int}$. Obviously, in thermal equilibrium, the total charge injected to the capacitor $Q_L$ is given with:
\begin{equation}
Q_L=Q_{in} - Q_{out}=0
\end{equation}
While there is no net charge transfer, the noise $\Delta Q$ can be estimated by means of Gaussian error propagation with:
\begin{equation}
\Delta Q_L= \sqrt{(\Delta Q_{in})^2 + (\Delta Q_{out})^2} = \sqrt{2 \cdot \Delta Q_{in/out}} 
\end{equation}
By using the relation between shot noise and leakage current from Eq. \ref{3TShotNoiseEquation}, one obtains:
\begin{equation}
\Delta Q_L=  \sqrt{2\, e \cdot I_l \cdot t_{int}} 
\label{Eqn-SB-ShotNoise}
\end{equation}
Consequently, one expects SB-pixels to show a by $\sqrt{2}$ higher shot noise then related 3T-pixels.

Unlike to 3T-pixels, the shot noise of SB-pixels does not increase toward infinity with increasing integration time.
This is as the signal clearing of the pixel acts also for the small ``signals'' of the noise fluctuations. In a rough
picture, one can therefore distinguish two regimes. In case $t_{int} \ll \tau$, Eq. \ref{Eqn-SB-ShotNoise} may be 
applied. In case $t_{int} \gg \tau$, the integration time of the pixel is effectively limited by the recharge process.
In this case, Eq. \ref{Eqn-SB-ShotNoise} becomes:
\begin{equation}
\Delta Q_L=  \sqrt{2\, e \cdot I_l \cdot \tau} 
\end{equation}
This equation predicts a saturation of the noise for very high leakage currents (determining a very fast $\tau$) and
long integration times. By using Eq. \ref{Eqn-SB-Tau} and assuming $I_S \ll I_L$, one may estimate the saturation value:
\begin{equation}
\Delta Q_{L} =  \sqrt{ 2\, n \cdot k_B \cdot T \cdot C}
\label{Eqn-SB-NoiseSaturation}
\end{equation}
 Unfortunately, the saturation of the pixel noise remains of few practical interest as it 
occurs by construction only if the clearing time becomes too short for effective charged particle detection.

As shown in Fig. \ref{fig:SB-PixelNoise}, this saturation effect is indeed observed. The figure shows the noise 
of a SB-pixel with standard diode and one with hardened diode as a function of the radiation dose and the integration time. The related measured values of 
$\tau$ can be extracted from Fig. \ref{fig:Mi11IonizingTau}. As suggested by Eq. \ref{Eqn-SB-NoiseSaturation}, the noise saturates at a level of $\sim 70~e ~ \rm ENC$, which suggests a plausible pixel capacitance of $\sim 10 \rm fF$. The full saturation is only reached after $t_{int}\approx 10 \tau$. The same observation is made for the radiation hardened diode, 
which saturates\footnote{The saturation of this pixel is more rigorously shown by additional measurements presented in \mbox{\cite{MichaelPhD}}, which also vary the temperature.} at the highest integration times shown thanks to its lower $I_L$ and longer $\tau$. The non-irradiated pixels do not reach saturation.

As expected, the previously described approaches for reducing the leakage current of the diodes of 3T-pixels also reduce the noise of the SB-pixel. In
the region of interest with less than few $\rm ms$ integration time, the difference is rather sizeable. This holds despite the fact that the radiation hardened pixels show a slightly higher noise before irradiation.

\subsubsection{Clamping pixel}

A number of studies were carried out in order to understand the tolerance of clamping pixels to ionizing radiation. A comparative study on the radiation tolerance of different sensing elements shows that a combination of the clamping structure with a 3T-sensing element turns into an unacceptable fixed pattern noise already after moderate radiation doses of $300~\rm krad$ \cite{PaperMichalPhd}. This may be as the clamping structure is not suited to compensate the leakage currents and, more importantly, the pixel-to-pixel fluctuations of this current. The increase of the fluctuations with increasing radiation dose turns into an increased fixed pattern noise. This feature was not observed for a pixel design relying on a SB-structure, which resisted reasonably well to doses of up to $300~\rm krad$. 

Radiation studies beyond this dose revealed that the clamping pixel shows a shift of the threshold voltages and once more a increase of fixed pattern noise \cite{PaperPhD-Dennis} in a pixel realized in a $0.35 \mum$ CMOS process. This effect could be reduced by means of thermal annealing. A potential origin of this effect is a radiation induced bulk to source leakage current of the clamping transistor, which discharges the clamping node. This effect was not observed in clamping pixels relying on a $0.18 \mum$ production process, which showed satisfactory performances up to $1.6~\rm Mrad$, which was the highest radiation dose applied
\cite{AugustFSBB}.

\subsection{Effect of ionizing radiation to the charge collection}

As ionizing radiation does in first order not create bulk damage, which would reduce the lifetime of signal electrons, one does not expect this radiation to deteriorate the charge collection efficiency of a sensor. However, an impact on the charge collection efficiency was observed in a specific case \cite{Su1-Paper} with a sensor manufactured in a $0.35~\rm \upmu m$ CMOS process with an only $\sim 4~\upmu \rm m$ thick epitaxial layer. The 3T-pixels had a pitch of  $\sim 20~\upmu \rm m$ and all transistors of the pixels were designed as enclosed transistors. Due to this feature and the use of an at least moderately deep sub-micron process, the sensor was hoped to be particularly radiation tolerant. Instead, a about 50\% drop of the CCE was observed after an irradiation of $400~\rm krad$. This effect was associated to a depletion of the p-well layer, which was intended to isolate the enclosed reset transistor from the sensing element. According to this interpretation, the depletion manifested itself once the fields of the \mbox{$3.3~\rm V$} potential of the outer ring of the transistor were complemented by the fields caused by radiation induced positive oxide charge. Due to this depletion, the transistor acted as parasitic collection diode. The issue was solved by inverting the layout of the transistor such that outer ring of the transistor was put on a lower potential. Other sensors with similar layout, which were produced in a different CMOS process, did not show this feature for doses of up to $1~\rm Mrad$  \cite{MichaelPhD}. The effect is therefore considered as specific to a, today obsolete, production process. However, it should be remembered as a warning, that ionizing doses may cause unexpected effects, which sometimes depend on undocumented features and modifications of the CMOS-processes used. 

\subsection{Results from beam test campaigns}

A detailed study comparing the performances of SB-pixels with standard and radiation hardened design in beam  was made with the sensor MIMOSTAR-2, which featured two matrices with each $128\times 64$ pixels with $30\times 30 \mum^2$ pitch \cite{ShabetaiPhD}.
The readout was done in serial analogue mode and the integration time was varied between $4.1$ and $0.8~\rm ms$ by changing the readout clock frequency. The test was part of the R\&D for the STAR PiXeL detector, which was believed to be exposed to a dose of up to $50~\krad$ at this time. Operating the sensors after this dose was complicated by the long integration times and the need for a high, typically $+30\C$ operation temperature, which created a relatively high leakage current and the related shot noise and signal clearing issues. It was observed that both, the standard and radiation hardened pixels reached a detection efficiency of >99\% if being operated at this unfavourable temperature and at a integration time of $4.1 ~\rm ms$. Once being irradiated to $20~\rm krad$, the standard sensors showed only a poor detection efficiency of <90\%. This performance drop was first of all created by an increase of the shot noise, which increased from initially $10~e~\rm ENC$ to slightly below $35~e~\rm ENC$.
The pixels with radiation hardened diode remained operational up to the highest dose applied. As expected from the above discussions on the properties of the SB-pixel, best performances were reached at short integration times and low temperatures as both steps reduce the shot noise and the pixel clearing. A better than 99\% detection efficiency was observed for the radiation hard pixel after applying a dose of $50~\rm krad$ at a temperature of $30\C$ and a integration time of $2~\rm ms$. For a temperature of $40\C$, this efficiency dropped to about 96.6\%. Despite this was not analysed explicitly, $^{55}$Fe spectra recorded for the same chip in \cite{MichalSzelezniakPhD} suggest that the primary failure mechanism of the standard pixel was a too fast signal clearing.
This effect was substantially alleviated by radiation hard design.

Results of a beam test with a MIMOSA-26 (see Sec. \ref{Sec:Performances} for technical data) with high resistivity epitaxial layer, which was irradiated with $150~\rm krad$ X-rays is reported in \cite{Mimosa26Ionizing}. The sensor was operated 
intentionally under unfavourable conditions ($T=+35\C$ and an integration time of $230\mus$ instead of the nominal $115 \mus$). This was done once more in the prospective of operating the sensor at the STAR PiXeL detector. The longer integration 
time was chosen to emulate the properties of the final sensor (MIMOSA-28/ULTIMATE), which was not yet available at the time. A detection efficiency for $120~\rm GeV/c$ pions of $99.8\%$ in combination with a dark rate of $10^{-5}$ was reported for the sensor flavour hosting a $15\mum$ thick $400~\Omega \cdot \rm cm$ high resistivity wafer. This exceeds the performances of the non-irradiated MIMOSA-26 relying on a low resistivity epitaxial layer.

The FSBB-M0 prototype (see Sec. \ref{Sec:Performances} for technical data), which integrated clamping pixels into the Tower $0.18 \mum$ process resisted to doses of up to $1.6 ~\rm Mrad$ in combination with a non-ionizing dose of $10^{13}\neqcm$ with 99\% detection efficiency and a dark hit rate of 
$10^{-5}$ according to a beam test \cite{AugustFSBB}.

\section{Effects of integrated non-ionizing radiation}
\label{SectionNonIonizingRadiation}

\subsection{Sensors with  $10~\Omega \cdot \rm cm$  epitaxial layer}

The first generation CPS relied on non-optimized CMOS processes, which provided a sensitive volume of about $10~\Omega \cdot \rm cm$ 
resistivity and a nominal bias voltage of 3 or $5~\rm V$. Most sensors were equipped with 3T-pixels.
The depletion depth of the related charge collection diodes is $\lesssim 1 \mum$ \cite{PaperMi25}.
Most of the signal charge is collected slowly by thermal diffusion. In this early phase, a number of different combinations of pixel pitches
and epitaxial layer thicknesses were explored. The typical values for the pixel pitch amounted $10 - 40\mum$, the thickness of the 
epitaxial layer started from $4\mum$. 

\subsubsection{Charge collection efficiency, gain and noise}
\label{Sec:LRCCE}

\begin{figure}[t]
  \centering
	%viewport=1.0cm 0cm 24cm 18cm, clip,
   \includegraphics[width=0.75\columnwidth]{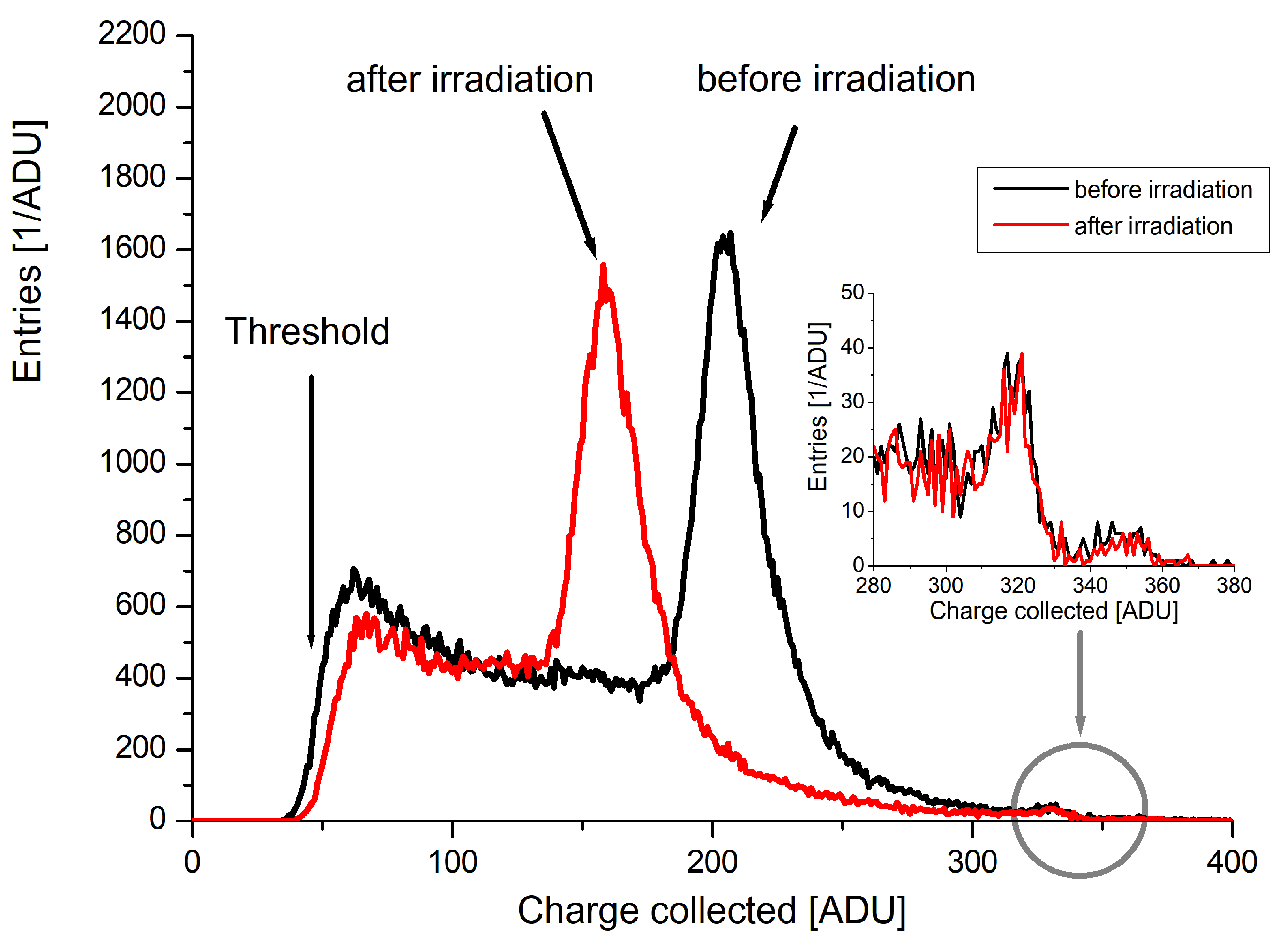}
    \caption{Illustrative picture of the impact of bulk damage on CPS with $10~\rm \Omega \cdot cm$ epitaxial layer.
		The $^{55}$Fe amplitude spectrum (sum signal of four pixels) of a virgin and a sensor irradiated with $3\times 10^{12} \neqcm$ neutrons is shown. The zoom shows the single pixel
		amplitude spectrum in the region of the calibration peak. From \cite{PaperNeutronRadMi2}. See text.}
    \label{fig:Fe55FourSpectrum1}
\end{figure}%https://www.overleaf.com/6458423ktgndt#

\begin{figure}[t]
\begin{center}
 \includegraphics[viewport=0cm 0cm 25.5cm 20cm, clip,width=0.8\columnwidth]{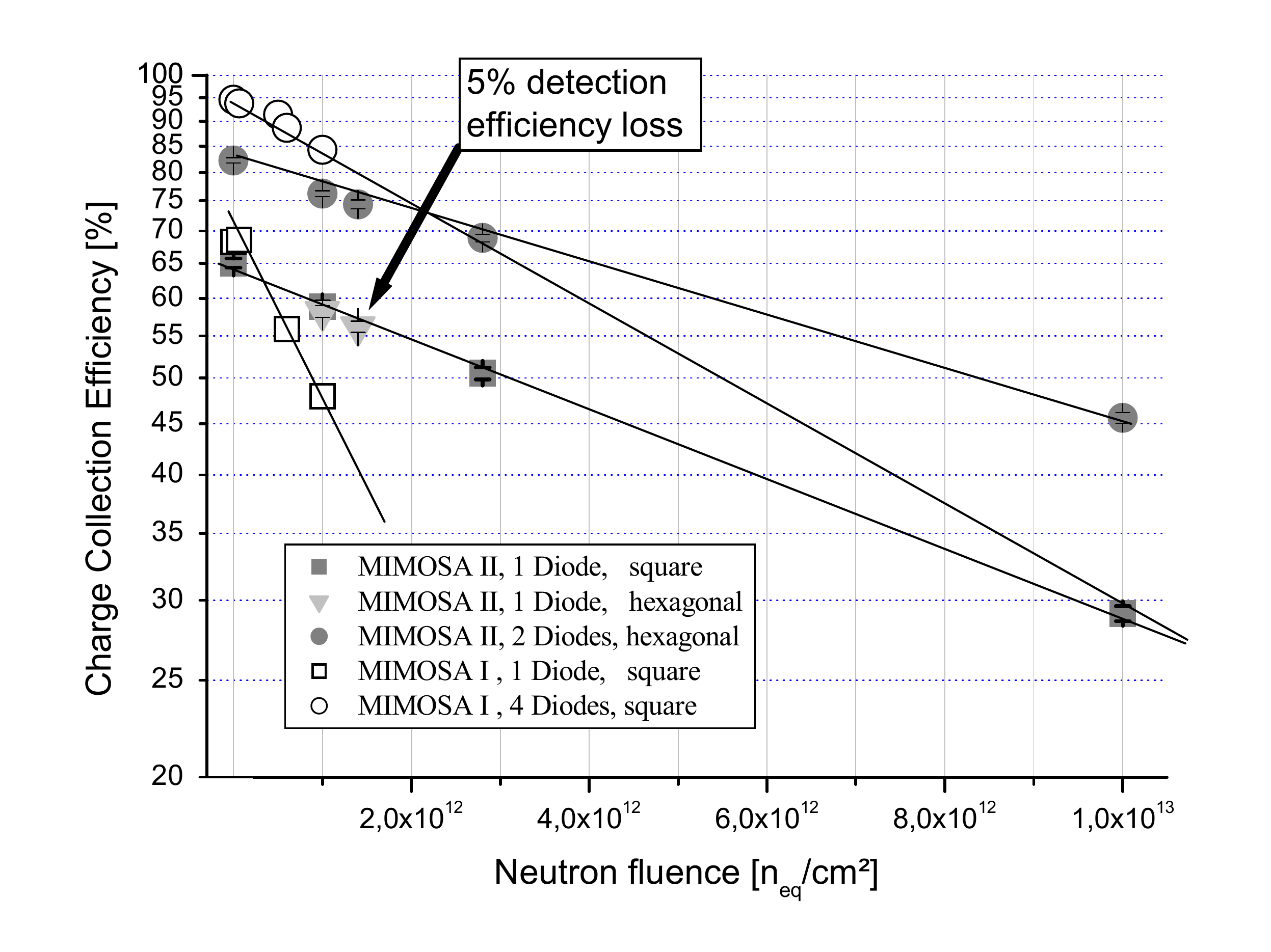}
\end{center}
 \caption{Charge collection efficiency for groups of 4 pixels as a function of the radiation dose. 
The data was taken with an $^{55}$Fe-source with MIMOSA-1 ($14 \mum$ epitaxial layer) and MIMOSA-2 ($4 \mum$ epitaxial layer).
Both sensors had a pixel pitch of $20\mum$. Some pixels hosted more than one diode, which were connected in parallel. Lines to guide the eye. From \cite{MichaelPhD}.}
    \label{fig:CCEMi2}
\end{figure}

First observations were made with non-optimized prototype sensors, which relied on 3T-pixels with $20\mum$ pitch \cite{PaperNeutronRadMi2}. The sensors were irradiated with $\sim 1 ~\rm MeV$ reactor neutrons and their charge collection efficiency (CCE) was measured with a $^{55}$Fe-source . For practical reasons, this CCE was expressed for groups of four pixels, which should not be confused with the CCE of the seed pixel. An illustrative example of the amplitude spectra obtained is shown in Fig. \ref{fig:Fe55FourSpectrum1}. One observes that the charge collection peak\footnote{See Sec. \ref{SectionFe55Spectrum} for an introduction on how to read $^{55}$Fe amplitude spectra.} found in the spectrum is shifted along x-axis, which is related to a reduced CCE in the non-depleted active volume in the medium.  The seed pixel spectrum shown in the zoom does not indicate any displacement or significant modification in terms of entries. This shows that the gain of the CPS did not change as consequence to the irradiation. Moreover, it suggests that the properties of the depleted volume of the charge collection diode was not modified and thus excludes, in accordance with measurements on the doping concentrations performed much later, significant modifications in the doping concentration of the epi-layer.

The sensors showed a deterioration of the charge collection efficiency (CCE) after being irradiated with reactor neutrons beyond  $1~\rm MeV$ equivalent fluences of $10^{11}~\rm n_{eq}/cm^2$. As shown in \mbox{Fig. \ref{fig:CCEMi2}}, this CCE$_4$ dropped to 30-50\% after $10^{13}~\rm n_{eq}/cm^2$. 
The functional radiation tolerance limit of this kind of devices was evaluated with a beam test and found to amount \mbox{$\sim10^{12}\neqcm$} \cite{MichaelPhD}. A S/N=15 (most probably value) was extracted as necessary requirement to obtain the 95\% detection efficiency for minimum ionizing particles, which were considered as minimum for considering the sensor as operational.

From the findings, it was suspected that, due to the radiation induced reduction of the minority charge carrier life time, the signal charge carriers generated in the epitaxial layer might have recombined before being collected. Indeed, Eq. \ref{Eq:ChargeCarrierLifeTime} suggests a charge carrier lifetime of $\sim 400~\rm ns$ for a radiation dose of $10^{12}\neqcm$. This value can be compared with a measurement on the charge collection time, which was carried out with a similar sensor relying on a $14~ \upmu \rm m$ thick epitaxial layer. The sensor was illuminated with pulses from a $1060~\rm nm$ IR laser and the delay of the pixel signal was measured. This delay was interpreted as the charge collection time of the pixel, which is reported to amount $\sim 100~\rm ns$ for pixels incorporating one diode and $\sim 60~\rm ns$ in case the pixel hosts four diodes, which are connected in parallel \cite{DoktorarbeitDeptuch, DeptuchSimulationAndLaserPaper}.  Accounting for this charge collection time, one expects to collect 86\% or 77\% of the initial charge after a dose of $10^{12} \neqcm$ for a pixel with four and one diode per pixel respectively. This simple assumption reproduces the data shown on the related sensor (MIMOSA-1 in Fig. \ref{fig:CCEMi2}) with reasonably good precision.

A much higher radiation tolerance was reported for a sensor built in a $0.25\mum$ CMOS process with an $8\mum$ epitaxial layer \cite{PaperVelthuis1e14}. It hosted different kinds of 3T-pixels with $15\mum$ pitch and diodes of either $3\times 3 \mum^2$ or $1.2 \times 1.2 \mum^2 $ surface. One structure was equipped with enclosed transistors and one with four parallel diodes per pixel. Some samples were irradiated with $24~\rm GeV/c$ protons at the CERN-SPS and intentionally not powered during irradiation. They were kept at $-20\C$ hereafter for both, test and storage. The test demonstrated that the sensors remained operational up to the highest dose applied ($10^{15}\neqcm$), which demonstrated CPS to tolerate very high non-ionizing doses before breaking down. Surprisingly, no significant increase of the leakage currents is reported. Based on the $S/N$ as measured with a $\beta$-source, the authors concluded that the sensors still operate reasonably well after a dose of $10^{14}\neqcm$. 

This controversial conclusion is supported somewhat by the fact that one irradiated structure does indeed indicate a S/N of 15, which one can consider as a reasonable limit for efficient MIP detection. On the other hand, the measurements presented do not necessarily follow a clear trend. For example an insufficient S/N of 7 is reported for the successful structure for the lower dose of $5 \times 10^{13}\neqcm$. Moreover, the predicted charge carrier life time for a dose of $10^{14}\neqcm$ amounts $\sim 4~\rm ns$, which is dramatically shorter than the above reported charge collection time\footnote{The $15\mum$-pixels discussed here would plausibly show a moderately faster charge collection time than the $20 \mum$-pixels used for measuring the charge collection time. However, this is unlike to change the picture.}.

It should be mentioned that measuring the small signal charge of CPS with $\beta$-rays alone creates room for a specific mistake. The charge deposit of those particles follows a broad Landau-fluctuation. Irradiated sensors may fail to detect $\beta$-rays contributing to the lower region of this distribution, which moves the indicated mean value up. In addition, early experiments computed an individual discrimination threshold for each pixel, which smears the lower edge of the remaining Landau-distribution such the the remaining tail appears as a good distribution. The biased measurement indicates therefore an apparently perfect Landau distribution, which is however reduced in number of entries and indicates a too high S/N. It is well possible that the authors of the study would have arrived at different conclusions if they have had the possibility to cross-check their findings with a $^{55}$Fe-source or a particle beam. 

\subsubsection{Leakage currents}
\label{Sec:NonIonizingLeakageCurrents}
\begin{figure}[t]
\begin{minipage}{0.55\textwidth}
\includegraphics[viewport=0cm 0cm 18cm 15cm, clip,width=\columnwidth]{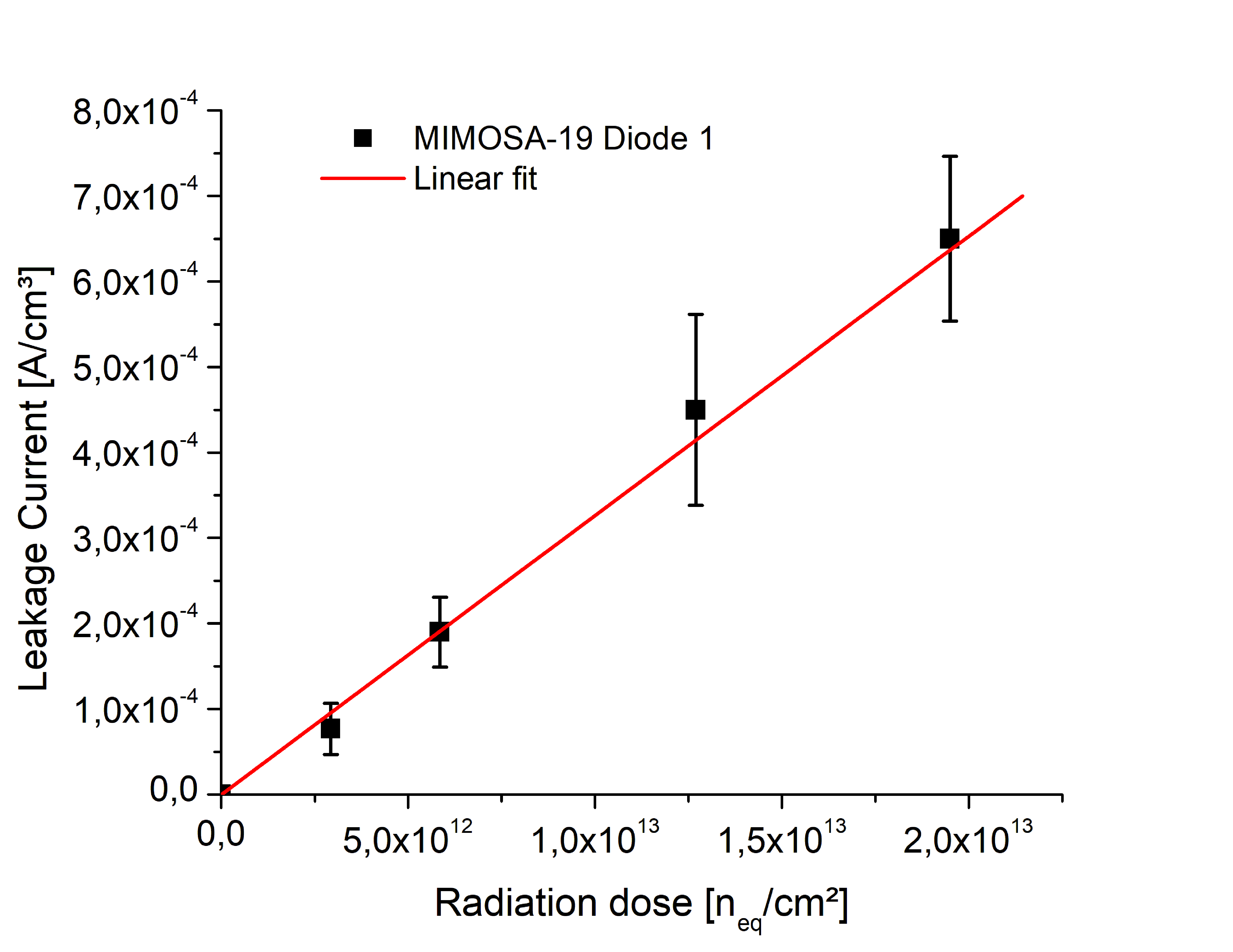}
\end{minipage}
\hspace{0.02\textwidth}
\begin{minipage}{0.41\textwidth}
 \caption{Measured increase of the radiation induced leakage current for a CPS with $12\mum \times 12\mum$ pixel pitch, a $39.6 \mum^2$ L-shaped diode and a $\sim 11 \mum$ thick, low resistivity epitaxial layer. The uncertainty on the dosimetry amounts \mbox{$\sim 10\%$}. The slope amounts \mbox{$\alpha(20^\circ{\rm C})=3.3 \times 10^{-17} \rm A/cm$}. Note that the standard annealing procedure was not respected, which introduces a systematic uncertainty of some 10\% (see text). Computed based on data from \cite{BudenbenderBachelor, YazgiliBachelor}.}
    \label{fig:LeakageCurrent}
\end{minipage}
\end{figure}

As introduced in Sec. \ref{SectionNonIonizingTheory}, the radiation induced bulk leakage current $\Delta I$ of a PN-junction which was irradiated with a fluence $\Phi$ of  \mbox{$1~\rm MeV$} equivalent reactor neutrons, is given with:
\begin{equation}
\Delta I= \alpha(T) \cdot \Phi \cdot V
\label{LeakageCurrentEquation}
\end{equation}
A value of $\alpha (20^{\circ}\rm C)= 4.0 \times 10^{-17} ~\rm A/n_{eq}/cm$ is given for  $T=20^{\circ}\rm C$ and an annealing of 80 min at $\rm 60^{\circ}C$ \cite{MollPhD}. For depleted silicon detectors with sizeable, depleted diodes, $V$ denotes the depleted volume of the diode. 
%For CPS relying on charge collection by diffusion, the equation states that electrons, which were thermally excited to the conduction band of the non-depleted epitaxial layer are not collected by the charge collection diodes. At the same time, this charge collection is obviously taking place for electrons, which were excited by a particle. It is a priori not obvious, why the charge collection process should be selective to the history of the electrons.
For CPS relying on diodes with tiny depleted volume, one may consider that an unknown amount of minority charge carriers from an undepleted volume nearby the junction might diffuse into its depleted volume and such generate a significant excess current as compared to the prediction of Eq. \ref{LeakageCurrentEquation}. 

An experiment aiming to test this hypothesis and the validity of Eq. \ref{LeakageCurrentEquation} was carried out with the CPS prototype MIMOSA-19, which hosts 3T-pixels with $12\mum \times 12\mum$ pixel pitch, a $39.6 \mum^2$ L-shaped diode and a $\sim 11 \mum$ thick, low resistivity epitaxial layer. The leakage currents of this sensor were reported in \cite{BudenbenderBachelor} and the data was combined with recent measurements \cite{YazgiliBachelor} on the depleted volume, which were performed according to the procedure discussed in Sec.~\ref{SectionFe55Spectrum}. No particular care for a controlled annealing (80 min at $60\C$) of the sensors was undertaken as the study aimed initially toward a different objective.  

The result of the study is displayed in Fig. \ref{fig:LeakageCurrent}, which was plotted under the assumption that $V$ is identified with the depleted volume of the photo-diode. One finds the expected linearity of the leakage currents and $\alpha(20^\circ{\rm C})=3.3 \times 10^{-17} \rm A/cm$, which confirms Eq. \ref{LeakageCurrentEquation}. The difference with respect to the literature value may be caused by the uncertainties of the measurement of the depleted volume. More likely, it reflects that the sensors under test were subject to a (too) long room temperature annealing of radiation damages prior to the experiment.  This is supported by the fact that the results match reasonably well the values reported for very long room temperature annealing as reported e.g. in \cite{MichaelPhD, WunstorfPhD}.

\subsubsection{Radiation hardening strategies for CPS with low resistivity active medium}

Radiation hardening strategies for CPS with low resistivity active medium aim to alleviate the impact of the losses in signal amplitude. The strategies presented were developed for a use with simple CMOS processes, which do not allow for depleting the pixel. This reflects the technology available in the early stage of the R\&D on CPS. Instead of depleting, it was tried to accelerate the charge collection by shortening the diffusion paths of the signal electrons. This was done by reducing the thickness of the epitaxial layer, the pixel pitch and/or using multiple diodes per pixel. Moreover, it was tried if extending the epitaxial layer may increase the initial number of charge carriers generated to an extent, which overcompensates the reduced charge collection efficiency.

\paragraph{Thickness of the active medium}
The latter approach was driven to its extreme by implementing a CPS on a CMOS wafer without epitaxial layer. Instead, the pixels with $20 \mum$ pitch used the few $100 \mum$ thick and moderately  \mbox{($\sim 10^{14}~\rm p/cm^3$)} doped substrate as active medium. The sensor showed a rather good initial S/N and detection efficiency, which was considered as a proof that CPS do not necessarily require an epitaxial layer \cite{Iouri-MapsTrackingPaper}. However, the width of the hit clusters was increased and a reduced spatial resolution for MIPs was observed.  The radiation induced drop of the signal charge was found exceed the one known from the standard sensors. According to beam tests, the MIP detection efficiency of sensors being previously exposed to a neutron dose of $10^{12} \neqcm$ dropped to only $\sim 90\%$. Therefore, the modified sensors showed a lower radiation tolerance than the standard sensors with epitaxial layer (see \cite{MichaelPhD}, P. 160). 

This weakness is plausibly related to the interface between the epitaxial layer and the substrate. In sensors with epitaxial layer, this interface creates a built in voltage, which hampers signal electrons from leaving the epitaxial layer and from getting lost in the deep substrate. Such, it increases the CCE of the sensor. In the absence of the interface, the CCE of the upper part of the active medium is reduced but the effective thickness of the active medium is increased. Consequently, one observes initially a higher signal charge. After irradiation, the life time of the electrons is reduced and electrons created in the deep substrate do not reach the charge collection diodes. Therefore, the disadvantage of the missing interface persists while the advantage of the larger effective thickness of the active medium vanishes. 

The limitations toward the edge of thinner epitaxial layers is discussed in a study \cite{PaperCCENeutronLowRes}, which compared the performances of sensors build on epitaxial layers with a thickness of $4 \mum$, $11 \mum$ and $14\mum$. The study estimates the charge signal and the S/N of sensors (based on the known thickness of the epitaxial layer and the CCE as measured with $^{55}$Fe) as a function of the radiation dose. In accordance with Fig. \ref{fig:CCEMi2}, it is found that for pixels with $20\mum$ pitch, the CCE of a $14\mum$ thick epitaxial layer deteriorates faster than the one of a $4\mum$ thick epitaxial layer. However, a higher signal is created by MIPs in the thicker epitaxial layer. This over-compensates the losses in CCE for all doses below the functional radiation limit of both devices. Therefore, the thicker epitaxial is found to be more radiation tolerant.

\paragraph{Pixels with multiple, parallel collection diodes}
\label{SectionMultipleDiodes}
The use of multiple diodes has repeatedly been considered as a mean to improve the charge collection performances of CPS. Therefore, the technology was implemented into various early sensor prototypes \cite{TurchettaPaper,DoktorarbeitDeptuch,PaperVelthuis1e14}. The obvious draw back of the approach consists in the additional leakage current and increased pixel capacitance, which can be considered to scale approximatively with the number of diodes connected in parallel. The additional capacitance decreases the charge-to-voltage conversion gain of the pre-amplifier of the pixels and increases in consequence the noise of the device. E.g. in \cite{TurchettaPaper}, a noise of $15 \e$ is reported for a pixel with $20\mum$ pitch and one diode with a surface of $3.1\times 3.1 \mum^2$. This compares to a noise of $31 \e$ for the related version of this pixel hosting four diodes. 
%However, the four diode pixel shows a substantially higher signal charge than the one with a single diode ($944~e$ instead of $458 ~e$, seed pixel, $^{55}$-Fe}). 
In \cite{PaperCCENeutronLowRes}, it is estimated, if the advantage of the better charge collection efficiency dominates the disadvantage caused by the higher noise. It is concluded that the pixels hosting multiple diodes show a higher signal and that their CCE shrinks slower than the one of the pixels with individual diode. However, due to the higher noise, this advantage does not turn into a relevant advantage in terms of S/N and radiation tolerance. Similar conclusions were drawn for pixels hosting diodes
with a very high surface \cite{BudenbenderBachelor}. 

\begin{figure}[t]
\begin{center}
 \includegraphics[viewport=0cm 0cm 35cm 25cm, width=0.8\columnwidth]{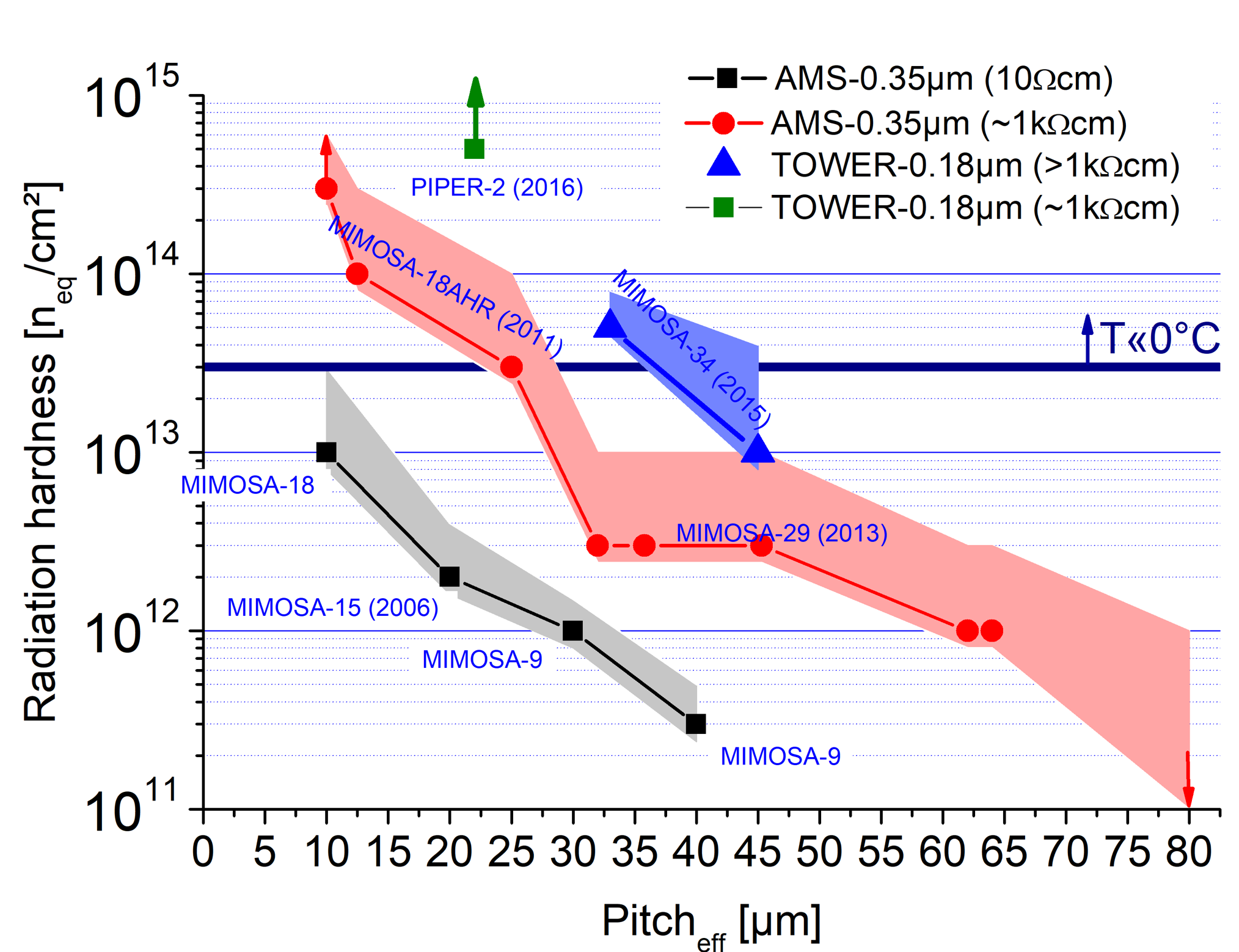}
\end{center}
 \caption{Compilation of the tolerance of CPS of the MIMOSA-series as a function of the effective pixel pitch and the initial resistivity/doping of the epitaxial layer. Cooling to $<0\C$ was required to control leakage currents while performing the underlying experiments at higher
radiation doses. See text for detailed explanation. Compiled from 
\cite{PaperPhD-Dennis, DennisMi18AHR, PaperCCENeutronLowRes, PaperRadHardnessOfMi18,  BachelorStefanStrohauer, LinnikPhD}.}
    \label{fig:RadTolCompilation}
\end{figure}

\paragraph{Pixel pitch} Decreasing the pixel pitch reduces the diffusion paths between the impact point of the particles and the collection diodes. Moreover, the relation between the surface of the charge collection diode and the pixel surface is improved. One expects both modifications to accelerate the charge collection process and thus the CCE of irradiated pixels. As the capacitance of the charge collection diodes remains unchanged, no increase of the pixel noise is to be expected. Indeed, a strong correlation between the pixel pitch and the radiation tolerance of the sensors was observed (see Fig. \ref{fig:RadTolCompilation}). This was first discussed in \cite{PaperCCENeutronLowRes} and \cite{PaperRadHardnessOfMi18} for SB-pixels relying on the AMS $0.35\mum$ OPTO process and a $11\mum$ thick epitaxial layer with a high doping ($\sim 10 ~\Omega \cdot \rm cm$) and confirmed with beam tests \cite{MichaelPhD}. As shown in the figure, the radiation tolerance of pixels with $40\mum$ was found limited to $3\times 10^{11}\neqcm$ while sensors with  $10\mum$ pixel pitch were found to reach a radiation tolerance above $10^{13}\neqcm$. By fitting the data (and with all caution given the uncertainties), one may state that the radiation tolerance $\Phi_{max}$ of the sensors as a function of the pixel pitch $10\mum \leq p \leq 40\mum$ is given with:
\begin{equation}
\Phi_{max} (10 ~ \Omega \cdot {\rm cm}) \approx 10^{15} \neqcm \cdot \left(\frac{p}{1 \mum} \right)^{-2.3}
\end{equation}
According to the equation, one may reach an impressive radiation tolerance by shrinking the pixels to a small pitch of few$\mum$, which is technologically possible in case a simple pre-amplifier and a sequential analogue pixel readout is employed. However, this approach turns into a large number of pixels per unit surface and thus a large number of pixels per readout line. As discussed in Sec. \ref{sec:RollingShutterReadout}, this slows down the readout. Sensors relying on pixels with a $p$ of few$\mum$ would show a rather slow time resolution, which is not acceptable for most applications.

\paragraph{Doping gradients}
\label{ChapterDopingGradients}
The Fermi energy of doped semiconductors depends slightly on the doping concentration of this semiconductor. Consequently, doping gradients may be used to modify the bands of silicon and to generate built in voltages also outside PN-junctions. To understand this, one remembers that the location of the Fermi energy $E_F$ with respect to the intrinsic energy $E_i$ of p-doped silicon is given (see \cite{BookSze}, P. 27) with:
\begin{equation}
dE=E_i- E_F = k_B \, T \log \left(\frac{N_A}{n_i} \right)
\end{equation}
In semiconductors, the distance between $E_i$ and the potential of the valence and the conduction band depends by definition not on the doping concentration. Moreover, in thermal equilibrium and in the absence of depletion fields, the absolute potential of $E_F$ is constant over the full semiconductor. A change in $N_A$ creates therefore a shift for the absolute potential of $E_i$ and consequently for the potential of the bands. The potential difference between silicon doped with the concentrations $N_{A_1}$ and $N_{A_2}$ is given with:
\begin{equation}
V=\frac{dE(N_{A_1})-dE(N_{A_2})}{e}=\frac{ k_B T}{e} \log \left(\frac{N_{A_1}}{N_{A_2}} \right)
\end{equation}
The related electric field $\mathcal{E}$ amounts:
\begin{equation}
\mathcal{E} = \frac{dV}{dx} = \frac{ k_B T}{dx \cdot e} \log \left(\frac{N_{A_1}}{N_{A_2}} \right)
\end{equation}
Here, $dx$ denotes the distance between the points with the related doping concentrations.
The drift speed of a signal electron exposed to a doping gradient is then given with:
\begin{equation}
v=\mu \cdot \mathcal{E} = \frac{\mu}{dx} \frac{ k_B T}{e} \log \left(\frac{N_{A_1}}{N_{A_2}} \right)
\end{equation}
At room temperature, the electron mobility $\mu$ is slightly above $1000~\rm cm^2/V/s$ and 
the built in voltage caused by doping gradient of one order of magnitude amounts about $60~\rm mV$. 

Assume a $\sim 14 \mum$ thick epitaxial layer with ideal exponential doping gradient. Be the doping
close to the diode for example of 
seven decades (e.g. from $\rm 10^{12}/cm^3$ to $\rm 10^{19}/cm^3$) between
the most lowly doped point of the epitaxial layer (near the diode) and the highly doped substrate,
this would turn into a sizable average field of about $30~\rm kV/m$. This drift field is sufficient
to move the electrons to a speed of $\sim 3~\rm \upmu m/ns$. This drift complements
the diffusion and allows for crossing the epitaxial layer within less than $\sim 5~\rm ns$,
which may speed up the charge collection by one order of magnitude with respect to pure diffusion. 
One may therefore consider, that fields introduced even by moderately big doping gradients may be
rather relevant. 

An explicite attempt to profit from this effect was made at IPHC with the sensor MIMOSA-21 (\cite{PaperMichalPhd}, P. 133). Two flavours of this sensor 
were manufactured in an $0.25 \mum$ BiCMOS process provided by an anonymous vendor. The sensor relied on a 
$9 \mum$ thick epitaxial layer with doping gradient. The precise doping profile and concentration was not documented. The process provided
deep p- and n-well implantations and a part of the pixels was covered with deep p-well. 
Various pixel designs with a pixel pitch of $10 \mum$ and $20 \mum$ were tested. Among the parameters, which were modified in
a systematic way, were the diode surface and the fraction of deep p-well in the pixel cell. In particular, it was tried to build a pixel,
which is surrounded by a deep p-well ring. This design had the conceptual potential to concentrate the signal charge in one
pixel diode and such to increase the S/N of the seed pixel.

The results of the study were somewhat disappointing. Most of the different pixel designs showed an insufficient charge collection efficiency. This is 
interpreted as a consequence of a layer with increased p-doping concentration between the epitaxial layer and the collection diode. Due to 
the above mentioned built in voltages generated, this layer has the potential to deflect the signal charge away from the diode. Moreover, most
pixels showed a sizeable leakage current, which was by significant factors above the ones reported for conventional pixels. A leakage current of 
several $\rm pA$ per pixel is reported for sensors, which were exposed to $500 \rm ~kRad$ and operated at room temperature. This high number
exceeds the one observed in standard CMOS processes by at least one order of magnitude and reflects plausibly
that the CMOS process was not optimized for building CPS.

However, an improved CCE was reported for a specific pixel with $2 \times 2 \mum^2$ pixel diode.  In this pixel, most of the pixel surface was
covered with ordinary p-well and deep p-well was arranged to a ring surrounding the pixel. This design had the conceptual potential to concentrate 
the signal charge in one pixel diode and such to increase the S/N of the seed pixel, which complements a potential effect of the doping gradients.
Indeed, an increase of the CCE from about $22\%$ (seed pixel, reference pixel manufactured in $0.35 \mum$ AMS Opto) to 34\% was reported. This CCE dropped
to about 20 after a dose of $6 \times 10^{12} \neqcm$, which appears by its own a rather good value. Unfortunately, the pixel showed a rather
high noise of $19~e~\rm ENC$ already prior to the irradiation. This, and the relatively thin epitaxial layer reduced the S/N of the pixel to an unacceptable
value (13 before irradiation, 7 after $6 \times 10^{12} \neqcm$). As by far better results were obtained with sensors relying on high resistivity epitaxial layers, the concept was abandoned. 

\subsection{Sensors with high resistivity $\sim{1~\rm k \Omega \cdot cm}$ epitaxial layer}

The tolerance of standard CPS to non-ionizing radiation is dominantly limited by the slow charge collection time of the sensor.
This process may be accelerated by extending the depleted volume of the charge collection diodes, which increases the geometrical ``capture cross-section'' of the diodes for diffusing charge carriers. As discussed in Sec. \ref{Sec:FundamentalDesign}, this extension may be reached by reducing the doping of the epitaxial layer. Doing so requires to process  wafers featuring lowly doped epitaxial layers with existing CMOS processes.  
Since this option became available, it forms  the standard technology
for building CPS. To the best of the author's knowledge, all CPS used in big
experiments of particle and heavy ion physics until nowadays, namely the STAR-PXL, the NA61-SAVD and the upcoming upgrade of the ALICE-ITS, rely on this technology.

\subsubsection{Charge Collection Efficiency}
\label{Sec:HRCCE}
The impact of using high resistivity (HR-) epitaxial layers as active medium of a CPS was first studied with TCAD  \cite{TCAD} simulations. A pixel, which bases on a $0.6 \mum$ CMOS-process with a $15\mum$ thick epitaxial layer was simulated assuming a standard low resistivity (LR, $\sim 10~\rm \Omega\, \cdot cm$) and a HR ($1~\rm k\Omega\, \cdot cm$) epitaxial layer \cite{PaperMi25}. Reducing the doping concentration is found to increase the thickness of the depleted layer of the collection diodes from $\sim 1 \mum$ to about $7 \mum$ at the available depletion voltage of $5 ~\rm V$. The lateral radius of the depleted zone is increased from essentially the diode radius of $\sim 3 \mum$ (LR) to $\sim 7 \mum$ (HR). The depleted volume of the diode is thus dramatically increased. However, it remains insufficient to deplete the simulated $20\mum \times 20 \mum$ pixels fully. 
The simulated sensor (MIMOSA-25) was realized. It exhibited an exceptionally high S/N for MIPs of about 50. After being irradiated to a dose of $3\times 10^{13} \neqcm$, the sensors showed a remaining S/N between 30 and 35, which is a dramatic improvement as compared to LR-sensors and sufficient for a good MIP detection efficiency \cite{PaperMi25}. 

A more quantitative comparison between LR- and HR-sensors became possible once the AMS $0.35\mum$ - OPTO process could be combined with both, a $10 \ohmcm$ and a $400 \ohmcm$ resistivity epitaxial layer. This allowed to realize sensors with both process flavours and to compare the performance of the otherwise identical devices \cite{PaperPhD-Dennis,DennisMi18AHR}.
\begin{figure}[tp]
 \centering
%1
\begin{subfigure}{.45\textwidth}
%\vspace{0.4cm}
  \centering
	   \includegraphics[width=1\columnwidth]{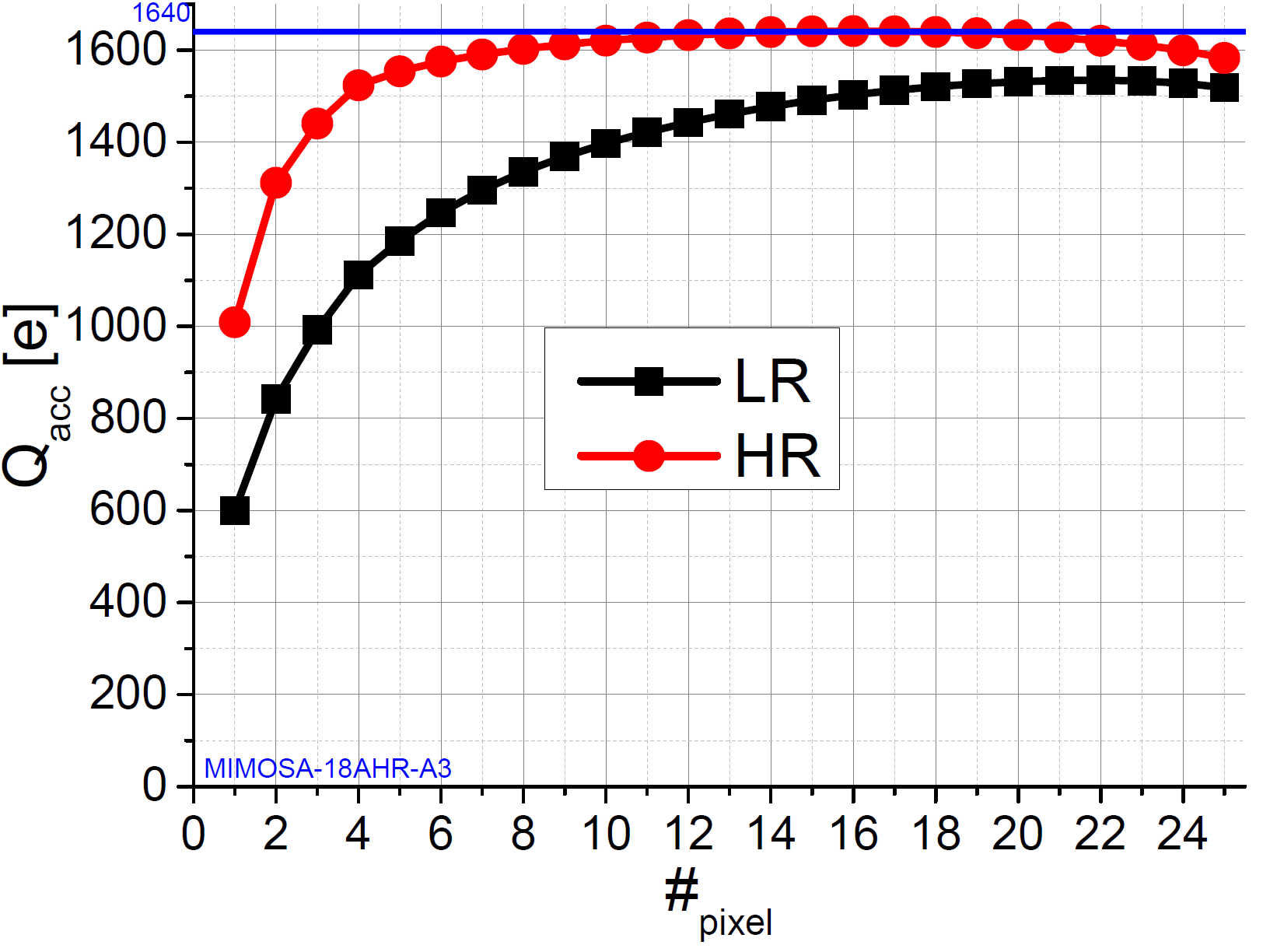}
    \caption{Average summed signal charge from groups of $N$ pixels. MIMOSA-18, $10\mum \times 10\mum$ pixels. From \cite{PaperPhD-Dennis}.  }
    \label{fig:AccumulatedChargeMi18}
\end{subfigure}%
\hspace*{0.05\textwidth}
%2
\begin{subfigure}{.45\textwidth}
\vspace{-0.2cm}
  \centering
	   \includegraphics[viewport=0cm 0cm 25.5cm 20cm, clip, width=1\columnwidth]{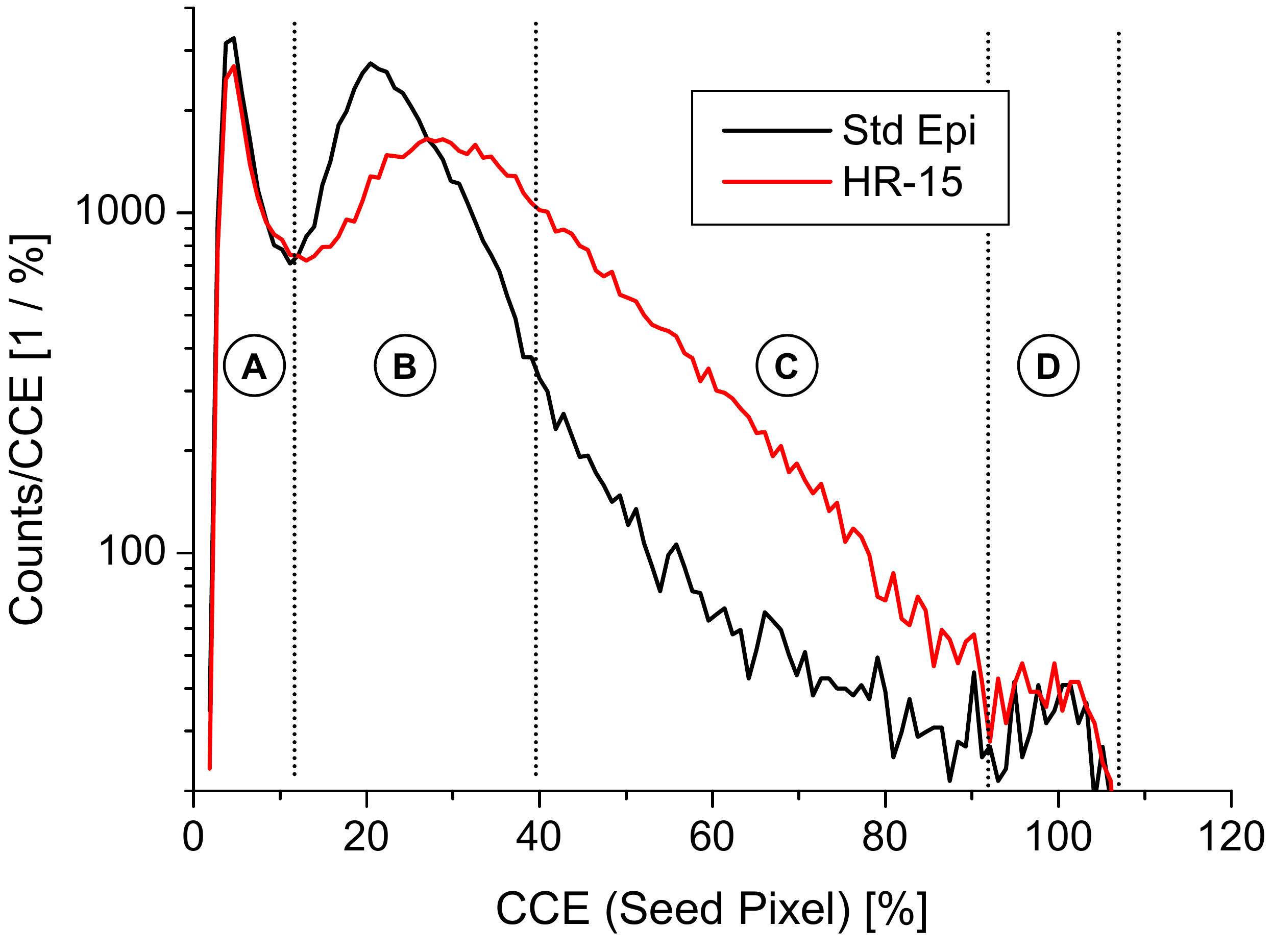}
    \caption{Amplitude spectra of the clamping pixel of MIMOSA-26 ($18.4\mum$ pitch, $V_{depl.}<1V$). From \cite{RadToleranceDev}.  }
    \label{fig:SpectrumFeMi26LRHR-2}
\end{subfigure}%

%3
\begin{subfigure}{.45\textwidth}
  \centering
	   \includegraphics[viewport=1cm 0cm 28cm 19.8cm, clip,width=1\columnwidth]{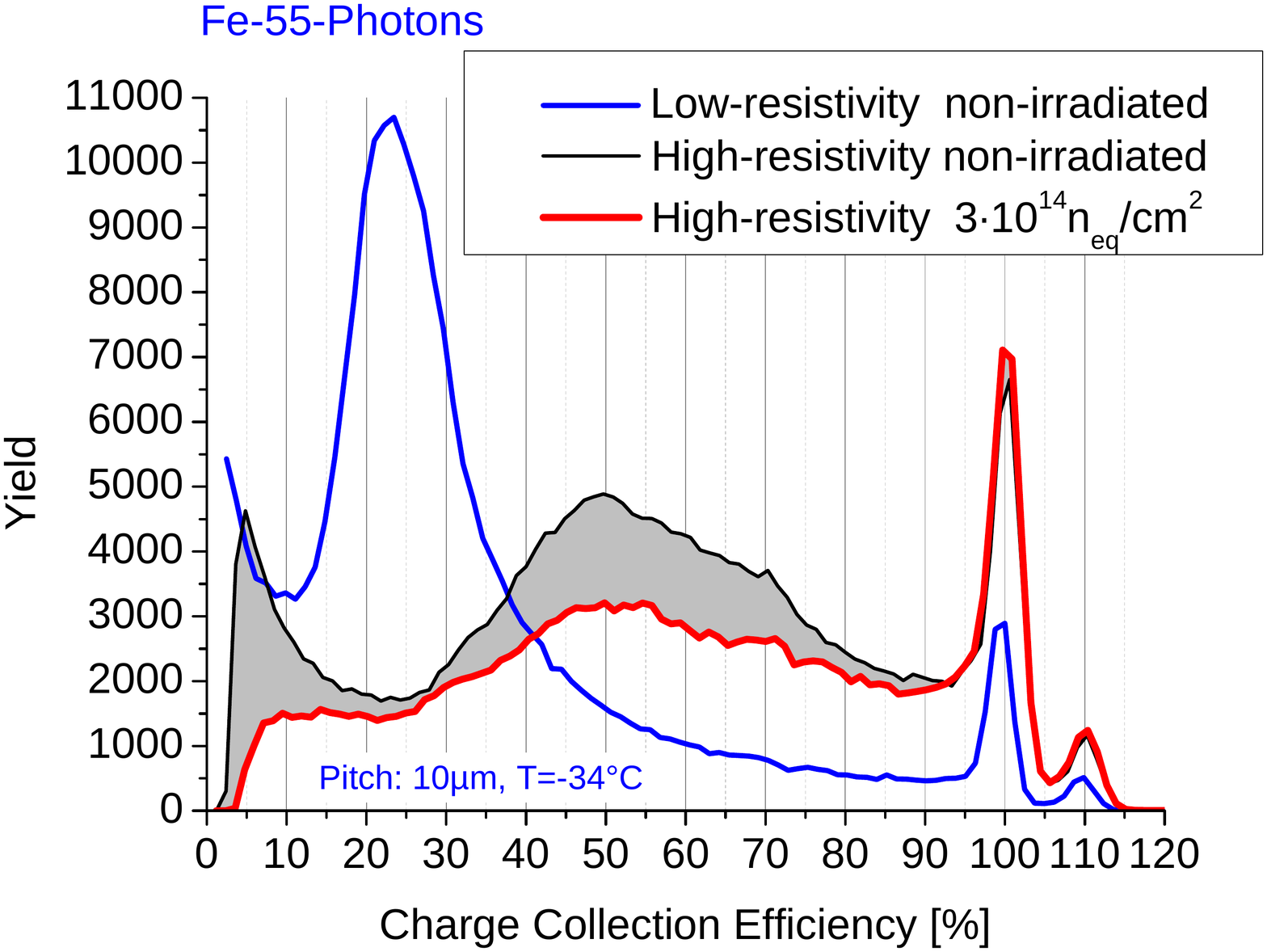}
    \caption{Seed pixel spectrum of MIMOSA-18 with \mbox{$10\times10 \mum^2$} pixels. From \cite{DennisMi18AHR}.}
    \label{fig:Fe55Mi18NeutronIrrad10mu}
\end{subfigure}%
\hspace*{0.05\textwidth}
%4
\begin{subfigure}{.45\textwidth}
  \centering
	   \includegraphics[width=1\columnwidth]{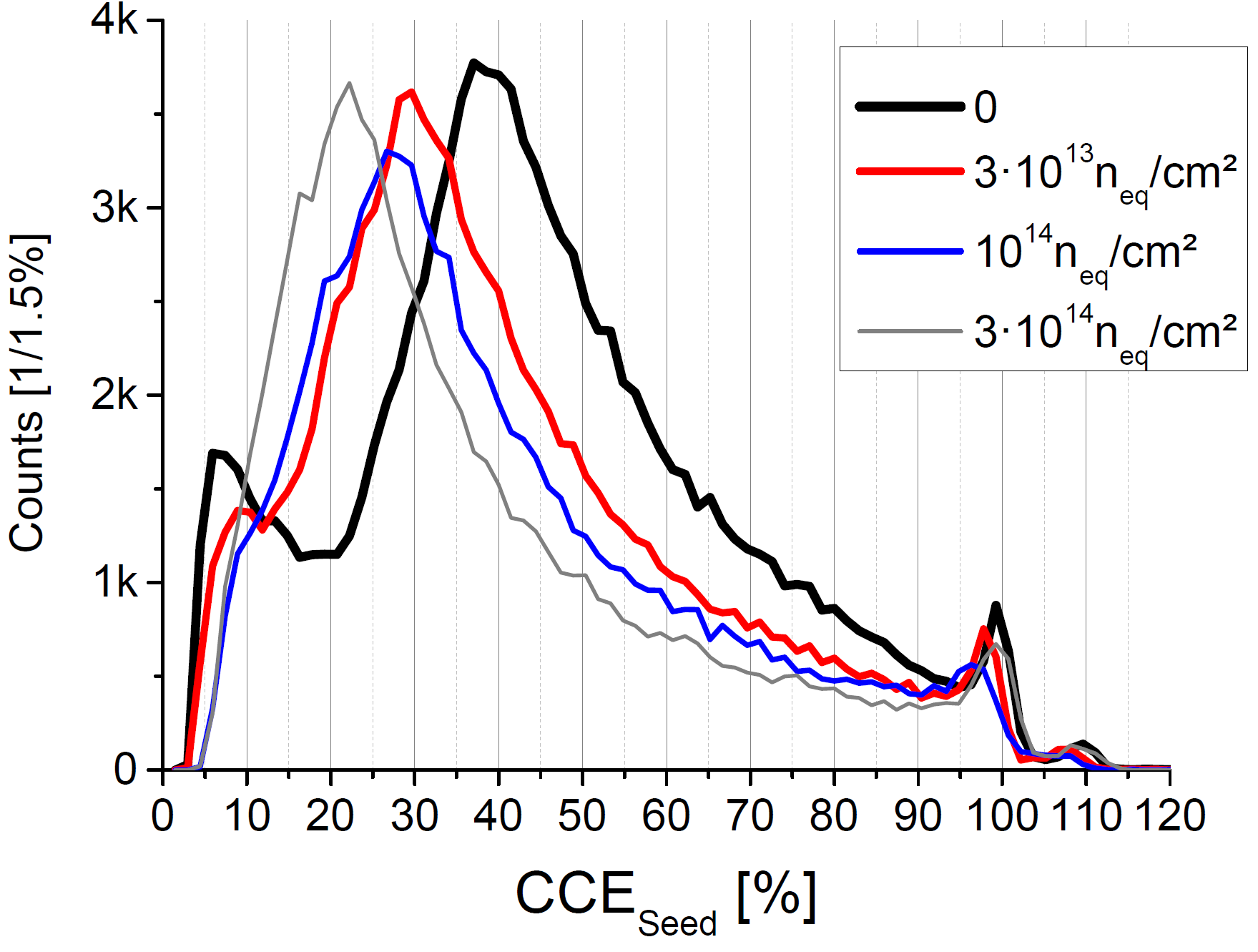}
    \caption{Seed pixel spectrum of MIMOSA-18 with \mbox{$25\times25 \mum^2$} pixels. From \cite{PaperPhD-Dennis}.}
    \label{fig:Fe55Mi18NeutronIrrad25mu}
\end{subfigure}%
\caption{Comparison amplitude specta recorded with an $^{55}$Fe-source for pixels with low (Std.) and high resistivity (HR) epitaxial layer.}
\label{fig:Fe55SpectrumCompareHRLR}
\end{figure}
According to those studies, the charge collection efficiency of HR-sensors does not really exceed the one of good LR sensors. This is as the latter approaches 100\% already. However, the charge is concentrated into the decisive seed pixel of the cluster. This is illustrated in Fig. \ref{fig:AccumulatedChargeMi18}, which shows the average accumulated charge of pixel clusters generated by X-rays from a $^{55}$Fe-source. The X-axis of the plot represents the number of pixels in the cluster considered for the charge measurement and the Y-axis the related charge. One observes that the seed pixel ($N=1$) collects a charge of 600 electrons in the LR-pixel and of slightly above 1000 electrons in the HR-pixel. The full signal charge is distributed to about 20 pixels for the LR-sensor but concentrated to roughly 4 pixels in the HR-sensor.  
The importance of this re-shaping of the clusters consists in the fact that the discrimination threshold of CPS, which is used for indicating the impact of one particle, acts mostly on the seed pixel. 

Going into detail, one finds that the impact of using HR-epitaxial layers depends on the sensor design.
A first comparison, was made with the popular MIMOSA-26 \cite{PaperChristineMi26,PaperPhD-Dennis,RadToleranceDev} prototype ($18.4 \times 18.4 \mum^2$ pixel pitch, clamping pixels with an effective depletion voltage $V_{depl} \lesssim 1\rm ~V$)  and the related $^{55}$Fe spectra are displayed in Fig.~\ref{fig:SpectrumFeMi26LRHR-2}. One observes that the number of entries in the calibration peak (region D in the figure) does not change significantly. However, the average charge collection efficiency is increased for the HR-sensor, which is in particular illustrated by the increased number of entries in region C. Therefore one observes an improved CCE as expected from enlarging the depleted volume of the charge collection diodes but one does not observe this enlargement. 

Additional studies were made with the MIMOSA-18AHR \cite{PaperPhD-Dennis,PaperRadHardnessOfMi18} prototype ($10 \times 10 \mum^2$ and $25 \times 25 \mum^2$, SB-pixels with  $V_{depl} \approx 3\rm ~V$) . The related spectra as recorded with the $10 \mum$ pixel of this chip are shown in Fig.~\ref{fig:Fe55Mi18NeutronIrrad10mu}. One observes that the number of entries in the ``calibration peak'' of the spectrum of the HR-pixel exceeds the one of the LR-pixel by factors, which reflects the expected enlargement of the depleted volume. Again, the shift of the big peak of the spectrum indicates the expected increase of the CCE of the non-depleted active volume of the sensor. 
One may thus conclude from the observations that the use of a HR-epitaxial layer may or may not enlarge the depleted volume of the charge collection diode but the CCE from the non-depleted epitaxial layer is improved anyway.

 A speculative explanation of those contradictory finding assumes that the p-doping concentration of the epitaxial layer increases nearby the wells, potentially due to a out-diffusion of p-dopands from the p-wells\footnote{The existence of such a kind of layer was suggested by the observation that an enlarged diode size is required in the AMS $0.35~\mum$ technology to connect the n-well and the epitaxial layer to a working charge collection diode, see \cite{MichaelPhD} for details.}. As this layer is created during the processing of the p-wells, its doping does not depend on the initial doping of the epitaxial layer. The n-well implantations of the diodes are embedded into this layer. Due to the small depletion voltage, the depleted volume of the charge collection diode of MIMOSA-26 remained within this layer. Therefore, the size of the depleted volume of the diode remains independent of the doping of the remaining epitaxial layer. The depletion voltage of MIMOSA-18AHR is sufficient to deplete the layer, to reach the lowly doped HR epitaxial layer and to generate a sizeable depleted volume there. The size of this depleted volume depends thus on the doping of the epitaxial layer. This does also improve the CCE from the epitaxial layer.

The reason for the improvement the CCE from the non-depleted volume of MIMOSA-26 remains speculative in this scenario. It might well be that the mentioned p-layer forms a shallow potential barrier, which hampers the signal electrons from entering the charge collection diode of the sensor and that this barrier is reduced/eliminated in the HR-sensor. Alternatively, one may speculate that the HR-sensor shows a beneficial doping gradient between epitaxial layer and substrate in analogy to what is discussed in Sec. \ref{ChapterDopingGradients}. Testing both hypothesis would require a detailed knowledge on the doping profile of the epitaxial layer, which is unfortunately not available.

As illustrated in Fig.~\ref{fig:Fe55Mi18NeutronIrrad25mu} for the MIMOSA-18AHR pixel with $25\mum$ pitch, the CCE of the HR-sensor decreases with increasing non-ionizing radiation. On a qualitative level, this is equivalent to the observations from LR-pixels but higher doses are required for reducing the CCE. The latter suggests that the ambitioned acceleration of the charge collection time was reached. It is also possible to build working pixels with a pitch above $40\mum$. Studies on the radiation tolerance of such pixels are found in \cite{PaperPhD-Dennis, BachelorStefanStrohauer} and cover pixels with a pitch of up to $80 \times 80 \mum^2$ as required for an accelerated rolling shutter readout. It is observed that the radiation tolerance of the pixels decreases in first order as a function of the surface of the pixels. A staggering of the pixels, which consists in shifting an elongated pixel by a half pitch, seems not to have significant impact on the radiation tolerance. The use of two or four diodes per pixel was also studied and the results confirm the mechanisms discussed in Sec. \ref{SectionMultipleDiodes}. However, a slight preference toward two diodes per pixel is observed for strongly elongated pixels. In pixels with four diodes, a region with good signal to noise and low charge sharing is indicated by a double peak in the amplitude spectrum. This region is likely located between the connected diodes at the center of the individual pixel. A region with worse performance is likely located in the periphery of the pixels and gives room for the creation of blind spots of poor CCE after moderate irradiation. 

As shown in Fig.~\ref{fig:Fe55Mi18NeutronIrrad10mu}, the specific HR-version of the $10 \mum$ pixel of MIMOSA-18AHR showed an unusual response to a neutron dose of $3\times 10^{14} \neqcm$. Instead of the expected drop in the average CCE of the sensor, one observes a reduced number of entries in the charge collection peak. This suggests that the size of the active volume was reduced while the charge collection capabilities of the remaining volume remained unchanged. A more detailed analysis carried out with X-rays and $\beta$-rays \cite{PaperRadHardnessOfMi18} suggests that the effective thickness of the active medium was reduced by about $20\%$ while most likely no blind spots are formed in the pixel. 

An overview on the maximum radiation tolerance of pixels realized in HR-processes as a function of the pitch is once more given 
in Fig. \ref{fig:RadTolCompilation}. Most of the underlying test were made in the laboratory. A significant cooling of the sensors exposed to doses above $10^{13}\neqcm$ was applied to control the leakage currents. A sensor was considered to resist to a given radiation dose if i) the S/N(MPW)$>15$ was observed for $\beta$-rays from a $^{90}$Sr-source, 
ii) the number of counts in the related amplitude spectra was not significantly reduced, iii) both, the  $^{90}$Sr and the $^{55}$Fe amplitude spectra did not show hints for local inefficiencies. 
For the points of MIMOSA-29, the effective pixel pitch is given. This effective pitch is defined as $p_{\rm eff}=\sqrt{A}$, where $A$ is the surface of the pixel. One finds that increasing the resistivity of the epitaxial layer from $10 \ohmcm$ to $400 - 1000 \ohmcm$ (AMS $0.35\mum$) increases the tolerance of the pixels by roughly one order of magnitude. 
Again with all caution, one may fit the data of the sensors produced in AMS $0.35\mum$ technology and state that those sensors show a reasonable radiation tolerance up to a dose of :
\begin{equation}
\Phi_{max} \approx 4.5\times 10^{17} \neqcm \cdot \left(\frac{p_{\rm eff}}{1 \mum} \right)^{-3.2}
\end{equation}

Increasing the resistivity substantially above $1000\ohmcm$ was studied with sensors based on the Tower/Jazz $0.18\mum$. The use of this resistivity does not change the qualitative properties of the sensors but yields moderate additional factor in terms of radiation tolerance. 

%It should be mentioned that the HR-sensors display a substantially faster increase of shot noise as a function of the radiation dose than LR sensors. In both cases, the additional shot noise could be suppressed by moderate cooling (see \cite{PaperPhD-Dennis}, P. 108). This can be explained in first order by the extended depleted volume of the charge collection diodes, which increases the leakage current with respect to the one observed in LR-sensors.

\subsubsection{Noise and leakage current}

\begin{figure}[tp]
 \centering
%1
\begin{subfigure}{.43\textwidth}
%\vspace{0.4cm}
  \centering
	   \includegraphics[width=1\columnwidth]{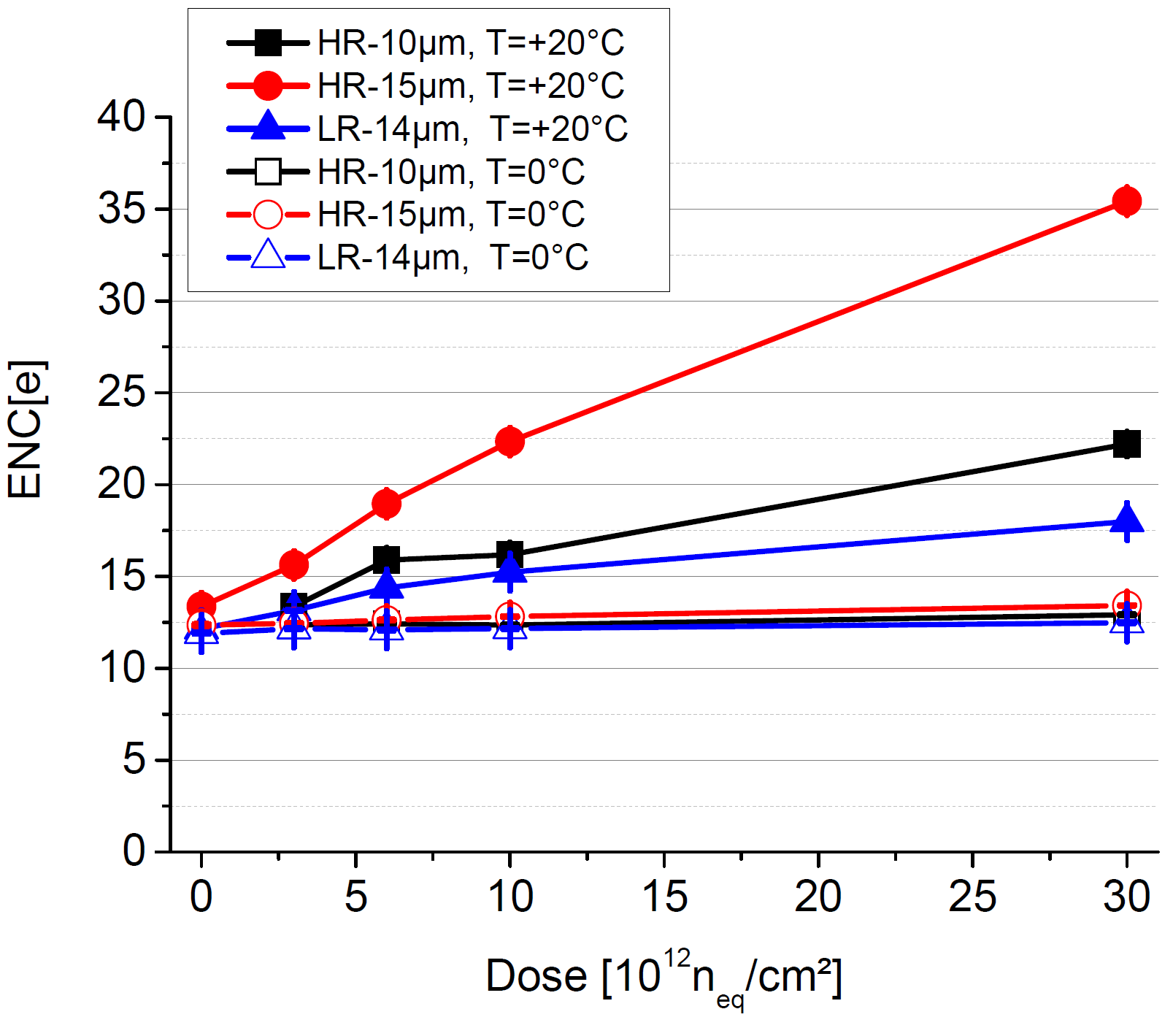}
    \caption{Comparison of the noise of MIMOSA-26 sensors with HR- and LR-pixels as a function of the temperature and the radiation dose.
		The standard high resistivity sensor MIMOSA-26AHR relies on the $15\mum$ thick HR epitaxial layer.
		From \cite{PaperPhD-Dennis}.  }
    \label{fig:ComparisonNoise}
\end{subfigure}%
\hspace*{0.05\textwidth}
%2
\begin{subfigure}{.47\textwidth}
\vspace{-.1cm}
  \centering
	   \includegraphics[width=1\columnwidth]{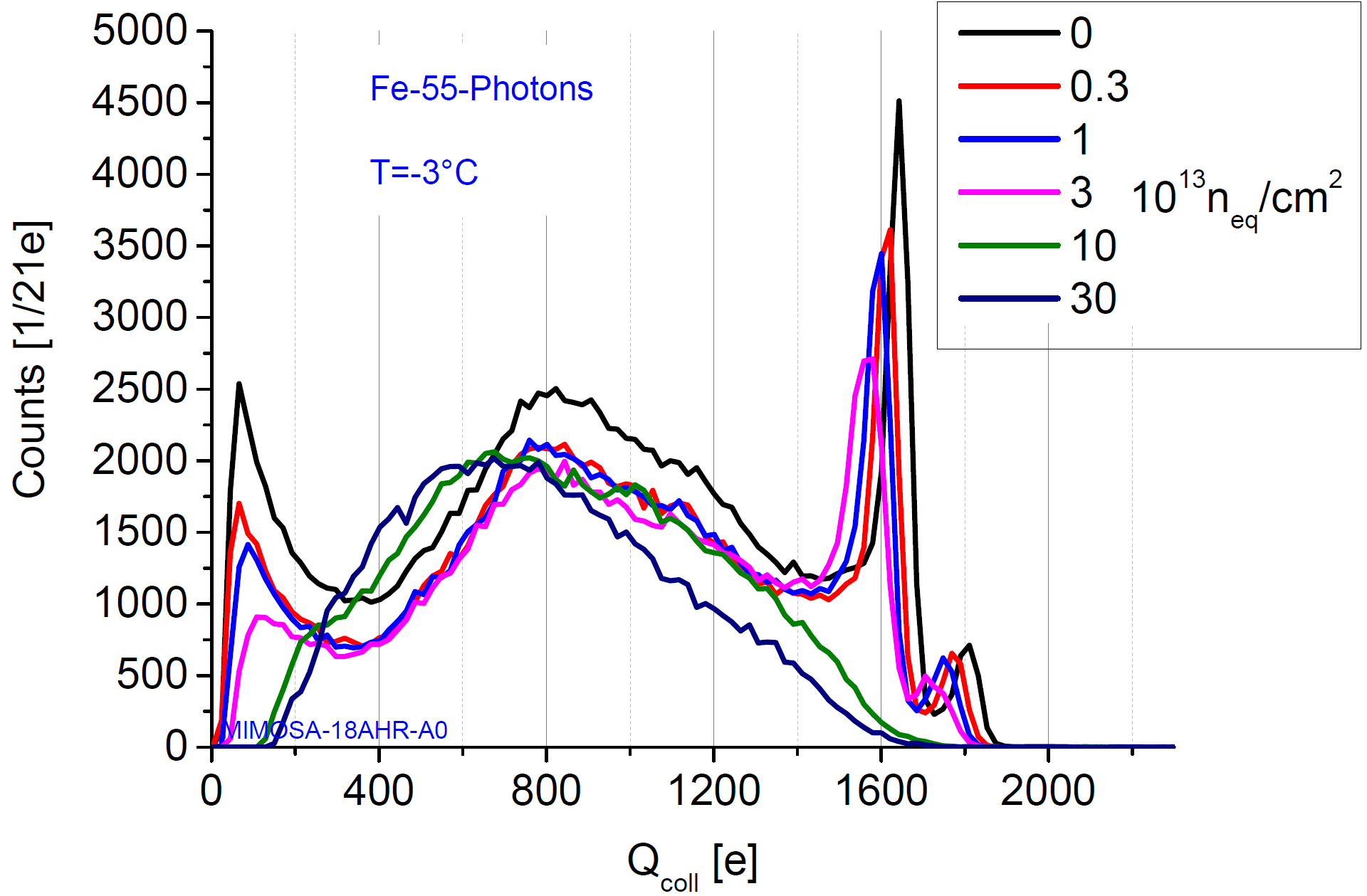}
		\vspace{-0.3cm}
    \caption{Amplitude spectrum of a $10\mum$ pixel of MIMOSA-18 as a function of the radiation dose.
			The increasing impact of the pixel clearing with increasing dose is visible \cite{PaperPhD-Dennis}.}
    \label{fig:ClearingMi18AHR-2}
\end{subfigure}%

\caption{Indirect consequences of the leakage current in CPS with HR-epitaxial layer.}
\label{fig:AcceleratedClearingLeakageCurrentHR}
\end{figure}

Only few systematic studies were performed in order to compare the leakage currents and the noise of irradiated LR- and HR-sensors of the MIMOSA-series.
This is on one hand as the sensors of this generation were not designed to allow for a quantitative access to those currents (SB-pixels without direct external access to the biasing lines of the diodes). On the other hand, the leakage current increases observed at the doses of interest ($\sim 10^{13}\neqcm$) were not considered as crucial as they remained on a level, which could be managed by means of moderate cooling and moderately fast readout. 

One study \cite{PaperPhD-Dennis} comparing the leakage current and noise properties of LR- and HR-pixels relied on the (shot-) noise as an indirect measure for the current. A result of the study is shown in Fig. \ref{fig:ComparisonNoise}, which displays the noise of MIMOSA-26 prototypes as a function of the epitaxial layer, the radiation dose and the temperature. Three epitaxial layers (low resistivity with $14\mum$ thickness, high resistivity with $10\mum$ and $15\mum$ respectively) were considered. At a coolant temperature of $0\C$, all options show the same noise. This suggests that the shot noise of the pixel diode is negligible at this temperature. For a temperature of $20\C$, the noise of the pixel relying on a $15\mum$ thick HR epitaxial layer (HR-15) exhibits the highest noise, while the noise of the HR-10 pixel and of the pixel with low resistivity silicon remains substantially lower. This noise increase is likely caused by shot noise and points to a related increase of the leakage currents. 

The increase of the leakage currents observed in the HR-10 and HR-15 sensors with respect to the reference sensor remains somewhat puzzling. On one hand, it is expected as one expects the HR-sensors to provide a more sizeable depleted volume and thus a higher radiation induced leakage current (see Sec. \ref{Sec:NonIonizingLeakageCurrents}). On the other hand, the related increase of the depleted volume was not observed in the related $^{55}$Fe amplitude spectra (see previous section). One might once more consider that the finding is driven by built-in-fields caused by doping gradients. As the doping profile of the different epitaxial layer was however not disclosed, this remains speculative and the sole conclusion of the study consists in the fact that the increase of the leakage currents of moderately irradiated sensors can be comfortably controlled by moderate cooling.

The limits of this strategy is illustrated by the amplitude spectrum of MIMOSA-18AHR pixels with $10\mum$ pitch, which is shown in Fig. \ref{fig:AcceleratedClearingLeakageCurrentHR}. 
One observes that the spectrum of the sensor operating at $t_{int}=6.6~\rm ms$ and a temperature of $T=-3\C$ is getting deformed with increasing radiation dose. This deformation becomes unacceptable above a dose of $3\times 10^{13} \neqcm$. It was associated with the clearing of the pixel and could be recovered by cooling the sensor to $T=-34\C$ up to the highest dose (see Fig.  \ref{fig:Fe55Mi18NeutronIrrad10mu}). 

Given that the charge collection of HR-pixels was significantly improved as compared to LR-pixels, the radiation induced leakage current and the related pixel clearing becomes the dominating issue limiting the tolerance of HR-CPS to radiation. The limit of the radiation tolerance is determined by the temperature dependent leakage current and the integration time of the sensor. As a good rule of thumb, one may state that room temperature operation might be possible up to $\sim 10^{13}\neqcm$. Above this limit, the use of a moderate cooling should be considered a loss in charge collection might set in for sizeable pixels. Starting from a dose of $10^{14}\neqcm$, either a massive cooling or a very fast sampling of the pixel signal is required to handle the leakage current and only small pixels may show a sufficient charge collection. 

%End of comment ===================================0

\subsection{Results from beam tests}

\begin{figure}[h]
\begin{center}   
 \begin{minipage}[c]{0.45\textwidth}
  \includegraphics[viewport=0cm 0cm 28cm 20cm, clip,width=\columnwidth]{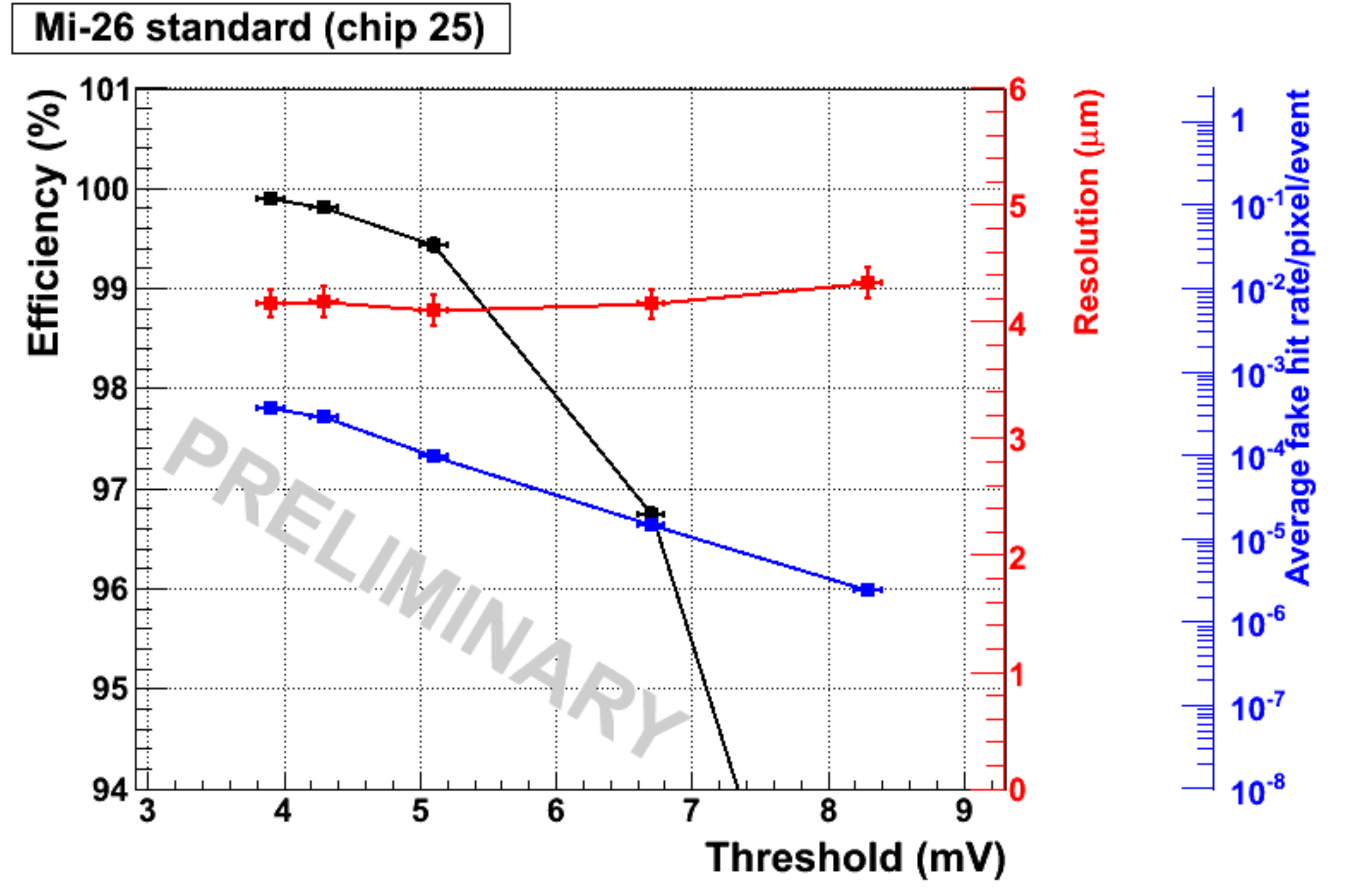}
     \end{minipage}
 \begin{minipage}[c]{0.45\textwidth}
  \includegraphics[viewport=0cm 0cm 28cm 20cm, clip,width=\columnwidth]{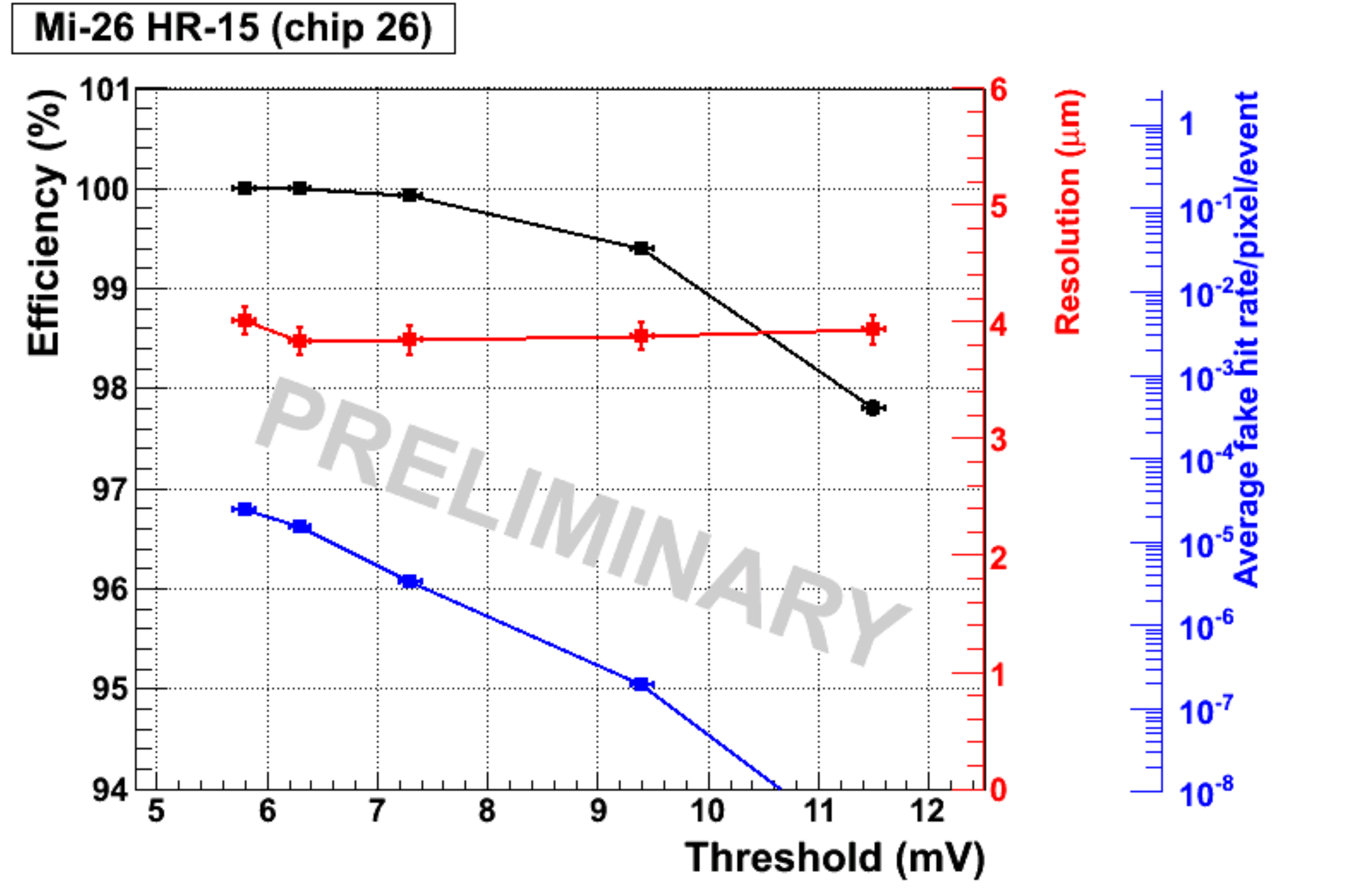}
     \end{minipage}
     
%\begin{minipage}[c]{0.44\columnwidth}
%}

 %    \end{minipage} 
         \begin{minipage}[c]{0.52\textwidth}
   \includegraphics[viewport=0cm 0cm 28cm 20cm, clip,width=1\columnwidth]{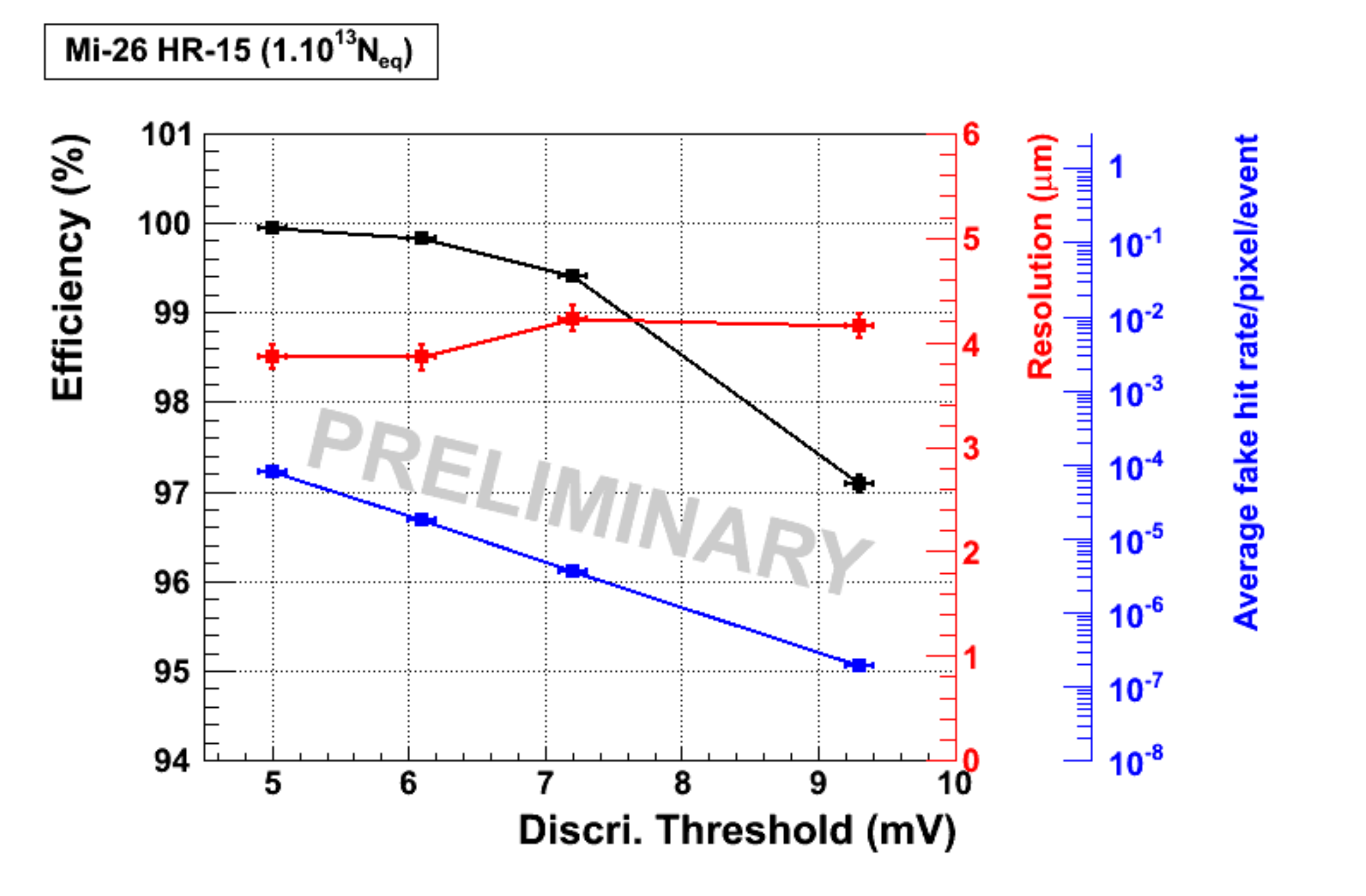}
     \end{minipage}     
		\hspace{0.5 \textwidth}
		
		   \caption{Beam test results for MIMOSA-26 with standard epi (upper left panel), HR-15 epi (upper right panel) and HR-15 epi irradiated with $10^{13}~\rm n_{eq}/cm^2$ (lower panel). From
	               \cite{RadToleranceDev}.}
       \label{fig:Beamtest}
\end{center} 
\end{figure}

Beam test with CPS being exposed to non-ionizing radiation were routinely undertaken but only a small subsection of the 
total results were published. Results shown in \cite{MichaelPhD} confirm the 
above discussed dependency of the limits of the radiation tolerance of sensors with low resistivity epitaxial layer and the pixel pitch
and demonstrate the radiation hardness limits of non-depleted CPS relying on high resistivity substrates.

A comparative study of the performances of MIMOSA-26 sensors (see Sec. \ref{Sec:Performances}) being implemented on a wafer 
with low and high resistivity epitaxial layers are shown in \cite{RadToleranceDev} and the related beam test results are displayed
in Fig. \ref{fig:Beamtest}. One observes that both sensor flavours create about the same spatial resolution of $\sim 4 \mum$ (a slighly 
better of $\sim 3.5 \mum$ is typically reached with a refined analysis). The ``HR-15'' sensors relying on the high resistivity epitaxial layer
exhibit a substantially better combination of high detection efficiency and low dark rate than their low resistivity counter part. The sensor
irradiated to $10^{13} \neqcm$, which was the highest dose applied in this study, shows superior detection capabilities than the non-irradiated
standard sensor. It is worth mentioning that the spatial resolution of the sensor is mostly not affected by the radiation damage, which 
may be considered as a general rule. 

In \cite{BaudotMi32BeamTestPaper}, beam test performances of a chip with SB- and clamping pixels of $20\times 20 \mum^2$
and $20\times 40 \mum^2$ are reported. The sensor named MIMOSA-32 relied on the Tower/Jazz 180 nm CMOS process and 
was shown to resist once more to the highest radiation doses applied ($1~\rm Mrad$ plus $10^{13} \neqcm$ at $T=+30\C$). Its 
tolerance to high leakage current effects was improved by the relatively fast integration time of $t_{int}=32 \mus$.
The sensor showed a good performance of $>99.5\%$ after irradiation with exception of the $20\times 40 \mum^2$ size pixel,
which showed a $98\%$ efficiency at $T=+30\C$ but recovered to $99.5\%$ once being cooled to $T=+15\C$. Once more, the
spatial resolution of the detector did mostly not change with radiation. A relatively high dark rate is reported, which was caused
by RTS in the entrance stage of the in-pixel amplifier (see Sec. \ref{SectionRTSinFET}). A sensor relying on clamping pixels
(FSBB-M0) resisted as well to a radiation dose of $10^{13} \neqcm$, which was the highest dose applied \cite{AugustFSBB}.

\section{Thermal Annealing}
\label{Sec:ThermalAnnealing}
Both, ionizing and non-ionizing radiation damage are known to be modified by annealing.
During this process, a part of the microscopic defects are recovered. In case of surface damage,
this may e.g. happen in case a hole trapped at the interface between Si and SiO$_2$ is filled by
a thermally excited electron. In the case of bulk damage,
an
interstitial atom may for example recombine with a vacancy in the crystal. On the other hand, 
some defects may bind to more complex defect molecules, which are sometimes
more damaging than the initial defect. In this case, the annealing generates
additional macroscopic damage and is referred to as reverse annealing \cite{Lutz}.

In a first study on annealing in CPS \cite{DoktorarbeitDeptuch}, charge collection diodes and
pixels implemented in a $0.25\mum$ CMOS process were irradiated with
$800~\rm krad$. During half the irradiation, the sensor was not powered.
The leakage currents were found to increase by a factor of 200 after irradiation
and decreased by a factor of three after three weeks of room temperature annealing.
24 hours of forced annealing at a temperature of $100^{\circ}\rm C$ reduced the leakage currents by another factor of five.

\begin{figure}[t]
\begin{minipage}{0.6\textwidth}
 \includegraphics[width=\columnwidth]{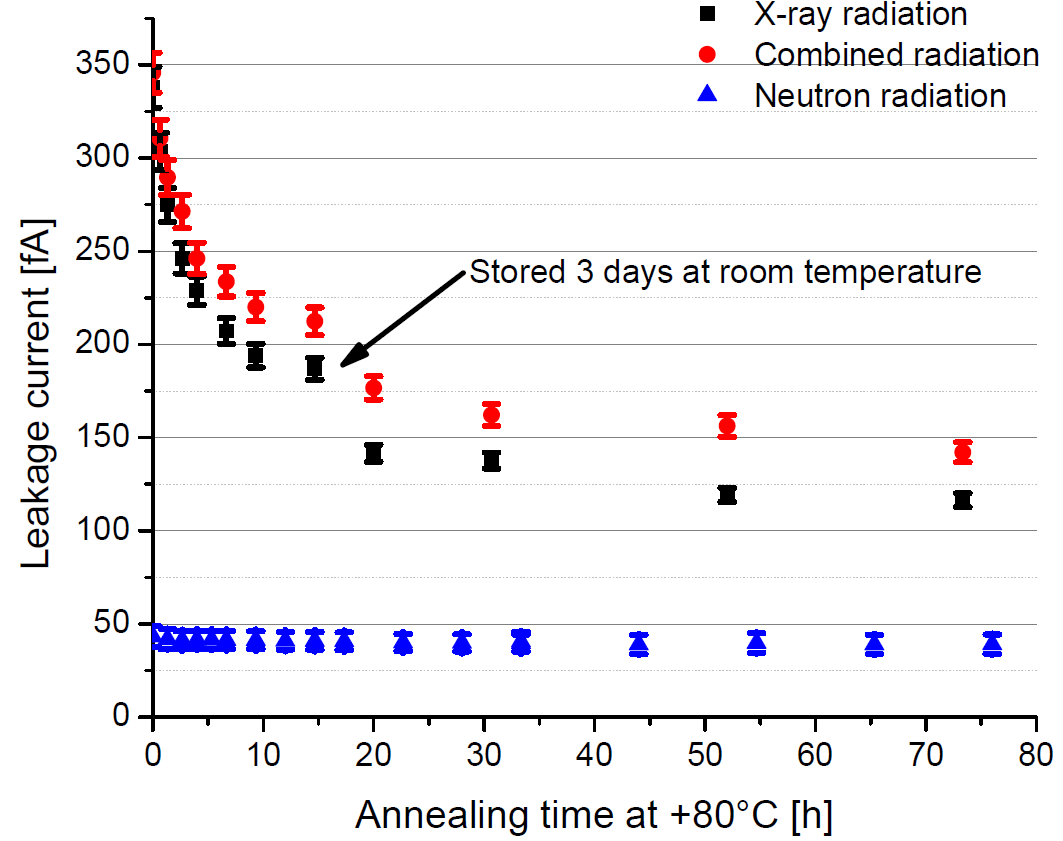}
\end{minipage}
\hspace{0.02\textwidth}
\begin{minipage}{0.36\textwidth}
 \caption{Leakage current of irradiated sensors as a function of the annealing time. See text. From \cite{DennisMaster}.}
    \label{fig:AnnealingLeakageCurrent}
\end{minipage}
\end{figure}

A more systematic study was carried out with a CPS named MIMOSA-19, which was based on an AMS 0.35 Opto process with $12\mum$ pixel pitch and a $14\mum$ low resistivity epitaxial layer \cite{DennisMaster,AnnealingStudyPaper}. The L-shaped diodes of the sensor have a surface of  $40\mum^2$ and are not hardened against ionizing radiation damage. The sensor was irradiated with $\sim 1~\rm MeV$ fission neutrons to a dose of $1.95 \times 10^{13} \neqcm$ and with soft X-rays to a dose of $200~\rm krad$. One sensor was exposed consecutively to both radiation doses. 

During and after the neutron irradiation, the sensors were stored for few weeks at room temperature until they were bonded and ready for test. Hereafter, their charge collection efficiency, gain, leakage current and noise were studied. This test was repeated after one year of room temperature storage. According to the test results, no significant modification of the charge collection efficiency and of the sensor gain were observed (within $\sim 5\%$ uncertainty). Concerning the leakage current and the noise, a slight but non-significant trend toward beneficial annealing was observed. 

Once the measurements were concluded, a virgin sensor and a sensor, which was previously irradiated with neutrons, were irradiated with $200~\rm krad$ soft X-rays. The irradiation was done at room temperature and the properties of the irradiated sensors was measured two hours after irradiation. As expected, the ionizing dose caused a strong increase of the leakage currents $I_L$  in both chips. The sensor, which was exposed to both, X-rays and neutrons, showed an $I_L$, which amounted about the sum of the leakage currents observed at sensors suffering from the individual radiation damages.

During the following 280h, the pixels were stored at room temperature and leakage current of their pixels was monitored. About  20\% of the total current was observed to vanish according to an exponential decay with a time constant of about $90\pm 50~\rm h$. Both sensors showed a similar behaviour within error bars, which suggests that the annealing acted on the surface damage. After the room temperature annealing went into saturation, a forced annealing was performed by heating the sensors to $T=80 {^\circ}\rm C$. As illustrated in Fig. \ref{fig:AnnealingLeakageCurrent}, the leakage currents of the X-ray irradiated sensors was reduced by more than a factor of two. The annealing procedure was also applied to a sensor, which was irradiated with $1.95\times 10^{13} \neqcm$ neutrons only. This sensor shows a 9\% decrease of the leakage currents after annealing, which is considered as significant w.r.t the uncertainties of the measurement. Moreover, a significant 5\% improvement of the charge collection efficiency are reported. 

Over all, one may conclude that thermal annealing is suited to reduce the leakage currents as caused by ionizing radiation. Beneficial effects on bulk damage have been seen but they are too small to be of practical relevance. More importantly, no reverse annealing was observed for sensors with low resistivity epitaxial layer. Note, that no statement is made on potential reverse annealing of sensors with high resistivity epitaxial layer.
\section{Specific radiation damage effects}
\label{Sec:SpecificRadDamage}
\subsection{Random Telegraph Signal}
\label{SectionRTS} \label{sectionRTS}
\subsubsection{Introduction}

\begin{figure}[t]
  \begin{center}
    \includegraphics[angle=0,width=0.75\textwidth]{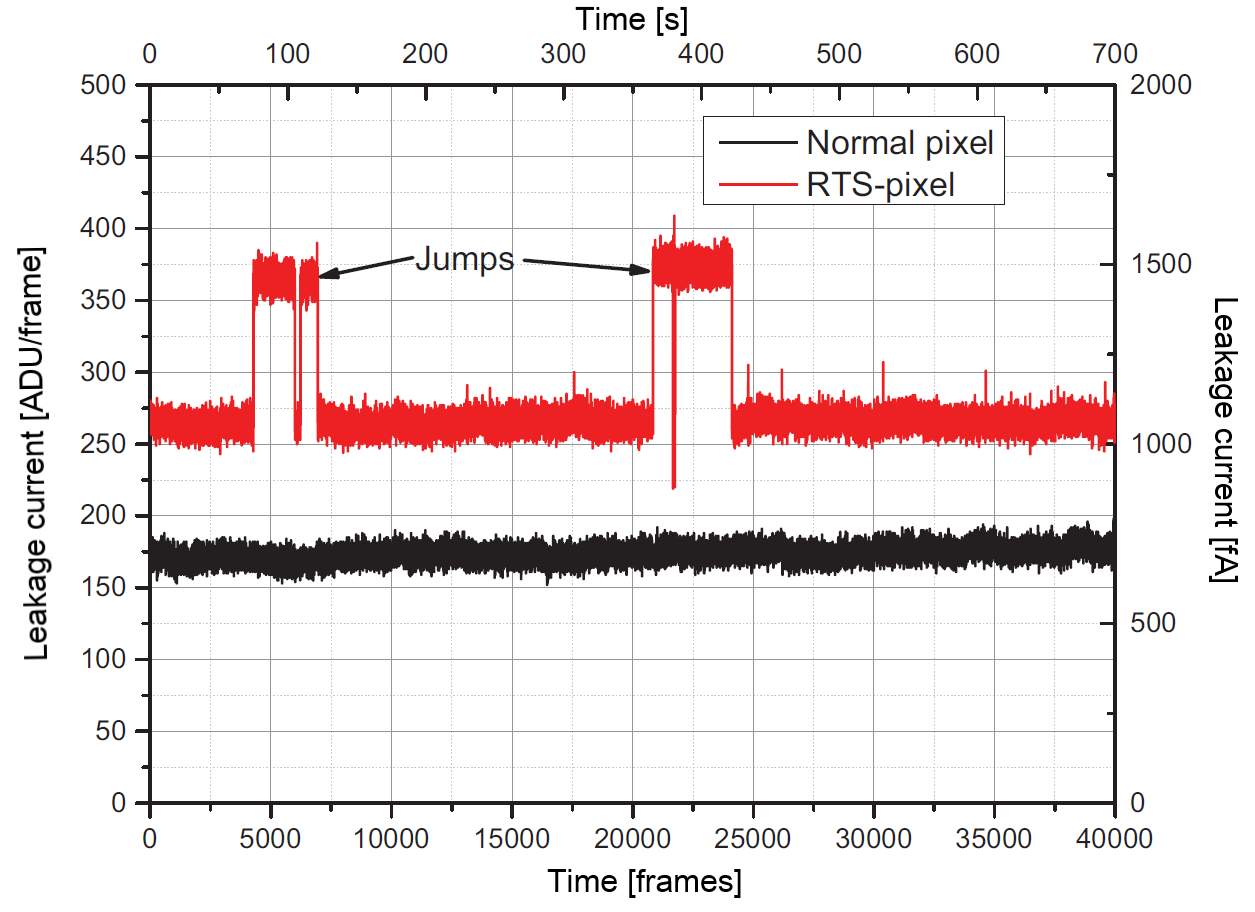} 
	\end{center}
    \caption{Random Telegraph Signal (RTS): The leakage current of a 3T-pixel showing RTS behaviour and of a similar 3T-pixel not showing this feature is shown.
		               In case the ``jumps'' exceed the threshold of the detector, numerous consecutive false hits are indicated by the RTS-pixel. From \cite{Paper:DeoringBachelor}, kindly reworked by the original author. }
		%This leakage current is denoted as Random Telegraph ''Signal`` by the original author, which does not correspond to a particle signal.
		
	%	with and without leakage current is denoted as Random Telegraph ''Signal`` by
		%the original author and given in raw units ADC (per frame) and in units of $\rm fA$. signal as the original authorupper (red) signal shows the leakage current of a 3T-pixel with RTS
   %   behaviour whereas the lower (black) line corresponds to the output of a stable pixel without RTS. The leakage is given
		%	in raw units ADC (per frame) and as Random Telegraph ''Signal`` in units of $\rm fA$.
   %   Once the amplitude of the ``jumps'' exceed the threshold of the detector, numerous consecutive false hits are indicated by the RTS-pixel. From \cite{Paper:DeoringBachelor}, kindly reworked by the author.}
    \label{fig:rts}
  
\end{figure}

Random Telegraph Signal is a rectangular modulation of the current passing reversed biased diodes or field effect transistors. An example of this modulation as recorded 
from the charge collection diode of a CPS with 3T-pixel is shown in Fig. \ref{fig:rts}. The signal is interpreted as the effect of the presence or absence of an individual electron
in a trap, which modulates the current with its charge. As the switching electron may
only be present or absent, the current is modulated between two well defined levels. RTS with more than two levels was observed and is usually explained as a superposition of multiple RTS-sources.
As RTS originates from defects located in an individual structure, it affects uniquely this structure. The defects may be created during chip production or by ionizing and non-ionizing radiation damage. 

Recognizing RTS signatures appears straight forward but becomes non-trivial as soon as its amplitude shrinks. Therefore, there is no commonly accepted definition of whether a pixel is affected by RTS.
In CPS for charged particle tracking, a pixel may be considered as affected in case the RTS amplitude is sufficient to exceed the signal discrimination threshold. If so, the pixel is converted into a 
``hot'' pixel and may dominate the dark rate of the CPS. However, as discussed more below, a number of detected RTS pixels 
according to this definition depends typically on the temperature and integration time of the pixel and the definition is not suited for CPS used in optical imaging.

\subsubsection{RTS in the leakage current of charge collection diodes}
\label{SectionRTSDiode}

%=============================
RTS in the leakage current is a property of the individual charge collection diode. The probability of a pixel to show RTS is not correlated to the state of the neighbouring pixels ~\cite{Paper:DeoringBachelor}, but increases for diodes showing a high absolute leakage current \cite{RTSPaper2}. For neutron irradiated sensors, this probability is reported to scale with the NIEL (see for example \cite{RTSPaper2,RTSPaper4, RTSPaper5}).
The amplitude of the RTS is reduced once the depletion voltage of affected diodes is reduced \cite{RTSPaper3}. 

A common feature of strategies to recognize diodes with RTS is that one compares the RTS-amplitude $I_{\rm RTS}$ with a certain threshold $I_t$.
For sensors with defined integration time $t_{int}$, this current may also be interpreted as a threshold charge $Q_{t}$, 
which may be chosen identical to the detector threshold. The latter is typically be chosen as a multiple $N_t$ of the temperature dependent pixel noise $Q_{noise}(T)$. 
Therefore, a pixel may be considered to be affected by RTS if:
\begin{equation}
I_{\rm RTS}> I_t (t_{int},T)=\frac{Q_{\rm t}}{t_{\rm int}} = \frac{N_t \cdot Q_{noise}(T)}{t_{int}}
\label{Eqn:RTSThreshold}
\end{equation}

As very different $t_{int}$ and $T$ are used in literature, it is hard to compare the results of the different pioneering studies \cite{RTSPaper2,RTSPaper3,FirstRTSPaper, Paper:MichaelRTS} on RTS
in a quantitative way. 

A deeper insight on the nature of RTS in diodes is found in \cite{RTSPaper4}, which relies on measurements with a very low $I_{t} (t_{\rm int}, T)$.
\iffalse %====================== Comment
\begin{figure}[t]
  \begin{minipage}{0.6\textwidth}
	\begin{center}

    \includegraphics[width=\textwidth]{./Figures/RTS-Amplitude.png} 
		%\includegraphics[viewport=0cm 0cm 19cm 12cm, clip,width=1.05\columnwidth]{Figures/RTS-Temperature.pdf}
	\end{center}
	\end{minipage}
	\hspace{0.05 \textwidth}
	\begin{minipage}{0.34\textwidth}
	\begin{center}
    \caption{Amplitude of RTS for differently irradiated imagers. From \cite{RTSPaper4}.}
    \label{fig:rts-amplitude}
  \end{center}
	\end{minipage}
\end{figure}
\fi %=================== end of comment
The probability to identify a pixel with a certain RTS amplitude
is reported to follow the distribution:
\begin{equation}
P(I_{\rm RTS})= k \cdot \lambda \cdot \exp{(-\lambda \cdot I_{\rm RTS})}
\label{Eqn:RTSProbability}
\end{equation}
Here, $k$ is a normalisation constant and $1/\lambda$ is given with $1/ \lambda \approx 110 \rm ~e/s$ 
for a sensor irradiated with $10~\rm Gy$ X-rays and with $1/\lambda \approx 1200 ~\rm e/s$
for a sensor irradiated with $1.8\times 10^{10} \neqcm$ neutrons. The distribution for a proton irradiated sensor
may be interpreted as a sum of the related ionizing and non-ionizing contribution. The very low amplitude
of the RTS caused by surface damage explains why this RTS was not observed in many early studies. 

\begin{figure}[t]
 \begin{center}
  \includegraphics[viewport=0cm 0cm 19cm 12cm, clip,width=0.8\columnwidth]{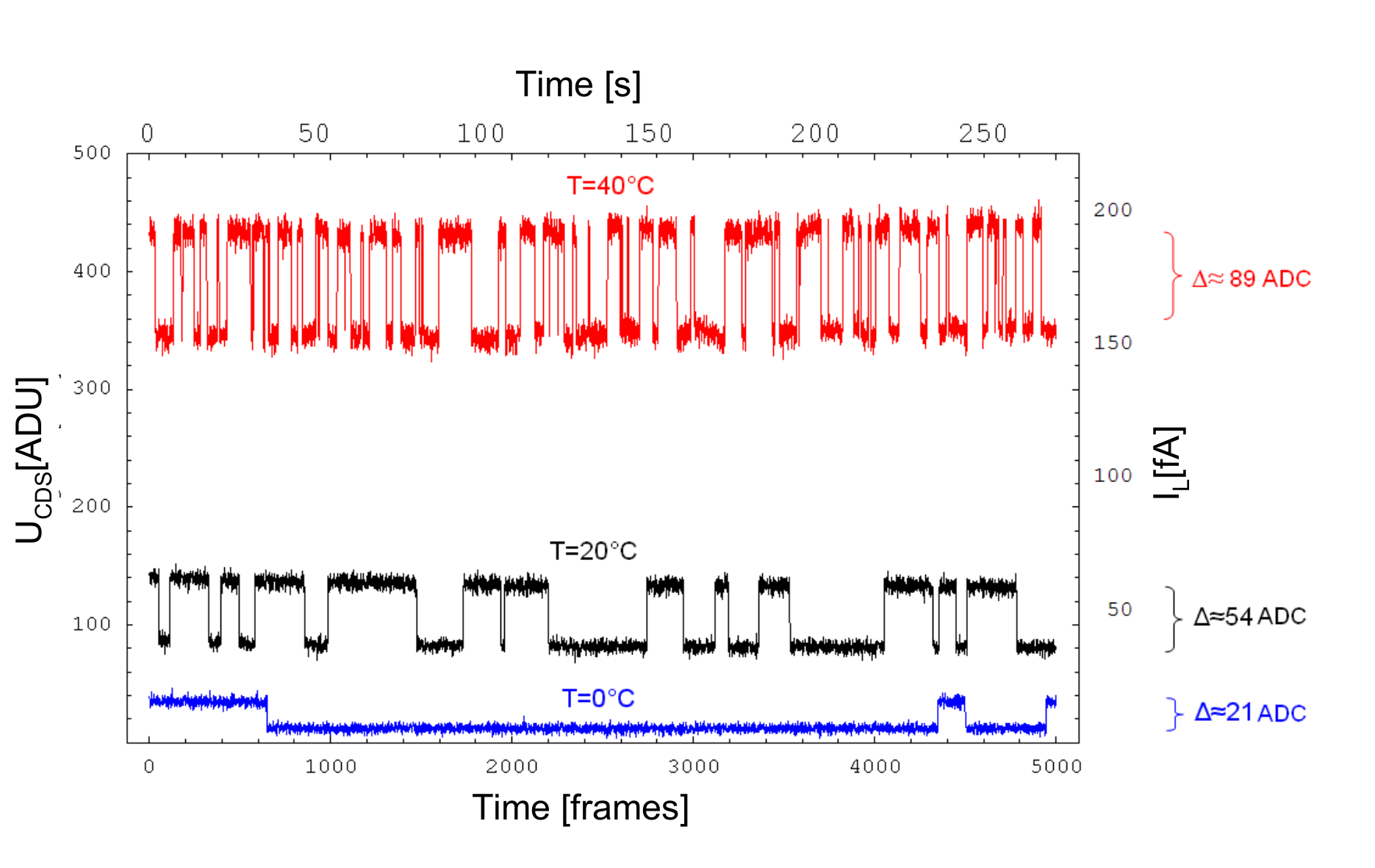}
   \caption{Temperature dependence of the RTS-amplitude and the RTS-frequency
  in a selected 3T-pixel of MIMOSA-19. From \cite{Paper:MichaelRTS}.
  }\label{fig:RTS-T}
 \end{center}
\end{figure}

As illustrated in Fig. \ref{fig:RTS-T}, the amplitude of the RTS is found to scale with temperature. 
A related parametrization is given in  \cite{RTSPaper4} with:
\begin{equation}
I_{RTS}  \propto \exp \left(- \frac{ E_a}{k_B T}\right)
\end{equation}
where $I_{RTS}$ is the transition maximum amplitude and $E_a$ is the activation energy. This energy is a property of the individual 
pixel and seems to vary over a significant range around the mid-gap energy.
The width of the RTS pulses spans a rather long range and an affected pixel may remain quiet for several minutes. For proton irradiated CPS, the probability of an pulse with a length of $t_{pulse}$ to occur 
is reported \cite{RTSPaper2} to follow:
\begin{equation}
P(t_{pulse}) = C_N \cdot  \exp\left(-\frac{t_{pulse}}{\tau} \right)
\end{equation}
Here, $C_N$ is a normalisation constant and $\tau$ is given with:
\begin{equation}
\tau^{-1} \propto \exp \left(-\frac{E_{a}}{k_B T} \right)
\end{equation} 
For temperatures between $23\C$ and $57\C$, \mbox{$E_{a}=0.58 ~\rm eV$} and $E_{a}=0.61 ~\rm eV$ are reported for the upper and lower state respectively \cite{RTSPaper4}.

Most studies on the relation between the number of RTS pixels and the dose are done for optical imaging under space conditions. In \cite{RTSPaper4}, CPS imagers with $128\times128$ pixels of $10 \mum$ pitch, which were manufactured in a $0.18\mum$ CMOS process, were used. The sensors were irradiated with ionizing and non-ionizing radiation and operated hereafter at a temperature of $T=22 \rm ^{\circ}C$. RTS was detected by scanning the leakage currents with a dedicated edge recognition technique discussed in \cite{RTSDetection} and a very low effective threshold of roughly $I_{t} (t_{\rm int}, T) \approx 10^{-17}\rm  A$. 
%Selected results of the study are displayed in Figure \ref{fig:rts-probability}. 
The number of pixels affected by RTS is observed to scale linearly with the non-ionizing dose and to reach $\sim 70\%$ of all pixels at the highest dose applied (a Displacement Damage Dose of \mbox{ $365 ~\rm TeV/g$}, which was applied by means of $14~\rm MeV$ neutrons and which is equivalent to $1.7 \times 10^{11}\neqcm$ in the wording of particle physics). For the ionizing dose, the number of RTS-pixels increases rapidly as well but the scaling shows a signature of a saturation. However, 100\% of all pixels are reported to show RTS after a dose of $100~\rm Gy = 10~\rm krad$. 

Both doses are small as compared to the radiation doses expected for modern vertex detectors. Therefore, one might expect RTS in all pixels already after a moderate operation time. However, CPS operating in vertex detectors may be cooled and are optimized for a short $t_{int}$, which reduces $I_{RTS}$ and increases $I_{t}$. A study on CPS with 3T-pixels \cite{Paper:MichaelRTS, Paper:DeoringBachelor}, which were operated with $t_{int}=3.6~\rm ms$ and at a temperature of $T_{coolant}=-20\C$ indicated less than 0.1\% RTS-pixels after a neutron dose of $2\times 10^{13} \neqcm$. At a temperature of $+40\C$, more than 40\% of all pixels indicated RTS. Besides the cooling, this result was caused by the high effective RTS threshold of $I_t=6.7 ~\rm fA$. 

\paragraph{Impact of the diode RTS on 3T-pixels}

\begin{figure}
\begin{minipage}{0.55\textwidth}
  \centering
	  \includegraphics[viewport=1.7cm 0.5cm 27cm 20cm, width=1\columnwidth]{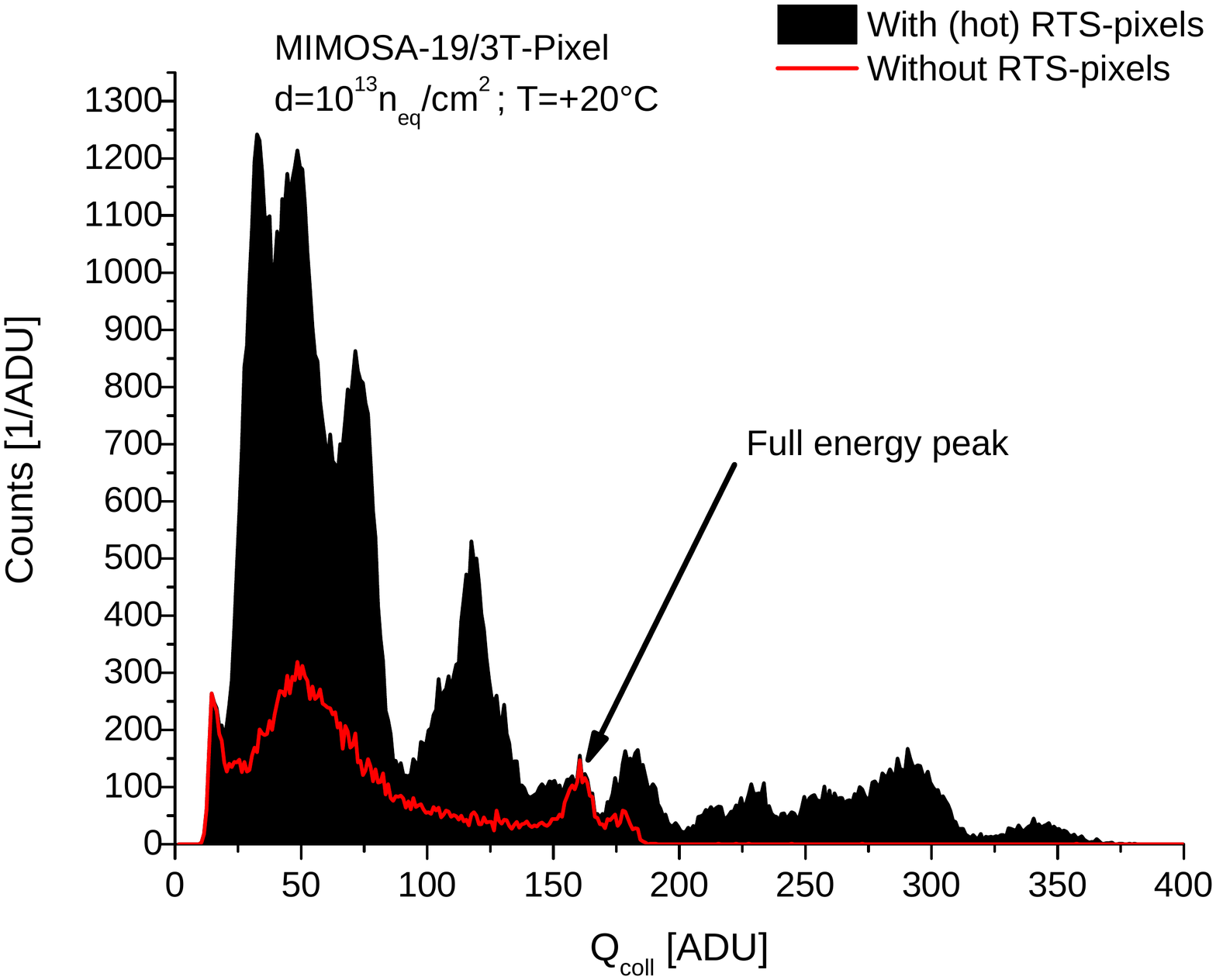}
\end{minipage}
\hspace*{0.05\textwidth}
 \begin{minipage}{0.35\textwidth}
    \caption{$^{55}$Fe amplitude spectrum of a 3T-pixel sensor showing massive RTS after being irradiated with neutrons. Individual pixels generate long trains of hits with similar amplitude, which creates peaks in the spectrum. After masking those pixels, the X-ray amplitude spectrum becomes visible. \mbox{From \cite{Paper:DeoringBachelor}, reworked.}}
    \label{fig:SpectrumFeMi2RTS-2}
\end{minipage}
\end{figure}%

CPS indicate a particle hit, whenever the charge collection diode collects an excess charge exceeding the discrimination threshold within
the integration time. 
In case the integrated amplitude of the RTS signal $Q_{RTS}=I_{RTS}\cdot t_{int}$ exceeds the discrimination threshold of the pixel, a series of hits is generated. All hits of this series will indicate the same signal charge and may
thus generate peaks in the amplitude spectra, which have to be removed before analysing them (see Fig.~\ref{fig:SpectrumFeMi2RTS-2}). 
The number of false hit indications generated by an RTS pulse is rather high. It is given by the number of frames recorded during the up-time of the pulse,
which may amount tens of seconds. Therefore, RTS in diodes dominates typically the dark occupancy of irradiated CPS \cite{Paper:DeoringBachelor, Paper:MichaelRTS}. 

\begin{figure}[t]
 \centering
%1
\begin{subfigure}{.45\textwidth}
  \centering
  \includegraphics[viewport=1.3cm 0cm 18cm 15cm,clip,width=1\columnwidth]{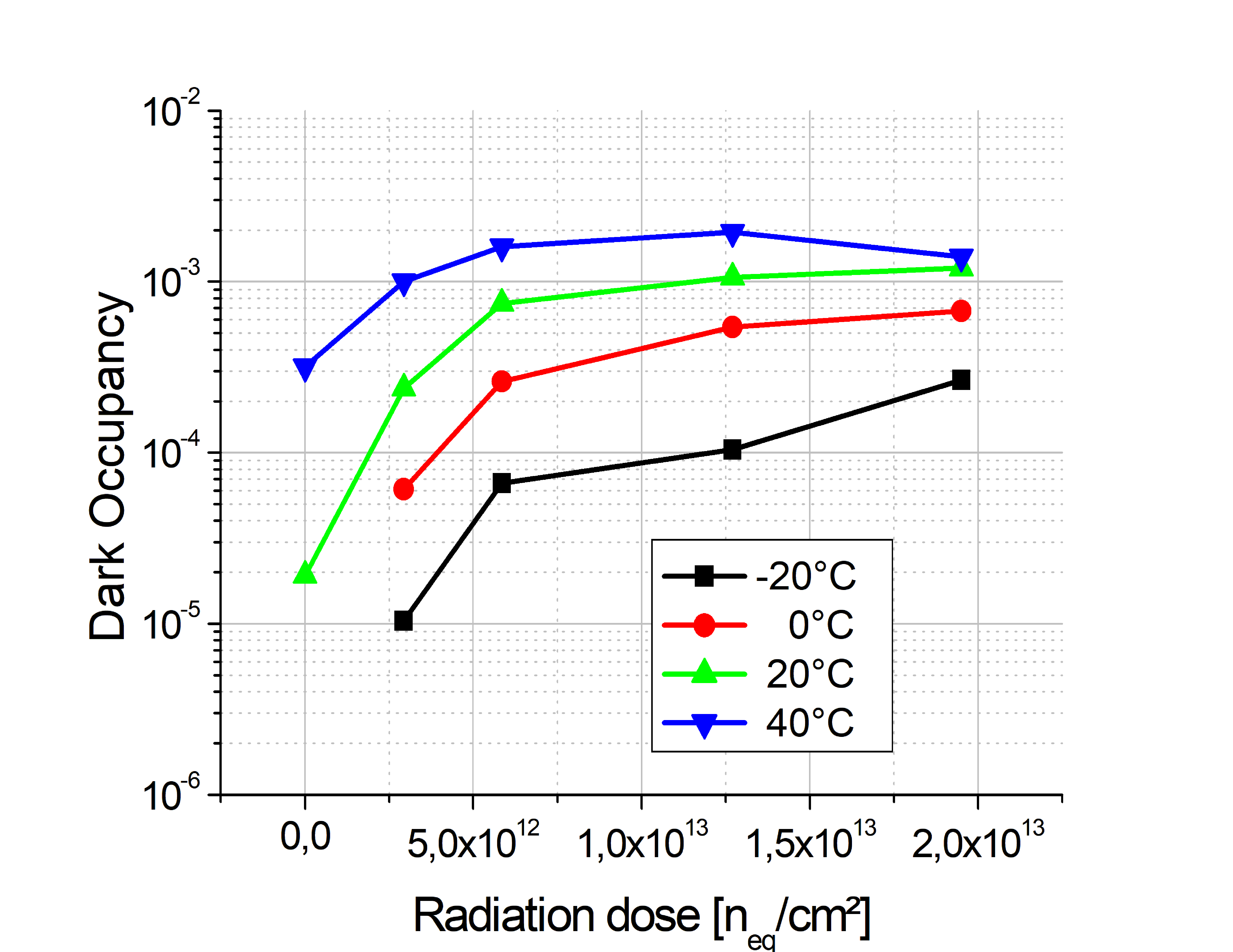}
    \caption{Fake hit rates of the 3T-pixels of MIMOSA-19. Fake hits from both, RTS-pixels and regular pixels are accounted for.}
    \label{fig:3TFakeHits}
\end{subfigure}%
\hspace*{0.05\textwidth}
%2
\begin{subfigure}{.45\textwidth}
  \centering
   \includegraphics[viewport=1.3cm 0cm 18cm 15cm,clip,width=1\columnwidth]{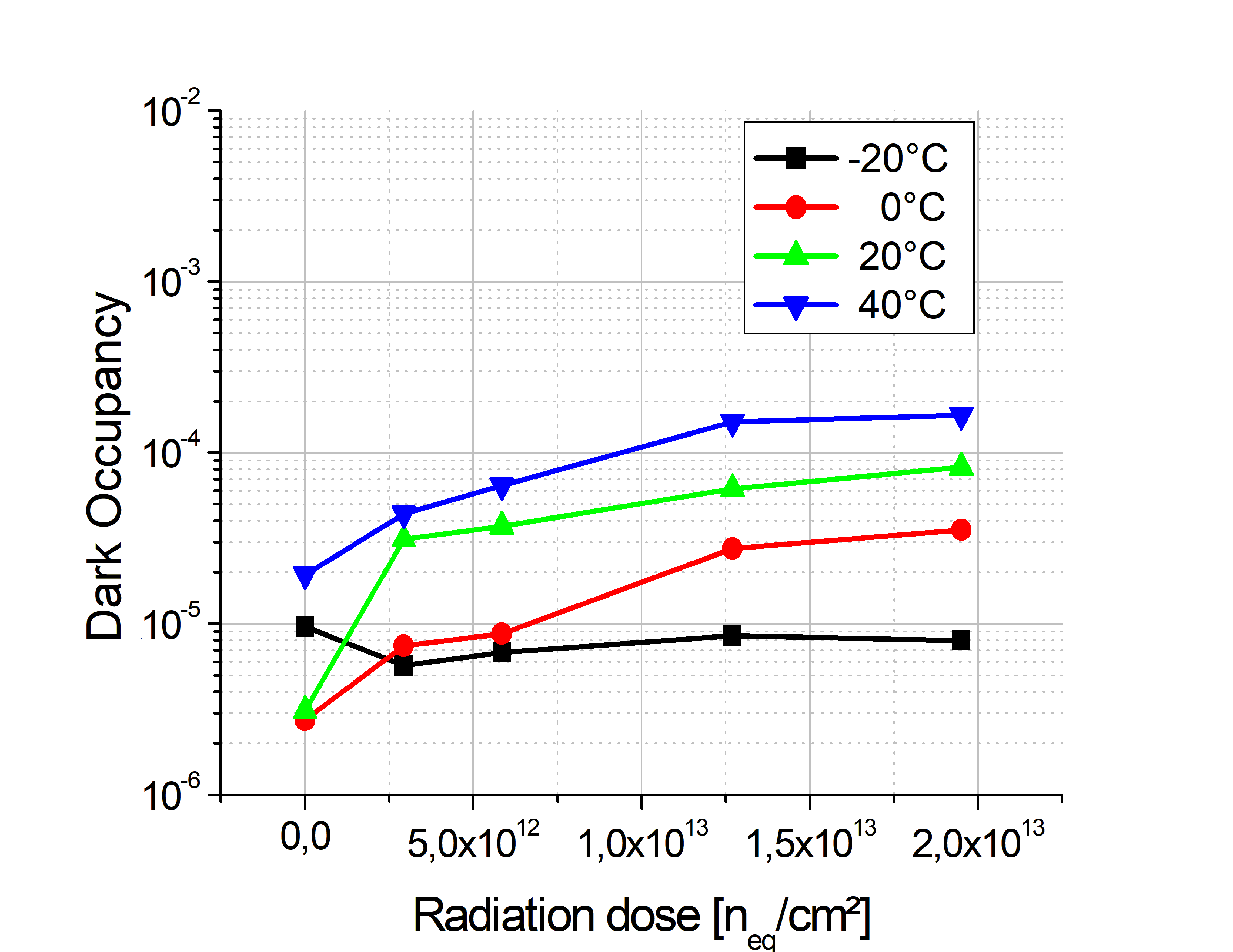}
    \caption{Fake hit rates of the SB-pixels of MIMOSA-18. Fake hits from both, RTS-pixels and regular pixels are accounted for.}
    \label{fig:SBFakeHits}
\end{subfigure}%https://www.overleaf.com/6458423ktgndt#
\caption{Fake hit rates for CPS with SB- and 3T-pixels. Based on data previously published in \cite{Paper:MichaelRTS}. 
See text.}
\label{fig:RTSFakeHits}
\end{figure}

A quantitative measurement of the related dark rate is shown in Fig. \ref{fig:3TFakeHits}. One observes that the related hit
rate increases with the radiation dose and even more importantly with the temperature. The latter effect is considered
as related to the temperature dependent increase of $I_{RTS}$ (see previous section). 
Besides the use of cooling, accelerating the integration time of the sensor forms an option to reduce the
number of false hit indications caused by RTS. This is as this measure may reduce the integrated charge related to 
the additional leakage current ($Q_{RTS}$) below the detection threshold. Finally, it may be appropriate
to mask some of the affected pixels.

\paragraph{Impact of diode RTS on SB-pixels}

The intrinsic leakage compensation of SB-pixels may adapt rapidly to the RTS induced steps of the leakage current. 
Fake hit indications are only expected at the rising edge of the RTS signal while the falling edge may, in theory, generate
a certain dead-time of the individual pixel. 
Rectangular RTS signatures are thus not visible
in the dark signal after CDS processing but may be seen in the raw dark signal of the pixels. Using this signal remains
however a rather qualitative tool. Due to the logarithmic response of the dark signal of the pixel to leakage currents,  
$I_{RTS}$ decreases with the increasing static 
leakage. Moreover, the detection concept is not suited to distinguish between RTS occurring in the diode 
and RTS occurring in the input transistor of the pre-amplifier of the pixel (see next section). Studies on the 
number of SB-pixels showing RTS were done \cite{Paper:MichaelRTS} but the results are not necessarily reliable.
The indication shows that most pixels exhibit RTS at low temperatures and prior to irradiation and that the fraction
of RTS affected pixels shrinks with higher dose and temperature. This is compatible with the consideration that
the indication is mostly driven by the increase of the diode leakage currents, which increase the detection threshold $I_{t}(t_{int},T)$ 
of the measurements.

However, as shown in Fig. \ref{fig:SBFakeHits}, SB-pixels show a lower dark rate than 3T-pixels. This is due to the 
above mentioned leakage compensation, which hampers the SB-pixels to generate longer series of hits. 

\iffalse
\begin{figure}[t]
  \begin{minipage}{0.65\textwidth}
	\begin{center}

    \includegraphics[viewport=0cm 2cm 25cm 19cm, clip,width=1\columnwidth]{Figures/ResponseSB.pdf}
		%\includegraphics[viewport=0cm 0cm 19cm 12cm, clip,width=1.05\columnwidth]{Figures/RTS-Temperature.pdf}
	\end{center}
	\end{minipage}
	\hspace{0.05 \textwidth}
	\begin{minipage}{0.3\textwidth}
	\begin{center}
    \caption{RTS-signatures in a SB-pixel. The dark voltage of the pixel $U=U_{1}$ and the indicated signal charge $U_{\rm CDS}$ are shown. Due to the dead time of the readout system, only the charge indication related to one edge of the RTS-signal was recorded.
    }\label{fig:SBResponse}
  \end{center}
	\end{minipage}
\end{figure}
\fi

%In \cite{Paper:MichaelRTS}, the number of RTS-pixels was measured as function of the radiation dose and the temperature for a sensor relying on LR epitaxial layer and a standard SB-pixel. According to the indicated results of the measurements, most SB-pixels show occasional signatures of RTS from the start. The fraction of indicated RTS-pixels decreases with high temperatures and radiation doses. This suggests that the above mentioned detection issues apply as both parameters increase the static leakage current. For this reasons, the authors of the paper consider the measurement as non-reliable. As displayed in Fig. \ref{fig:SBFakeHits}, SB-pixels exhibit indeed a by about one order of magnitude lower dark hit rate than about similar 3T-pixels. 

\subsubsection{RTS in MOSFET-transistors}
\label{SectionRTSinFET}

Besides in diodes, RTS can be found in FET transistors. Here, RTS modulates the source-drain current of the transistor. This RTS is among others caused by traps in the conductive channel nearby the gate of the transistors. The charges of those individual traps add to the gate voltage and thus modulate the electric current passing the device. The amplitude of the RTS depends on the precise position of the trap in the FET, which turns into a wide and continuous range of possible amplitudes. The superposition of multiple RTS sources is considered to cause the so-called 1/f - noise \cite{RTSinFET1}. 
%
%\paragraph{Properties of the RTS}
The typical current amplitude of the RTS signal is given in \cite{RTSFETPaper2} with:
\begin{equation}
\frac{\Delta I_D}{I_D} = \eta \cdot \frac{g_m}{I_D} \cdot \frac{q}{W\,L\:\:C_{ox}} \cdot \left (1-\frac{x_t}{t_{ox}} \right ) 
\label{Eqn:RTSFETAmplitude}
\end{equation}
Here, $I_D$ is the drain current, $\Delta I_D$ the related RTS-amplitude, $\eta$ a process constant, $g_{m}$ the transconductance of the MOSFET, $q$ the elementary charge, $W$ and $L$ the gate dimensions, $C_{ox}$ the MOSFET gate oxide capacitance, $x_t$ the distance between the the trap and the Si/SiO$_2$ interface and $t_{ox}$ the gate oxide thickness. From the equation, one concludes that the RTS amplitude increases with shrinking length and width of the transistor gate. This can be intuitively understood as a consequence of the limited range of the fields caused by the individual traps in the semiconductor. In \cite{RTSinFET1} it is reported that this amplitude scales typically with $1/L^2$, where $L$ denotes the length of the transistor gate. This is compatible with Equation \ref{Eqn:RTSFETAmplitude} as $C_{ox}$ scales with $L$.

The amplitude of RTS in transistors is simulated in \cite{RTSinFETSimulation}. It becomes particularly important for transistors working in weak inversion at drain currents of $\lesssim 1~\rm \upmu A$. The predicted RTS-amplitude may scale from $\sim 5\%$ to 40\% of the drain current for transistors with  $100 \times 100 ~\rm nm^2$ and $30 \times 30 ~\rm nm^2$. Besides of the distance between the Si/SiO$_2$ interface and the trap, the location of the trap along length of the gate is a decisive parameter for the RTS amplitude. A maximum amplitude is reached by traps in the region of the minimum electron density in the middle of the gate.  

MOSFET RTS affects the input stage of the on-pixel pre-amplifier of CMOS pixels \cite{RTSinFET3}, which is the source follower transistor for 3T-pixels and SB-pixels. A naive design of this transistor aims for a low drain current and tiny gate sizes, which reduces the power consumption and capacitive noise of the device respectively. Reducing $t_{ox}$ increases in first order the tolerance to ionizing radiation. However, each of those ``optimizations'' increases the RTS amplitude.

\begin{figure}[t]
\begin{minipage}{0.59\textwidth}
\includegraphics[viewport=1cm 0cm 26.2cm 20cm, clip, width=\textwidth]{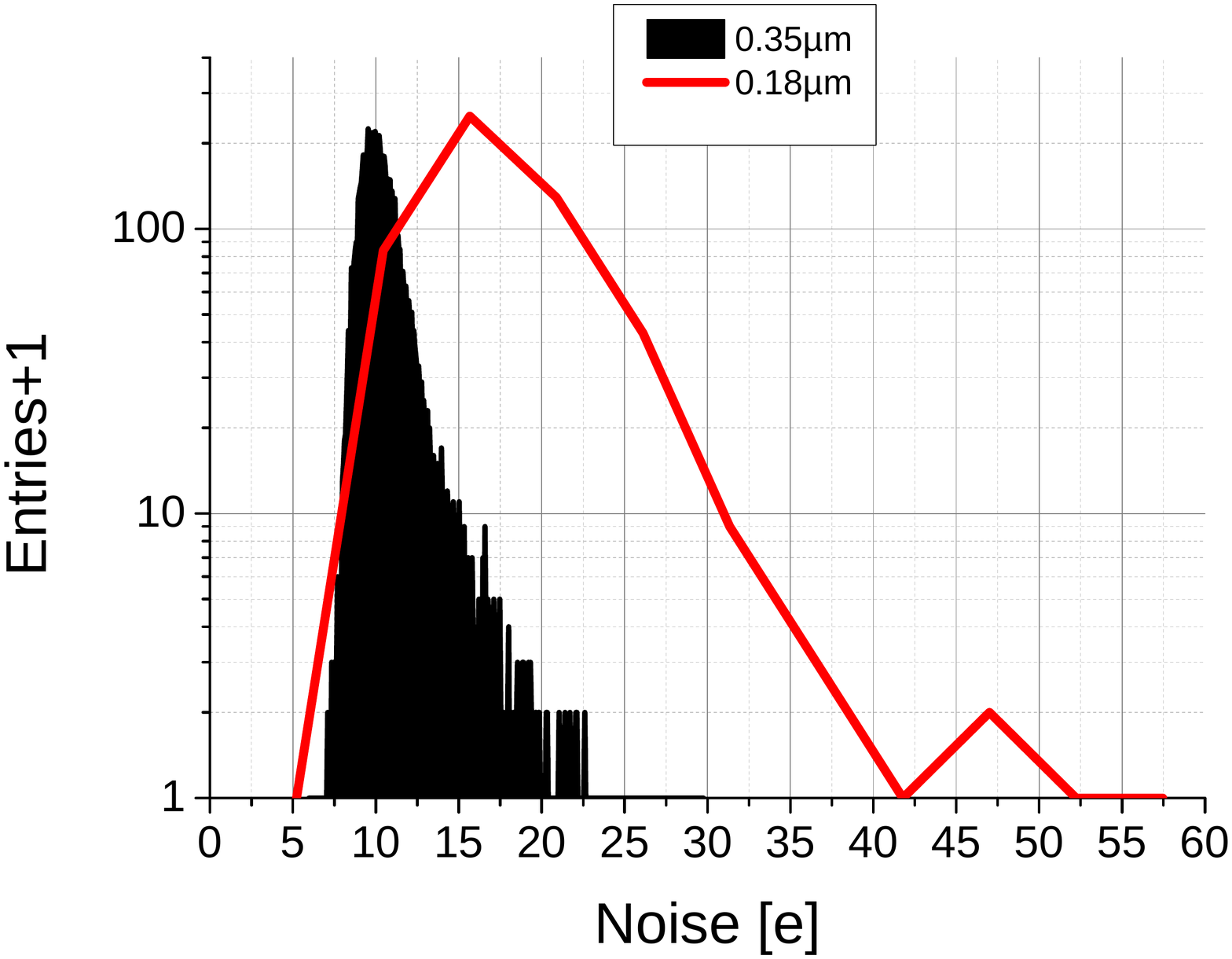}
\end{minipage}
\hspace{0.02\textwidth}
\begin{minipage}{0.35\textwidth}
\caption{Noise distribution (noise vs. number of pixels showing this noise) of two CPS with SB-pixel engineered in a $0.35\mum$ and in an $0.18\mum$ CMOS process.
Both imagers (MIMOSA-18AHR and MIMOSA-32ter) were equipped with collection diodes with $11\mum^2$ surface. The gate length of the SF-transistors was $0.35\mum$ and $0.2\mum$ respectively. From \cite{PaperPhD-Dennis}.}
\label{fig:Mi18Mi32Mi34Noisevergleich}
\end{minipage}
\end{figure}

The consequence of this behaviour is shown in Fig. \ref{fig:Mi18Mi32Mi34Noisevergleich}. The figure shows a histogram of the pixel noise of two reasonably similar SB-pixel architectures, which were implemented in a $0.18\mum$ and a $0.35\mum$ CMOS process, respectively. The gate length of the SF transistors was essentially chosen to amount this minimum feature size, which turns for a smaller pixel capacitance of the "$0.18\mum$-pixels". This lower capacitance improves the charge-to-voltage conversion gain of the pixel and thus was expected to amplify the signal while keeping the noise (in units of mV) constant. Based on this consideration, one expects a drop in the noise (in units of $e ~\rm ENC$). This advantage can indeed be seen for some of the "$0.18\mum$-pixels". However, due to RTS, most of those pixels exhibit a substantially higher noise than the  "$0.35\mum$-pixels". This reduces the overall performance of the device as CPS use typically a common threshold, which has to be set according to the noise of the most noisy pixels. It was shown, that increasing the surface of the transistor gate of the SF-transistor in SB-pixels alleviates the issue \cite{PaperMi32RadHardness}. Increasing the gate surface of CPS manufactured in $0.18\mum$ technology was found to reduce the pixel noise decisively \cite{Paper:AliceItsTDR}. 

Little is known on the impact of radiation damage on the RTS noise of MOSFETs. A pioneering study carried out with a n-channel poly-SI gate MOSFET and an effective gate size of $L=1.2 \mum / W=0.5 \mum$ is presented in \cite {RTSRadDamageFET1}.  The study reports an appearance of multiple RTS states after applying an X-ray dose of up to \mbox{$\sim 1 ~\rm Mrad$}. Simultaneously, the disappearance of some RTS states existing prior to the irradiation was observed. Most of the radiation effects could be recovered after annealing the device at $T=400^{\circ}C$.

A first study on a commercial CPS manufactured in a $0.18~\mum$ CMOS process was done by exposing the device to up to $2.19~\rm Mrad$ X-rays \cite{RTSinFET4}. The sensor was operated with a integration time of $2\mus$ and a mean noise of $194~\rm \upmu V$ is reported. About 0.4\% of the 3T-pixels of the sensor exhibited a significant MOSFET-RTS prior to irradiation. This number was increased starting from a dose of few $100~\rm krad$ and by up to 50\% after the highest dose applied. As for the previous study, modifications of the RTS amplitude and frequency are reported. 

\subsection{Effects of thermal and cold neutrons}
\label{SectionBorIrradiation}

\begin{figure}[t]
\begin{minipage}{0.59\textwidth}
\includegraphics[width=\textwidth]{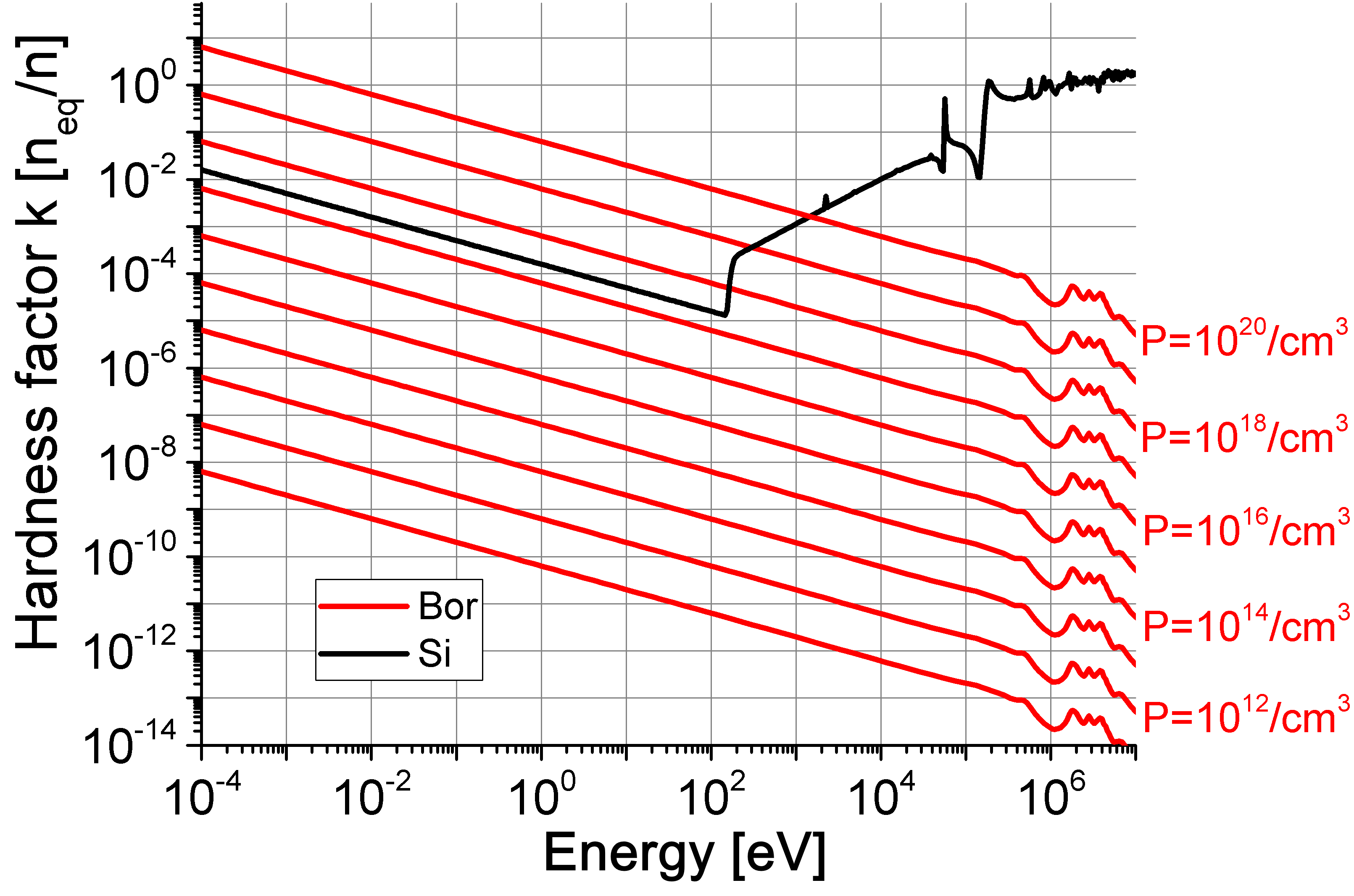}
\end{minipage}
\hspace{0.02\textwidth}
\begin{minipage}{0.35\textwidth}
\caption{Estimated hardness factor of neutrons impinging pure silicon (black, data from \cite{NIEL-compilation}), predicted damage due to boron fission (red). Warning: Model predictions do not match experimental data, see text. From \cite{PaperThermalNeutrons}.}
\label{fig:hardnessfactorwithdoping}
\end{minipage}
\end{figure}

As discussed in Sec. \ref{SectionNonIonizingTheory}, the NIEL-model neglects by construction effects of the device chemistry and possible reactions of trace elements found in the irradiated silicon. This was suspected to generate a bias in the radiation tolerance studies of CPS, as their p-doped epitaxial layer is typically doped with boron. Being exposed to cold and thermal neutrons, the dopants show a huge cross-section for the neutron induced fission \mbox{(n + ${}^{10}$B $\rightarrow ^7$Li + $^4$He + 2.8~MeV)}
and thus add a radiation energy, which is ignored by the NIEL-model. 

Studies addressing the question on whether this fission may generate significant radiation damage were carried out with with n-p-n transistors realized in the DMILL process \cite{BulkDamageInDMILL, RadiationDamageBipolar}. The transistors were irradiated with fast and thermal neutrons and the radiation damage was compared. It was observed that thermal neutrons damage the transistors stronger than 1 MeV neutrons, which is in strict contradiction to the standard NIEL tables.  

The question, if this effect may generate additional radiation damage in CPS, was addressed in a theoretical and experimental pilot study \cite{PaperThermalNeutrons}.
The NIEL caused by the fission process was estimated for natural boron based on the energy dependent cross-section of the neutron capture process causing it.
The fission energy of 2.8~MeV was distributed by means of energy and mass conservation to the ions. The frequent generation of a $\gamma$-ray in the process was neglected. The movement of the ions in the silicon and the bulk damaged
was simulated with the dedicated software SRIM \cite{PaperSRIM}. The hardness factor of the ions was estimated by comparing the number of vacancies indicated by
this software with the one of protons simulated with the same device. The results of the study are displayed in Fig. \ref{fig:hardnessfactorwithdoping}. 
%
\iffalse
\begin{figure}[t]
   \centering
  \includegraphics[width=0.5\linewidth]{Figures/haertefaktor_v2.png}
\caption{ Estimated hardness factor of neutrons impinging pure silicon (black, data from \cite{NIEL-compilation}), predicted damage according due to boron fission (red). Warning: Model predictions do not match experimental data, see text. From \cite{PaperThermalNeutrons}.
 \label{fig:hardnessfactorwithdoping}}
\end{figure}
\fi
%
They suggest that the effect of boron fission may provide a significant contribution to the over-all NIEL starting from doping concentrations above $P=10^{17}\rm/cm^3$. 
This value is not reached in the sensitive volumes of CPS but e.g. in the substrate of the sensors and in the P-Wells / P-Diffs used for implementing transistors. Therefore,
one does not expect the sensitive volume of CPS to take particular damage in first order. However, according to the simulation, the fission products show a certain range ($6 \mum$ for $^{4}$He and $3 \mum$ for $^{7}$Li). Therefore, ions created in the highly doped structures may create indirect damage in the more moderately doped structures.
\begin{figure}[tp]
\centering
%1
\begin{subfigure}{.45\textwidth}
    \vspace{-0.1cm}
	\centering
  \includegraphics[width=\textwidth]{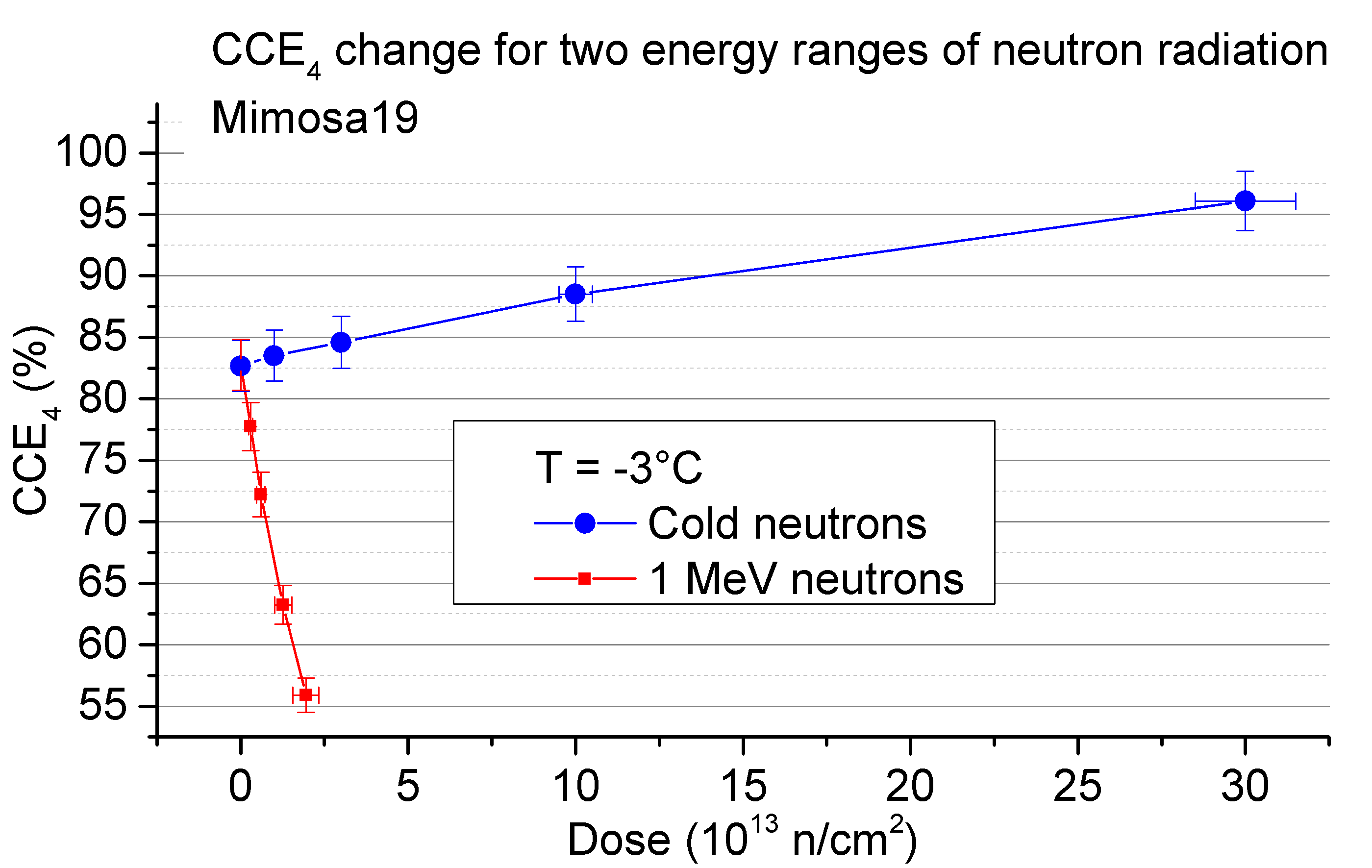}
  \caption{The charge collection efficiency of MIMOSA-19 for the four most significant pixel in a 5x5 cluster as a function of the 
	neutron dose and neutron energy. From \cite{PaperThermalNeutrons}.}
  \label{fig:CCEthermal}
\end{subfigure}%
\hspace*{0.05\textwidth}
%2
\begin{subfigure}{.45\textwidth}
  \centering
  \vspace{-0.5cm}
	\includegraphics[width=.9\textwidth]{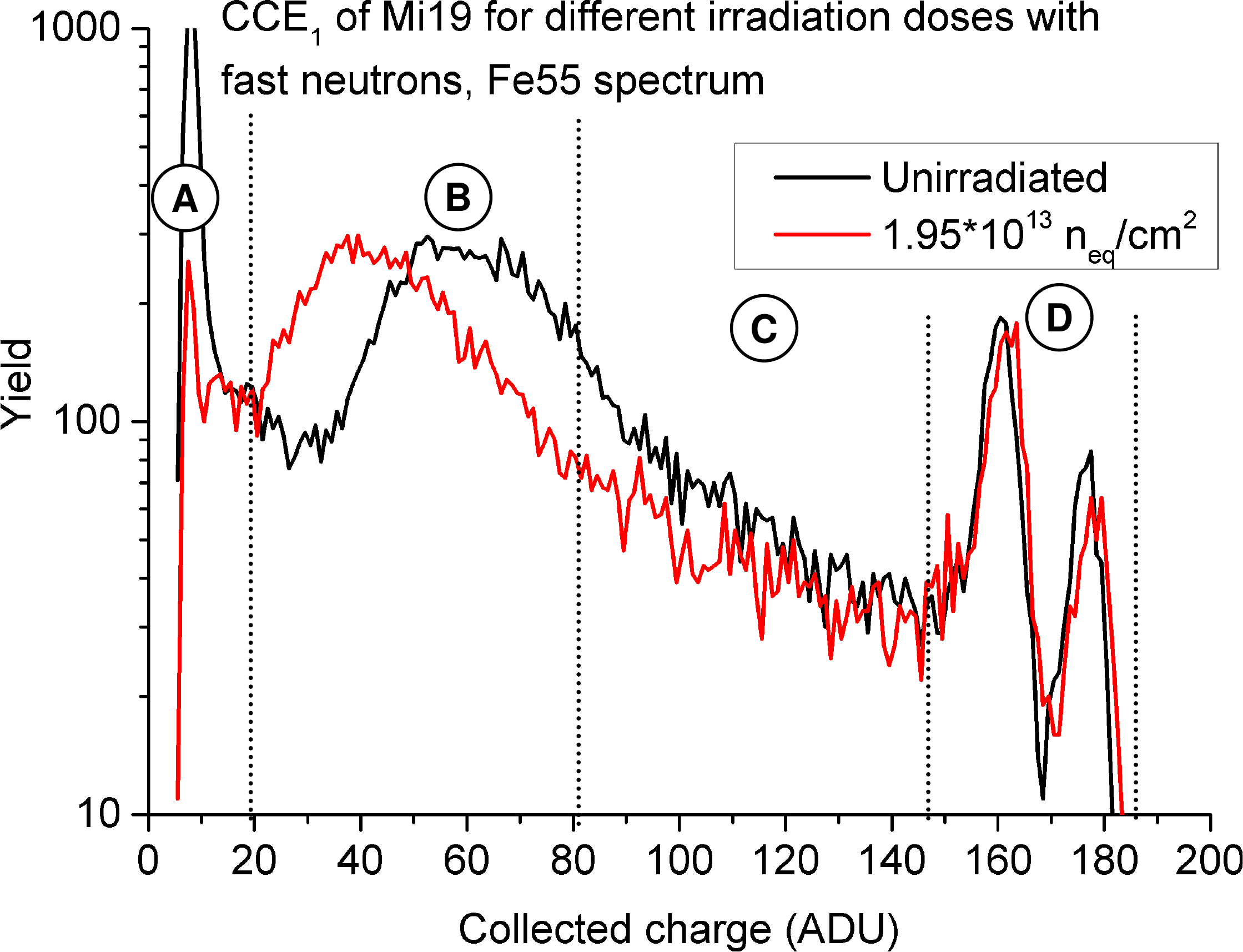}
  \caption{Amplitude spectrum of a non-irradiated MIMOSA-19 and of a MIMOSA-19 sensor after irradiation with $1~\rm MeV$ neutrons. 
	From \cite{PaperThermalNeutrons}.
  \label{fig:CCE1Fast}}
\end{subfigure}%https://www.overleaf.com/6458423ktgndt#
\newline
%3

\begin{subfigure}{.45\textwidth}
  \centering
	\vspace{0.6cm}
  \includegraphics[width=.9\textwidth]{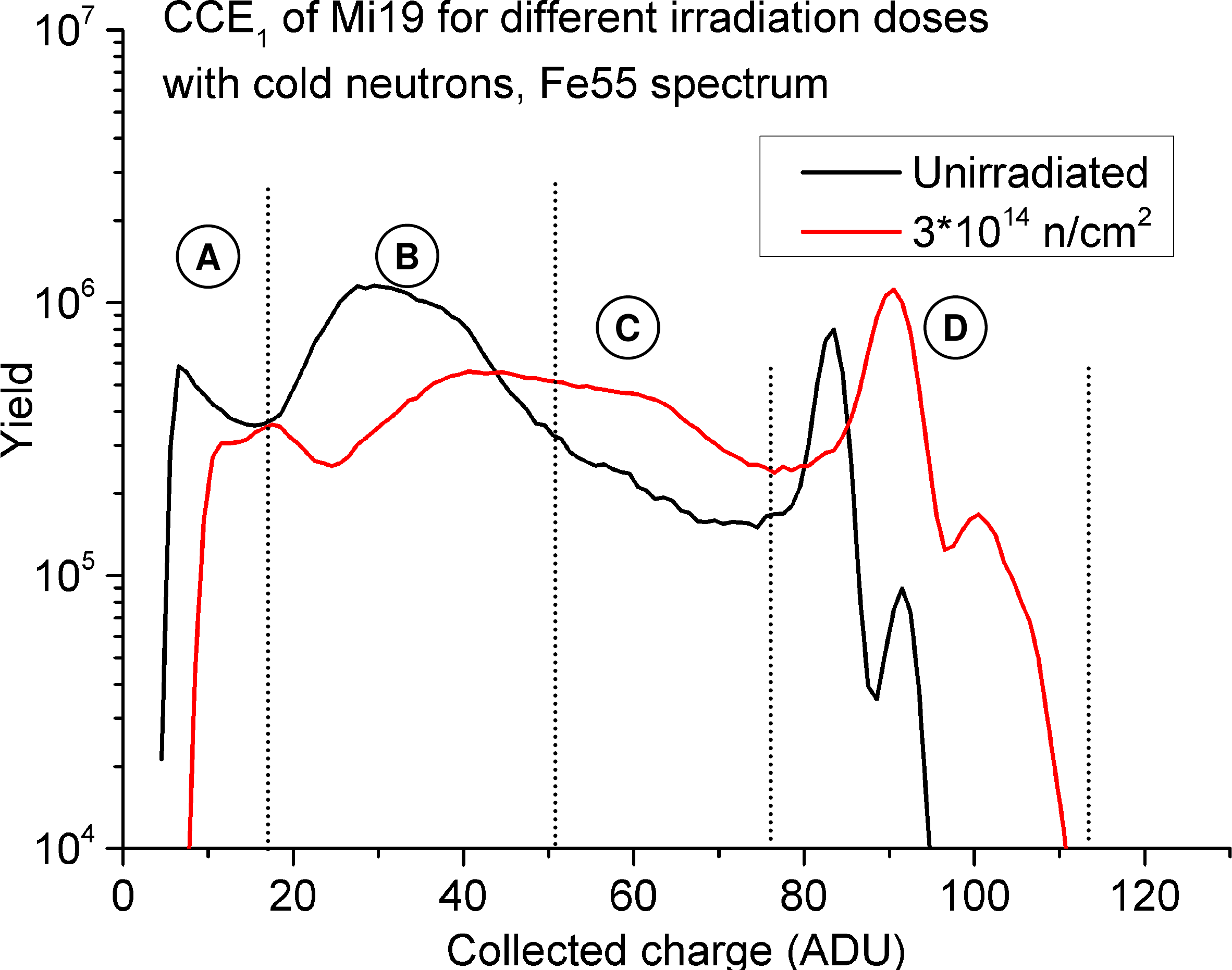}
%	\vspace{-0.7cm}
  \caption{Amplitude spectrum of a reference sensor and a sensor irradiated with cold neutrons. From \cite{PaperThermalNeutrons}.
  \label{fig:CCE1Thermal}}
\end{subfigure}
\hspace*{0.54\textwidth}
%4
\iffalse
\begin{subfigure}{.45\textwidth}
	\vspace{0.6cm}
  \centering
	   \includegraphics[viewport=1cm 0cm 28cm 19.8cm, clip,width=1\columnwidth]{Figures/RESMDD-Fe.pdf}
    \caption{Seed pixel spectrum of MIMOSA-18 (AMS $0.35\mum$, SB-pixels with  \mbox{$10\times10 \mum^2$} pitch)
			with low resistivity ($\sim 10 \ohmcm$) and high resistivity ($\sim 1000 \ohmcm$)
		epitaxial layer before irradiation and after irradiation with $1~\rm MeV$ neutrons. From \cite{DennisMi18AHR}.}
    \label{fig:Fe55Mi18NeutronIrrad10mu2}
\end{subfigure}%
\fi
	
	\caption{Charge collection and $^{55}$-Fe spectra of sensors exposed to cold ($1.8~\rm meV$) and fast ($\sim 1 ~\rm MeV$) neutrons. 
	Mind the scale while comparing the spectra.}

\end{figure}

The experimental part of the study was carried out with MIMOSA-19 sensors, which rely on a 14$\mum$ thick, LR epitaxial layer with an anticipated P-doping of few $10^{15}/\rm cm^3$. The $196\times196$ 3T-pixels with 12x12$\mum^2$ pitch have L-shaped diodes with 39.6$\mum^2$ surface. The anticipated doping of the substrate of the P-Well is $\sim10^{19}\rm/cm^3$. 
The chips were irradiated at the FRM II research reactor in Garching. The MEDAPP beam line~\cite{MEDAPP} provided fast neutrons (a direct fission spectrum with $E_{\rm neutron}=2~\rm MeV$ (peak) and a hardness factor of $k\approx 1$ (1 MeV neutron equivalent per neutron), $E_{\rm neutron}>100~\rm keV$ for 99\% of all neutrons). Unwanted $\gamma$-rays caused an ionizing dose of \mbox{$<100~\rm krad$} per $10^{13} ~\rm n/cm^2$. Other sensors were irradiated at the PGAA beam line \cite{PaperPgaa} with cold neutrons ($E_{\rm neutron} = 1.8\times 10^{-3}~\rm eV$ (peak), $k\approx$0.003 N$_{\rm eq}$) and an unknown ionizing dose. The irradiation was done at room temperature. Despite the sensors remained unpowered during irradiation, the ionizing radiation damage might determine the leakage currents after irradiation.

The sensors were bonded, illuminated with X-ray sources ($^{55}$Fe$\rightarrow 5.9~\rm keV$, $^{109}$Cd$\rightarrow \rm 22.1~keV$) and their amplitude spectra were recorded. 
%according to our standard protocol (see \cite{MichaelPhD}). The summed amplitude spectrum of groups four pixels of X-ray clusters are known to provide a good relative measure of the charge collection efficiency (CCE) of neutron irradiated sensors. The CCE of MAPS with 3T-pixel was previously found not to be significantly modified by ionizing radiation damage (see e.g. \cite{MichaelPhD}, section 5.3.3 and 5.3.4.3). Comparing CCE measurements done with $^{55}$Fe and $^{109}$Cd was confirmed to be appropriate due to the linear response of the sensor.
The results of the study are displayed in Fig. \ref{fig:CCEthermal}. As expected, the sensors irradiated at MEDAPP display a \emph{decrease} of the average CCE with increasing radiation dose. Unexpectedly, the sensors irradiated at PGAA show an \emph{increase} of the CCE. The slope of the CCE does thus depend on the neutron energy, what cannot be explained with a trivial quantitative modification of the NIEL scaling and thus also contradicts the theoretical results displayed in Fig. \ref{fig:hardnessfactorwithdoping}. 

An analysis of the related $^{55}$Fe amplitude spectra suggest an explanation for this finding. As illustrated in Fig. \ref{fig:CCE1Fast}, the sensors exposed to the fast neutrons show the usual response to the radiation load (compare Sec. \ref{Sec:LRCCE}), which is a reduced charge collection efficiency and an unchanged calibration peak. The latter suggests that the properties of the photo-diode including the doping of the epitaxial layer was not changed. The response of the sensor, which was irradiated with cold neutrons, is qualitatively different. As displayed in Fig. \ref{fig:CCE1Thermal}, the calibration peak of the amplitude spectrum gains in number of entries after irradiation and its position is displaced. Moreover, one observes an improved CCE in the epitaxial layer, which manifests itself by a displacement of the big charge collection peak toward higher values. Those signatures cannot be explained by simple effects like e.g. gain modifications of the related amplifier. Instead, they match the 
findings on sensors with high resistivity active volume as illustrated in Fig. \ref{fig:Fe55Mi18NeutronIrrad10mu} (note that the shift of the calibration peak is suppressed in this figure by construction). This suggests that 
the initially high p-doping of the LR-epitaxial layer of the MIMOSA-19 studied might have been substantially reduced. This would expand the depleted volume of the diode, and such simultaneously reduce its capacity (as indicated by the shift of the calibration peak $\rightarrow$ higher gain $\rightarrow$ less $C$), increase the number of hits recorded in the depleted volume (more entries in the calibration peak) and increase the CCE of the epitaxial layer. 

The number of entries of the calibration peak increases by slightly more than a factor two after irradiation. This suggest a the doping to drop by a rough factor of five (assuming flat PN-junctions, higher values may hold for the small charge collection diodes of MIMOSA-19). A similar radiation induced acceptor removal was seen before (see Sec. \ref{Sec:Neff}). However, to match the observations from MIMOSA-19 with literature, one would have to assume that cold neutrons show a hardness factor of about unity. This fits the findings on the DMILL transistors reported earlier but is two to three orders of magnitude above the predictions of the NIEL-model. 
Despite the details remain to be understood, it is safe to state that the damage caused by thermal neutrons in CPS is not predicted correctly by the NIEL-model. Dedicated tests are thus recommended before operating CPS in a harsh radiation field caused by thermal neutrons.
 
%----------------------------------------------------------------------------------------------------------------------------------------------

\subsection{Radiation damage caused by relativistic heavy ions}

Vertex detectors are typically installed very close to the beam axis. This holds in particular for fixed target experiments,
where the distance between beam axis and sensors amounts as few as $3~\rm mm$ (realized in the SAVD of NA61/SHINE) or $5.5 ~\rm mm$
(foreseen for the MVD of CBM). Due to this small distance, the sensors are exposed to primary beam particles
from the so-called beam halo. Moreover, they may suffer from impacts of the full beam during beam tuning or in case
the beam steering system fails. This is particularly worrying in heavy ion experiments as each beam ion has the potential to created
irreversible radiation damage. Obvious methods for protecting the sensors consist in collimating the beam, adding an emergency
stop system detecting beam displacements to the accelerator and installing the vertex detector on a moving table. The latter
allows to move the detector out during beam tuning but it is typically too slow to protect the device in case of an unwanted beam displacement.

%The width of the beam halo depends on the properties of the individual accelerator and beam line and must be estimated by measurement or based on input
%of the accelerator team. In any case, the issue should be taken serious if the distance between beam centre and sensors is below the order of few $\rm cm$.
%Reducing the width of the beam spot by collimators located close to the experiment is non-trivial. Besides of absorbing the ions, the collimators may generate unwanted nuclear fragments and particle showers. Moreover, their borders may scatter additional beam ions out of their initially acceptable path into the halo region.

According to the Bethe-Bloch-equation, the energy deposit caused by a beam ion with a 
charge $Z$ scales with $Z^2$. Consequently, the energy deposit of a Pb-ion in silicon exceeds the one of a proton
by close to four orders of magnitude. Depending on the structure of the beam halo, the total ionizing and non-ionizing 
doses induced by the beam ions may provide even a dominant contribution to the total radiation exposure of the sensors. 
 Moreover, due to the high charge deposit, even individual ions may generate damaging effects in a CMOS device.
Those are referred to as single event upsets (SEU). Relevant SEUs in CPS are changes of the logic state of a memory cell
(bit flip) and the generation of a meta-stable short circuit in the device (latch-up). While the bit flip may be tolerated in most cases,
latch-ups may destroy CPS irreversibly and call for protection measures.

\subsubsection{Single event upsets and latch-up}

SEUs occur, if a particle deposits a sufficient amount of charge at a, typically small, sensitive point of a CMOS device. 
The probability of obtaining an SEU is typically expressed as a cross-section, which depends on the nature and
energy of the impinging particle. 

The cross-section for the occurrence of single bit flips caused by ionizing particles was studied during the preparation
of the the ALICE Inner Tracking System \cite{Paper:AliceItsTDR}. Various memory cells manufactured in $0.18 \mum$ CMOS technology
were exposed to slow (25-250 MeV) protons, which show a charge deposit similar to the one of relativistic heavy ions. 
The cross-section for bit flips was found to amount in the order of $10^{-13}~\rm cm^2/bit$. 
This value is not considered as particularly worrying for data memories as individual bit errors in the data stream of the sensors can typically be tolerated. 
More care is needed for the registers storing the information required for the internal control of the sensors. A bit flip in those registers may disturb
the sensor operation, which has to be detected and to be cured by restarting the device. In order to reduce the down-time of the sensors, 
the ALICE collaboration decided to protect those registers. This can for example be done by storing each bit in three memory cells and by deciding
about the correct state by majority voting. A bit flip in individual cells can then be tolerated and cured on the fly.

\begin{figure}[t]

  \centering
  \includegraphics[width=0.5\textwidth]{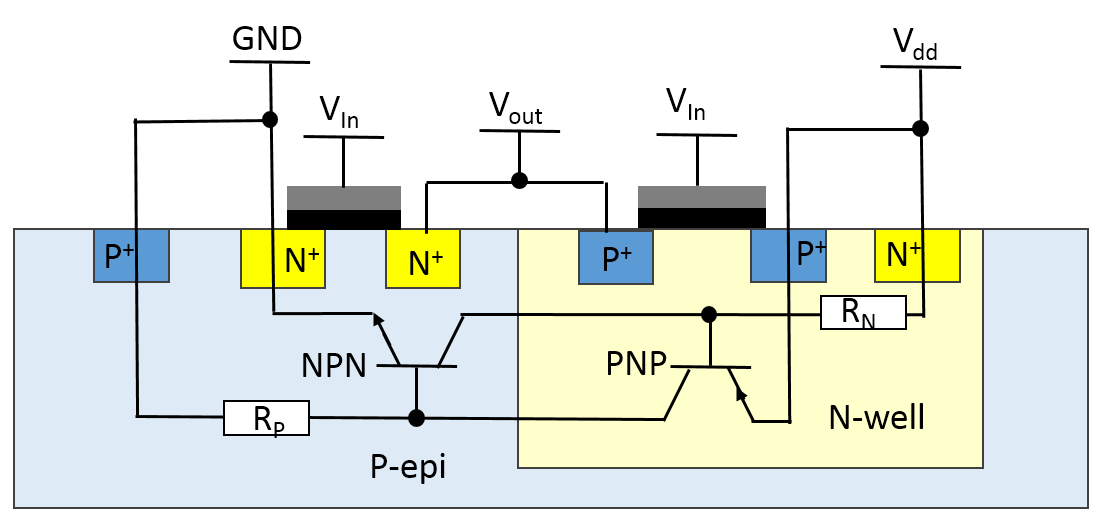}
	\vspace{0.6cm}
	
	\caption{
CMOS inverter and bipolar transistors making the structure vulnerable for latch-up. After an illustration by "Cepheiden" found on Wikipedia.}
     \label{fig:LatchUp}

\end{figure}

Latch-ups occur in vulnerable CMOS structures as for example in the CMOS inverter displayed in \mbox{Fig. \ref{fig:LatchUp}}. 
This inverter is composed from a PMOS and an NMOS field effect transistor, which are realized by N+ implantations and an N-well with P+
implantations respectively. As shown in the figure, this CMOS structure realizes unintentionally a combination of two bipolar transistors,
which are switched of in the normal state of operation.
This may change in case a major current is generated in the reverse biased P-epi/N-well junction. In this case, the resistivity of the 
silicon material (represented by $\rm R_P$ and $\rm R_N$ in the figure) generates a voltage drop. This voltage drop may be sufficient 
to switch one of the parasitic transistors into a conductive state. In this case, the voltage drop at the opposite resistor is increased
and the second bipolar transistor becomes conductive, which increases the initial current further. Finally, both bipolar transistors are 
stabilized in the conductive state. Therefore, rather large currents may pass between $\rm V_{dd}$ and $\rm GND$ and the ohmic losses
at $\rm R_P$ and $\rm R_N$ may heat up the structure until it is thermally destroyed. The conductive state may be switched back to the
initial, non-conductive state by switching off the power of the device and restarting it. If this action is undertaken with sufficient speed, no
damage is to be expected. 

The probability of latch-up was studied \cite{MichalSzelezniakPhD}  with the MIMOSTAR-2 sensor (a sensor relying on SB-pixels) in the prospective of using the
device for the PXL detector of STAR. No latch-up was observed for LET values up to $8~\rm MeV \cdot cm^2/g$. Hereafter, a sharp
increase of the related cross section was observed and a cross section of roughly $5\times 10^{-4} \rm cm^2$ is reported for a LET above
$30~\rm MeV \cdot cm^2/g$. According to the study, also soft upsets were observed. The related cross section was  $2\times 10^{-7}$ to
$7\times 10^{-7} \rm cm^2$ at the highest LET applied ($13~\rm MeV \cdot cm^2/g$), which is a small factor above the related latch-up cross 
section.

Latch-up was found to be the major source of losses of sensors in the STAR-PXL detector, which was carried out with ULTIMATE CMOS sensors \cite{PaperUltimateSensor,Paper:StarPixel}.
The sensors were protected against damage caused by latch-up by a dedicated power supply, which was measuring the current consumed by a ladder of 10 sensors and by
performing a power cycle, if this value exceeded a certain threshold value. However, in a first run carried out in 2014, 14\% of the active surface of the detector was lost during the first two weeks of operation \cite{Paper:STAR-Latchup}. The damage was concentrated to the most inner layer of the detector. After reducing the activation threshold of the latch-up protection system from initially $400~\rm mA$ to $120~\rm mA$ above the nominal current, only four additional sensors were lost during the 10 remaining weeks of operations. Operating the sensors with an over-current protection threshold of $80~\rm mA$ above the nominal current consumption reduced the number of lost sensors to 5 during each, the RHIC run in 2015 and 2016.

A follow-up study was carried out by bombarding individual sensors and ladders with protons and heavy ions produced by the 88" Cyclotron BASE Facility at LBNL. Current limited latch-up states with a typical increase of the digital current of $300~\rm mA$ were observed. A permanent damage could only be reproduced in sensors, which were previously thinned to $50 \mum$. The $700 \mum$ thick sensors studied were found to be reasonably robust, maybe because the additional material reduces the thermal stress by conducting the heat away from the location of the latch-up. The location of the damage structures was identified through infrared camera inspection. The substrate and the epitaxial layer of the damaged sensors were removed in the region of interest. Electron microscope pictures of the damaged region revealed that the metal lines in the damaged regions were melted down.

\subsubsection{Integrated displacement damage caused by heavy ions}

While the total ionizing dose caused by relativistic heavy ions in silicon can be estimated based on the Bethe-Bloch-equation, little is known on  the related integrated NIEL. 
Computations on the NIEL caused by up to $1~\rm GeV$ (not AGeV) heavy ions \cite{SlowIonRadDamagePaper} suggest that the radiation damage caused by one
slow gold ion may exceed $\sim 10^6\neqcm$. The numbers are in reasonable agreement with the experimental results reported in \cite{SlowIonMeasurements}. A recent experimental study on the damage caused by $40.5~\rm AMeV$  $^{40}$Ar-ions relying on depleted $\rm p^+-n-n^+$ silicon detectors \cite{IonArgonPaper} suggests that those ions generate a non-ionizing radiation damage equivalent to $\sim 10\neqcm$. As stated in the paper, this result is not compatible with the previously 
mentioned theoretical calculations and observations, which remains to be understood. In any case, the studies do not cover the ion energy of primary beam particles of relativistic heavy ion physics. 

An estimate for the NIEL
caused by relativistic ions is provided in \cite{xapsos2004niel}. In accordance with the focus of the work, the natural ion spectrum of cosmic radiation, this study limits itself to 
energies of $2~\rm A GeV$ to atoms up to $Z=26$ (Fe). The authors underline that nuclear fragmentation is to be considered at high ion energies. For light ions, namely for protons, the fragmentation of the target atom provides a significant contribution to the overall NIEL. For atoms with high atomic numbers $Z$, the NIEL scales approximately with $Z^2$. For $\gg 1~\rm AGeV$ energy range of interest for heavy ion physics, the NIEL seems to decrease with increasing particle energy, which would
be consistent with the NIEL tables for protons. A naive extrapolation of the values toward higher $Z$ values suggests, that the NIEL of a Pb-ion with an energy above $2~\rm AGeV$ should be below few $100 \neqcm$. 

The validity of this rough assumption was cross checked with an experiment \cite{Paper:NA61Annual2016}. 
Four CPS (\mbox{MIMOSA-34}) were exposed to primary 
$30~\rm AGeV$ Pb-ions at the CERN-SPS. Those sensors were developed by the 
IPHC Strasbourg and manufactured in a TOWER/JAZZ $0.18 \mum$ CMOS process with $18~\rm \upmu m$ high resistivity epitaxial layer.
They host SB-pixel with various pixel pitches.  The tolerance of those pixels to $\sim 1~\rm MeV$ neutrons was measured
independently, which allowed to compare the radiation damage caused by the heavy ions with the one caused by the neutrons.

The sensors were intentionally not powered but wrapped into aluminium foil and mounted on a holding structure made from plastic material. 
They were placed $200~\rm m$ upstream the target in
the beam line of NA61/SHINE. Dosimetry was provided by an ion sensitive, $4 \times 4 ~\rm mm^2$
small scintillator. This scintillator together with its light guide and a suited, small PMT were placed on the support upstream the sensors. 
The signals of the PMT were forwarded to the NA61/SHINE counting house and the dosimetry was performed by means of single ion 
counting. 
During two weeks of operation, a maximum ion flux of $1.2 \times 10^{10} \rm ~ Pb ~ ions/cm^2$ was integrated. Hereafter, the sensors were bonded at the IPHC
Strasbourg and illuminated with X-rays from a $^{55}$Fe-source. No significant drop of CCE was observed, which suggests that the 
equivalent radiation damage caused by the ions was $\lesssim 3 \times 10^{12} ~\rm n_{eq}/cm^2$.  The noise of the pixel was also checked 
and again no significant noise increase due to irradiation was observed. This preliminary
finding provides an upper limit for the hardness factor of $30~\rm AGeV$ Pb ions, which is given with $k_{ion} \lesssim 300~ \rm n_{eq}$.
This upper limit is compatible with the above mentioned theoretical predictions. From the practical point of view, it allows to conclude that the integrated
non-ionizing radiation damage caused by relativistic heavy ions in CPS will typically be dominated by the related ionizing damage.
\section{Outlook: Toward depleted CPS (DMAPS)}
 \label{CPSOtherGroups}

\subsection{Depleting CPS, fundamental considerations}

Depleting the active medium of a CPS aims to collect the signal charge with drift fields instead of thermal diffusion.
Doing so accelerates the charge collection and focuses the signal charge to one central pixel, which both increases
the tolerance of the device to non-ionizing radiation damage.
The particular challenge of depleting the active medium of CPS consists in the fact that the epitaxial layer of the sensors
is thin as compared to the pixel pitch. Therefore, depleting a sensing element equipped with the traditional tiny collection diodes
is rather an issue of lateral depletion than of depleting
into the depth. Moreover, applying a classical back bias is hampered by the fact that the 
p-sub/p-epi/p-well structure of the standard CPS sensing element is conductive unless a full depletion is reached.

Attempts to generate fully depleted sensors follow currently two major streams. One consists in replacing the initially 
small charge collection diode by a big diode, which approaches the size of the pixel surface. This solves straight forwardly
both, the side depletion and the conductivity issue and forms flat junction with a depletion depth scaling with $\sqrt{U_{depl}}$. If combined with a thick
active medium, this sensor geometry may reach impressive depletion depths. However, the big diodes come with sizeable capacitances, 
which reduce the pixel gain and thus complicate to amplify a particle signal above the noise level.

Alternatively, it is being tried to apply a high voltage to the conventional, tiny diodes. This yields very low diode
capacities but depleting the pixel, namely to the side, is harder. This is as the field lines of the small diodes follow 
rather a radial than a parallel geometry. As a consequence, the depletion depth reached scales with $\sqrt[n]{U_{depl}}$.
The precise value of $n>2$ is so far not established and the equation will in any case provide a rough approximation
of the complex geometry only. One expects $n=3$ for an ideal point-like diode and values 
 as high as $n \approx 6$ were reported to fit measured data \cite{PaperPipper} (details see below).

The R\&D on fully depleted CPS targets a radiation tolerance, which is only exploited by sensors with very fast readout. 
The latter is reached by adding a discrimination logic into the pixel itself and by replacing the rolling shutter with
fast, asynchronous readout schemes. In simple words, the latter skip pixels, which are not hit and thus don't loose
time on reading them. Both, depletion and on-pixel signal discrimination cannot be done with 
classical double well CMOS processes. The progress in the field rely therefore on the impressive optimizations, 
which are introduced by the CMOS industry in order to serve the market of high-end imaging products. 

\begin{figure}[t]

  \centering
	   \includegraphics[width=\textwidth]{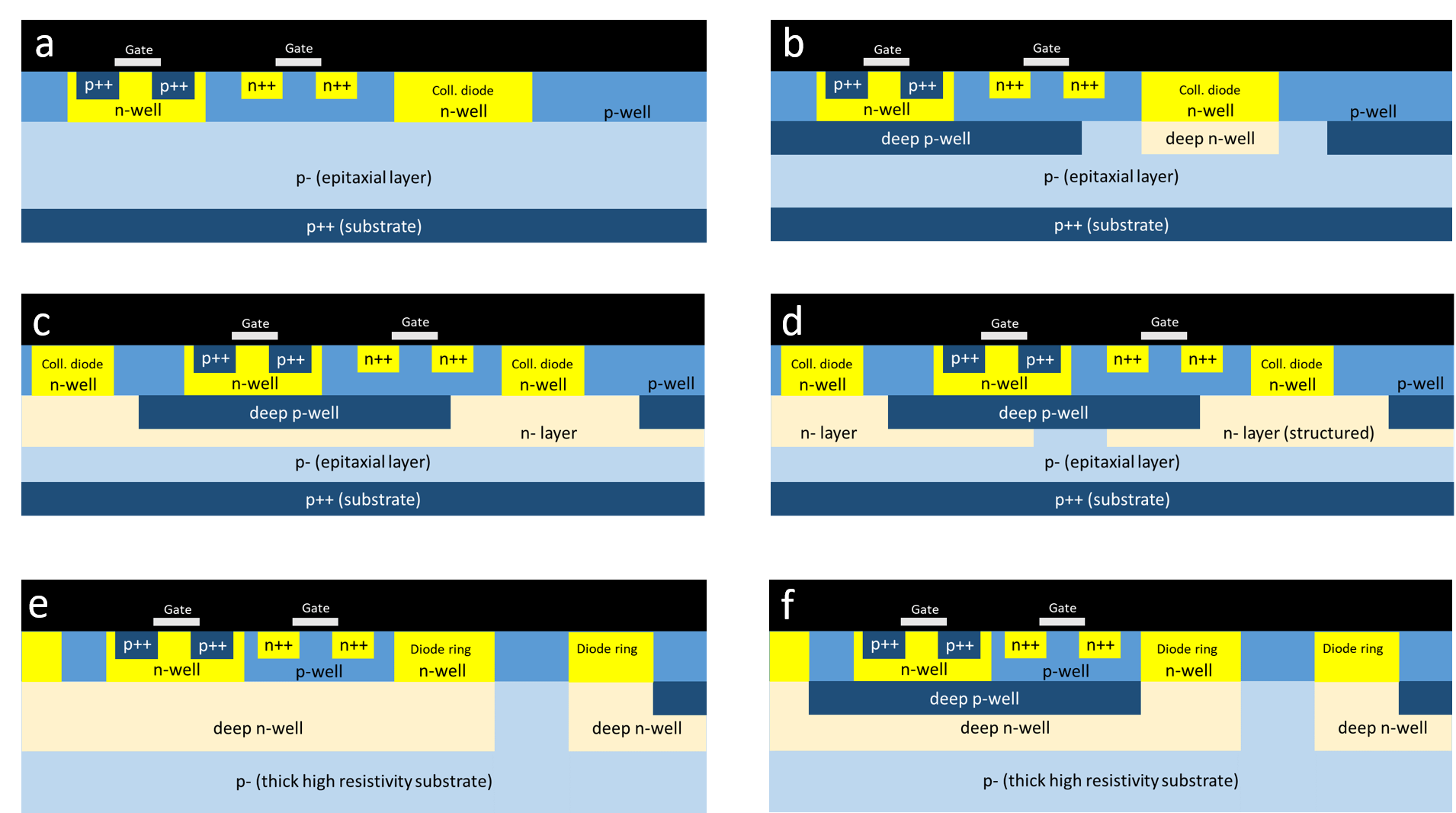}

    \caption{Features of different CMOS processes. The standard double layer process (a) features p- and n-wells only.
		The n-wells of PMOS transistors, as shown at the left, act as parasitic charge collection diode. Quad well processes
		(b) feature an additional deep p-well, which may separate transistor and epitaxial layer. The ``modified'' quad well
		process (c) adds a n-layer to the epitaxial layer, which eases depletion. The ``double modified'' process (d) 
		structures this n-layer, the lateral p/n-junction is predicted to provide lateral drift fields and to accelerate the charge 
		collection. Triple well processes (e) feature a deep n-well and PMOS transistors may be embedded into a sizeable collection
		diode made with this implantation. A more modern version of this pixel (f) uses again a deep p-well to separate the PMOS transistor bulk and the diode.}
		
    \label{fig:Processes}
\end{figure}

Improved CMOS processes come with the high resistivity epitaxial layer required for
obtaining a deep depletion. Moreover, as shown in Fig. \ref{fig:Processes}, they provide additional structured silicon layers 
beyond the standard diffusions and wells. As shown in Fig. \ref{fig:Processes} (b), the first additional layer (so-called
deep p- and n-well) allows for isolating the n-well of PMOS transistors and the epitaxial layer, which
allows for the use of full CMOS electronics in the pixel area and such to build fast pixels with
integrated discriminators. Recently, a CMOS process\footnote{This process is currently referred to as modified or double modified process in the 
jargon of the community.} providing an additional, fourth N-layer was designed (see Fig. \ref{fig:Processes} (c,d)). This layer is of particular interest for
depleting pixels as it is used for generating
planar and sizeable diodes within the epitaxial layer.

High voltage (HV)-CMOS processes are being used to bypass the limitation of standard CMOS in terms of maximum operation
voltage. The processes are intrinsically optimized for voltages of several $10 ~\rm V$, which allows for applying sizeable bias
voltages to the charge collection diode in a rather straight forward way.  In parallel, solutions for applying an up to $40~\rm V$
depletion voltage to pixels manufactured in standard CMOS processes were identified.

%The precise value of $n$ is not fully established. First observations and related analytic assessments \cite{LinnikPhD} suggest that $n$ might scale between
%$n=3$ (ideal point like diode) and $n=6$. 

From the experimental point of view, one has to consider that the
classical C-V test for measuring the depletion depth, as known from planar diodes, may not be reliable. This is as the underlying assumption of a flat junction is not fulfilled. Mapping out the 
active volume with an $^{55}$Fe-source (see Sec. \ref{SectionFe55Spectrum}) appears as a reasonable complement for estimating
the depleted volume from the data stream of the sensor. Useful additional information may be obtained from the leakage currents of mildly neutron irradiated samples (see Sec. 
 \ref{Sec:NonIonizingLeakageCurrents}) but first exploratory tests of the approach question its reliability (see details below). A saturation of the the depleted volume as indicated by C-V test, $^{55}$Fe-photons
and leakage currents with increasing depletion voltage does not necessarily indicate a full depletion. Complex diode/field geometries  may feature other reasons why additional voltage has mostly
no further power to extend the depleted volume. Two photon absorption TCT (TPA-TCT) \cite{TPA-TCT-Paper} is being established for performing a reliable direct measurement
of the shape of the depleted zone relying on laser illumination but requires additional instrumentation.

The development of next generation CPS is at present driven by applications in relativistic heavy ion physics (namely ALICE and CBM) and by
the vision to upgrade existing LHC experiments (namely ATLAS) with those sensors. The first application requires a tolerance of 
in the order of  $\sim 1 ~\rm MRad$ and $10^{13}$ to $10^{14} \neqcm$, which seems in reach of an advanced partial depletion. The targeted radiation
tolerance for ATLAS amounts rather $\sim 100 ~\rm Mrad$ and $>10^{15}\neqcm$, which clearly calls for full depletion.

In the following, some current R\&D projects are introduced and the results of the ongoing research is summarized.

\subsection{R\&D projects relying on small diodes}

\subsubsection{LePix}
A  first attempt to build fully depleted CPS was carried out within the LePix project \cite{LePixPaper1}. The sensor was intended 
for a use in LHC detectors, which requires a fast sensor readout. As no quad-well processes were available yet, no in-pixel discrimination could
be carried out. Instead, it was planned to connect each pixel with an individual discriminator located aside the pixel matrix.
This requires a high number of metal lines, which determined the use of a \mbox{90 nm} CMOS technology and a rather high pixel pitch of $50\mum$.
The pixels were modified 3T-pixels, which stored their output signal into two in-pixel capacitors. Those analogue memory cells were suited to 
hold the signal of both frames until the analogue readout was completed. The reset and the source follower transistor were implemented as PMOS transistors, which were embedded
into a deep N-well collection diode. The diode is biased with +0.6 V and the $\rm \sim 400~  \Upomega \cdot cm$ high resistivity substrate, which served as active medium. The latter 
was back-biased with a voltage of some \mbox{-10~\rm V}.

During laboratory tests, the diodes exhibited an unusually high leakage current of $3.5~\rm pA$. By operating the device with an integration time of $10 \mus$, the related shot noise was limited to a reasonable value of $\sim 15~ e ~\rm ENC$. However, a disappointing detection efficiency of around 70\% was observed in a beam test carried out with $300~\rm MeV$ pions. Based on the results of additional studies, it was concluded that some peripheral regions of the pixels show an insufficient CCE.

The radiation tolerance of LePix was also studied \cite{LePixPaper2}. Based on measurements of a $770 \mum^2$ diode, which was irradiated up to $2 \times 10^{15} \neqcm$, an radiation induced increase of the bulk current according to $\alpha = 3.18 \times 10^{-16} ~\rm A/cm$ is reported. This value exceeds the literature value for traditional detector grade silicon (see Sec. \ref{SectionNonIonizingTheory}) by an order of magnitue. This was interpreted as a consequence of not annealing the irradiated sensor prior to the measurement. The radiation induced increase of the doping concentration of the bulk material from $N_{\rm eff} \approx \rm 10^{13}/cm^3$ to few $\rm 10^{14}/cm^3$ is reported. This turns into an increase of the capacitance of the charge collection diodes, which became relevant starting from a dose of $10^{14} \neqcm$. The pixels were exposed to up to $10~\rm Mrad$ of soft X-rays and showed significant shifts of their response starting from $500~\rm krad$. For high doses, the sensor failed completely but recovered after 1 week of room temperature storage. The authors conclude that the CMOS process used is suited for producing sensors with a tolerance to $\sim 10^{14} \neqcm$ and $\sim 10~\rm Mrad$. As the non-irradiated device showed issues in particle detection, this potential was unfortunately not exploited by the LePix prototype itself.

\iffalse

\subsubsection{APSEL}

A first attempt  to bypass the limitations of standard double well CMOS processes and to realize pixels with a full \mbox{amplifier-,} shaper-, discriminator-chain
the pixels was carried out with the APSEL series \cite{RizzoMapsPaper1}. The sensors were designed with a triple well process, which provided a deep
n-well. This allows for placing NMOS and PMOS transistors into the deep n-well of the collection diode of the pixel. However, the bulk of the PMOS transistors 
is floating with this diode. A deep p-well as required for fully isolating the PMOS transistors was not available.

The $25 \times 25 \mum^2$ and $50 \times 50 \mum^2$ pixels CPS were equipped with 
large deep N-well diodes, which covered more than 85\% of their surface. The N-wells of the few
PMOS transistors required for building the discriminator were not explicitly separated from the epitaxial layer and acted thus as parasitic collection diode. However, it was hoped 
that their unwanted charge collection would remain marginal as compared to the charge collection of the main diode. The performance of the device was initially found to be 
limited by the high capacitance of the collection diode, which turned into a high capacitive noise. This issue could be solved by optimization and a S/N of
25 was reached. However, the detection efficiency for minimum ionizing particles remained limited to $\sim 90\%$ \cite{DNW-MAPS-Paper1}. This is likely
because the parasitic charge collection of the PMOS transistors could not be overcome in this technology. 

\fi

\subsubsection{ALPIDE}
\label{Sec:ALPIDE}
The ALPIDE sensor \cite{ALPIDE1,ALPIDE2} was developed for the upgrade of the ALICE Inner Tracking System \cite{Paper:AliceItsTDR}. The fully monolithic sensor relies on a 180nm quad-well process with high resistivity epitaxial layer and integrates rather granular pixels with a small diode and a pitch of $29.2 \times 26.9 \mum^2$. Each pixel integrates a discriminator. The readout is done via a priority encoder and may operate in a triggered and a self-triggered mode. Thanks to this innovative design, ALPIDE reaches a time resolution of $\sim 10\mus$, which exceeds the time resolution of elder sensors with rolling shutter by roughly one order of magnitude.

The sensor separates the electronics and the sensitive volume by means of a deep P-well implantation (see Fig. \ref{fig:Processes}(b)). It offers the feature to put the deep P-well and the epitaxial layer to a moderate depletion voltage of -6V. A second feature, which allows for a higher depletion depth, consists in separating 
the n-well implantation of the diode and the surrounding p-well structures. Like this, an intermediate layer of lowly p-doped silicon is generated aside the n-well. By doing so, the effective width of the charge collection diode is increased while its capacitance is reduced. 

As required by the ALICE-ITS, ALPIDE was designed to resist to $1.7 \times 10^{13} \neqcm$. A detection efficiency of far better than 99\% in combination with a very low dark hit rate of below $10^{-9}$/pixel/event was observed in beam tests after irradiating the sensor to this dose with neutrons. Unfortunately, no tests at higher doses were performed. The unusually low dark hit rate was reached by optimizing the input stage of the pre-amplifyer for low random telegraph signal (RTS, see Sec. \ref{SectionRTSinFET}). 

A remaining weak point of ALPIDE consists in its limited tolerance to ionizing doses, which is guaranteed to amount $\sim 500\krad$ for fast ($\gtrsim 100 ~\rm krad/h$) irradiations \cite{Hillemanns}. The radiation damage affects parameters like the internal threshold and the internal DACs generating reference voltages. Moreover, a drop in current consumption is observed after a dose of $300~\rm krad$. Most of the radiation effects were observed to recover after a few days. Their occurrence depends thus on the dose rate of the irradiation. Following up this observation, the radiation tolerance of ALPIDE is being restudied by irradiating the sensor with doses of $\sim 100 ~\rm krad$ about once per month. Preliminary results show that the sensors recover after each irradiation step, which was so far confirmed up to a dose of $2.5-2.7~\rm Mrad$. This does not hold for the observed radiation induced shifts of the discrimination threshold, which may however be compensated by the use of appropriate sensor settings. As a next step of the ongoing study, it is foreseen to validate the performances of the irradiated sensors by means of a dedicated beam test.

%In December 2017, an ALPIDE sensor was exposed $\sim 50\krad$ of highly relativistic Xe-ions at the CERN-SPS. The sensor was reported to remain in stable operation during this test and no crucial single event effects were observed \cite{PaoloPersonal}. The sensitivity of the irradiated device to minimum ionizing particles remains to be confirmed. 

\subsubsection{PIPPER-2}
\label{sec:pipper2}
\begin{figure}[t]

  \centering
	   \includegraphics[width=6cm]{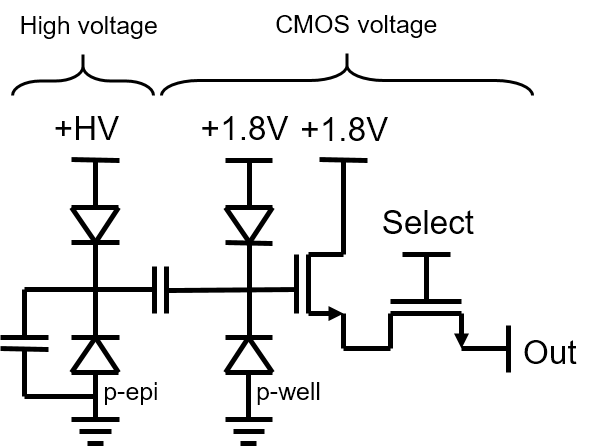}

    \caption{Schematics of the AC-coupled pixel of PIPPER-2.}
    \label{fig:AC-SB-PixelLayout}
\end{figure}

Pipper-2 \cite{PipperPaper} relies on the same Tower/Jazz 180 nm quad-well CMOS process as ALPIDE. It was developped for X-ray imaging
but also tested for radiation tolerance in the context of CBM. The sensor is equipped with SB-diodes (see Fig. \ref{fig:AC-SB-PixelLayout}), which are separated from the pre-amplifier with a 
high voltage tolerant capacitor. This allows to apply a depletion voltage of up to 40 V to the diodes while the sensitive transistors operate within their 1.8 V voltage limit.
The DC-working point of the gate of the source follower transistor is set by a second pair of diodes at the low voltage side. The related reversed biased diode 
does not reach the epitaxial layer and collects therefore no signal charge.
The chip hosts four submatrices of pixels with eight columns of 128 pixels. The pixel size is $22 \times 22 \mum^2$. The chip was realized on a $> 1\rm k \Omega \cdot cm$
epitaxial layer with a thickness of $18 \mum$ and on a wafer with high resistivity ($>600~\rm  \Omega \cdot cm$) CZ bulk substrate.

The sensor was operated with depletion voltages of up to 30 V and the depletion depth was estimated by measurements with X-rays and compared with simulations
done with TCAD. Simulation and measurement match and it is concluded that a depletion layer with a thickness of $12\mum$ is created for the sensors relying on the epitaxial layer. Below this 
depth, the doping concentration increases consecutively to substrate doping, which hampered reaching a further depletion.  For the sensors based on
the CZ-silicon, the depletion depth is extended to $17 \mum$. An analytical equation predicting the depletion depth to scale with $\sqrt[6]{U_{depl}}$ is reported to 
fit the data. 

A first pilot study on the tolerance of PIPPER-2 relying on the $18\mum$ thick epitaxial layer to non-ionizing radiation is being carried out with sensors being irradiated to $10^{13} \neqcm$ and
$5 \times 10^{14} \neqcm$ %\cite{MichaelProgressReport2017}
. Besides of obtaining a first insight on radiation effects, the project aimed to test procedures for
measuring the thickness of the depleted zone of the sensor. This was done by means of capacity measurement, leakage current measurement and mapping out the sensor with a $^{55}$Fe and a $^{109}$Cd source
relying on Eqn. \ref{Eqn:Depletion}.
$^{109}$Cd was used as its $22 ~\rm keV$ photons penetrate Si with little attenuation. Therefore, no correction of the data for attenuation effects is required. 
The results of the ongoing study are considered as preliminary but some selected points should be brought to the attention of the reader. Final results of the ongoing work are scheduled for publication in \cite{LinnikPhD,BennyGSI2017}.

\begin{figure}[t]
 \centering
%1
\begin{subfigure}{.45\textwidth}
%\vspace{0.4cm}
  \centering
	   \includegraphics[viewport= 1cm 1cm 45cm 29cm, clip, width=1\columnwidth]{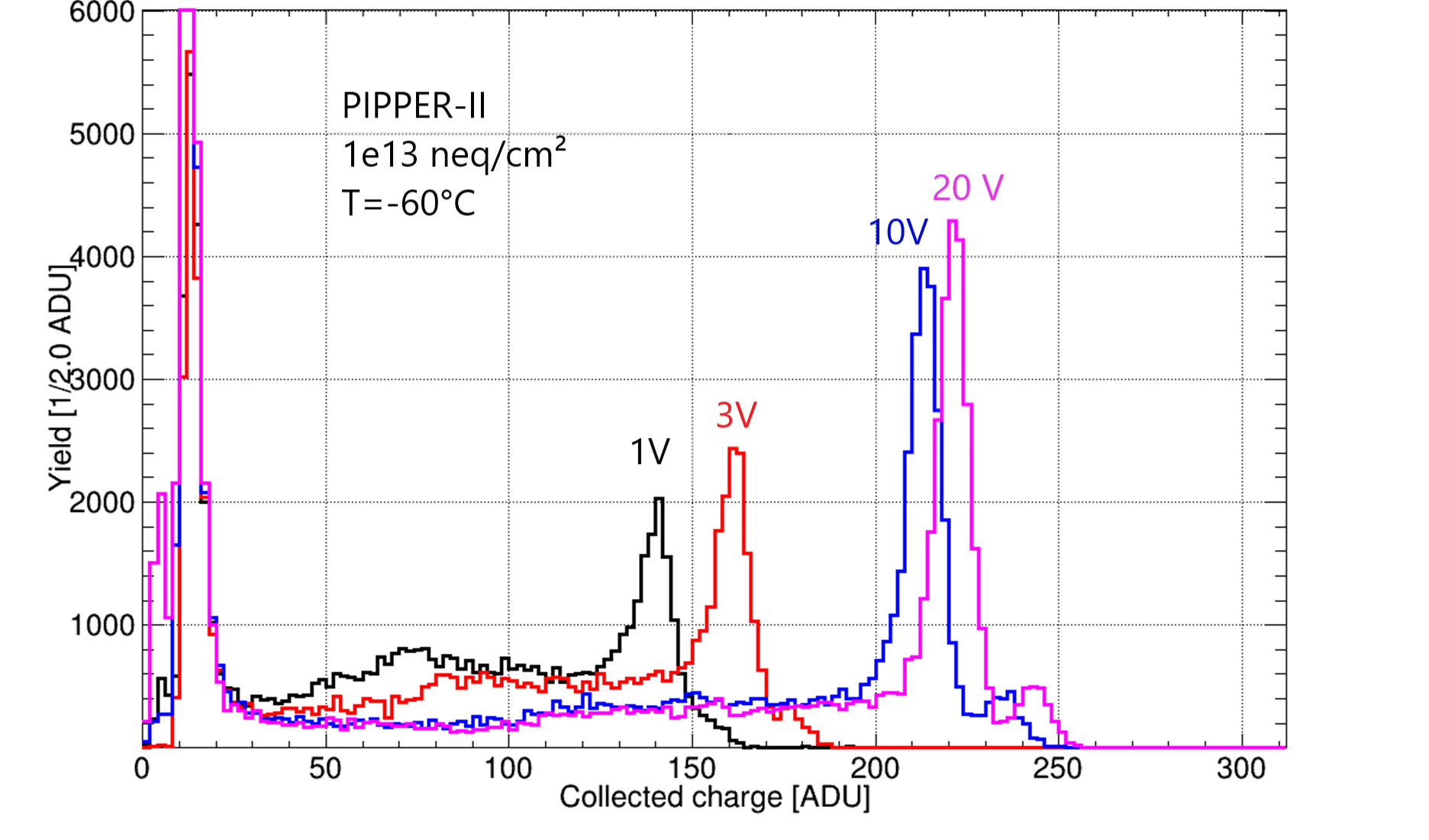}
    \caption{$^{55}$Fe spectrum of a PIPPER-II prototype, which was irradiated with $10^{13} \neqcm$ as a function of the bias voltage applied.  }
    \label{fig:Pipper2CCE1e13}
\end{subfigure}%
\hspace*{0.05\textwidth}
%2
\begin{subfigure}{.45\textwidth}

  \centering
	   \includegraphics[viewport= 3cm 1cm 47cm 29cm, clip,width=1\columnwidth]{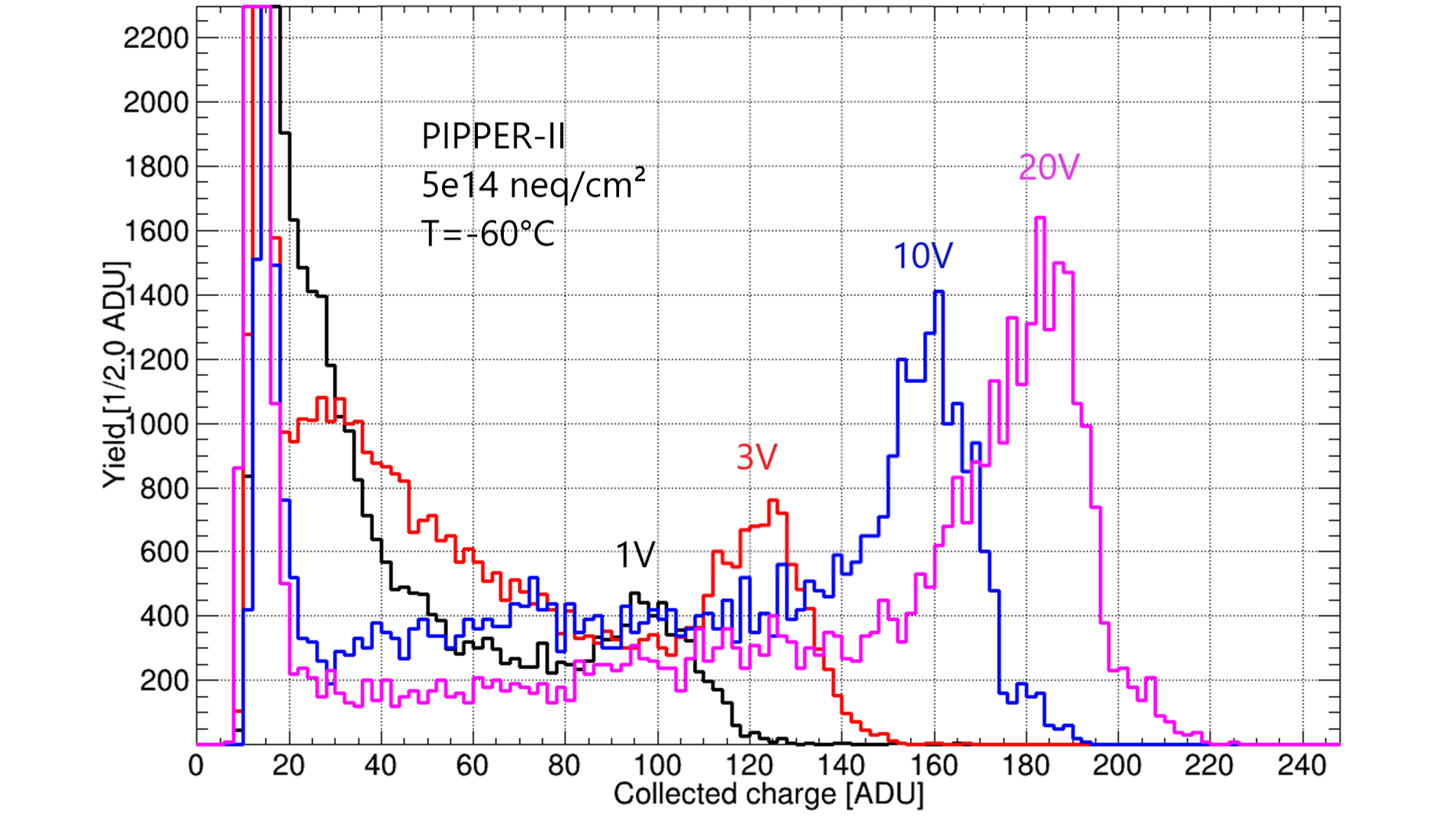}
	
    \caption{$^{55}$Fe spectrum of a PIPPER-II prototype, which was irradiated with $5 \times 10^{14} \neqcm$ as a function of the depletion voltage applied.
		}
    \label{fig:PipperResults}
\end{subfigure}%

\caption{Preliminary results on the charge collection efficiency of the PIPPER-II prototype as a function of the depletion voltage and radiation dose. Mind the scale. 
%From \cite{BennyWuhan}.
}
\label{fig:PIPPER-2CCE}
\end{figure}

$^{55}$Fe spectra of the irradiated sensors are shown in Fig. \ref{fig:PIPPER-2CCE}. As expected for full depletion, the charge collection peak 
mostly vanishes for $U_{depl}\gtrsim 10\rm~V$, while the calibration peak grows and shifts to a higher position ($\rightarrow$ more gain $\rightarrow$ less capacity). 
Few modifications are seen for higher $U_{depl}$, which holds to a reduced extent for sensor exposed to $5 \times 10^{14} \neqcm$.

% A preliminary measurement of the depleted volume indicates a volume of $\sim 6000\mum^3$, which corresponds
%to thickness of the depleted layer of $12\mum \pm 1 \mum$. Given that the instrumental approach differs in its details, this finding is in good accordance with the results reported in \cite{PaperPipper}.

\begin{figure}[t]
 \centering
%1
\begin{subfigure}{.45\textwidth}
%\vspace{0.4cm}
  \centering
	   \includegraphics[width=1\columnwidth]{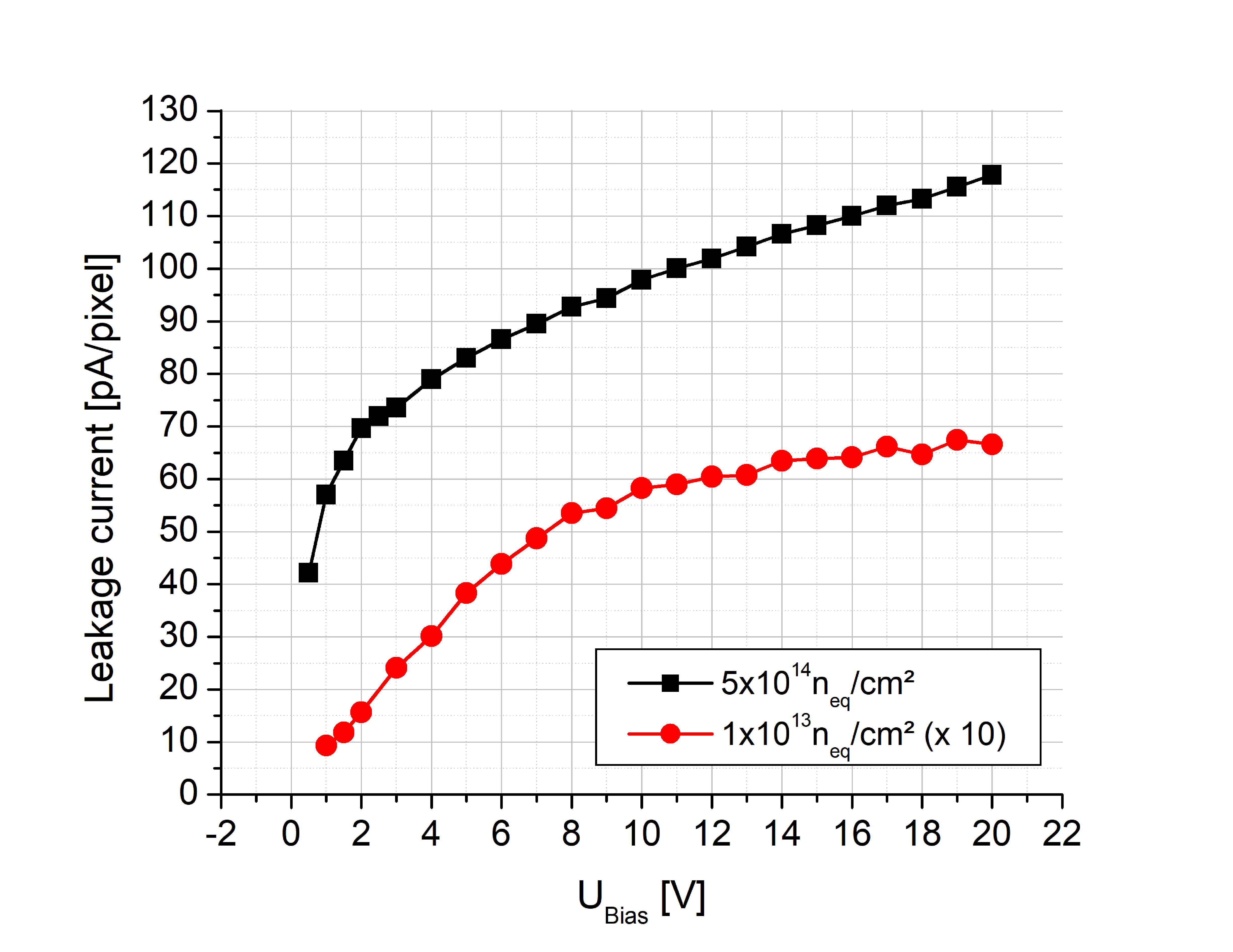}
    \caption{Combined leakage current of the three pixel matrices of PIPPER-2 at a temperature of $20\C$. The leakage current of the sensor irradiated with $10^{13}\neqcm$ was multiplied
			with a factor of 10 for clarity.}
    \label{fig:PipperLeakageCurrentAbsolute}
\end{subfigure}%
\hspace*{0.05\textwidth}
%2
\begin{subfigure}{.45\textwidth}
\vspace{-0.9cm}
  \centering
	   \includegraphics[width=1\columnwidth]{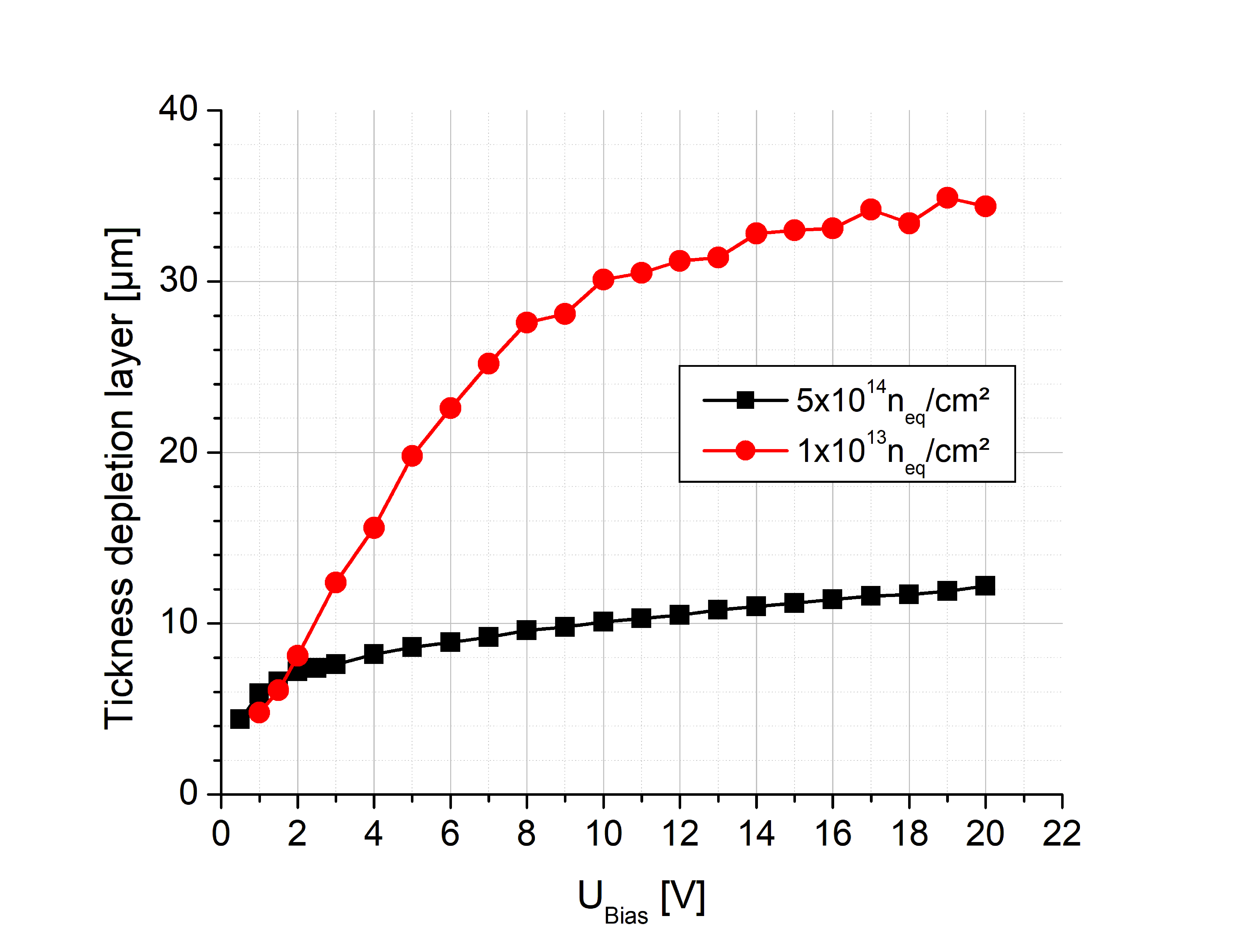}
	
    \caption{Thickness of the depleted layer of PIPPER-2 as indicated by the leakage current.
		}
    \label{fig:PipperLeakageCurrentThicken}
\end{subfigure}%

\caption{Preliminary results on the leakage current of neutron irradiated PIPPER-2 sensors. %Data from \cite{BennyWuhan}.
}
\label{fig:PipperLeakageCurrentFig}
\end{figure}

Results from measurements of the leakage current of PIPPER-2 are displayed in Fig. \ref{fig:PipperLeakageCurrentFig}. 
Fig. \ref{fig:PipperLeakageCurrentAbsolute} indicates the absolute leakage current per pixel as measured with an ammeter from the biasing line of the
diodes. Note that the data was taken after some months of room temperature annealing. Moreover, the results are somewhat biased by the fact that, 
due to hardware constraints, only the summed leakage current of three pixel matrices could be measured.
Among those matrices, one has 
four charge collection diodes, which is expected to reduce the voltage required for reaching full depletion. This complicates the quantitative interpretation of the results but should show few effect on the results as soon as full depletion is reached. One observes that the leakage current per diode reaches values of $\sim 6 ~\rm pA$
after a dose of $10^{13}\neqcm$ and $120 ~\rm pA$ after a dose of $5\times 10^{15} \neqcm$ at $T=20\C$. Those unusually high currents
reflect the depleted volume and created issues with shot noise and signal clearing. Cooling the sensors to temperatures of $-60\C$
 helped to control both effects but signal clearing remained an issue. 

The leakage current was used to estimate the volume and thus the thickness of the depletion layer in accordance with Eq. \ref{Eqn:VolumeCurrent}. 
The indicated result of this procedure are displayed in Fig. \ref{fig:PipperLeakageCurrentThicken}. While
the highly irradiated sensor reproduces the previously reported thickness of the depleted layer, the indication of the sensor irradiated to $10^{13}\neqcm$
overshoots and converges to an indicated thickness of $34\mum$. This points to the presence of a non-negligible source of leakage current, 
which might be formed from the surface damage generated by the $\upgamma$-ray background of the neutron source used for radiation. 
For the higher dose, this surface damage might have been saturated and thus become negligible w.r.t the bulk damage. Annealing effects
appear as a less plausible cause of the effect as both sensors had mostly the same annealing history. Until the effect is clarified,
leakage currents measurements should be used with caution only for estimating the depletion depth.

%This can be interpreted as a signature of full depletion for the highly irradiated chip
%while the observation for the sensor irradiated to $10^{13}\neqcm$ remains to be understood.

\begin{figure}[t]
 \centering
%1
\begin{subfigure}{.45\textwidth}
%\vspace{0.4cm}
  \centering
	   \includegraphics[width=1\columnwidth]{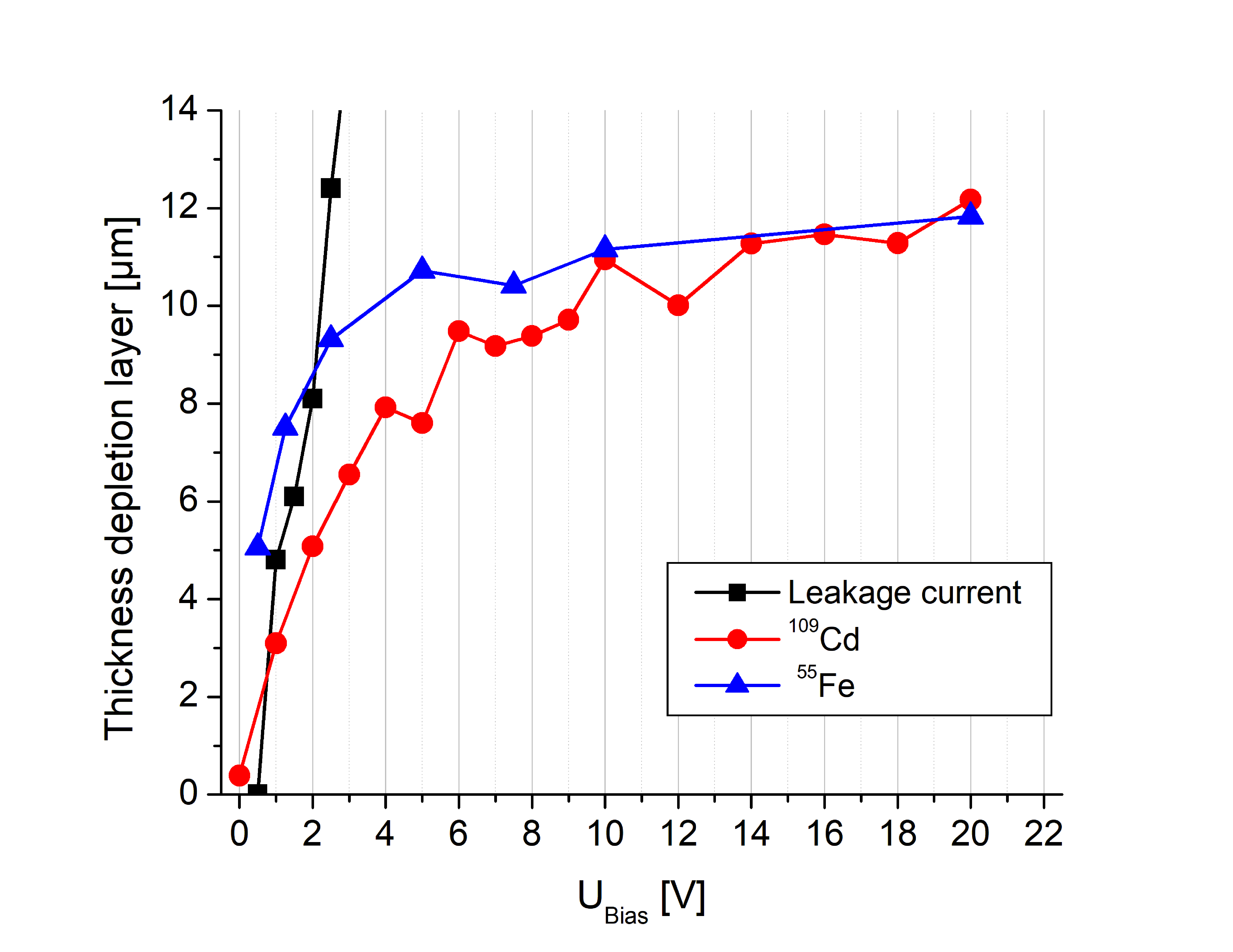}
    \caption{Thickness of the depletion layer of PIPPER-2 as function of the bias voltage for a dose of $10^{13}\neqcm$.}
    \label{fig:PipperCCE55109for1e13}
\end{subfigure}%
\hspace*{0.05\textwidth}
%2
\begin{subfigure}{.45\textwidth}
%\vspace{-0.9cm}
  \centering
	   \includegraphics[width=1\columnwidth]{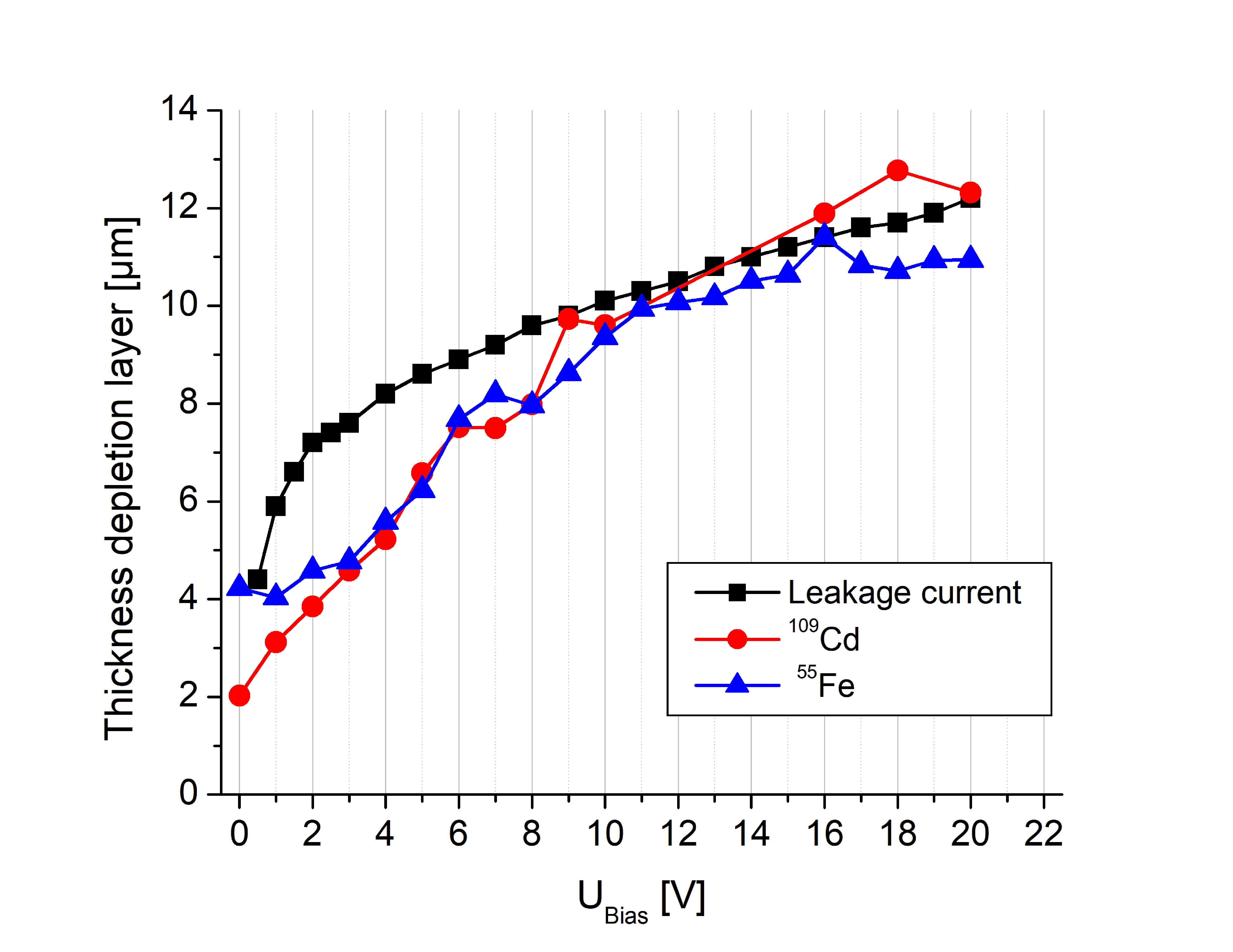}
	
    \caption{Thickness of the depletion layer of PIPPER-2 as function of the bias voltage for a dose of $5\times 10^{14}\neqcm$.
		}
    \label{fig:PipperCCE55109for5e14}
\end{subfigure}%

%Piotr Gasic

\caption{Preliminary results on the thickness of the depletion layer of PIPPER-2 as obtained from amplitude spectrum and leakage current measurements. %Data from \cite{BennyWuhan}.
}
\label{fig:PipperCCE55109}
\end{figure}

Preliminary results on the thickness of the depleted layers as obtained from mapping the sensor with a $^{109}$Cd and a $^{55}$Fe source are shown in Fig. \ref{fig:PipperCCE55109}. 
The plots show the estimated thickness of the depleted layer as a function of the voltage applied to the diode. The indications of the leakage current measurements are shown for comparison.
The results displayed in Fig. \ref{fig:PipperCCE55109for1e13} hold for the sensor irradiated with $10^{13}\neqcm$. The amplitude spectra indicate a layer thickness of $\sim 12\mum$ at
a voltage of $20~\rm V$, which can be considered as full depletion within the known limits of the sensor. The results match the expectation and contrast the results obtained from the leakage current
measurement mentioned above. 

For the sensor irradiated to  $5\times10^{14}\neqcm$, one observes all three measurements to show similar indications as soon as a reasonable degree of depletion is reached. 
The indication suggests that the sensor depletes slower than the $10^{13}\neqcm$ sensor but a full depletion is reached once the highest voltage is applied. 

It should be mentioned that the sensor irradiated with $10^{13}\neqcm$ shows a noise of $23 ~e~\rm ENC$ at a temperature of $-55\C$ and an integration time of $12.5 \mus$. The most probable charge collected from the $\upbeta$-rays of a $^{90}$Sr source was found to amount $1290 ~ e$ at a depletion voltage of $20~\rm V$, which turns into a S/N of 55.  For the sensor irradiated with $5\times 10^{14} \neqcm$, a noise of $27  ~e~\rm ENC$  and a most probable charge of $\gtrsim 860 ~ e$ was seen.
The high remaining S/N of 31 and the apparent full depletion of the sensing element suggest that the irradiated device will show a good detection efficiency for minimum ionizing particles, which remains to be confirmed.

\subsubsection{Investigator (modified process)}
\label{Sec:Investigator}

Traditional CPS rely on charge collection diodes, which are made from a tiny n-well/p-epi junction. This junction geometry turns into a small capacitance but complicates 
depleting the sensor. An alternative option became recently available thanks to a modification of the $0.18\mum$ TowerJazz process. As shown in Fig. \ref{fig:Processes} (c), this modification
consists in adding a thin, lowly doped n-layer on top of the p-doped epitaxial layer. By doing so, one creates a planar pn-junction inside the active medium.
This modified doping profile was expected to ease depleting the epitaxial layer quite substantially. To some extent, this approach realizes for a first time a theoretical
concept, which was discussed in \cite{Turchetta-VDNwellPaper} but reached far beyond the technology available at the time.

A first systematic study of CPS relying on the modified process was carried with the Investigator-prototype \cite{InvestigatorPaper,VeryDeepNwellMapsPaper},
which was developed aiming for ATLAS. Investigator hosts pixels with a pitch
between $25 \times 25 \mum^2$ and $50 \times 50 \mum^2$. A spacing of $3\mum$ between the n-well implantation forming the core of the diode
 and the deep p-well layers protecting the electronics was implemented. For the $50 \times 50 \mum^2$ pixels, this spacing was extended to $18.5 \mum$.

According to the tests, the modification increases the capacitance of the pixels by about $30\%$ once a (backside) depletion voltage of $-6~\rm V$ is applied.
After correcting for the different gain, the charge collection of the pixel carried out in the modified process is reported to be slightly worse than the one 
observed on a similar pixel build with the non modified high resistivity epitaxial layer. Moreover, the signal rise time of the modified $50 \mum$ pixel is slightly slower
than the one of the standard pixel ($16.6~\rm ns$ instead of $15.9~\rm ns$). However, the modification reduces the spread of the charge collection time
from $4.6$ to $1.9~\rm ns$.

After an irradiation of $10^{14} \neqcm$, the gain of the $50\mum$ standard pixel is reduced by 14\% and the most probable charge collected from $\upbeta$-rays of a $^{90}$Sr 
source shrinks by 10\%. For a dose of  $10^{15} \neqcm$, a loss of the pixel sensitivity is reported. This loss was not observed for the modified pixel,
which provides a by 17\% reduced gain but a mostly unchanged signal charge even after $10^{15} \neqcm$. The numbers hold for a depletion voltage of $-6~\rm V$ and a 
operation temperature of $-30\C$. Beam tests of the modified pixels with a $180~\rm GeV/c$ pions of the CERN-SPS suggest a S/N of 39 ($25\mum$ pitch), of 33 ($30 \mum$ pitch) and 13 ($50\mum$ pitch). The restricted S/N of the latter pixel is explained with its limited gain. While the smaller pixels show charge sharing, mostly no charge sharing is observed for the $50\mum$ pixel.

For the non-irradiated sensors, an efficiency of $98.5\% \pm 0.5\% (stat) \pm 0.5\% (syst)$ is reported for the modified $50\mum$ pixel. After a dose of $10^{15} \neqcm$, the $30\mum$ pixel reached an efficiency of $97.5\% \pm 1.5\% (stat) \pm 0.6\% (syst)$. 

\subsubsection{Malta}
The MALTA sensor \cite{ProceedingsMalta, PaperMalta,TalkMalta} combines the sensing elements of the above mentioned Investigator with
a fast readout electronics. It is equipped 
with $512\times 512$ pixels of $36.4\times 36.4 \mum^2$ pitch. The charge collection diode is formed from a small, $\sim 2 \mum$ n-well implantation with a capacity of $< 5~\rm fF$, which is embedded into the lowly n-doped layer covering the 
p-doped epitaxial layer. The sensor may be biased to -6 V at the deep p-well and to -20 V at the substrate. 

The hit discrimination is performed on the pixel by means of a fast analogue front end, which was derived from the one of ALPIDE (see above) but modified to provide a time resolution better than $25~\rm ns$ as required for ATLAS. 
According to simulations, the modified front-end was expected to provide threshold level of $300~ e$ with a dispersion of $10~e$ and an ENC of $10~ e$. The sensor is designed to operate at a very low energy consumption of
$80~\rm mW/cm^2$.

First test of the device show that the sensor features a non-negligible non-gaussian tail in the noise distribution, which was attributed to the presence of RTS \cite{TalkMalta} in some pixels. As those pixels could not be masked individually,
defining a good threshold setting was problematic. The lowest threshold setting was found to amount $230 ~e$ with a dispersion of $36 ~e$ and an ENC of $ 7~e$. This threshold increased to  $500 ~e$ with a dispersion of $70 ~e$ and an ENC of $ 12~e$
after irradiating the sensor with $5 \times 10^{14} \neqcm$ fast reactor neutrons.
A beam test with $180~\rm GeV/c$ pions carried out at the CERN SPS
indicated a 96\% detection efficiency for minimum ionizing particles for the non-irradiated device, which was operated with a bias of -6V at the p-well and moreover with a back-bias of -15V. A refined analysis of the data revealed that the
pixels are reasonably efficient in their centre while the efficiency in the corners of the pixels was found to be reduced to 90\%. This is among others as, due to the RTS, a relatively high threshold had to be used. The sensor was irradiated to 
$10^{15} \neqcm$ and once more tested in the beam. The results indicate a 89\% efficiency in the centre of the pixel and an efficiency as low as 56\% at their edge. 

The origin of the weakness, which was revealed in this first design attempt, was spotted by means of TCAD simulation  \cite{DoubleModifiedProcessSimulationPaper}. It was found that the electric field in the fully depleted active volume
is rather non-homogeneous: As the lateral component of the drift fields generated by the diodes of neighbouring pixels cancel out at the at the pixel borders, a depleted, but essentially
field free volume is generated in this region. Signal electrons being attracted to this field minimum drift only slowly and may easily be lost. According to the simulation, this field minimum may be mostly eliminated 
if the n-layer of the pixel is interrupted by a p-doped region located at the edge of the pixels. This generates a lateral p/n junction and thus a lateral field. Two approaches for realizing this, adding a gap in the n-layer (see Fig. \ref{fig:Processes} (d)) and
adding explicitly a very deep p-implantation, are studied. Both are considered to be compatible with the Tower/Jazz process and referred to as ``double modified'' process. It is concluded that the first approach is most efficient and may accelerate the charge collection from the pixel corner from few 10 ns to few ns, which is predicted to increases the charge collection efficiency from this region by a rough factor of four.

\subsubsection{TJ-Monopix}

TJ-Monopix \cite{Monopix} uses the same CMOS process and similar pixels and pre-amplifiers as the above mentioned MALTA sensor but uses a different digital front end. 
The $2\times 1 ~\rm cm^2$ prototype hosts $224 \times 448$ pixels with $36\times 40 \mum^2$ pitch. The pixels receive a $40~\rm MHz$ time information. This allows to use time-over-threshold
to obtain an analogue charge information. The readout of the pixels is done by a column drain architecture.
During tests, the sensor demonstrated a gain of $\sim 400{~\rm  \upmu V}/e$ \cite{LF-Monopix2}. The threshold before irradiation was $350~e$ with a dispersion of $35~e$ and a noise of $16 ~e ~\rm ENC$. 
After being irradiated with $10^{15} \neqcm$ fast reactor neutrons, a threshold of $570~e$ with a disperson of $65~e$ and a noise of $23 ~e \rm ~ ENC$ was observed. As the same sensing elements were used,
TJ-Monopix shared the sensing issues of MALTA as discussed in the previous section. The detection efficiency reported for the two pixel flavours was $97.1\%$ and $93.7\%$ respectively before irradiation and 
dropped by $\sim 28\%$ and $\sim 43\%$ after irradiation.

\subsubsection{MIMOSIS}

The MIMOSIS project \cite{MichaelVCI2019} aims the Micro Vertex Detector of the CBM experiment. The sensor will host small ($26.88\times 30.24 \mum^2$) pixels as required to reach the ambitioned very good spatial resolution of $\sim 5 \mum$.
The targeted minimum radiation tolerance is $\gtrsim 3\times 10^{13} \neqcm$ and $\gtrsim 3~\rm Mrad$ at $T=-20\C$ before replacement
at the end of a run. The front end electronics of MIMOSIS is inspired by the above mentioned ALPIDE sensor. Modifications include the adaptation
of the digital front end to an continuous readout with a peak counting rate of $70~\rm MHz/cm^2$ and at a time resolution of $5 \mus$. Moreover, it is being considered to replace the initially DC-coupled pixel  by a AC-coupled version,
which is realized in analogy to the one of the PIPPER sensor. This allows to top-bias the pixel node with voltages of up to 40 V. As MIMOSIS is produced in the Tower/Jazz 180nm technology, the AC coupled diode can be combined with the use of the modified and double modified process, which would presumably allow for reaching full depletion.

A first prototype named MIMOSIS-0 was realized with the standard Tower/Jazz 180nm process with non-modified epitaxial layer. The prototype hosts DC- and the novel AC-coupled pixels and their priority encoder readout. Preliminary measurements
based on pulse injection suggest that a low threshold of $<150~e$ may be used for both pixels. For the DC- coupled pixel a threshold dispersion of $15 ~e$ and a noise of $\lesssim 9~e$ is reported.
The threshold disperson of the AC-coupled pixel amounts $\lesssim 7~e$ with a noise of $\lesssim 3~e$. 

\subsection{R\&D projects relying on large fill factor diode}

\subsubsection{HV-MAPS/CCPD}
\label{sec:HV-MAPS-1}
HV-MAPS \cite{HV-MAPS1} relied initially on high-voltage CMOS technologies as used for devices for automotive and industrial applications. The triple-well processes used provide a lowly doped, deep N-well, which may embed PMOS and NMOS transistors. The N-well is optimized to separate the potential of those transistors from the one of the P-doped substrate of the chip. Therefore, the embedded transistors may operate normally while a depletion voltage of typically more than $50~\rm V$ toward the substrate is applied. HV-MAPS use this feature by implementing the amplification electronics into a large deep N-well diode. This combines a high depletion voltage with the availability of full CMOS, which allows to perform signal discrimination on the pixel. However, the large diodes show a significant capacitance and thus a high noise. Moreover, the related processes provided initially only low resistivity substrates, which limits the depletion depth to $\sim 10\mum$. 

The above mentioned design concept is followed up with ``capacitive coupled hybrid pixel detectors'' (CCPD) \cite{HVMaps2}.
CCPDs are composed from a dedicated sensor chip and a dedicated
readout chip as known from classical hybrid pixels. However, instead of bump bonding both chips, they are glued to each other with a very thin glue 
layer. The electrical communication between the chips is done by means of capacitive coupling. To do so, opposing bonding pads of both 
chips are used as capacitor plates. This approach allows to combine CPS with existing, highly integrated readout chips used for hybrid pixel detectors.
At the same time, the cost intense bump bonding is avoided. CCPDs relying on a CMOS process with high resistivity epitaxial layer became available 
meanwhile and are sometimes labelled as HR-MAPS.

Recent results of different radiation tolerance studies on HV-MAPS is summarized in \cite{HV-MapsRadiationSummary}: 
CCPDs realized in the AMS H18 process with $10 \Omega \cdot \rm cm$ active volume were exposed to neutron fluences of up to $2 \times 10^{16} \neqcm$ \cite{HVRadiation1,
HVRadiation2,HVRadiation3}. The sensor chip integrates a full discriminator chain in each pixel, which provides a digital output signal. Three CMOS pixels send their output
signal via a common bump-bond pad to a pixel cell of a FE-14 readout chip. The signal amplitude of the three pixels is intentionally different, which allows the charge
sensitive readout cell to identify the firing pixel based on the signal amplitude. Radiation hardness studies showed the beneficial effect of acceptor removal on the charge amplitude of the detector, which however vanished after exposing the device to $\sim 10^{16} \neqcm$. During a beam test, the device showed a detection efficiency of 99.7\% after a dose of $10^{15} \neqcm$. For $5 \times 10^{15} \neqcm$, this value remains above 97\% \cite{HVRadiation5}.  

\subsubsection{LF-MonoPix}

LF-Monopix is a depleted MAPS (DMAPS) prototype program \cite{LF-Monopix-Twepp, Monopix, LF-Monopix2} aiming for ATLAS.
The concept of the sensor relies on sizeable $250 \times 50 \mum$ pixels
with high fill factor diodes. This pixel size is sufficient to match the requirements of ATLAS in terms of spatial resolution.
The high diode capacitance of $\sim 400 ~\rm fF$ creates noise levels of $200~e ~\rm ENC$, which was however
considered to be acceptable as a sizeable depletion depth above $100 \mum$ was ambitioned. 
The LF-Monopix-01 produced with the 150~nm LFoundry CMOS process is relying on a $> 2 \rm ~k \Omega \cdot cm$ high resistivity p-type wafer.
The LFoundry process provides quad well features and an additional very deep n-well implantation (see Fig. \ref{fig:Processes} (f)). The latter is used to manufacture the high fill factor charge collection electrode, 
which also embeds the pixel electronics. Deep p-well implantations available in the process are used to isolate the n-wells of
PMOS transistors from the deep n-well of the diode. The pixels hold a 4-bit DAC for threshold tuning and receive 8-bit time stamps with $40~\rm MHz$ to match the $25 ~\rm ns$ time resolution
required for the intended use in an upgraded ATLAS inner tracker. The collected charge may be measured by the time-over-threshold, which amounts $\sim 2\mus$ ($\sim 80$ clock cycles) for full depletion signals.
The readout of the pixels is done by a column drain architecture.
 
The sensor was thinned to  $100\mum$ and $200 \mum$ and processed at its backside. Hereafter, it was found to withstand depleting voltages above 260 V. A full depletion is reported to occur at
a voltage of 60~V for the $200 \mum$ thick, non-irradiated sensor. Radiation test results from LF-Monopix and its precurser LF-CPIX are reported in \cite{LF-Monopix3}. The LF-CPIX, which hosts similar
analog pixel structures but misses the complex readout structure, was exposed to up to 50 Mrad soft X-rays. After irradiation, a mild increase of the threshold dispersion from $370 ~e$ to $420 ~e$ was reported.
This disperson could be reduced to $55~e$ before and $76~e$ after irradiation, which is negligible as compared to the readout noise of $\sim 200~e~\rm ENC$. The gain of the pixel amplifiers is found to 
decrease by $\sim 5\%$ and the noise increases by $20\%$ for the highest dose applied. Sensors irradiated with $10^{15} \neqcm$ fast reactor neutrons were tested in beam. A strong decrease of  a rough
factor of five of the most probable value of the indicated charge is observed. However, the remaining $4.5~ ke$ is found to be sufficient for reliable particle detection. At a chip temperature of $T<-40\C$,
an average hit efficiency of $99.6\%$ and $98.9\%$ are observed. The background rate remained below $10^{-7}$ after individual noisy pixels were switched off.

\section{Summary and conclusion}

CPS form an emerging technology for charged particle tracking. Their advantage with respect to other silicon pixel detector technologies consist
in an excellent spatial resolution, a light material budget, a reliable particle detection and a low price. Their initially modest tolerance to radiation
was dramatically improved during the last two decades. 

The tolerance of the sensors to non-ionizing radiation is limited
mostly by two radiation effects. Those are the radiation induced reduction of the lifetime of the minority charge carriers in the silicon, which tends
to absorb signal electrons before they are collected; and the radiation induced increase of leakage currents. For sensors using traditional, mostly
non-depleted active volumes, the first effect dominates and the typical radiation tolerance reached amounts $\lesssim 10^{13} \neqcm$. Depleting the
active volume assures a good charge collection up to doses of $\gtrsim 10^{15} \neqcm$. However, the leakage current of the diode, which scales 
with its depleted volume and the radiation dose, may increase to uncomfortable values. If so, it creates shot noise and, in the so-called SB-pixels, 
an accelerated clearing of the pixel signal in the pre-amplifier. Both effects may be alleviated by cooling and by accelerating the readout/shaping
time of the pixel, where the second measure is required independently to obtain a good time resolution and counting rate capability. However,
implementing both measures may be limited by external system aspects like constraints concerning operation temperature and the maximum acceptable power dissipation
of a fast on-pixel electronics. 

Reducing the pixel size was found to improve the charge collection properties of all kinds of neutron irradiated CPS with small charge collection diodes
dramatically. Moreover, it may reduce the leakage current per pixel and the related radiation effects of depleted CPS as the depleted pixel volume is reduced.
The measure is limited by the minimum size of the on-pixel pre-amplifiers. Connecting more than one collection diode in
 parallel to a common pre-amplifier was tried but the unwanted additional capacitive noise is found to dominate the improved charge collection efficiency. 
For non, or partially, depleted CPS, the optimal thickness of the active volume (epitaxial layer) amounts $\sim 15-25 \mum$. Using a high resistivity substrate
instead of an epitaxial layer reduces the radiation tolerance of non-depleted pixels while it is somewhat a prerequisite for reaching 
very thick depletion layers with sensors following the high fill factor diode approach. 

Ionizing radiation increases the leakage currents of collection diode of the device. Badly designed reset transistor of 3T pixels
may form a dominating additional source of leakage currents. Again, most related effects may be controlled by fast readout and cooling.
Moreover, the currents may be reduced by design by adding dedicated guard rings to the diodes. Those 
push vulnerable, thick SiO$_2$ away from the diode and cut conductive paths. The use of CMOS-processes with small feature size help to increasing
the radiation tolerance of transistors but does not necessarily help to increase the radiation tolerance of the diodes. This is as the thinner and thus more radiation
hard gate oxide provided by those processes does not improve the properties of diodes, which are surrounded by other, thick oxide structures.
The tolerance of CPS to ionizing radiation depends significantly on the specific design of the device and may exceed $10~\rm Mrad$.

Both, ionizing and non-ionizing radiation are found to create RTS in the charge collection diodes of CPS. This noise manifests itself as a fluctuation of the 
dark signal of the pixel between two or more well-defined levels. It is suited to generate a substantial number of false hit indications and to transform a pixel
into a hot pixel. While the definition on whether a pixel shows RTS is not precise in literature, it is appropriate to state that all pixels show this feature after
moderately high radiation doses. At the time given, no strategy to reduce the RTS of diodes by design is known. However, the dark occupancy caused by this effect may be reduced
by means of cooling the sensors and accelerating the readout. Moreover, SB-pixels show by design a substantially lower RTS induced dark rate than 3T-pixels.
RTS is also observed in transistors and may generate false hit signatures in case the input stage of the pixel pre-amplifiers is affected. Increasing the surface
of the related transistor gate is a known and suited approach for reducing the amplitude of the RTS signal below the detector threshold. The additional pixel
capacitance generated by this step is found to increase the average noise of the pixels but to improve the overall performance of the device.

CPS are formed from p-doped silicon. Cold and thermal neutrons may induce a nuclear reaction of the dopants \mbox{(n + ${}^{10}$B $\rightarrow ^7$Li + $^4$He + 2.8~MeV)}. The decay products may generate a substantial NIEL, which is not accounted for in the standard NIEL model. According 
to theoretical estimate within the framework of the NIEL model, this contribution dominates as soon as the doping concentration exceeds a value of few $p=10^{17}\rm/cm^3$. Tests performed with sensors, which were irradiated with cold ($1.8~\rm meV$) neutrons, showed a qualitatively different picture. It is observed that the neutron radiation generates an unexpectedly
strong acceptor removal, which is about equal to the one expected for fast $1~\rm MeV$ neutrons and thus exceeds the predictions of the standard NIEL model by three orders of magnitude. 

Heavy-ions may generate a so-called latch-up in CPS, which is a reversible short circuit in the devices.
Unless this short circuit is extinguished by means of a rapid power cycle, it may generate a permanent thermal destruction of the sensor. $50\mum$ thin sensors were found substantially more vulnerable to this effect than the standard $700\mum$ sensors. The integrated NIEL created by relativistic ions is badly known. A related study suggests that a fully stripped $30~A\rm GeV$ Pb-ions creates an equivalent damage of $\lesssim 300 \neqcm$.

Over all, the performances of CPS was substantially improved during the last two decades. This includes an improvement of their radiation tolerance, which amounts
 two to three orders of magnitude. Thanks to those progresses and to the novel design options provided by next generation CMOS processes, the technology is now ready for a by far extended number of applications in detectors of heavy ion and particle physics. 

%\acknowledgments

%Acknowledgments.

\section{Acknowlegements}
The author wishes to thank J. Stroth (Goethe University Frankfurt) and M. Winter (PICSEL group, IPHC Strasbourg) as much
as the members and students of their research groups for their continuous support during almost two decades of common research on CPS. I 
would also like to thank the anonymous reviewer of this work for the efforts and many valuable comments.
This work was supported by BMBF (05P15RFFC1), GSI and HIC for FAIR. The author was supported by the MainCampus Stipendiatenwerk
der Stiftung Polytechnische Gesellschaft Frankfurt am Main.

%\include{AppendixNeu}

%\bibliographystyle{unsrt}
%\bibliography{bibliothek.bib}

%\printbibliography

\end{document}